%% file: main.tex
\pdfoutput=1
\documentclass[12pt,a4paper]{article}

\usepackage{ifthen} 
\newboolean{pdflatex}
\setboolean{pdflatex}{true} 

\newboolean{articletitles}
\setboolean{articletitles}{true} 

\newboolean{uprightparticles}
\setboolean{uprightparticles}{false} 

\usepackage{tabularx,booktabs}
\newcolumntype{Y}{>{\centering\arraybackslash}X}

\usepackage{subcaption}
\def\mkpisq     {{\ensuremath{m_{K\pi}^{2}}}\xspace}
\def\phitentwenty {\ensuremath{\phi{(1020)}}\xspace}
\def\mB         {{\ensuremath{m_B}}\xspace}

\def\phih       {{\ensuremath{\phi}}\xspace}

\def\BdToJpsiKst {\decay{\Bd}{\jpsi\Kstarz}}
\def\BdToPsitwosKst {\decay{\Bd}{\psitwos\Kstarz}}

\DeclareRobustCommand{\optmybar}[1]{\shortstack{{\miniscule (\rule[.3ex]{0.65em}{.18mm})}
  \\ [-.7ex] $#1$}}
\def\bnbar#1  {\kern \thebaroffset\optmybar{\kern -\thebaroffset \ensuremath{#1}}\xspace}

\def\paperauthors{LHCb collaboration} 
\def\paperasciititle{Comprehensive analysis of local and nonlocal amplitudes in the B to Kstar mu mu decay} 
\def\papertitle{Comprehensive analysis of local and nonlocal amplitudes in the \decay{\Bz}{\Kstarz\mup\mun} decay} 
\def\paperkeywords{{High Energy Physics}, {LHCb}, {Penguin decays}} 
\def\papercopyright{\the\year\ CERN for the benefit of the LHCb collaboration} 
\def\paperlicence{CC-BY-4.0 licence}
\def\paperlicenceurl{https://creativecommons.org/licenses/by/4.0/}

\input{preamble}
\usepackage{longtable} 
\usepackage{cleveref}

\begin{document}

\renewcommand{\thefootnote}{\fnsymbol{footnote}}
\setcounter{footnote}{1}

\input{title-LHCb-PAPER}

\renewcommand{\thefootnote}{\arabic{footnote}}
\setcounter{footnote}{0}

\cleardoublepage


\pagestyle{plain} 
\setcounter{page}{1}
\pagenumbering{arabic}


\input{body}
\input{acknowledgements}

\input{appendix}

\input{appendix-fit-parameters}
\input{appendix-plots}
\input{appendix-projections}
\addcontentsline{toc}{section}{References}
\bibliographystyle{LHCb}
\bibliography{main,standard,LHCb-PAPER,LHCb-CONF,LHCb-DP,LHCb-TDR}

\newpage
\input{Authorship_LHCb-PAPER-2024-011}
\end{document}

%% file: preamble.tex
\usepackage[top=1in, bottom=1.25in, left=1in, right=1in]{geometry}

\usepackage{microtype}
\usepackage{lineno}  
\usepackage{xspace} 
\usepackage{caption} 

\usepackage{graphicx}  
\usepackage{color}
\usepackage{colortbl}
\graphicspath{{./figs/}} 

\usepackage{amsmath} 
\usepackage{amssymb}
\usepackage{amsfonts}
\usepackage{upgreek} 

\newcommand*\patchAmsMathEnvironmentForLineno[1]{%
\expandafter\let\csname old#1\expandafter\endcsname\csname #1\endcsname
\expandafter\let\csname oldend#1\expandafter\endcsname\csname
end#1\endcsname
 \renewenvironment{#1}%
   {\linenomath\csname old#1\endcsname}%
   {\csname oldend#1\endcsname\endlinenomath}%
}
\newcommand*\patchBothAmsMathEnvironmentsForLineno[1]{%
  \patchAmsMathEnvironmentForLineno{#1}%
  \patchAmsMathEnvironmentForLineno{#1*}%
}
\AtBeginDocument{%
\patchBothAmsMathEnvironmentsForLineno{equation}%
\patchBothAmsMathEnvironmentsForLineno{align}%
\patchBothAmsMathEnvironmentsForLineno{flalign}%
\patchBothAmsMathEnvironmentsForLineno{alignat}%
\patchBothAmsMathEnvironmentsForLineno{gather}%
\patchBothAmsMathEnvironmentsForLineno{multline}%
\patchBothAmsMathEnvironmentsForLineno{eqnarray}%
}

\usepackage{hyperref}
\usepackage{hyperxmp}
\hypersetup{pdftex,
            pdfauthor={\paperauthors},
            pdftitle={\paperasciititle},
            pdfkeywords={\paperkeywords},
            pdfcopyright={Copyright (C) \papercopyright},
            pdflicenseurl={\paperlicenceurl}}

\usepackage[colorinlistoftodos,textsize=scriptsize]{todonotes}

\usepackage[bottom,flushmargin,hang,multiple]{footmisc}

\usepackage[all]{hypcap} 

\input{lhcb-symbols-def} 

\usepackage{cite} 
\usepackage{mciteplus}

%% file: lhcb-symbols-def.tex
\usepackage{xspace} 
\usepackage{upgreek}

\def\lhcb   {\mbox{LHCb}\xspace}
\def\atlas  {\mbox{ATLAS}\xspace}
\def\cms    {\mbox{CMS}\xspace}

\def\babar  {\mbox{BaBar}\xspace}
\def\belle  {\mbox{Belle}\xspace}
\def\belletwo {\mbox{Belle~II}\xspace}

\def\cdf    {\mbox{CDF}\xspace}

\def\MagUp {\mbox{\em Mag\kern -0.05em Up}\xspace}

\ifthenelse{\boolean{uprightparticles}}%
{

 \def\Pmu         {\ensuremath{\upmu}\xspace}

 \def\Ppi         {\ensuremath{\uppi}\xspace}                 
                  
 \def\Prho        {\ensuremath{\uprho}\xspace}                 
                  
 \def\Ptau        {\ensuremath{\uptau}\xspace}                 
                  
 \def\Pphi        {\ensuremath{\upphi}\xspace}

 \def\Ppsi        {\ensuremath{\uppsi}\xspace}                 
 \def\Pomega      {\ensuremath{\upomega}\xspace}                 

 \def\PDelta      {\ensuremath{\Delta}\xspace}                 
 \def\PXi         {\ensuremath{\Xi}\xspace}                 
 \def\PLambda     {\ensuremath{\Lambda}\xspace}                 
 \def\PSigma      {\ensuremath{\Sigma}\xspace}                 
 \def\POmega      {\ensuremath{\Omega}\xspace}                 
 \def\PUpsilon    {\ensuremath{\Upsilon}\xspace}
 \let\oldPi\Pi
 \def\PPi         {\ensuremath{\oldPi}\xspace}

 \def\PB      {\ensuremath{\mathrm{B}}\xspace}                 
                  
 \def\PD      {\ensuremath{\mathrm{D}}\xspace}

 \def\PJ      {\ensuremath{\mathrm{J}}\xspace}                 
 \def\PK      {\ensuremath{\mathrm{K}}\xspace}

 \def\Pb      {\ensuremath{\mathrm{b}}\xspace}                 
 \def\Pc      {\ensuremath{\mathrm{c}}\xspace}

 \def\Pi      {\ensuremath{\mathrm{i}}\xspace}

 \def\Pp      {\ensuremath{\mathrm{p}}\xspace}                 
 \def\Pq      {\ensuremath{\mathrm{q}}\xspace}                 
                  
 \def\Ps      {\ensuremath{\mathrm{s}}\xspace}                 
 \def\Pt      {\ensuremath{\mathrm{t}}\xspace}

 \def\thebaroffset{0.0em}
}
{

 \def\Pmu         {\ensuremath{\mu}\xspace}

 \def\Ppi         {\ensuremath{\pi}\xspace}                 
                  
 \def\Prho        {\ensuremath{\rho}\xspace}                 
                  
 \def\Ptau        {\ensuremath{\tau}\xspace}                 
                  
 \def\Pphi        {\ensuremath{\phi}\xspace}

 \def\Ppsi        {\ensuremath{\psi}\xspace}                 
 \def\Pomega      {\ensuremath{\omega}\xspace}                 
 \mathchardef\PDelta="7101
 \mathchardef\PXi="7104
 \mathchardef\PLambda="7103
 \mathchardef\PSigma="7106
 \mathchardef\POmega="710A
 \mathchardef\PUpsilon="7107
 \mathchardef\PPi="7105
                  
 \def\PB      {\ensuremath{B}\xspace}                 
                  
 \def\PD      {\ensuremath{D}\xspace}

 \def\PJ      {\ensuremath{J}\xspace}                 
 \def\PK      {\ensuremath{K}\xspace}

 \def\Pb      {\ensuremath{b}\xspace}                 
 \def\Pc      {\ensuremath{c}\xspace}

 \def\Pi      {\ensuremath{i}\xspace}

 \def\Pp      {\ensuremath{p}\xspace}                 
 \def\Pq      {\ensuremath{q}\xspace}                 
                  
 \def\Ps      {\ensuremath{s}\xspace}                 
 \def\Pt      {\ensuremath{t}\xspace}

 \def\thebaroffset{0.18em}
}
\newcommand{\offsetoverline}[2][\thebaroffset]{\kern #1\overline{\kern -#1 #2}}%

\makeatletter
\ifcase \@ptsize \relax
  \newcommand{\miniscule}{\@setfontsize\miniscule{4}{5}}
\or
  \newcommand{\miniscule}{\@setfontsize\miniscule{5}{6}}
\or
  \newcommand{\miniscule}{\@setfontsize\miniscule{5}{6}}
\fi
\makeatother

\DeclareRobustCommand{\optbar}[1]{\shortstack{{\miniscule (\rule[.5ex]{1.25em}{.18mm})}
  \\ [-.7ex] $#1$}}

\def\mup        {{\ensuremath{\Pmu^+}}\xspace}
\def\mun        {{\ensuremath{\Pmu^-}}\xspace} 

\def\mumu       {{\ensuremath{\Pmu^+\Pmu^-}}\xspace}

\def\tauon      {{\ensuremath{\Ptau}}\xspace}
\def\taup       {{\ensuremath{\Ptau^+}}\xspace}
\def\taum       {{\ensuremath{\Ptau^-}}\xspace}

\def\ellell     {\ensuremath{\ell^+ \ell^-}\xspace}

\def\quark     {{\ensuremath{\Pq}}\xspace}
\def\quarkbar  {{\ensuremath{\overline \quark}}\xspace}

\def\squark    {{\ensuremath{\Ps}}\xspace}

\def\cquark    {{\ensuremath{\Pc}}\xspace}
\def\cquarkbar {{\ensuremath{\overline \cquark}}\xspace}

\def\bquark    {{\ensuremath{\Pb}}\xspace}
\def\bquarkbar {{\ensuremath{\overline \bquark}}\xspace}

\def\tquark    {{\ensuremath{\Pt}}\xspace}

\def\pion   {{\ensuremath{\Ppi}}\xspace}
\def\piz    {{\ensuremath{\pion^0}}\xspace}
\def\pip    {{\ensuremath{\pion^+}}\xspace}
\def\pim    {{\ensuremath{\pion^-}}\xspace}

\def\rhomeson {{\ensuremath{\Prho}}\xspace}
\def\rhoz     {{\ensuremath{\rhomeson^0}}\xspace}

\def\kaon    {{\ensuremath{\PK}}\xspace}
\def\Kbar    {{\ensuremath{\offsetoverline{\PK}}}\xspace}

\def\KorKbar {\kern \thebaroffset\optbar{\kern -\thebaroffset \PK}{}\xspace}
\def\Kz      {{\ensuremath{\kaon^0}}\xspace}

\def\Kp      {{\ensuremath{\kaon^+}}\xspace}
\def\Km      {{\ensuremath{\kaon^-}}\xspace}

\def\KS      {{\ensuremath{\kaon^0_{\mathrm{S}}}}\xspace}

\def\Kstarz  {{\ensuremath{\kaon^{*0}}}\xspace}
\def\Kstarzb {{\ensuremath{\Kbar{}^{*0}}}\xspace}

\def\Kstarp  {{\ensuremath{\kaon^{*+}}}\xspace}

\newcommand{\phiz}{\ensuremath{\Pphi}\xspace}
\newcommand{\omegaz}{\ensuremath{\Pomega}\xspace}

\def\Dbar    {{\ensuremath{\offsetoverline{\PD}}}\xspace}
\def\D       {{\ensuremath{\PD}}\xspace}
\def\Db      {{\ensuremath{\Dbar}}\xspace}
\def\DorDbar {\kern \thebaroffset\optbar{\kern -\thebaroffset \PD}\xspace}

\def\Dp      {{\ensuremath{\D^+}}\xspace}
\def\Dm      {{\ensuremath{\D^-}}\xspace}

\def\DpDm    {\ensuremath{\Dp {\kern -0.16em \Dm}}\xspace}
\def\Dstar   {{\ensuremath{\D^*}}\xspace}
\def\Dstarb  {{\ensuremath{\Dbar{}^*}}\xspace}

\def\B       {{\ensuremath{\PB}}\xspace}
\def\Bbar    {{\ensuremath{\offsetoverline{\PB}}}\xspace}

\def\BorBbar {\kern \thebaroffset\optbar{\kern -\thebaroffset \PB}\xspace}
\def\Bz      {{\ensuremath{\B^0}}\xspace}

\def\Bd      {{\ensuremath{\B^0}}\xspace}
\def\Bdb     {{\ensuremath{\Bbar{}^0}}\xspace}
\def\BdorBdbar {\kern \thebaroffset\optbar{\kern -\thebaroffset \Bd}\xspace}
\def\Bu      {{\ensuremath{\B^+}}\xspace}

\def\Bp      {{\ensuremath{\Bu}}\xspace}

\def\Bs      {{\ensuremath{\B^0_\squark}}\xspace}

\def\BsorBsbar {\kern \thebaroffset\optbar{\kern -\thebaroffset \Bs}\xspace}

\def\jpsi     {{\ensuremath{{\PJ\mskip -3mu/\mskip -2mu\Ppsi}}}\xspace}
\def\psitwos  {{\ensuremath{\Ppsi{(2S)}}}\xspace}

\def\Y#1S{\ensuremath{\PUpsilon{(#1S)}}\xspace}

\def\proton      {{\ensuremath{\Pp}}\xspace}

\def\Lz          {{\ensuremath{\PLambda}}\xspace}

\def\LorLbar     {\kern \thebaroffset\optbar{\kern -\thebaroffset \PLambda}\xspace}

\def\Lb           {{\ensuremath{\Lz^0_\bquark}}\xspace}

\def\BF         {{\ensuremath{\mathcal{B}}}\xspace}

\newcommand{\decay}[2]{\ensuremath{#1\!\to #2}\xspace} 

\def\to                 {\ensuremath{\rightarrow}\xspace}

\def\qsq       {{\ensuremath{q^2}}\xspace}

\def\CP                {{\ensuremath{C\!P}}\xspace}

\def\BdToKstmm    {\decay{\Bd}{\Kstarz\mup\mun}}

\def\bsll     {\decay{\bquark}{\squark \ell^+ \ell^-}}

\def\AT#1     {\ensuremath{A_{\mathrm{T}}^{#1}}\xspace}           

\def\ctl       {\ensuremath{\cos{\theta_\ell}}\xspace}
\def\ctk       {\ensuremath{\cos{\theta_K}}\xspace}

\def\C#1      {\ensuremath{\mathcal{C}_{#1}}\xspace}                       
\def\Cp#1     {\ensuremath{\mathcal{C}_{#1}^{'}}\xspace}                    
\def\Ceff#1   {\ensuremath{\mathcal{C}_{#1}^{\mathrm{(eff)}}}\xspace}        
\def\Cpeff#1  {\ensuremath{\mathcal{C}_{#1}^{'\mathrm{(eff)}}}\xspace}       
\def\Ope#1    {\ensuremath{\mathcal{O}_{#1}}\xspace}                       
\def\Opep#1   {\ensuremath{\mathcal{O}_{#1}^{'}}\xspace}                    

\newcommand{\nospaceunit}[1]{\ensuremath{\text{#1}}}       
\newcommand{\aunit}[1]{\ensuremath{\text{\,#1}}}       

\newcommand{\tev}{\aunit{Te\kern -0.1em V}\xspace}
\newcommand{\gev}{\aunit{Ge\kern -0.1em V}\xspace}
\newcommand{\mev}{\aunit{Me\kern -0.1em V}\xspace}
\newcommand{\kev}{\aunit{ke\kern -0.1em V}\xspace}
\newcommand{\ev}{\aunit{e\kern -0.1em V}\xspace}
 
\newcommand{\mevc}{\ensuremath{\aunit{Me\kern -0.1em V\!/}c}\xspace}
\newcommand{\gevc}{\ensuremath{\aunit{Ge\kern -0.1em V\!/}c}\xspace}
\newcommand{\mevcc}{\ensuremath{\aunit{Me\kern -0.1em V\!/}c^2}\xspace}
\newcommand{\gevcc}{\ensuremath{\aunit{Ge\kern -0.1em V\!/}c^2}\xspace}
\newcommand{\gevgevcccc}{\ensuremath{\gev^2\!/c^4}\xspace} 

\def\mum  {\ensuremath{\,\upmu\nospaceunit{m}}\xspace}

\def\fb   {\ensuremath{\aunit{fb}}\xspace}
\def\invfb   {\ensuremath{\fb^{-1}}\xspace}

\def\deriv {\ensuremath{\mathrm{d}}}

\def\gsim{{~\raise.15em\hbox{$>$}\kern-.85em
          \lower.35em\hbox{$\sim$}~}\xspace}
\def\lsim{{~\raise.15em\hbox{$<$}\kern-.85em
          \lower.35em\hbox{$\sim$}~}\xspace}

\def\sPlot{\mbox{\em sPlot}\xspace}

\def\pt         {\ensuremath{p_{\mathrm{T}}}\xspace}

\def\ptot       {\ensuremath{p}\xspace}

\def\mrad{\aunit{mrad}\xspace}

\def\evtgen     {\mbox{\textsc{EvtGen}}\xspace}

\def\geant      {\mbox{\textsc{Geant4}}\xspace}

\def\photos     {\mbox{\textsc{Photos}}\xspace}

\def\pythia     {\mbox{\textsc{Pythia}}\xspace}

\def\tell1  {TELL1\xspace}
\def\ukl1   {UKL1\xspace}

\newcommand{\eg}{\mbox{\itshape e.g.}\xspace}
\newcommand{\ie}{\mbox{\itshape i.e.}\xspace}

\newcommand{\lhcborcid}[1]{\href{https://orcid.org/#1}{\hspace*{0.1em}\raisebox{-0.45ex}{\includegraphics[width=1em]{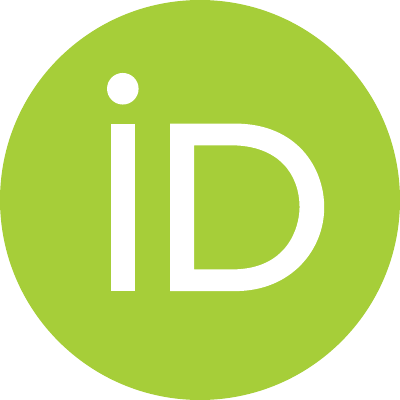}}}}


%% file: title-LHCb-PAPER.tex

\begin{titlepage}
\pagenumbering{roman}

\vspace*{-1.5cm}
\centerline{\large EUROPEAN ORGANIZATION FOR NUCLEAR RESEARCH (CERN)}
\vspace*{1.5cm}
\noindent
\begin{tabular*}{\linewidth}{lc@{\extracolsep{\fill}}r@{\extracolsep{0pt}}}
\ifthenelse{\boolean{pdflatex}}
{\vspace*{-1.5cm}\mbox{\!\!\!\includegraphics[width=.14\textwidth]{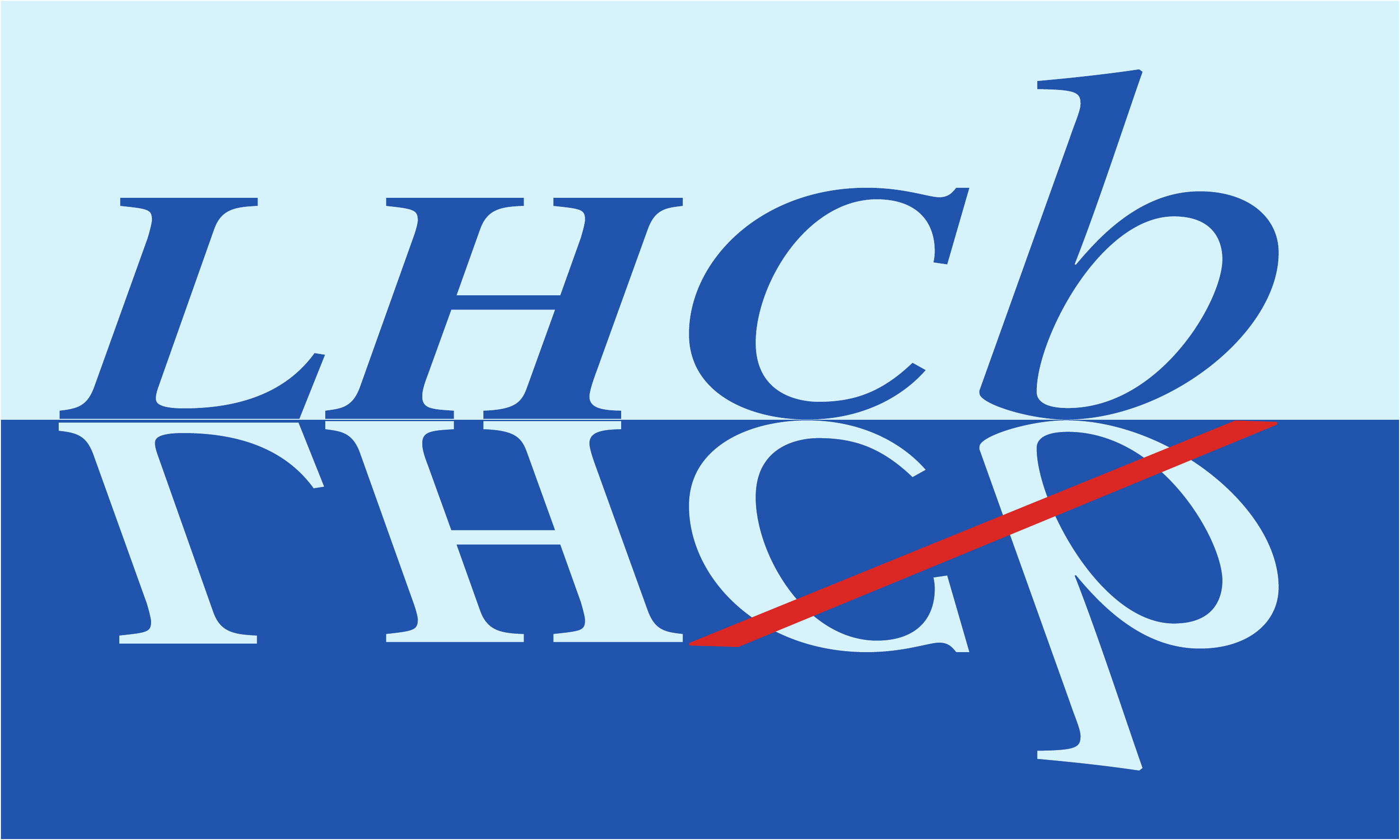}} & &}%
{\vspace*{-1.2cm}\mbox{\!\!\!\includegraphics[width=.12\textwidth]{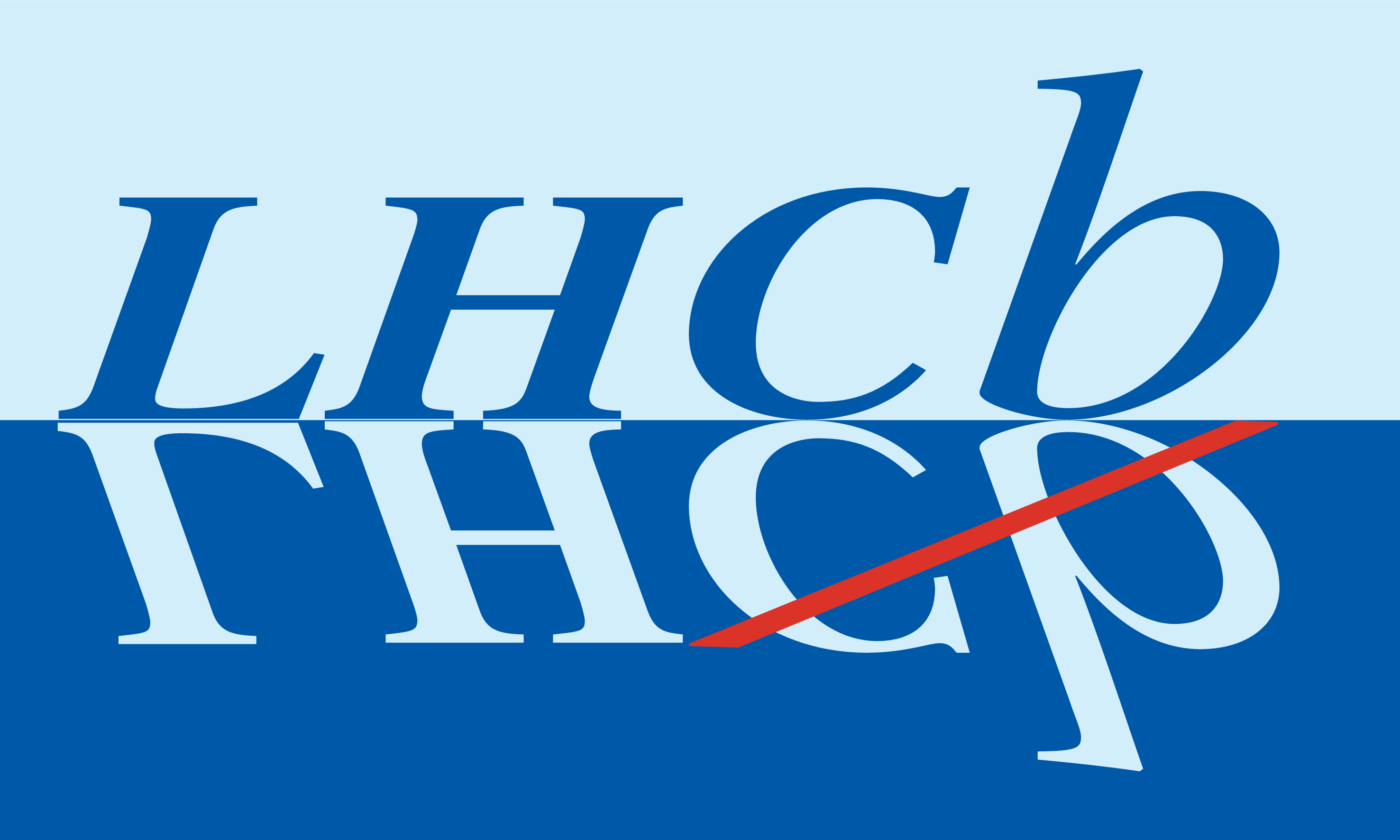}} & &}%
\\
 & & CERN-EP-2024-122 \\  
 & & LHCb-PAPER-2024-011 \\  
 & & 10 September 2024 \\ 
 & & \\
\end{tabular*}

\vspace*{2.0cm}

{\normalfont\bfseries\boldmath\huge
\begin{center}
  \papertitle 
\end{center}
}

\vspace*{1.0cm}

\begin{center}
\paperauthors\footnote{Authors are listed at the end of this paper.}
\end{center}

\vspace{\fill}

\begin{abstract}
  \noindent
  A comprehensive study of the local and nonlocal amplitudes contributing to the decay $B^0 \to K^{*0}(\to K^+ \pi^-) \mu^+ \mu^-$ is performed by analysing the phase-space distribution of the decay products. The analysis is based on $pp$ collision data corresponding to an integrated luminosity of 8.4\,$\rm{fb}^{-1}$ collected by the LHCb experiment. This measurement employs for the first time a model of both one-particle and two-particle nonlocal amplitudes, and utilises the complete dimuon mass spectrum without any veto regions around the narrow charmonium resonances. In this way it is possible to explicitly isolate the local and nonlocal contributions and capture the interference between them. 
  The results show that interference with nonlocal contributions, although larger than predicted, only has a minor impact on the Wilson Coefficients determined from the fit to the data. For the local contributions, the Wilson Coefficient $\mathcal{C}_9$, responsible for vector dimuon currents, exhibits a $2.1\sigma$ deviation from the Standard Model expectation. The Wilson Coefficients $\mathcal{C}_{10}$, $\mathcal{C}^{'}_9$ and $\mathcal{C}^{'}_{10}$ are all in better agreement than $\mathcal{C}_9$ with the Standard Model and the global significance is at the level of $1.5\sigma$. 
  The model used also accounts for nonlocal contributions from $B^{0}\to K^{*0}\left[\tau^+\tau^-\to \mu^+\mu^-\right]$ rescattering,
  resulting in the first direct measurement of the  $b s\tau\tau$ vector effective-coupling $\mathcal{C}_{9\tau}$.
\end{abstract}


\begin{center}
  Published in
   JHEP 09 (2024) 026
\end{center}

\vspace{\fill}

{\footnotesize 
\centerline{\copyright~\papercopyright. \href{\paperlicenceurl}{\paperlicence}.}}
\vspace*{2mm}

\end{titlepage}


\newpage
\setcounter{page}{2}
\mbox{~}

%% file: body.tex
\section{Introduction}
\label{sec:Introduction}

The decay \BdToKstmm has attracted significant attention in recent years, both theoretically and experimentally.\footnote{The inclusion of charge-conjugate processes is implied throughout, unless otherwise specified.} From the theoretical side, the combination of the loop suppressed flavour-changing neutral current \bsll Standard Model~(SM) amplitudes and the rich information arising from the multitude of \Kstarz polarisation amplitudes makes the decay an ideal channel to search for New Physics~(NP) contributions~\cite{Kruger:1999xa}. From the experimental side, the \BdToKstmm mode stands out as the corresponding \decay{\Bz}{\Kstarz e^+e^-} mode is harder to reconstruct at a hadron collider and the \decay{\Bz}{\Kstarz \tau^+\tau^-} mode is so difficult to reconstruct at any experiment that it has not been observed yet. Previous \lhcb results of angular analyses of the \BdToKstmm decay have consistently deviated from SM predictions at the level of three standard deviations ($3\sigma$)~\cite{LHCb-PAPER-2013-019,LHCb-PAPER-2013-037,LHCb-PAPER-2015-051,LHCb-PAPER-2020-002, LHCb-PAPER-2023-033}. 
The decay has also been studied by \atlas~\cite{ATLAS:2018gqc}, \babar~\cite{BaBar:2006tnv,BaBar:2015wkg}, \belle~\cite{Belle:2009zue,Belle:2016fev}, \cdf~\cite{CDF:2011tds}, and \cms~\cite{CMS:2015bcy,CMS:2017rzx}.

The matrix element for the \BdToKstmm decay has components related to both local and nonlocal contributions. The local contributions correspond to an energy scale well above the beauty hadron masses and are where NP might manifest itself. The leading nonlocal contributions are due to narrow charmonium resonances in the dimuon mass spectrum of the \BdToKstmm decay but influence all parts of the phase space. The subleading contributions come from two-particle amplitudes at dimuon masses above the open charm threshold. By explicitly excluding the charmonium resonance regions, previous analyses relied on theoretical estimates of the remaining nonlocal contributions. A novel feature of the analysis presented in this paper is to include the full phase space, including the narrow charmonium resonances, and determine the local and nonlocal contributions simultaneously. In this way the theoretical model dependence related to the nonlocal contributions is reduced as they are determined directly from the data.

Theoretically, the decay \BdToKstmm is described in the Weak Effective Theory (WET) framework (see \eg  Ref.~\cite{Altmannshofer:2008dz}), which is encapsulated by the Hamiltonian
\begin{align}
    \mathcal{H}_\text{WET} = \frac{-4 G_F}{\sqrt{2}} V^*_{\tquark\squark}V_{\tquark\bquark} \sum_i \C{i} ^{(')} (\mu) \Ope{i} ^{(')} (\mu),
\end{align}
where $G_F$ is the Fermi constant and $V_{q_{k}q_{j}}$ are elements of the Cabibbo-Kobayashi-Maskawa (CKM) matrix corresponding to the $q_j \to q_k$ quark transition. The effective operators $\Ope{i} ^{(')}$ describe all possible interactions between ingoing and outgoing particles, while the Wilson Coefficients $\C{i} ^{(')}$ are the corresponding effective coupling constants. Lastly, $\mu$ defines the energy scale. 

In the WET framework, the $\Ope{i} ^{(')}$ operators arise from integrating out all heavy particles at and above the electroweak mass scale $M_W\sim 80\gevcc$, and the Wilson Coefficients are calculated by matching results for physical observables calculated in the full and effective theories. The values of the Wilson Coefficients can then via renormalisation be evolved to the relevant energy scale for $\bquark$-hadron decays, given by the mass of the $b$ quark $m_\bquark$~\cite{EOSAuthors:2021xpv}. The dominant effective local operators for \bsll decays are defined using the standard notation as
\begin{equation}
    \begin{split}
        \Ope {7} &= \frac{e}{16 \pi^2} m_\bquark \left( \bar{\squark}_L \sigma^{\mu \nu} \bquark_R \right) F_{\mu \nu}, \\
        \Ope {9\ell} &= \frac{e^2}{16 \pi^2} \left( \bar{\squark}_L \gamma_\mu \bquark_L \right) \bar{\ell} \gamma^\mu \ell, \\ 
        \Ope {10\ell} &= \frac{e^2}{16 \pi^2} \left( \bar{\squark}_L \gamma_\mu \bquark_L \right) \bar{\ell} \gamma^\mu \gamma_5 \ell,
    \end{split}
    \quad\quad\quad
    \begin{split}
        \Opep {7} &= \frac{e}{16 \pi^2} m_\bquark \left( \bar{\squark}_R \sigma^{\mu \nu} \bquark_L \right) F_{\mu \nu}, \\
        \Opep {9\ell} &= \frac{e^2}{16 \pi^2} \left( \bar{\squark}_R \gamma_\mu \bquark_R \right) \bar{\ell} \gamma^\mu \ell, \\ 
        \Opep {10\ell} &= \frac{e^2}{16 \pi^2} \left( \bar{\squark}_R \gamma_\mu \bquark_R \right) \bar{\ell} \gamma^\mu \gamma_5 \ell,
    \end{split}
    \label{eqn:eftOpe}
\end{equation}
and in this analysis, the SM values for the Wilson Coefficients are obtained from Ref.~\cite{EOS, EOSAuthors:2021xpv}. The operators $\mathcal{O}_{1-6,8}$ mix through renormalisation with operators $\mathcal{O}_{7,9}$ resulting in observables sensitive to effective operators $\mathcal{O}_{7,9}^\text{eff}$. Further discussion on these effective operators and their corresponding Wilson Coefficients can be found in Sec.~\ref{sec:NonLocalModel}. The Wilson Coefficients for the primed operators, which correspond to the opposite chirality processes, are predicted to be highly suppressed in the SM or to be zero, but are considered here as potential sources of NP.

The dilepton operators and the corresponding Wilson Coefficients carry a lepton flavour index (\eg $\C{9\ell} $) implying that NP might not be lepton flavour universal. Wilson Coefficients related to muons are implied in this paper if no explicit lepton index is given. Possible NP contributions from (pseudo)scalar or tensor operators are not taken into account in this analysis since they are strongly constrained through measurements of other \bsll processes~\cite{Altmannshofer:2008dz,Descotes-Genon:2012isb}. The level of \CP violation in the \BdToKstmm decay is insignificant in the SM and, in this analysis, any NP contribution is assumed to be \CP conserving. In this case all the Wilson Coefficients are real and unless otherwise noted, the short hand notation \C{i} will be used to denote the real part of the Wilson Coefficient. 

Global analyses of the \BdToKstmm decay and others involving \bsll transitions indicate that the data are better described by models with one or more Wilson Coefficients differing from the SM predictions~\cite{Altmannshofer:2014rta,Capdevila:2017bsm,Beaujean:2013soa,Descotes-Genon:2013wba,Alguero:2021anc,Alguero:2023jeh}. Interestingly, the introduction of a NP contribution to the single parameter \C{9}, describing the effective \decay{\bquark}{\squark \ellell} vector coupling, is sufficient to explain the tension observed with the data. Even better descriptions of the data can be achieved by modifying the axial-vector coupling \C{10} as well. Such contributions arise naturally in scenarios involving $Z'$ bosons or leptoquarks, as discussed in \eg Ref.~\cite{London:2021lfn}.

Unfortunately, no robust conclusion regarding NP can be drawn at this point since both experimental measurements and theoretical predictions of decays involving the \bsll transition remain limited in precision. For the \BdToKstmm decay, the phase-space distribution depends on calculations of various nonperturbative hadronic matrix elements. The dominant contributions are proportional to matrix elements of the \decay{b}{s} quark current entering the local operators of Eq.~\ref{eqn:eftOpe}. The corresponding local form factors are calculable with good precision in lattice QCD (LQCD)~\cite{Horgan:2013hoa,Horgan:2015vla}  and light-cone sum rules (LCSR)~\cite{Bharucha:2015bzk,Gubernari:2018wyi}. However, additional contributions arise from the four-quark operators \Ope{1,2}, which produce a dimuon pair through a subsequent coupling to the electromagnetic current~\cite{Khodjamirian:2010vf}. These contributions are dominated by the so-called charm-loop operator describing a \decay{\bquark}{\squark \cquark\cquarkbar(\decay{\cquark\cquarkbar}{\decay{\gamma^\ast}{\mup\mun}})} process. Such processes are at an energy scale below the cutoff in the WET framework and lead to so-called nonlocal form factors. These objects are less well understood and reliable methods for their calculation with controlled uncertainties are not yet well-developed. The size of the nonlocal contributions has been the source of some debate in recent years~\cite{Khodjamirian:2010vf,Korchin:2012kz,Lyon:2014hpa,Descotes-Genon:2013wba,Ciuchini:2015qxb,Ciuchini:2022wbq,Gubernari:2020eft,Gubernari:2022hxn} and no consensus has yet been reached regarding their influence.

The nonlocal contributions manifest as resonant and nonresonant amplitudes in the $\qsq \equiv m^2_{\mu\mu}$ spectrum. The most strongly affected regions are those in the vicinity of the narrow quarkonia resonances, \ie the \phitentwenty, \jpsi, and \psitwos. The method traditionally employed to avoid these nonlocal contributions is to omit regions in \qsq around the resonances from the experimental analyses. 
However, the nonlocal contributions can have significant effects far away from the resonances through interference, both between the various nonlocal components and with the local component~\cite{Khodjamirian:2010vf}. Crucially, such effects lead to a shift in \C9 that can potentially be large enough to resolve the observed tensions in the angular observables without requiring any NP affecting the local contributions~\cite{Lyon:2014hpa,Ciuchini:2015qxb,Ciuchini:2020gvn,Blake:2017fyh}. 

Recently, an analysis was carried out in which the nonlocal contributions were parameterised following Refs.~\cite{Bobeth:2017vxj,Gubernari:2020eft,Gubernari:2022hxn} using a truncated series expansion designed to exploit the analytic properties of the hadronic matrix elements in the region $\qsq \leq m^2_\psitwos$. The coefficients of the series expansion were determined experimentally by combining \lhcb data from the low-\qsq ($1.1 \leq \qsq \leq 8.0\gevgevcccc$) and inter-resonance ($11.0 \leq \qsq \leq 12.5\gevgevcccc$) regions with independently obtained measurements of the polarisation amplitudes and strong-phase differences at the \jpsi pole~\cite{LHCb-PAPER-2023-033}. Based on the same dataset (2011--2012 and 2016) as prior \lhcb analyses it obtained a result indicating that the nonlocal contributions did not affect prior measurements in a significant way. However, the nonlocal model used in Ref.~\cite{LHCb-PAPER-2023-033} did not consider the effects of light hadron resonances below $\qsq = 1.1 \gevgevcccc$ or broad charmonium resonances and multibody states above $\qsq = m^2_\psitwos$, nor did it consider finite width effects of the \jpsi and \psitwos resonances.

In this paper, \lhcb data are fitted with a model that combines the local and nonlocal amplitudes across the \qsq spectrum in the range $0.1 \leq \qsq \leq 18.0 \gevgevcccc$. The model includes all known vector resonances coupling to muons, as well as two particle contributions from $\D^{(\ast)}\Db{}^{(\ast)}$ and $\taup\taum$ loops. It thereby simultaneously determines the nonlocal contributions and the Wilson Coefficients, \C9, \C{10}, \Cp9, \Cp{10} and $\C{9\tau} $ that describe the local contributions. The analysis is performed using proton-proton (\proton\proton) collision data corresponding to an integrated luminosity of 8.4\invfb collected during the years 2011--2012 (Run~1) and 2016--2018 (Run~2). 

The paper is organised as follows: in Sec.~\ref{sec:AmplitudeModel}, the decay amplitude model is described, followed by a description of the experimental considerations in Sec.~\ref{sec:ExperimentalModel}. The strategy for performing the measurement is detailed in Sec.~\ref{sec:Strategy} and the main contributions to systematic uncertainties are described in Sec.~\ref{sec:Systematics}. The results of the fit to data are then presented in Sec.~\ref{sec:Results}, which is followed by a discussion and concluding remarks in Secs.~\ref{sec:Discussion} and~\ref{sec:Conclusion}, respectively.

\section{Theoretical signal model}
\label{sec:AmplitudeModel}

\subsection{Angular definitions}
\label{sec:AngularDefinitions}

In a four-body final state, the $\Bd\to K^+\pi^-\mumu$ differential decay rate can be fully described by a five-dimensional phase space. It is parameterised in terms of three helicity angles \ctl, \ctk and \phih, along with \qsq and $m_{K\pi}^2$, which denote the mass squared of the dimuon and $K^+\pi^-$ systems, respectively. The angle $\theta_\ell$ is defined as the angle between the direction of the \mup (\mun) in the dimuon rest frame and the direction of the dimuon in the \Bd (\Bdb) rest frame. The angle $\theta_K$ is defined as the angle between the direction of the kaon in the \Kstarz (\Kstarzb) rest frame and the direction of the \Kstarz (\Kstarzb) in the \Bd (\Bdb) rest frame. The angle $\phi$ is the angle between the plane containing the dimuon pair and the plane containing the kaon and pion from the \Kstarz meson. The angular basis used in this paper is identical to that defined in Ref.~\cite{LHCb-PAPER-2013-019}, and is defined such that the \Bd and \Bdb angular distributions are described by the same set of angular functions.

Within the range $0.796 < m_{K\pi} < 0.996\gevcc$ considered in this analysis, the decay $\Bd\to K^+\pi^-\mumu$ receives amplitude contributions from P-wave $K^{*0}(892)$ transitions, henceforth referred to simply as \BdToKstmm, and S-wave $\Bd\to K^{*0}_{0}(700)\mumu$ transitions. Given previous measurements of higher \Kp\pim partial waves in \mbox{$\Bd\to K^+\pi^-\mu^+\mu^-$} and \mbox{$\Bd\to \psi^{(')}K^+\pi^-$} transitions~\cite{LHCb-PAPER-2016-025,Belle:2014nuw,Belle:2013shl}, contributions from higher partial wave \Kp\pim states in the $m_{K\pi}$ range of this analysis are considered to be negligible.

\subsection{Decay rate and angular observables}
\label{sec:DecayRate}

The five-dimensional differential decay rate of the $B^0\to K^+\pi^-\mumu$ process is expressed in terms of spherical harmonic angular coefficients $f_i(\cos{\theta_\ell},\cos{\theta_K},\phi)$ multiplied by the corresponding set of \qsq-dependent functions $\bnbar{J}_i (q^2)$ and $m_{K\pi}^{2}$-dependent functions $g_i(m_{K\pi}^{2})$. This is written compactly as  
\begin{align}
\label{eqn:fivediffdecayrate}
    \frac{{\rm d}^5\bnbar{\Gamma} (\decay{\Bz}{K^+\pi^-\mup\mun})}{{\rm d} q^2\,{\rm d}\vec{\Omega}\,{\rm d}m_{K\pi}^{2}} &= \frac{9}{32\pi} \sum_i \bnbar{J_i} (q^2)f_i(\cos{\theta_\ell},\cos{\theta_K},\phi) g_i(m_{K\pi}^{2}),
\end{align}
where $\Gamma$ and $\overline{\Gamma}$ indicate the \decay{\Bdb}{\Kstarzb\mumu} and \BdToKstmm decay rates, respectively, and likewise for the $J_i$ and $\bar{J}_i$ functions. The $\bnbar{J}_i (q^2)$ functions are built up from bilinear combinations of the transversity amplitudes, described in Sec.~\ref{sec:TransversityAmplitudes}. Their explicit forms are given in Appendix~\ref{app:DecayRateFunctions}, 
along with the definitions of $f_i(\cos{\theta_\ell},\cos{\theta_K},\phi)$. The functions $g_{i}(\mkpisq)$ in Eq.~\ref{eqn:fivediffdecayrate} represent bilinear products of the line shape models for the $\Kp\pim$ system. The P-wave line shape is modelled using a relativistic Breit--Wigner function and the S-wave line shape is modelled using the LASS parameterisation, both given in Ref.~\cite{LHCb-PAPER-2023-032}. The systematic uncertainty on the parameters reported in this paper due to the choice of the $K^+\pi^-$ line shape is found to be negligible. Note that the \qsq dependence of the $g_{i}(\mkpisq)$ factors is suppressed. This is a good approximation given the \qsq and \mkpisq ranges in which the analysis is performed. For this analysis, the differential decay rate of Eq.~\ref{eqn:fivediffdecayrate} is integrated over the selected region in $\mkpisq$ leading to
\begin{align}
\label{eqn:diffdecayrate}
    \frac{{\rm d}^4\bnbar{\Gamma} (\decay{\Bz}{K^+\pi^-\mup\mun})}{{\rm d} q^2\,{\rm d}\vec{\Omega}} &= \frac{9}{32\pi} \sum_i \bnbar{J_i} (q^2)f_i(\cos{\theta_\ell},\cos{\theta_K},\phi) G_i,
\end{align}
where $G_i$ are given by
\begin{align}
    G_i &= \int_{\mkpisq=0.796^2\gevgevcccc}^{\mkpisq=0.996^2\gevgevcccc}g_{i}(\mkpisq)\deriv\mkpisq ~. 
    \label{eqn:MkpiFactors}
\end{align}

\subsection{Transversity amplitudes}
\label{sec:TransversityAmplitudes}
Following the convention described in  Ref.~\cite{Altmannshofer:2008dz}, the seven transversity amplitudes for a P-wave \Kstarz state are given by
\begin{align}
    \begin{split}
        \mathcal{A}_0^{L,R}(q^2) =& N_{0}\bigg\{ \left[ \left( \C{9} ^{\text{(eff)},0}(q^2)-\Cp{9} \right) \mp \left( \C{10} -\Cp{10} \right) \right] A_{12}(q^2) \\ &\hspace{55mm}+ \frac{m_b}{m_B + m_{K^*}} \left( \C7 ^{\text{(eff)},0} - \Cp{7} \right) T_{23}(q^2)\bigg\}, 
    \end{split}
    \label{eqn:A0amplitude}
    \\[3mm]
    \begin{split}
         \mathcal{A}_\parallel^{L,R}(q^2) =& N_{\parallel}\bigg\{ \left[ \left( \C{9} ^{\text{(eff)},\parallel} (q^2)-\Cp{9} \right) \mp \left( \C{10} -\Cp{10} \right) \right] \frac{A_1(q^2)}{m_B - m_{K^*}} \\ &\hspace{55mm}+ \frac{2m_b}{q^2} \left( \C7 ^{\text{(eff)},\parallel} - \Cp{7} \right)T_2(q^2) \bigg\}, 
    \end{split}
    \label{eqn:APamplitude}
    \\[3mm]
    \begin{split}
       \mathcal{A}_\perp^{L,R} (q^2) =& N_{\perp} \bigg\{ \left[ \left( \C{9} ^{\text{(eff)},\perp}(q^2)+\Cp{9} \right) \mp \left( \C{10} + \Cp{10} \right) \right] \frac{V(q^2)}{m_B + m_{K^*}} \\ &\hspace{55mm}+ \frac{2m_b}{q^2} \left( \C7 ^{\text{(eff)},\perp} - \Cp{7} \right) T_1(q^2)  \bigg\},
    \end{split}
    \label{eqn:ATamplitude} 
    \\[3mm]
    \mathcal{A}_{t}(q^2) =& N_t\left\{ 2\left[ \C{10} - \Cp{10} \right]A_{0}(\qsq)\right\},
    \label{eqn:Atamplitude} 
\end{align}
where the functions $V$, $A_0$, $A_1$, $A_{12}$, $T_1$, $T_2$, and $T_{23}$ are the local hadronic form factors, while the nonlocal matrix elements are absorbed in the \qsq and polarisation-dependent effective Wilson Coefficients $\mathcal{C}_9^{\text{(eff)}, \mathcal{\lambda}} (\qsq)$. For the  $\mp$ symbol,  the minus and plus signs correspond to the $L$ and $R$ superscripts, respectively. Further discussion of the effective Wilson coefficients is provided in Sec.~\ref{sec:NonLocalModel}. The symbols $m_{B}$ and $m_{K^*}$ denote the known masses of the $B^0$ and $K^{*0}$ mesons~\cite{PDG2022}. Equations~\ref{eqn:A0amplitude}--\ref{eqn:Atamplitude} are referred to as the longitudinal, parallel, perpendicular, and timelike transversity amplitudes, respectively. 

The S-wave $K^{*0}_{0}$ state leads to two additional transversity amplitudes, given by~\cite{Becirevic:2012dp}
\begin{align}
    \begin{split}
        \mathcal{A}_{00}^{L,R}(q^2) =& N_{00}\bigg( (\C{9} ^{\text{(eff)},00} \mp \C{10} ) F_{1}(q^2) + \frac{2m_b}{m_B + m_{K^{*0}_{0}(700)}} \C7 ^{\text{(eff)},00}F_{T}(q^2)  \bigg),
    \end{split}
    \label{eqn:A00amplitude}
\end{align}
where $F_i$ are the corresponding local S-wave form factors as defined in Ref.~\cite{Doring:2013wka}. The symbol $m_{K^{*0}_{0}}(700)$ denotes the mass of the $K^{*0}_{0}$ state~\cite{PDG2022}. The choice of the $K^{*0}_{0}(700)$ mass leads to a systematic uncertainty discussed later in this paper. The timelike amplitude contribution for the S-wave $K^{*0}_{0}$ is ignored in this analysis owing to the smallness of the S-wave contribution as a whole in the \mkpisq range considered and the lepton-mass suppression of the timelike amplitudes.  
The various normalisation factors appearing in the transversity amplitudes are given by
\begin{align}
N_{0}=&\frac{-8N m_B m_{K^*}}{\sqrt{q^2}}, \nonumber \\
N_{\parallel} =&- N \sqrt{2}(m^2_B - m^2_{K^*}), \nonumber \\
\label{eqn:N00}
N_{\perp}=&N\sqrt{2\lambda},\\
N_{t}=&\frac{N \sqrt{\lambda}}{\sqrt{\qsq}}, \nonumber \\
N_{00}=&-N\frac{\lambda_{K_{0}^{*0}}}{\sqrt{q^2}}, \nonumber 
\end{align}
where the triangle (K\"all\'en) function $\lambda$ is given by
\begin{equation}
\label{eqn:kallen}
\lambda=m_{B}^{4}+m_{K^*}^{4}+q^4-2(m_{B}^{2}m_{K^*}^{2}+m_{K^*}^{2}q^{2}+m_{B}^{2}q^{2}),
\end{equation}
and the constant $N$ is
\begin{equation}
N=V_{tb}V_{ts}^{*}\sqrt{\frac{G_{F}^{2}\alpha_{\text{em}}^{2} q^2\lambda^{1/2}\beta_{\mu}}{3\times2^{10}\pi^5m_{B}^{3}}},
\end{equation}
with $\beta_\mu = \sqrt{1 - 4m_{\mu}^{2}/\qsq}$.
The $K_{0}^{*0}$ subscript appearing in the expression of $N_{00}$ in Eq.~\ref{eqn:N00} indicates the $m_{K^*} \to m_{K_{0}^{*0}(700)}$ replacement.
The angular coefficients of Eq.~\ref{eqn:diffdecayrate} can be fully constructed from combinations of the transversity amplitudes, as shown in Appendix~\ref{app:DecayRateFunctions}. Consequently, the parameters of the signal model for the fit to the \BdToKstmm differential decay rate are those that feature in Eqs.~\ref{eqn:A0amplitude}--\ref{eqn:A00amplitude}. These include the Wilson Coefficients $\mathcal{C}^{(')}_{7,9,10}$, and the parameters that describe the local and nonlocal hadronic contributions. The CKM elements, particle masses, and coupling constants are all taken to be known and are fixed in the fit. In the following subsections, specific parameterisations of the local and nonlocal amplitudes are presented, which allow both the hadronic parameters and the Wilson Coefficients to be determined from a fit to data.

\subsection{Local form factors}

The local $B^0\to \Kstarz$ form factors $F_i \in \{V, A_0, A_1, A_{12}, T_1, T_2, T_{23}\}$ are parameterised using a series expansion~\cite{Bharucha:2015bzk},
\begin{equation}
 \label{eq:formfactors}
F_i(q^2)=\frac{1}{1-q^2/m_{R,i}^2}\sum_{k=0}^{2}\alpha^{i}_{k}[z(q^2)-z(0)]^k,
\end{equation}
where the $\alpha^{i}_{k}$ coefficients are parameters to be determined, and $m_{R,i}$ is the mass of the lowest lying \bquarkbar\squark resonance with $J^P$ quantum numbers matching those of the form factor $F_i$ for $b\to s\ell^+\ell^-$ transitions. The values of $m_{R,i}$ used in this analysis are the same as those in Table 3 of Ref.~\cite{Bharucha:2015bzk}. The $z$ function in Eq.~\ref{eq:formfactors} is defined as, 
\begin{equation}
 \label{eq:zexpansion}
 z(t) = \frac{\sqrt{t_{+} -t}- \sqrt{t_{+}-t_{0}}}{\sqrt{t_{+} -t}+ \sqrt{t_{+}-t_{0}}} \, ,
\end{equation}
with $t_{\pm} = (m_{B} \pm m_{K^*})^{2}$ and $t_{0} = t_{+} (1- \sqrt{1- t_{-}/t_{+}})$. This analysis uses the results from Ref.~\cite{Gubernari:2022hxn} that rely on a combination of LCSR~\cite{Gubernari:2018wyi} and LQCD~\cite{Horgan:2013hoa,Horgan:2015vla} computations to constrain the coefficients $\alpha^{i}_{k}$. This constraint is propagated to the fit of the \mbox{\BdToKstmm} differential decay rate through a multivariate Gaussian likelihood factor. The behaviour of the fit under alternative LCSR and LQCD local form-factor calculations presented in Ref.~\cite{Bharucha:2015bzk} is also investigated.

Recent LCSR computations of $B^0\to \Kstarz$ form factors that account for the finite \Kstarz width have shown that narrow width $B^0\to \Kstarz$ form factors can be scaled by a global factor of 1.1 to account for the finite \Kstarz width~\cite{Descotes-Genon:2019bud,Gubernari:2022hxn}. This correction factor has only been demonstrated to work in the large recoil (low \qsq) region and further theoretical work is required in order to establish a finite width effect to $B^0\to \Kstarz$ form factors across the entire \qsq range. Therefore, the form-factor parameters used in this analysis, provided in Ref.~\cite{Gubernari:2022hxn}, implicitly account for this factor in the region $\qsq<8\gevgevcccc$ but not elsewhere.

\subsubsection{S-wave local form factors}

As no reliable S-wave local form-factor predictions exist as yet, in this analysis the S-wave amplitudes are treated as nuisance parameters. This means decoupling the local parameters appearing in $\mathcal{A}_{00}^{L,R}$ from those in the P-wave transversity amplitudes. Moreover, an estimation of the $B^0\to K^{*0}_{0}$ contribution using data is employed by adopting an effective form factor in the S-wave amplitude given by
\begin{align}
\label{eqn:effective_swave}
    \mathcal{A}^{L,R}_{00} &= -N\frac{\lambda_{K_{0}^{*0}}}{\sqrt{q^2}} \bigg[ F_{\text{eff}} (q^2)  \left( \mathcal{C}^{ {\rm(eff)}S}_{9} \mp \mathcal{C}^S_{10} + \frac{2 m_\bquark}{m_\B + m_{K^{*0}_{0}(700)}} \mathcal{C}^S_{7} \right) \bigg],
\end{align}
where the parameters $\mathcal{C}^{ {\rm(eff)}S}_{9}$, $\mathcal{C}^S_{10}$ and $\mathcal{C}^S_{7}$ are independent of the Wilson Coefficients \C9, \C10 and \C7.
The effective S-wave form factor $F_{\text{eff}}$ is in turn given by
\begin{align}
	\label{eqn:effective_swaveFF}
    F_{\text{eff}} (\qsq) &= \frac{F(0)}{1 + 2.15 (\alpha_F - 1) \left( \frac{\qsq}{m^2_B} \right) + 1.055 (1 - 2 \alpha_F + \beta_F) \left( \frac{\qsq}{m^2_\Bd} \right)^2},
\end{align}
with the parameters $\alpha_F$ and $\beta_F$ both positive. This form-factor parameterisation describes a very similar shape in $q^2$ to that used in Ref.~\cite{Doring:2013wka} but instead incorporates a polynomial with restricted parameters in the denominator. In this way, the fit can accommodate the wide range of $\Bd\to K^{*0}_{0}$ form factors in the literature while at the same time staying well behaved. Given Eqs.~\ref{eqn:effective_swave} and~\ref{eqn:effective_swaveFF}, a degeneracy exists between the decoupled S-wave Wilson Coefficients $\mathcal{C}_{7,9}^S$ and the effective form factor normalisation parameter $F(0)$. In the fit, the latter is allowed to vary along with $\mathcal{C}_{10}^S$, while $\mathcal{C}_{7}^S$ and $\mathcal{C}_{9}^S$ are fixed to the SM values for \C7 and \C9 , respectively. The $\alpha_F$ and $\beta_F$ form factor parameters are also highly correlated, requiring one of them ($\alpha_F$) to be fixed for a reliable fit.

\subsection{Nonlocal form factors}
\label{sec:NonLocalModel}
In Eqs.~\ref{eqn:A0amplitude}--\ref{eqn:A00amplitude}, the nonlocal form factors are absorbed into the Wilson Coefficients, which highlights the fact that the dominant effect of these contributions is a \qsq- and helicity-dependent shift in the effective \C9 value, given by
\begin{align}
        \mathcal{C}_9^{\text{(eff)}, \lambda} &= \C9 + Y_{\quark \quarkbar,\lambda} (\qsq).
    \label{eq:C9eff}
\end{align}
It should be noted that even if \C9 is real, $\mathcal{C}_9^{\text{(eff)}, \lambda}$ will take on complex values. The approach followed in this analysis begins with expressing the nonlocal term $Y_{\quark \quarkbar,\lambda} (\qsq)$ as a subtracted hadronic dispersion relation~\cite{Khodjamirian:2012rm,Bordone:2024hui}
\begin{align}
    Y_{\quark \quarkbar,\lambda} (\qsq) &= Y_{\quark \quarkbar\,\lambda} (q^2_0) + \frac{\left( \qsq - q^2_0 \right)}{\pi} \int^\infty_{4m_\mu^2} \frac{\rho_{\quark \quarkbar, \lambda} (s)}{(s - q^2_0) (s - \qsq - i \epsilon)} \deriv s, \label{eqn:dispersionRelation}
\end{align}
where $q^2_0$ is the subtraction point discussed below and the spectral density function $\rho_{\quark \quarkbar, \lambda}$ contains information on the hadronic intermediate states that contribute to the \BdToKstmm decay. The nonlocal model adopted in this analysis does not account for rescattering contributions such as those of $\Bd\to D^{*}D_{s}\to \Kstarz\mu^+\mu^-$ or $\Bd\to DD^{*}_{s}\to \Kstarz\mu^+\mu^-$~\cite{Khodjamirian:2010vf,Ciuchini:2022wbq} transitions. Recent estimates~\cite{Isidori:2024lng} show that such contributions for the similar \decay{\Bz}{\Kz\mup\mun} decay can manifest as \qsq independent shifts to \C9 far from the charmonium states at the level of a few percent of the value of $\mathcal{C}_{9}$.\footnote{This pre-print appeared after the submission of our paper.}

The expression for the nonlocal contributions given by Eq.~\ref{eqn:dispersionRelation} exploits the fact that $Y_{\quark \quarkbar,\lambda} (\qsq)$ is perturbatively calculable via an operator product expansion in the region $\qsq \lesssim 0$~\cite{Khodjamirian:2010vf}. If one performs such a calculation, thereby fixing the subtraction term $Y_{\quark \quarkbar,\lambda} (q^2_0)$, the second term in Eq.~\ref{eqn:dispersionRelation} then provides a means of extrapolating the result to the region $\qsq>4m_{\mu}^{2}$ by integrating the spectral density function $\rho_{\quark \quarkbar, \lambda} (s)$. The analytic structure of $Y_{\quark \quarkbar, \lambda} (\qsq)$, and therefore $\rho_{\quark \quarkbar, \lambda} (s)$, is determined by the set of possible on-shell intermediate states~\cite{Bobeth:2017vxj}; therefore, as described in the next subsection, the spectral density function can be decomposed into a sum over contributions from known intermediate states in the \BdToKstmm decay.

\subsubsection{Parameterisation of specific nonlocal contributions}\label{sec:nonlocalParam}

In this analysis, a parameterisation of the spectral density function $\rho_{\quark \quarkbar, \lambda} (s)$ in Eq.~\ref{eqn:dispersionRelation} is adopted based on that of Refs.~\cite{Cornella:2020aoq,Bordone:2024hui}. In particular, $\rho_{\quark \quarkbar, \lambda} (s)$ is decomposed into a sum of parametric contributions from all relevant one-particle (1P) and two-particle (2P) intermediate states,
\begin{align}
    \rho_{q\bar{q},\lambda}(\qsq) &= \rho_{q\bar{q},\lambda}^{1\text{P}}(\qsq) + \rho_{q\bar{q},\lambda}^{2\text{P}}(\qsq),
\end{align}
and correspondingly, the nonlocal terms are modified as
\begin{equation}
    Y_{\quark \quarkbar, \lambda}(\qsq) = Y_{\quark \quarkbar, \lambda} (q^2_0) + \Delta Y^{1\text{P}}_{\quark \quarkbar, \lambda} (\qsq) + \Delta Y^{2\text{P}}_{\quark \quarkbar, \lambda} (\qsq) + Y_{\tau\tau,\lambda} \, .
\end{equation}
The vector resonances $j=\{ \rho(770), \omega(782), \phitentwenty, \jpsi, \psitwos, \psi(3770), \psi(4040), \psi(4160) \}$ are considered for the 1P states, while for the 2P states the various open-charm meson pairs $k=\{ \D\Dbar, \Dstar\Dbar, \Dstar\Dstarb\}$ are considered. The spectral density function therefore formally becomes,
\begin{equation}
\label{eqn:spectral_dens_1p}
\rho_{q\bar{q},\lambda}^{1\text{P}}(\qsq) \propto \sum_{j}\mathcal{M}_{j}^{\lambda}(B\to\Kstarz V_j)\delta(\qsq-m_{j}^{2}), 
\end{equation}
\begin{equation}
\label{eqn:spectral_dens_2p}
\rho_{q\bar{q},\lambda}^{2\text{P}}(\qsq) \propto \sum_{k}\int\frac{\deriv p_{k}^{2}}{16\pi^2}
   \delta(\qsq-p_{k}^{2}) \int\frac{\deriv^{3}\vec{p}_{k_1}}{E_{k_1}}\frac{\deriv^{3}\vec{p}_{k_2}}{E_{k_2}} \mathcal{M}_{k}^{\lambda}(\decay{\B}{\Kstarz M_{k_1}M_{k_2}})\delta^{4}(p_{k}-p_{k_1}-p_{k_2}),
\end{equation}
where the $j$ index represents the different 1P states and $k$ represents the 2P states.

In Eq.~\ref{eqn:spectral_dens_1p}, the 1P states are treated as stable particles. In order to account for the one-particle resonance widths, the dispersive integral of $\rho_{q\bar{q},\lambda}^{1\text{P}}(\qsq)$ is modelled using the expression
\begin{align}
\label{eqn:dy1p}
    \Delta Y^{1\text{P}}_{\quark \quarkbar, \lambda} (\qsq) &= \sum_{j} H^{\lambda}_{j} \frac{\left( \qsq - q^2_0 \right)}{m^2_j - q^2_0}BW_j (\qsq) \equiv\sum_{j} |H^{\lambda}_{j}| e^{i \delta^\lambda_j} \frac{\left( \qsq - q^2_0 \right)}{m^2_j - q^2_0} BW_j (\qsq),
\end{align}
with each
\begin{align}
    BW_j (\qsq) &= \frac{m_j \Gamma_j}{\left( m^2_j - \qsq \right) - i\,m_j \Gamma_j},
\end{align}
describing a relativistic Breit--Wigner distribution. The pole masses $m_j$ and natural widths $\Gamma_j$ are set to their world-average values~\cite{PDG2022}. The widths are fixed in the fit for all resonances, while the pole masses are fixed for all except for the \jpsi and \psitwos resonances. The parameters $|H_{j}^{\lambda}|$ appearing in Eq.~\ref{eqn:dy1p} are normalised according to the branching fraction of the processes $B^0\to V_j(\to\mumu) K^{*0}$, where $V_j$ denotes a \mbox{$J^{PC}=1^{--}$} one-particle state, such that
\begin{equation}
    |H_{j}^{\lambda}|=|A_{j}^{\lambda}|\sqrt{\frac{\hbar(m^2_j - q^2_0)^{2}}{\displaystyle\tau_B\int\left |\mathcal{N}^\lambda (\qsq - q^2_0) BW_j (\qsq) F_{\rm vec}^{\lambda}(\qsq) \right |^{2}\deriv\qsq}},
\end{equation}
where $F_{\rm vec}^{\lambda}\in (V/(m_{B}+m_{\Kstarz}),A_1/(m_{B}-m_{\Kstarz}),A_{12})$, and \mbox{$\mathcal{N}_{\lambda}\in(N_{0},N_{\perp},N_{\parallel})$} as defined in Sec~\ref{sec:TransversityAmplitudes}. The $|A^\lambda_j|$ parameters are thus defined such that
\begin{equation}
	\label{eqn:1PMagDef}
	|A^\lambda_j|^2 = f^\lambda_j \BF{(\decay{\Bd}{V \Kstarz})}\BF{(\decay{V}{\mumu})},
\end{equation}
where $f^j_\lambda$ represents the corresponding polarisation fraction, $|A^\lambda_j|^2/\left(\sum_{\lambda'} |A^{\lambda'}_j|^2\right)$.
The $|A^\lambda_j|$ and $\delta^\lambda_j$ parameters, to be determined from data, are the relative magnitudes and phases of each resonance. The phase convention used here defines the longitudinal phases, $\delta^0_j$, relative to \C{9}, while the phases for the other polarisation components $\delta^\parallel_j$ and $\delta^\perp_j$ are defined relative to the longitudinal component. 

This analysis constitutes the first measurement of the phase differences between the local and nonlocal amplitudes in the \BdToKstmm decay in the range \mbox{$0.1<\qsq<18~\gevgevcccc$}. 
Existing measurements~\cite{LHCb-PAPER-2013-023,BaBar:2007rbr,Belle:2014nuw} of the relative phase differences between the helicity components of $\BdToJpsiKst$ and $\BdToPsitwosKst$ provide a cross-check of the parameterisation. Previous measurements of the polarisation amplitudes for the decays \decay{\Bd}{\rhoz\Kstarz} and \decay{\Bd}{\omegaz\Kstarz} from Ref.~\cite{LHCb-PAPER-2018-042}, and \decay{\Bd}{\phiz\Kstarz}~\cite{LHCb-PAPER-2014-005} are used in combination with the measured branching fractions~\cite{PDG2022} to fix the magnitudes and relative phases for these contributions such that only the overall phase relative to \C9 is measured for each. A different phase convention is used in this analysis, which amounts to shifting the previously measured phases by $+\pi$.

The relativistic Breit--Wigner approximation is a good description of well-separated narrow states. For the broad overlapping resonances above the open-charm region and below the $\phi(1020)$ meson, the modelling of the one-particle amplitudes constitutes an approximation that has been shown to be valid given the relatively small amount of signal in the open-charm region of this rare decay~\cite{LHCb-PAPER-2016-045}.

Following the recipe of Ref.~\cite{Cornella:2020aoq}, the two-particle amplitudes $\mathcal{M}_{k}^{\lambda}(\decay{\B}{\Kstarz M_{k_1}M_{k_2}})$, appearing in Eq.~\ref{eqn:spectral_dens_2p}, are described using the two-body phase-space function for a state with centre-of-mass energy $\sqrt{s}=\sqrt{q^2}$ decaying into the state $M_{k_1}M_{k_2}$, characterised by masses $m_{k_1}$ and $m_{k_2}$ with relative orbital angular momentum $L$ set to the lowest partial wave allowed by angular momentum conservation. For the set of two-particle states $k = \{\D\Dbar,\Dstar\Dbar,\Dstar\Dstarb\}$, the spectral density therefore has the form
\begin{equation}
\label{eqn:spectral_dens_phasespace}
    \rho_{q\bar{q}}^{2P}(\qsq) = \sum_{k}\left[\frac{\lambda(\qsq,m^{2}_{k_1},m^{2}_{k_2})}{\qsq}\right]^{\frac{2L+1}{2}},
\end{equation}
with $L=0$ for $\Dstar\Db$, and $L=1$ for $\D\Dbar$ and $\Dstar\Dstarb$, resulting in two-particle terms given by

\begin{align}
    \label{eq:Y2P}
    \Delta Y^{2\text{P}}_{q\bar{q},\lambda} (\qsq) = & A^{\lambda}_{\Dstar\Db} h_S (m_{\Dstar\Db}, \qsq) + \sum_{n=\D\Db,\Dstar\Dstarb} A_{n}^{\lambda} h_P (m_{n}, \qsq) ,\nonumber
    \\ \equiv& |A^\lambda_{\Dstar\Db} | e^{i \delta^\lambda_{\Dstar\Db}} h_S (m_{\Dstar\Db}, \qsq) + \sum_{n=\D\Db,\Dstar\Dstarb} |A^\lambda_{n}| e^{i \delta^\lambda_{n}} h_P (m_{n}, \qsq),
\end{align}
\noindent where the functions $h_S$ and $h_P$ are defined in Ref.~\cite{Cornella:2020aoq} and describe the amplitude as a function of \qsq. The quantity $m_{k_1k_2} = (m_{k_1} + m_{k_2})/2$ is an  effective mass for the two-body state state, while $|A_{k_1k_2}^{\lambda}|$ and $\delta_{k_1k_2}^{\lambda}$ are its magnitude and phase. A Gaussian constraint is placed on the open-charm components
relating the size of the real and imaginary parts for each polarisation of the three open-charm contributions. A systematic uncertainty is assigned for potential biases in the parameters due to this constraint, described in Sec.~\ref{sec:opencharm_constraint_systematic}.

Lepton flavour universality violating $b\to s\tau^+\tau^-$ transitions, with subsequent \mbox{$\tau^+\tau^-\to\gamma^*\to \mu^+\mu^-$} rescattering, introduces $\mathcal{C}_{9\tauon}$ contributions to \BdToKstmm decays via a nonlocal two-particle amplitude. Therefore, following Ref.~\cite{Cornella:2020aoq},
\begin{equation}
    Y_{\tauon \bar{\tauon}} (\qsq) = -\frac{\alpha_{\rm{em}}}{2 \pi} \mathcal{C}_{9\tauon} \left[ h_S (m_\tauon, \qsq) - \frac{1}{3} h_P (m_\tauon, \qsq) \right].
\end{equation}

Finally, for the S-wave, the nonlocal amplitudes give rise to an effective \C9 coefficient as in Eq.~\ref{eq:C9eff}. The only considered S-wave nonlocal contributions are those arising from $\Bd\to \jpsi K^{*0}_{0}$ and $\Bd\to \psitwos K^{*0}_{0}$ amplitudes, contributing to the $\mathcal{C}_{9}^{\rm (eff),00}$ coefficient. The S-wave amplitude of other nonlocal contributions is expected to be subdominant compared to their already relatively suppressed P-wave counterpart and is ignored.

\subsubsection{Subtraction constant}
\label{sec:SubtractionConstant}
The dispersion relation shown in Eq.~\ref{eqn:dispersionRelation} requires knowledge of the subtraction constant $Y_{q\bar{q}}(q^{2}_{0})$, which is in principle different for charmed and light-quark hadronic states. In this analysis, a subtraction point of $q^2_{0} = - 4.6\gevgevcccc$ is chosen for the subtraction constant $Y_{c\bar{c}}(q^{2}_{0})$ whose value 
is taken from the two-loop calculation in an operator product expansion of the dominant nonlocal contributions 
presented in Ref.~\cite{Asatrian:2019kbk}. 

As the light-quark contributions are CKM suppressed, the same subtraction constant is used for both the charm- and light-quark hadronic dispersion relation by default. A systematic uncertainty for this approach is assessed by studying the behaviour of the fit under an unsubtracted dispersion relation for light-quarks and is found to make a negligible difference to the fit results. 

\subsubsection{Further empirical components}
\label{sec:delta_c7}
Global fits to $\Bd\to \Kstarz\gamma$, $\Bd\to\Kstarz e^+e^-$ and $\Bs\to\phi\gamma$ measurements have placed stringent constraints on NP contributions to the Wilson Coefficients \C{7} and \Cp{7} \cite{Paul:2016urs,LHCb-PAPER-2020-020}. As such, in this analysis, the \C{7} and \Cp7 values are fixed to their SM predictions~\cite{EOSAuthors:2021xpv}. Instead, a helicity-dependent shift to the \C{7} Wilson Coefficient is introduced, encoded as $\mathcal{C}_{7}^{ {\rm (eff)},\lambda} = \mathcal{C}_{7}+\Delta\mathcal{C}_{7}^{\lambda}$, where $\Delta\mathcal{C}_{7}^{\lambda}$ are three complex parameters to be determined from data. Such a parameterisation allows for the presence of an additional helicity-dependent complex phase that is constant across \qsq~\cite{Blake:2017fyh}.

In the amplitude fits, the parameters $\Delta\mathcal{C}_{7}^{\parallel,0}$ are degenerate with the tensor form-factor coefficients $\alpha_{0}^{T_{2}}$, $\alpha_{0}^{T_{23}}$. Therefore, the choice is made to fix the parameters $\alpha_{0}^{T_{2}}$ and $\alpha_{0}^{T_{23}}$ to their values provided in Ref.~\cite{Gubernari:2022hxn}. In order to assess the level of compatibility between the entire set of baseline and postfit form-factor coefficients, a separate fit is performed where the parameters $\Delta \mathcal{C}_{7}^{\parallel,0}$ are instead fixed and the coefficients $\alpha_{0}^{T_{2}}$, $\alpha_{0}^{T_{23}}$ are allowed to vary in the fit.

\section{Experimental model of the signal}
\label{sec:ExperimentalModel}

In order to accurately describe the data, the theoretical \BdToKstmm decay rate must be augmented with a model for the detector response. A description of the \lhcb detector and simulation framework is provided in this section, along with an explanation of the event selection requirements. Using simulation, a model for the total efficiency of the event reconstruction and selection is developed, along with a model for the \qsq resolution of the detector. The final form and implementation of the signal model is described in Sec.~\ref{sec:FullSigPDF}.

\subsection{Detector and simulation}
\label{sec:Detector}
    
The \lhcb detector~\cite{LHCb-DP-2008-001,LHCb-DP-2014-002} is a single-arm forward spectrometer covering the \mbox{pseudorapidity} range $2<\eta <5$, designed for the study of particles containing \bquark or \cquark
quarks. The detector includes a high-precision tracking system consisting of a silicon-strip vertex detector surrounding the $pp$ interaction region, a large-area silicon-strip detector located upstream of a dipole magnet with a bending power of about $4{\mathrm{\,Tm}}$, and three stations of silicon-strip detectors and straw drift tubes placed downstream of the magnet. The tracking system provides a measurement of the momentum, \ptot, of charged particles with a relative uncertainty that varies from 0.5\% at low momentum to 1.0\% at 200\gevc. The minimum distance of a track to a primary $pp$ collision vertex (PV), the impact parameter (IP), is measured with a resolution of $(15+29/\pt)\mum$, where \pt is the component of the momentum transverse to the beam, in\,\gevc. Different types of charged hadrons are distinguished using information from two ring-imaging Cherenkov detectors. Photons, electrons and hadrons are identified by a calorimeter system consisting of scintillating-pad and preshower detectors, an electromagnetic and a hadronic calorimeter. Muons are identified by a system composed of alternating layers of iron and multiwire proportional chambers. The information from each of the particle identification (PID) detectors is used as input to several multivariate classifiers, each trained to identify a certain species of particle.

The reconstruction and selection of events is performed by a trigger~\cite{LHCb-DP-2012-004,LHCb-DP-2019-001}, which consists of a hardware stage based on information from the calorimeter and muon systems, followed by a software stage, which applies a full event reconstruction. In the software stage, trigger signals are associated with reconstructed particles and can be queried offline.

Simulation is required to develop and model the effects of the selection requirements on the signal angular distribution and to assess the impact of certain sources of background and potential systematic uncertainty. In the simulation, $pp$ collisions are generated using \pythia~\cite{Sjostrand:2007gs} with a specific \lhcb configuration~\cite{LHCb-PROC-2010-056}. Decays of unstable particles are described by \evtgen~\cite{Lange:2001uf}, in which final-state radiation is generated using \photos~\cite{davidson2015photos}. The interaction of the generated particles with the detector and its response are implemented using the \geant toolkit~\cite{Allison:2006ve, *Agostinelli:2002hh} as described in Ref.~\cite{LHCb-PROC-2011-006}. In a subset of the simulated datasets, the underlying $pp$ interaction is reused multiple times, with an independently generated signal decay for each~\cite{LHCb-DP-2018-004}. In order to ensure agreement between the simulation and real \lhcb data, independent samples are used to calibrate the simulation and correct for potential discrepancies.

\subsection{Signal candidate selection}
\label{sec:Selection}

The $\decay{\Bd}{\Kstarz\mumu}$ signal candidates are first required to pass the hardware trigger, which selects events containing at least one muon with high transverse momentum $\pt$. The minimum $\pt$ threshold varies between $1.36\gevc$ and $1.8\gevc$ for single muons, depending on the year of data taking. For pairs of muons, a threshold is placed on the product of their \pt, which varies between $1.68\gev^2/c^2$ and $3.24\gev^2/c^2$, depending on the year of data taking. In the subsequent software trigger, at least one of the final-state particles is required to have $\pt > 1.7\gevc$, unless the particle is identified as a muon in which case $\pt>1.0\gevc$ is required. The final-state particles that satisfy these transverse momentum criteria are also required to have an IP larger than $100\mum$ with respect to all candidate PVs in the event to reject prompt particles produced directly in \proton\proton collisions. Finally, a dedicated trigger line is employed to select multibody \B meson candidates based on the topology of the decay products. This trigger requires that the tracks of two or more of the final-state particles form a vertex that is significantly displaced from any PV. At all stages of the trigger, it is required that the particles composing the signal candidate are directly responsible for the trigger decision, as opposed to other particles in the event.

In the offline selection, signal candidates are formed from a pair of oppositely charged tracks that are identified as muons, combined with a \Kstarz meson candidate. The \Kstarz candidate is formed from two oppositely charged tracks that are identified as a kaon and a pion. The four tracks of the final-state particles are required to have a significant IP with respect to all PVs in the event and form a good-quality common vertex. The impact parameter of the \Bd candidate with respect to one of the PVs is required to be small and the decay vertex of the \Bd candidate is required to be significantly displaced from the same PV. The angle between the reconstructed \Bd momentum and the vector connecting the PV to the reconstructed \Bd decay vertex is required to be small. Candidates are required to have reconstructed \Bd mass, denoted as $m(\Kp\pim\mumu)$, in the range \mbox{$4800 < m(\Kp\pim\mumu) < 6500 \mevcc$}. Finally, the reconstructed mass of the \Kp\pim system, denoted as $m(\Kp\pim)$, is required to be in the range $796 < m(\Kp\pim) < 996 \mevcc$.

A significant background contribution arises from candidates formed by the random combination of kaons, pions, and muons originating from different parent particles or from the \proton\proton collision itself (referred to as combinatorial background). To reduce the level of combinatorial background, a Boosted Decision Tree~(BDT)~\cite{Breiman,AdaBoost} classifier is trained to discriminate between signal and background based on a set of input variables corresponding to reconstructed particle information. The BDT algorithm is trained entirely using data, with background subtracted \decay{\Bd}{\jpsi(\to \mumu)\Kstarz} events used as a signal proxy, and \BdToKstmm candidates with a mass above 5500\mevcc used as a background proxy. For the background proxy, events with dimuon masses close to the \phitentwenty, \jpsi, and \psitwos resonance masses are excluded to avoid biasing the training with many events that contain real resonant dimuons. The background subtraction is achieved using the \sPlot technique~\cite{Pivk:2004ty} where the weights are obtained from a fit to the $m(\Kp\pim\mumu)$ distribution. A total of thirteen training variables are used, and the ones found to provide the most discriminating power include various kinematic properties of the \B meson, along with PID and isolation variables of the daughter particles. The requirement on the BDT output is chosen to optimise the signal significance $S/\sqrt{S+B}$, where $S$ and $B$ and the expected signal and background yields, respectively.
The BDT classifier achieves a signal efficiency of approximately 87\% and 90\% in Run~1 and Run~2, respectively, whilst maintaining a consistent background rejection rate of greater than 98\%.

\subsection{Modelling of the detector efficiency and response}
\label{sec:AcceptanceResolution}

The reconstruction and selection of signal candidates sculpt the phase space of the signal decay. This effect can be accounted for in the fit to the data via an acceptance function $\epsilon(\ctl, \ctk, \phih, \qsq)$, which includes the effects of the detector geometry, triggering, reconstruction, and selection of events.

\begin{figure}
    \centering
    \begin{subfigure}{0.49\linewidth}
        \centering
        \includegraphics[width=0.9\linewidth]{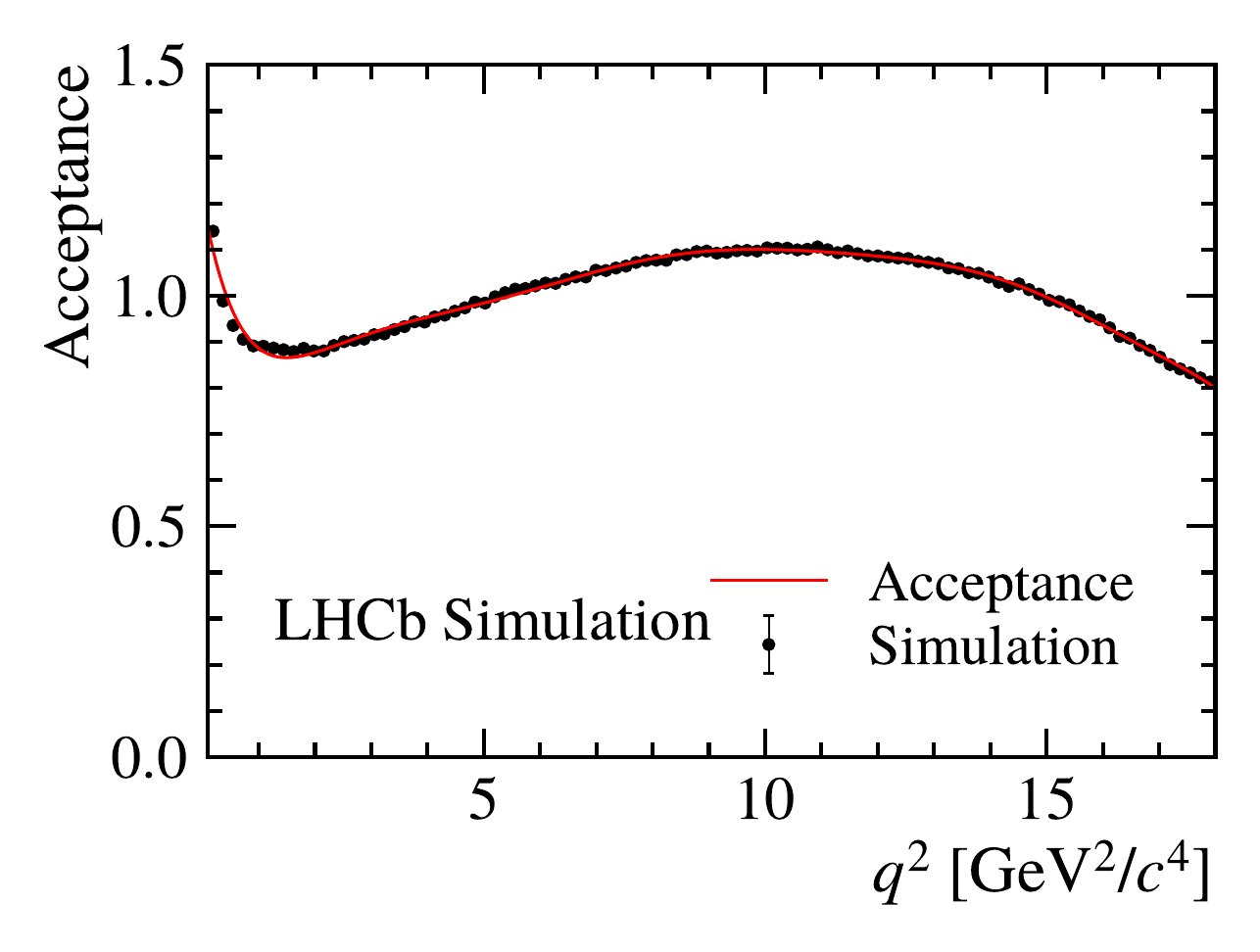}
    \end{subfigure}
    \begin{subfigure}{0.49\linewidth}
        \centering
        \includegraphics[width=0.9\linewidth]{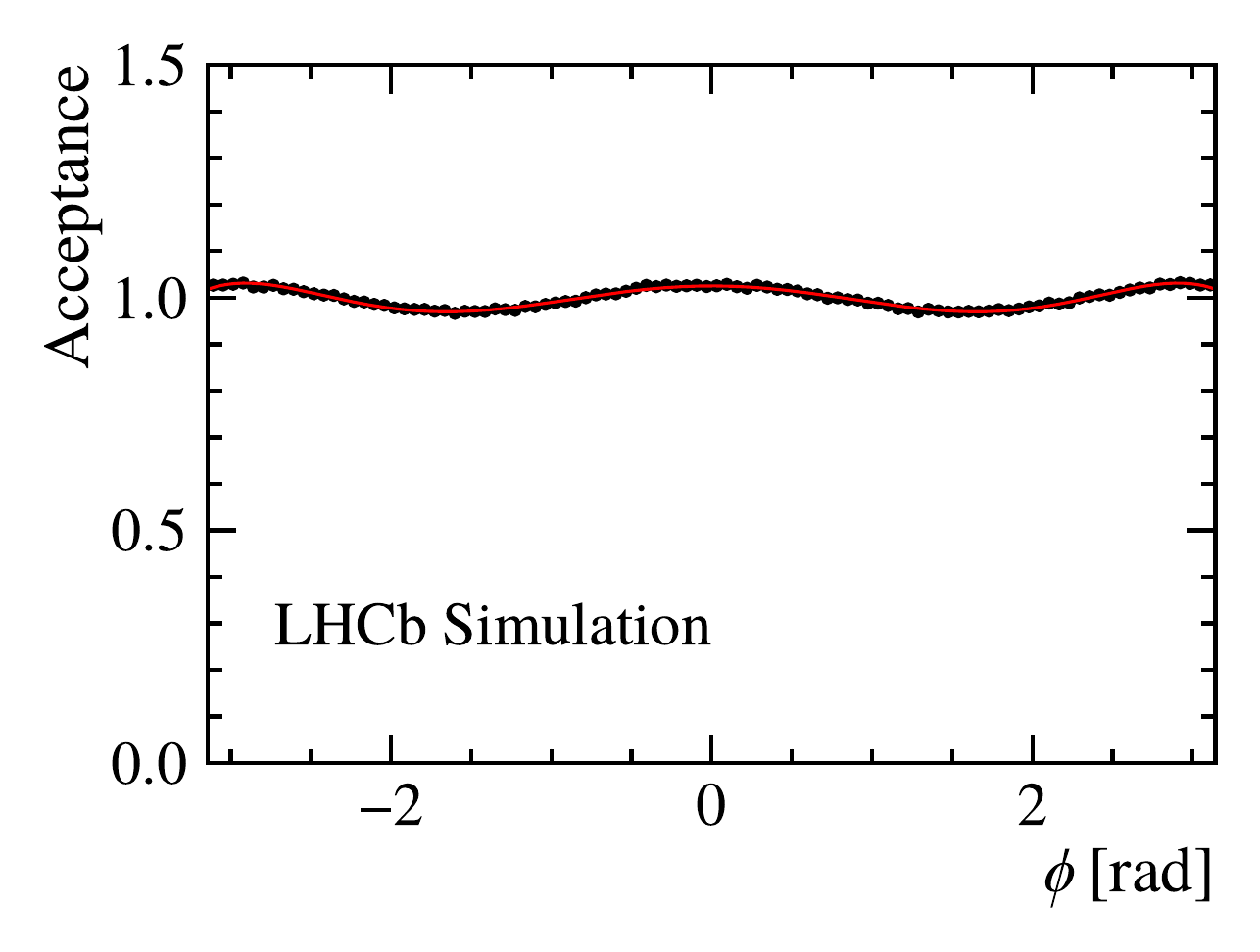}
    \end{subfigure}
    
    \begin{subfigure}{0.49\linewidth}
        \centering
        \includegraphics[width=0.9\linewidth]{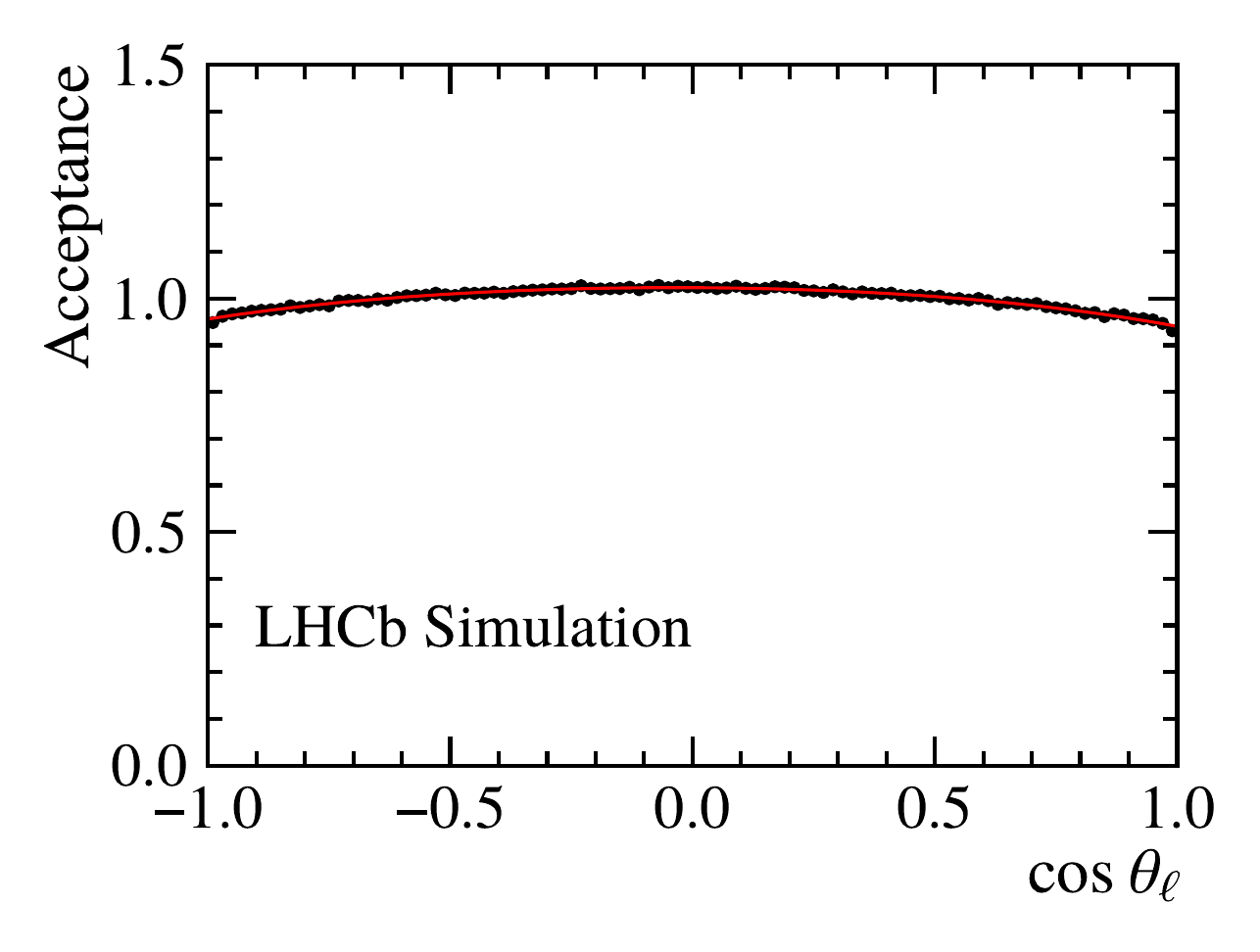}
    \end{subfigure}
    \begin{subfigure}{0.49\linewidth}
        \centering
        \includegraphics[width=0.9\linewidth]{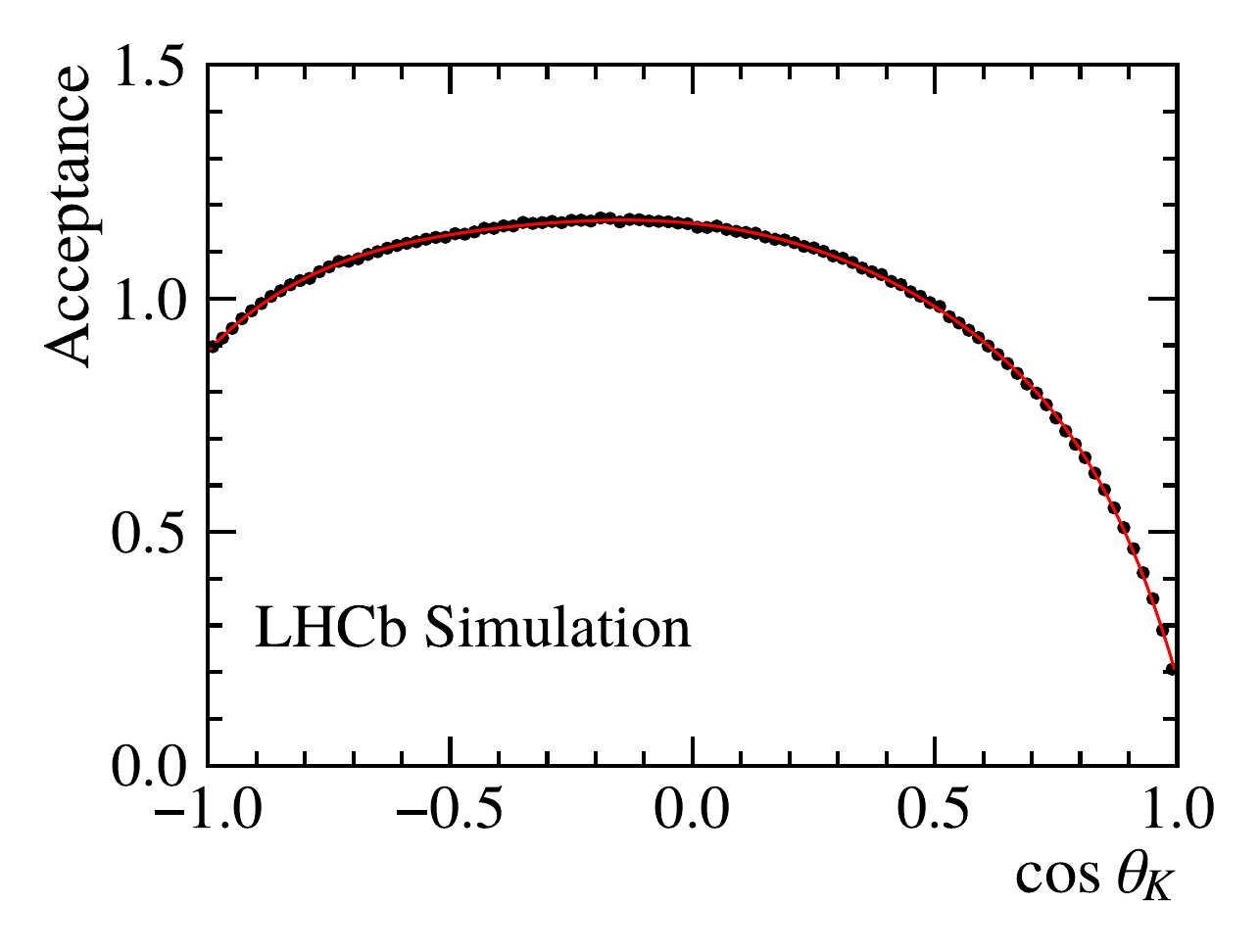}
    \end{subfigure}
    
    \caption{One-dimensional projections of the acceptance function determined from simulation.}
    \label{fig:acceptance}
\end{figure}

The acceptance function is determined in the four-dimensional phase space described by \ctl, \ctk, \phih, and \qsq, using simulation generated with a uniform distribution in each dimension. The simulated events are run through the complete reconstruction and selection chain, and the resulting deviation from uniformity is taken to quantify the acceptance. The acceptance function is modelled using Legendre polynomials, \ie 
\begin{align}
    \epsilon(\ctl, \ctk, \phih, q^2) = \sum_{ijkl} c_{ijkl} P_i(\ctk) P_j(\ctl) P_k(\phih) P_l(q^2), \label{eqn:Acceptance}
\end{align}
where $P_{m}$ refers to the Legendre polynomial of order $m$. The maximum polynomial order for each dimension is chosen empirically to give the set of lowest orders which are sufficient to model the acceptance well, leading to the choice of a ninth order polynomial for the \qsq dimension, seventh order for the \ctk dimension, fourth order for the \ctl dimension, and sixth order for the \phih dimension. The resulting acceptances are shown in Fig.~\ref{fig:acceptance} and a systematic uncertainty is assigned to choice of Legendre polynomial orders as described in Sec.~\ref{sec:AcceptanceSyst}.

In addition to the acceptance, a resolution model, given by $R (q^2 - q^2_\text{true})$, is implemented to account for the smearing of reconstructed dimuon masses relative to their true values. This effect is the combined result of the finite resolution in each subdetector involved in the reconstruction of muons. 
The \qsq resolution model is built up from the sum of a Gaussian function $G(\Delta\qsq; \mu_, \sigma)$, and two Crystal Ball (CB) functions~\cite{Skwarnicki:1986xj} $C_{l,u}(\Delta \qsq; \mu,\sigma,\alpha_{l,u},n_{l,u})$, with power law tails on opposite sides, 
\begin{equation}
\begin{split}
    R (\Delta q^2) =& f_{G} G(\Delta q^2; \mu, \sigma_{G}) \\ &\hspace{5mm}+
    (1 - f_{G}) \frac{1}{N_C}
    \left[ C_u(\Delta q^2; \mu, \sigma_{C},\alpha_{u},n_{u}) + 
    C_l(\Delta q^2;\mu, \sigma_{C},\alpha_{l},n_{l}) \right],
    \end{split} \label{eqn:resolution_function}
\end{equation}
where $\Delta \qsq \equiv q^2 - q^2_\text{true}$, $f_G$ is the Gaussian fraction, and $N_C$ is a normalisation factor for the CB sum component. The optimal parameters of the resolution model are determined using either simulation or data, depending on the \qsq region. The resolution varies depending on the value of \qsq itself in a nonlinear fashion, and ranges from approximately 0.01\gevgevcccc at \qsq = 1\gevgevcccc to 0.04\gevgevcccc at \qsq = 13.6\gevgevcccc.
The final fit to data is performed simultaneously in three \qsq regions, allowing variations in the resolution model and background composition to be accurately modelled. 
These three regions, referred to as low-, mid-, and high-\qsq, are chosen to have the three narrow \quark\quarkbar resonances, the \phitentwenty, \jpsi, and \psitwos in separate regions and are defined according to Table~\ref{tab:WCFitCategories}. 

\begin{table}[t]
\caption{Three $q^2$ regions defining the simultaneous fit categories when determining the Wilson Coefficients.}
\begin{center}
    \begin{tabular}{r  r}
        \hline
        \multicolumn{1}{c}{Category} & \multicolumn{1}{c}{Region} \\
        \hline
        Low-\qsq & $\phantom{0}0.10 \leq \qsq < \phantom{0}3.24 \gevgevcccc$ \\
        Mid-\qsq & $\phantom{0}3.24 \leq \qsq < 11.56 \gevgevcccc$ \\
        High-\qsq & $11.56 \leq \qsq \leq 18.00 \gevgevcccc$\\
        \hline
    \end{tabular}
\end{center}
\label{tab:WCFitCategories}
\end{table}

\begin{figure}
    \centering
    \begin{subfigure}{0.49\linewidth}
        \centering
        \includegraphics[width=0.9\linewidth]{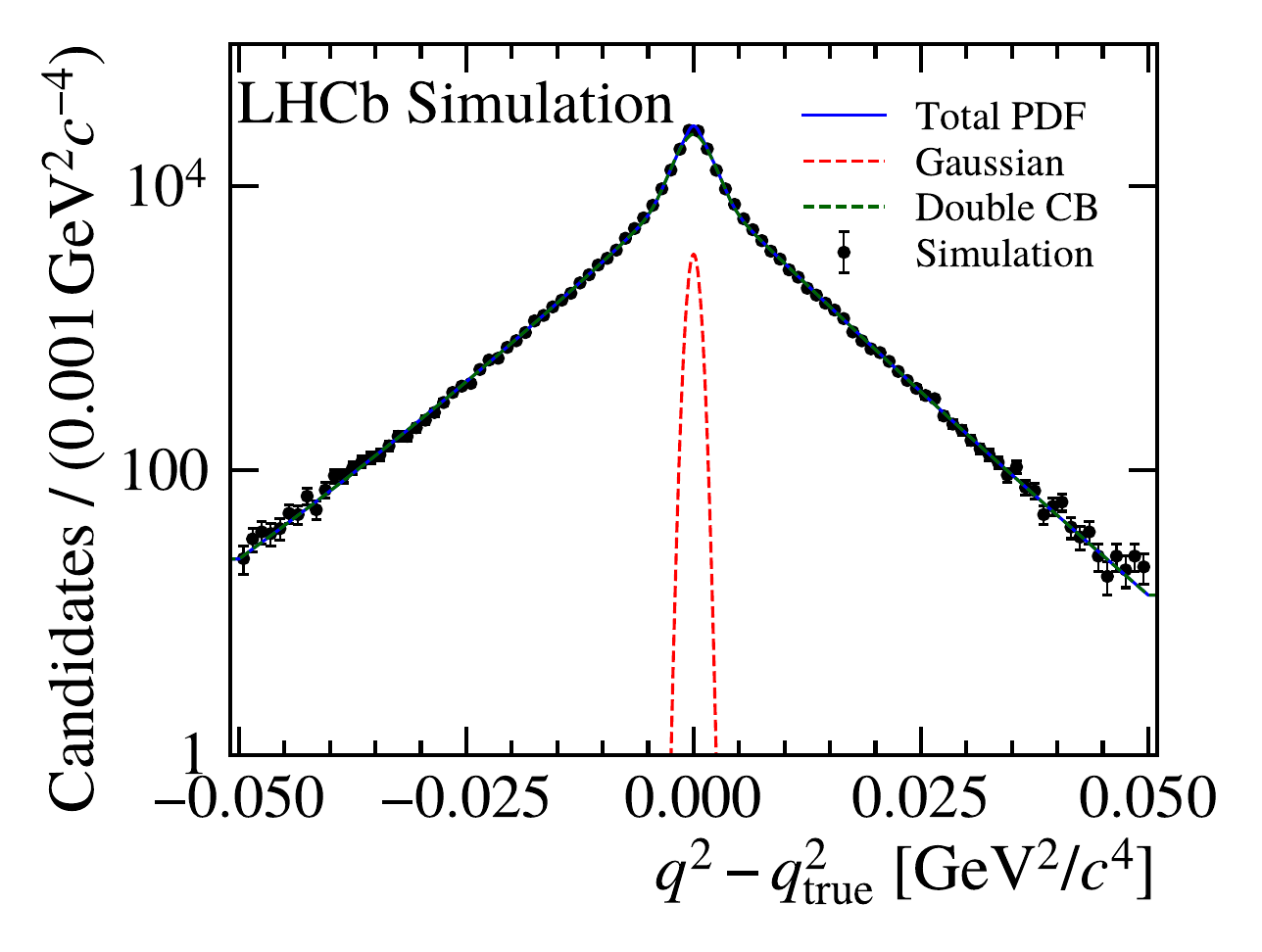}
        \caption{}
        \label{subfig:lowq2Res}
    \end{subfigure}\\
    \begin{subfigure}{0.49\linewidth}
        \centering
        \includegraphics[width=0.9\linewidth]{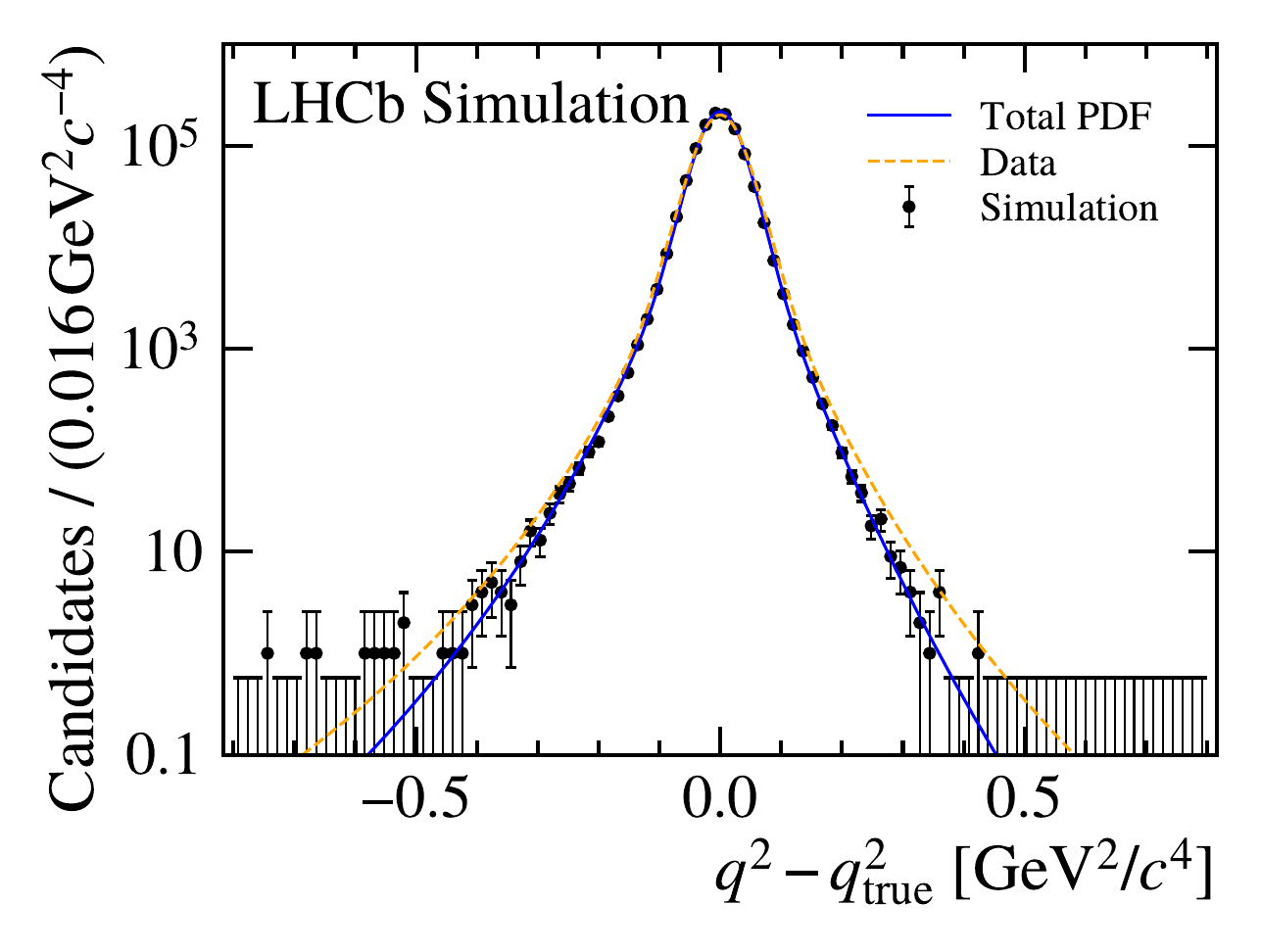}
        \caption{}
        \label{subfig:midq2Res}
    \end{subfigure}
    \begin{subfigure}{0.49\linewidth}
        \centering
        \includegraphics[width=0.9\linewidth]{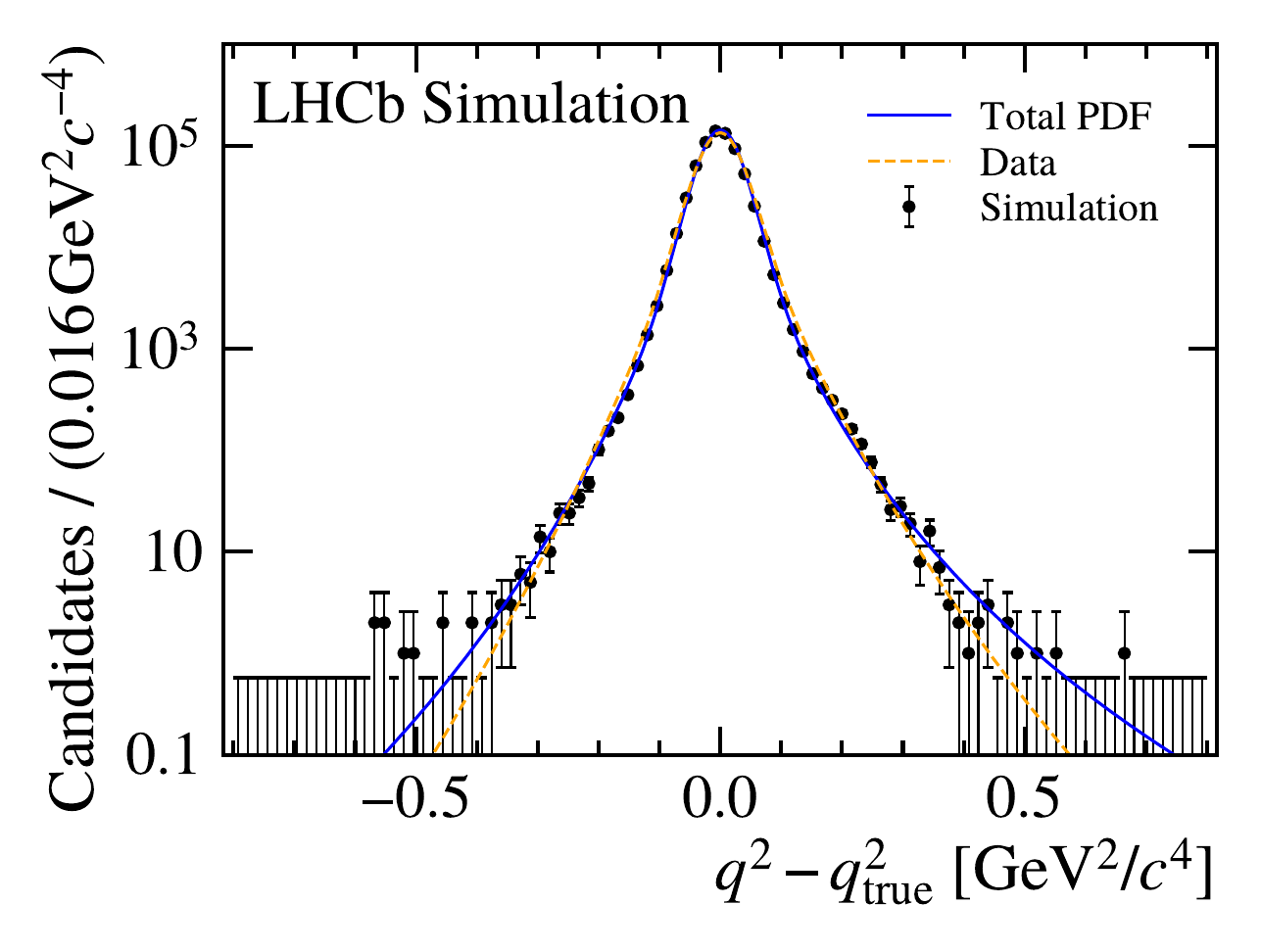}
        \caption{}
        \label{subfig:highq2Res}
    \end{subfigure}
    \caption{Distributions of reconstructed $q^2$ resolution in LHCb simulation, overlaid with the results of fitting the resolution function of Eq.~\ref{eqn:resolution_function}.~\subref{subfig:lowq2Res} shows the fit to the low-$q^2$ region $0.1 < q^2 < 3.24 \,\rm{GeV}^2/c^4$ for simulated $B^{0} \to K^{*0}\mu^+\mu^-$ events (blue), along with the Gaussian core (red) and double CB (green) contributions separately. ~\subref{subfig:midq2Res} shows the fit to the mid-$q^2$ region (blue) $3.24 < q^2 < 11.56 \,\rm{GeV}^2/c^4$ for simulated $B^{0} \to J/\psi K^{*0}$ events and~\subref{subfig:highq2Res} shows the fit to the high-$q^2$ region (blue) $11.56 < \qsq < 18.0 \,\rm{GeV}^2/c^4$ for simulated $B^{0} \to \psi(2S) K^{*0}$ events. 
    In the latter two plots, the dashed orange curves show the final resolution shape after the fit to data, which agrees well with the results from simulation.}
    \label{fig:ResolutionMC}
\end{figure}

In the low-\qsq region,
the resolution parameters are obtained through an unbinned maximum likelihood fit to the distribution of \qsq reconstruction errors in simulated \mbox{\BdToKstmm} decays. The $\alpha_{l,u}$ parameters of the CB tails are symmetrised in the low-\qsq region to improve stability, while the remaining parameters of the resolution model are allowed to vary freely in this fit. The model provides an excellent description of the resolution in simulation, and the results of this fit are shown in Fig.~\ref{subfig:lowq2Res}. The low-\qsq resolution parameters are fixed in the fit to data since the number of signal candidates in this region is insufficient to allow them to vary. 

In the mid-\qsq and high-\qsq 
regions, all of the resolution parameters are allowed to vary freely in the fit to data. Fits to the \jpsi and \psitwos peaks in simulated \BdToJpsiKst and \BdToPsitwosKst decays are shown in Figs.~\ref{subfig:midq2Res}, and~\ref{subfig:highq2Res}, respectively, along with a comparison to the final resolution shape obtained from the fit to data. Excellent agreement between the resolution models obtained from data (orange) and simulation (blue) is observed in all areas except for the far tails of the resonance peaks, giving additional confidence in the accuracy of the simulations and therefore also in the low-\qsq resolution model.

\subsection{Full signal probability density function}
\label{sec:FullSigPDF}

The full signal Probability Density Function (PDF) has the form,
\begin{align}
	\begin{split}
	\mathcal{P}_{\text{Sig},i}(\ctl, \ctk, \phih, \qsq) = \frac{1}{\mathcal{N}} \bigg[ \frac{\deriv^4\left( \Gamma + \bar{\Gamma}\right) (\decay{\Bz}{\Kstarz\mup\mun})}{\deriv q^2_\text{true}\deriv\vec{\Omega}} \otimes R_i (q^2 - q^2_\text{true}) \bigg] \\ 
    \hspace{15mm}\times \epsilon(\ctl, \ctk, \phih, q^{2} ),
	\end{split}
	\label{eqn:fullSignalPDF}
\end{align}
where the $\otimes$ symbol indicates a convolution, and the index $i$ labels the \qsq region. The angular resolution of the detector is not accounted for in the signal model. Based on simulation, the angular resolution is around 40\mrad for $\theta_l$ and $\theta_K$, and around 100\mrad for \phih, with little dependence on \qsq. The angles and \qsq used in the determination of the acceptance in Eq.~\ref{eqn:Acceptance} refer to the true values in the simulation and not the reconstructed ones. However, here in Eq.~\ref{eqn:fullSignalPDF} it is used for the reconstructed ones. As the variation in efficiency is very slow compared to the resolution in all dimensions, this difference only leads to a negligible systematic uncertainty on the results.  The \CP-averaged \BdToKstmm theoretical decay rate contains all parameters of interest, including the Wilson Coefficients and all parameters describing both the local and nonlocal hadronic form factors. The expression is built up by constructing the angular coefficients of Eq.~\ref{eqn:diffdecayrate} from the transversity amplitudes given in Eqs.~\ref{eqn:A0amplitude}--\ref{eqn:A00amplitude} (see Appendix~\ref{app:DecayRateFunctions} for details). The acceptance function is fixed from simulation, whilst the \qsq resolution adds a small number of nuisance parameters to the signal model which are either allowed to vary in the fit to data or are fixed from simulation, as already described in Sec.~\ref{sec:AcceptanceResolution}. 

To improve the \qsq resolution, a constraint is applied when determining the value of \qsq. The constraint involves performing a kinematic refit of the decay chain using a Kalman filter~\cite{Hulsbergen:2005pu} to vary the four-momenta of the final-state particles within their uncertainties such that the reconstructed mass is constrained to the known \Bd mass~\cite{PDG2022}. Unless otherwise stated, the use of \qsq throughout this paper always refers to the constrained value.

\subsection{Background composition}
\label{sec:BkgComposition}

The model requires to describe the contribution from processes other than the signal decay which contaminate the final sample. The combinatorial background, discussed already in Sec.~\ref{sec:Selection}, is the only contribution which remains significant after the full selection and is modelled as described in Sec.~\ref{sec:SidebandFit}. Beyond this, several physical background sources are identified, referred to as peaking backgrounds, which are suppressed using a combination of dedicated vetoes and machine-learning techniques.

The \decay{\Bu}{\Kp\mumu} mode mimics the signal decay when a random \pim is combined with the daughters of the true decay. This background is vetoed by removing all candidates with $m(\Kp\pim\mumu) > 5380\mevcc$ in which the mass of the \Kp\mumu combination is also compatible with the known \Bu mass. The \decay{\Bs}{\Kp\Km\mumu} decay forms a peaking background when one of the kaons is misidentified as a pion. The dominant contribution arises from \mbox{\decay{\Bs}{\phi(1020)\mumu}} decays followed by the transition \mbox{\decay{\phi(1020)}{\Kp\Km}}. Several vetoes are applied to remove this contribution, accounting for both the resonant and nonresonant parts of the $m(\Kp\Km)$ spectrum. Candidates are first reconstructed, assigning the kaon mass to the pion. For the resonant $\phi(1020)$ channel, strict pion PID requirements are applied to those candidates with reconstructed \B and \Kp\Km masses that are compatible with the known \Bs and $\phi(1020)$ masses. For the nonresonant mode, the requirement that the reconstructed \Kp\Km mass is compatible with the $\phi(1020)$ mass is removed, and slightly modified PID cuts are applied. The decay \decay{\Bd}{\pip\pim\mumu} forms a peaking background in a similar way if one of the pions is misidentified as a kaon. In this case, the dominant contribution comes via the \decay{\Bd}{\rhoz(\decay{}{\pip\pim})\mumu} resonant decay. Analogous PID requirements are applied to remove these decays after assigning a pion mass hypothesis to the reconstructed kaon. Backgrounds stemming from the double misidentification of the final-state particles in signal decays, \eg when the \pim (\Kp) of the \Kstarz meson is misidentified as a \Km (\pip) and vice versa, are highly suppressed due to PID requirements on the final state particles.

Several more peaking background sources arise from the \decay{\Lb}{\proton\Km\mumu} and \mbox{\decay{\Lb}{\proton\pim\mumu}} decays, which mimic the signal if 
one or both hadrons are misidentified and are reconstructed as a \Kstarz decay. These backgrounds are removed by reconstructing decays under the alternative mass hypotheses, and requiring that the final-state hadrons satisfy strict PID criteria if the mass is close to the known \Lb mass. 

Double hadron misidentification leads to peaking backgrounds that originate from true resonant signal decays, \decay{\Bd}{\jpsi\Kstarz} and \decay{\Bd}{\psitwos\Kstarz}, with two of the final-state particles swapped, \ie the \pim (\Kp) is misidentified as a \mun (\mup) and vice versa. These decays are vetoed by assigning the muon mass to the pion (kaon), and removing events for which the \pim \mup (\Kp \mun) combination has a mass close to either the known \jpsi or \psitwos mass, and the \pim (\Kp) also fails to satisfy stringent PID criteria. 

An additional peaking background can be formed from \decay{\Bu}{\Kstarp\mumu} decays with either \decay{\Kstarp}{\KS\pip} or \decay{\Kstarp}{\Kp\piz} states and the charged hadron from these decays is combined with a random charged pion or kaon from elsewhere in the event to create the \Kstarz candidate. These events are less trivial to separate from the signal; hence, two BDT classifiers are trained using simulation for the purpose of discriminating between \BdToKstmm decays and \decay{\Bu}{\Kstarp\mumu} decays in each of the two \Kstarp decay modes. A total of fourteen variables are used to train the BDT algorithm including various kinematic and isolation variables, with the highest discriminating power provided by the significance of the impact parameter with respect to the PV of the randomly charged hadron used to create the \Kstarz candidate, \ie the \Km in the BDT classifier trained to reject \decay{\Bu}{\Kstarp(\decay{}{\KS\pip})\mumu} decays, and the \pim in the BDT algorithm trained to reject \decay{\Bu}{\Kstarp(\decay{}{\Kp\piz})\mumu} decays. 

\section{Data analysis}
\label{sec:Strategy}

The primary aim of the analysis is to determine the Wilson Coefficients of the \bsll WET Hamiltonian as well as to obtain a full description of the nonlocal amplitudes. The measurement is performed by fitting the \BdToKstmm angular distribution of Eq.~\ref{eqn:diffdecayrate}, which provides sensitivity to the WET parameters through the \qsq dependent angular observables. The latter are parameterised in terms of a set of theoretical amplitudes, which depend directly on the \bsll Wilson Coefficients and various parameters describing the nonlocal contributions. This approach of explicitly modelling the signal in the \qsq dimension including the charmonium resonance regions is the main new feature with repect to previous angular analyses of the \BdToKstmm decay mode. 

A full description of the signal model is provided in Sec.~\ref{sec:AmplitudeModel}. The signal decay rate is modelled in five dimensions, \ie the three helicity angles \ctl, \ctk, and \phih, along with \qsq, and \mB masses. The signal shape in the mass of the $K\pi$ system is integrated out as mentioned in Sec.~\ref{sec:DecayRate}. This is done in order to simplify the already very complex model.
The model is ultimately used to perform an unbinned maximum-likelihood fit to the data, simultaneously in the \ctl, \ctk, \phih, and \qsq dimensions, within the range $0.1 < \qsq < 18.0 \gevgevcccc$. In order to constrain the background, two separate fits are performed in different ranges of the \Bd mass as described in Sec.~\ref{sec:BMassFit} and~\ref{sec:SidebandFit}.

\subsection{Determination of the signal fraction}
\label{sec:BMassFit}
\begin{table}[t]
    \centering
    \caption{The signal fraction in the full mass range $5220 \leq m(\Kp\pim\mumu) \leq 5840 \,\rm{MeV}/c^2$ determined in five $q^2$ regions chosen to isolate different combinatorial background contributions.}
    \begin{tabular}{r  c  l}
          \hline
          Category & $q^2$ region [\gevgevcccc] & Signal fraction ($f_{\text{Sig},i}^\text{full}$) \\
          \hline
         Low-$q^2$  & $[0.10, 3.24]$ & $0.9196 \pm 0.0088$ \\
         Fully combinatorial mid-$q^2$ & $[3.24, 8.20] \cup [10.6, 11.56]$ & $0.8045 \pm 0.0093$ \\ 
         Resonant mid-$q^2$ & $[8.20, 10.6]$ & $0.9934 \pm 0.0002$ \\ 
         Fully combinatorial high-$q^2$ & $[11.56, 12.40] \cup [14.40, 18.00]$ & $0.8656 \pm 0.0088$  \\ 
         Resonant high-$q^2$ & $[12.40, 14.40]$ & $0.9862 \pm 0.0010$\\
         \hline
    \end{tabular}
    \label{tab:BMassFitCategories}
\end{table}
One-dimensional fits to the $m(\Kp\pim\mumu)$ distribution are performed in the range $5220 \leq m(\Kp\pim\mumu) \leq 5840 \mevcc$, to determine the fraction of signal events $f_{\text{Sig}}^\text{full}$ relative to the background in the full mass range. The signal fraction is determined simultaneously in the five separate \qsq regions given in Table~\ref{tab:BMassFitCategories}. These regions correspond to the same three regions as those in Table~\ref{tab:WCFitCategories}, but with the mid- and high-\qsq regions further subdivided.
This is done in order to capture the fact that the combinatorial background composition differs depending on whether the \qsq value is close to one of the \jpsi or \psitwos resonances, or away of them. In particular, within the resonance regions (labelled resonant mid- and high-\qsq in Table~\ref{tab:BMassFitCategories}), the dominant contribution comes from true resonant dimuon candidates combined with a random \Kp\pim combination, resulting in a strongly peaking \qsq distribution. Outside the resonance regions, fully random combinations of \Kp\pim\mumu are the dominant contribution with no peaking structure, henceforth referred to as fully combinatorial.

\begin{figure}[tbp]
    \centering
   \begin{subfigure}{0.45\linewidth}
        \includegraphics[width=\linewidth]{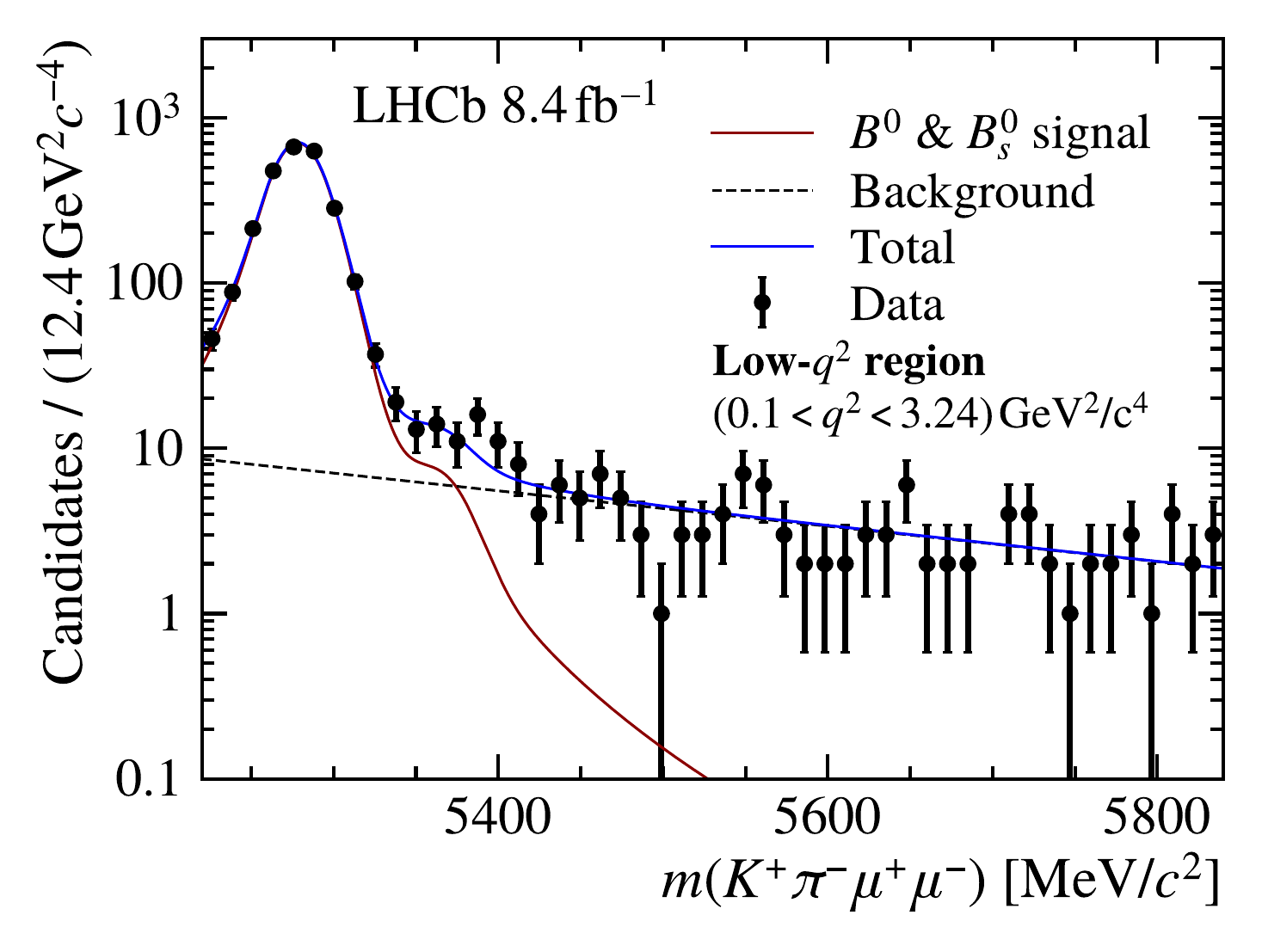}
    \label{fig:B0_mass_fit_low_q2}
    \end{subfigure}\\
    \begin{subfigure}{0.45\linewidth}
        \includegraphics[width=\linewidth]{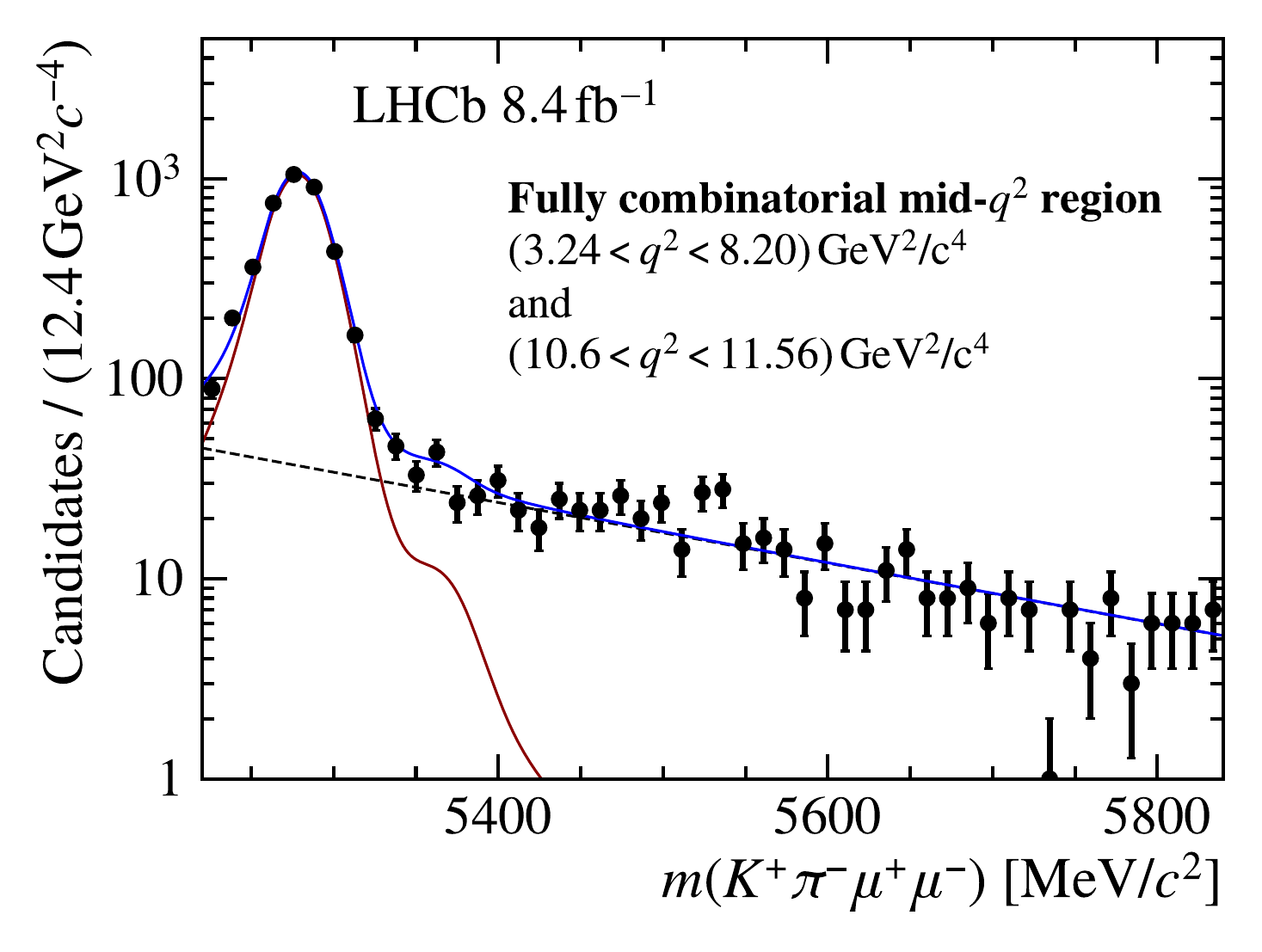}
    \label{fig:B0_mass_fit_mid_q2_comb}
    \end{subfigure}
    \begin{subfigure}{0.45\linewidth}
        \includegraphics[width=\linewidth]{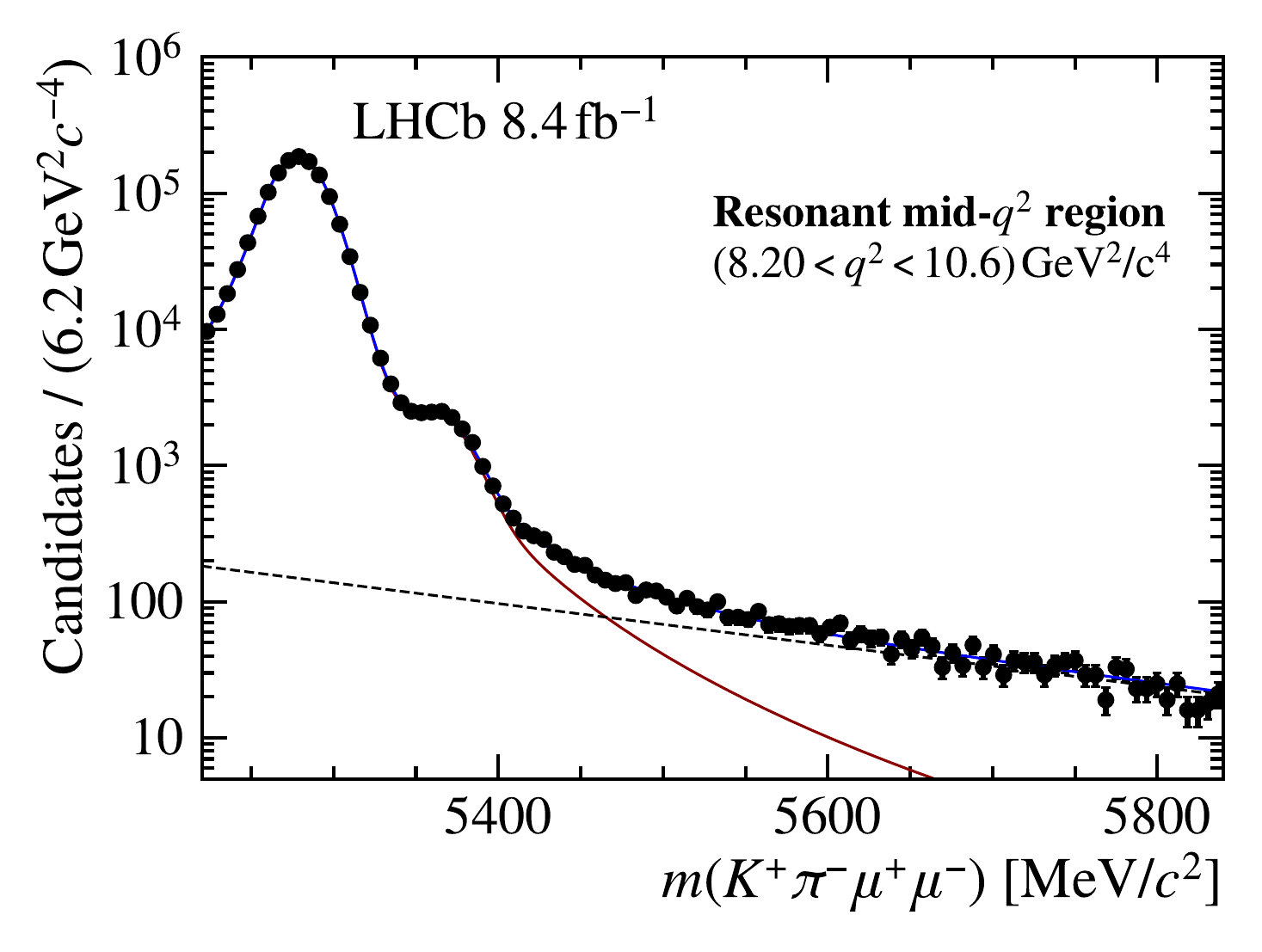}
    \label{fig:B0_mass_fit_mid_q2_reso}
    \end{subfigure}
    \begin{subfigure}{0.45\linewidth}
        \includegraphics[width=\linewidth]{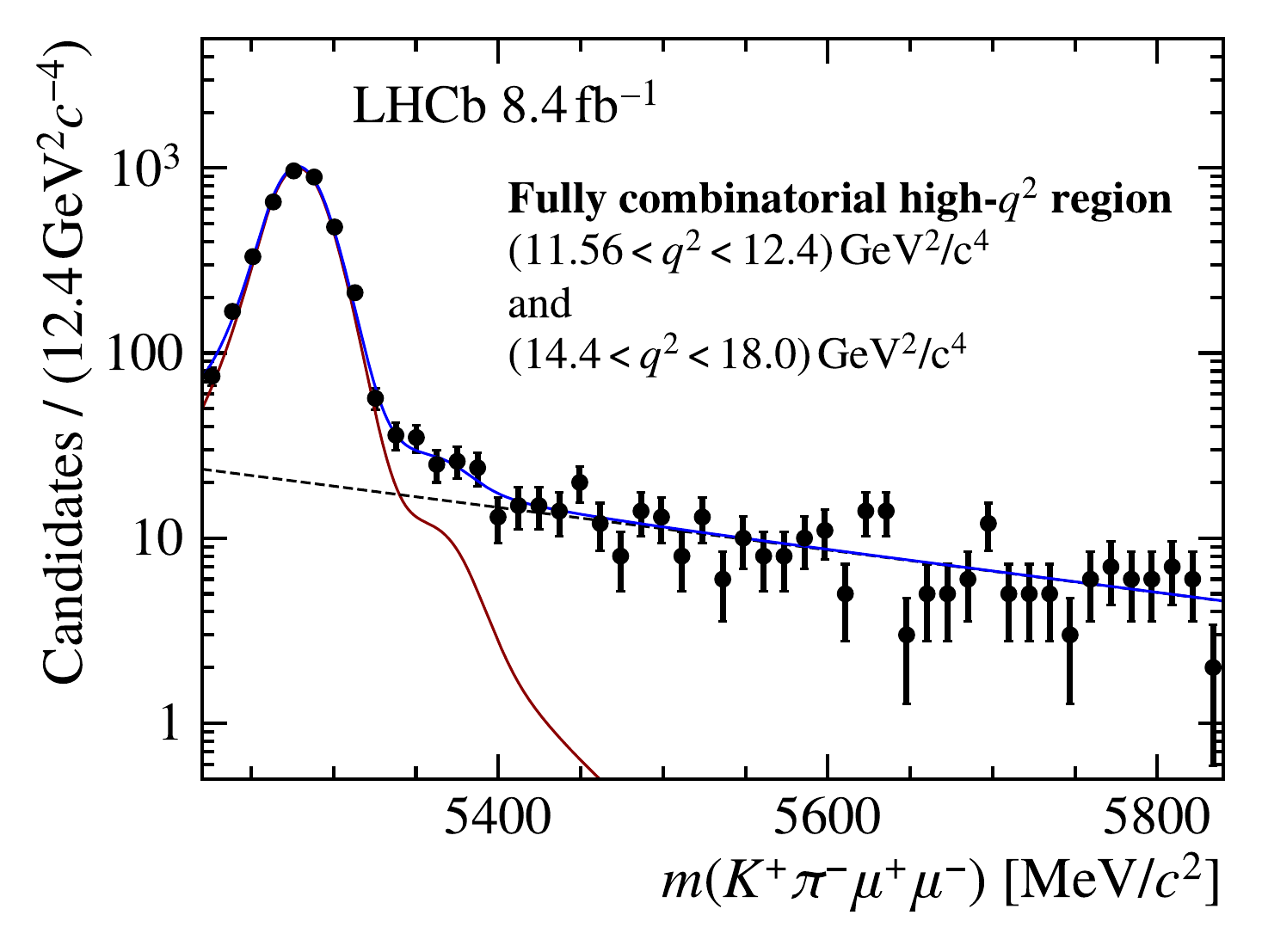}
    \label{fig:B0_mass_fit_high_q2_comb}
    \end{subfigure}
    \begin{subfigure}{0.45\linewidth}
        \includegraphics[width=\linewidth]{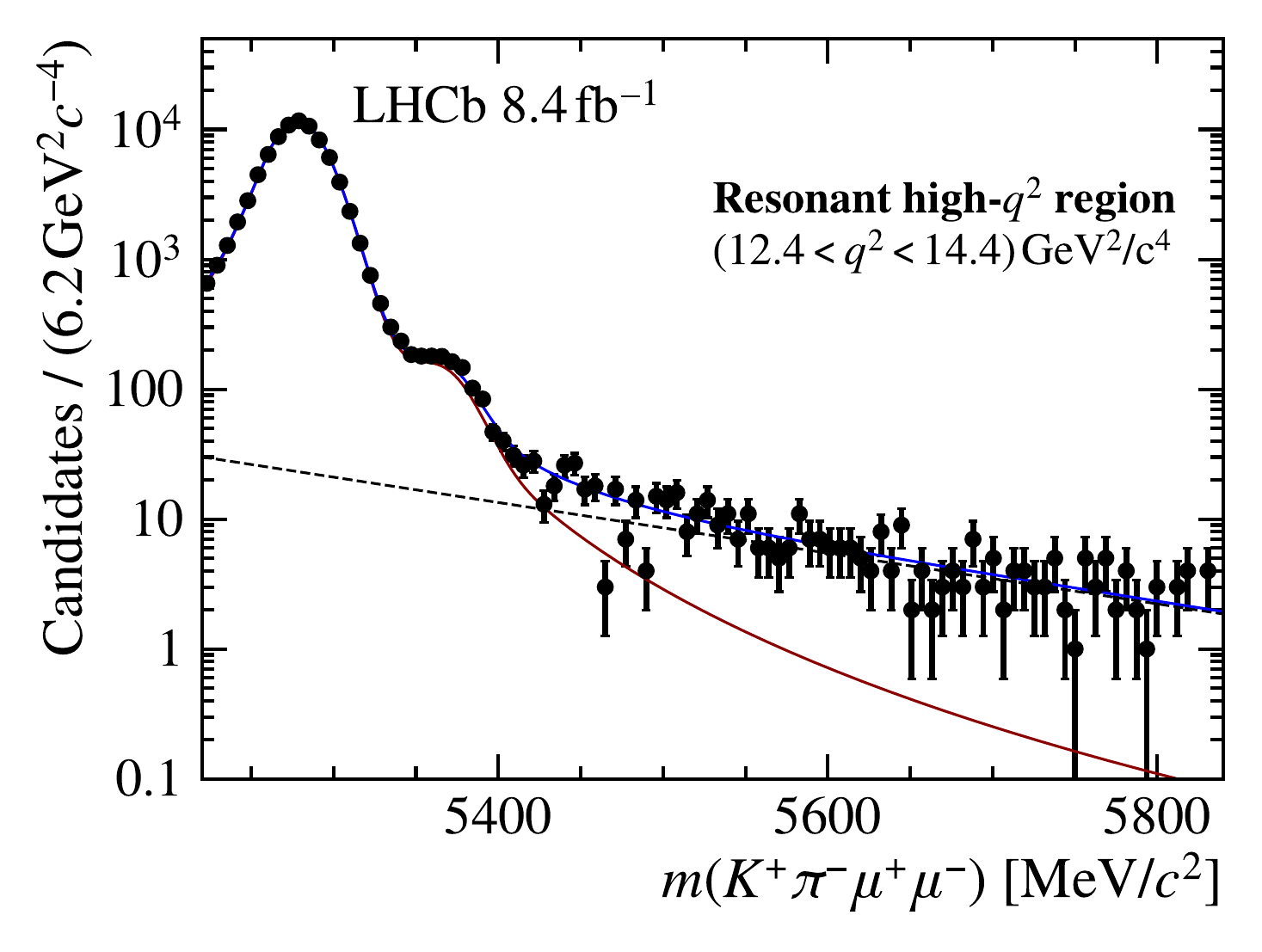}
    \label{fig:B0_mass_fit_high_q2_reso}
    \end{subfigure}
    \caption{The mass distribution $m(\Kp\pim\mumu)$ of candidates in the data in five separate $q^2$ regions. The data is overlaid with the results of a simultaneous fit to determine the signal fractions.
    }
    \label{fig:mBmassFits}
\end{figure}

Similar to previous \lhcb analyses of \BdToKstmm decays~\cite{LHCb-PAPER-2020-002,LHCb-PAPER-2015-051}, the shape of the signal mass is modelled using the sum of two single-sided CB functions,
\begin{align}
    \mathcal{P}_\Bd = \frac{1}{N}
    \left[ f_1 C(m; \mu, \sigma_{1},\alpha,n) + 
    (1-f_1) C(m;\mu, \sigma_{2},-\alpha,n) \right],
\end{align}
where $f_1$ represents the fraction of the first CB component, $N$ is normalisation constant, and $m$ represents the reconstructed \Bd mass $m(\Kp\pim\mumu)$. The best description of the data is obtained with symmetric CB tails on opposite sides of the Gaussian core. Due to large correlations between the CB tail parameters, only the $\alpha$ parameter is allowed to vary in the fit, whilst the $n$ parameter is fixed to the value obtained in Ref.~\cite{LHCb-PAPER-2015-051}. For the signal, both a \Bd and a \Bs component with the same shape are included with a fixed peak offset given by the known difference in the \Bs and \Bz masses $\Delta m \equiv m(\Bs) - m(\Bz)=  87.19 \mevcc$~\cite{PDG2022}. The fraction $f_\Bs$ of the \Bs component relative to the \Bd component is allowed to vary. 
The combinatorial background is modelled with an exponential function, leading to a total PDF of the form
\begin{align}
    \mathcal{P}_{\text{Total},i}(m) = f_{\text{Sig},i}^\text{full} \left[ (1-f_{B_s^0}) \mathcal{P}_{\Bz}(m) + f_{B_s^0}\mathcal{P}_{\Bs}(m) \right] + (1-f_{\text{Sig},i}^\text{full})\mathcal{P}_{\text{Bkg},i}(m),
    \label{eqn:MassFitPDF}
\end{align}
where the index $i$ labels the \qsq region. The fits to the different \qsq regions can be seen in Fig.~\ref{fig:mBmassFits}, and the values of $f_{\text{Sig},i}^\text{full}$ obtained for each region are listed in Table~\ref{tab:BMassFitCategories}. The signal fractions of Eq.~\ref{eqn:MassFitPDF} are used to calculate the number of background events per \qsq region contained in the signal region, which is defined as a 40\mevcc region around the \Bd mass peak, $5259.58 \leq m(\Kp\pim\mumu) \leq 5299.58 \mevcc$.

\subsection{Background fit in the upper {\boldmath \Bd} mass sideband}
\label{sec:SidebandFit}

\begin{figure}[t]
    \centering
    \begin{subfigure}{0.49\linewidth}
        \includegraphics[width=\linewidth]{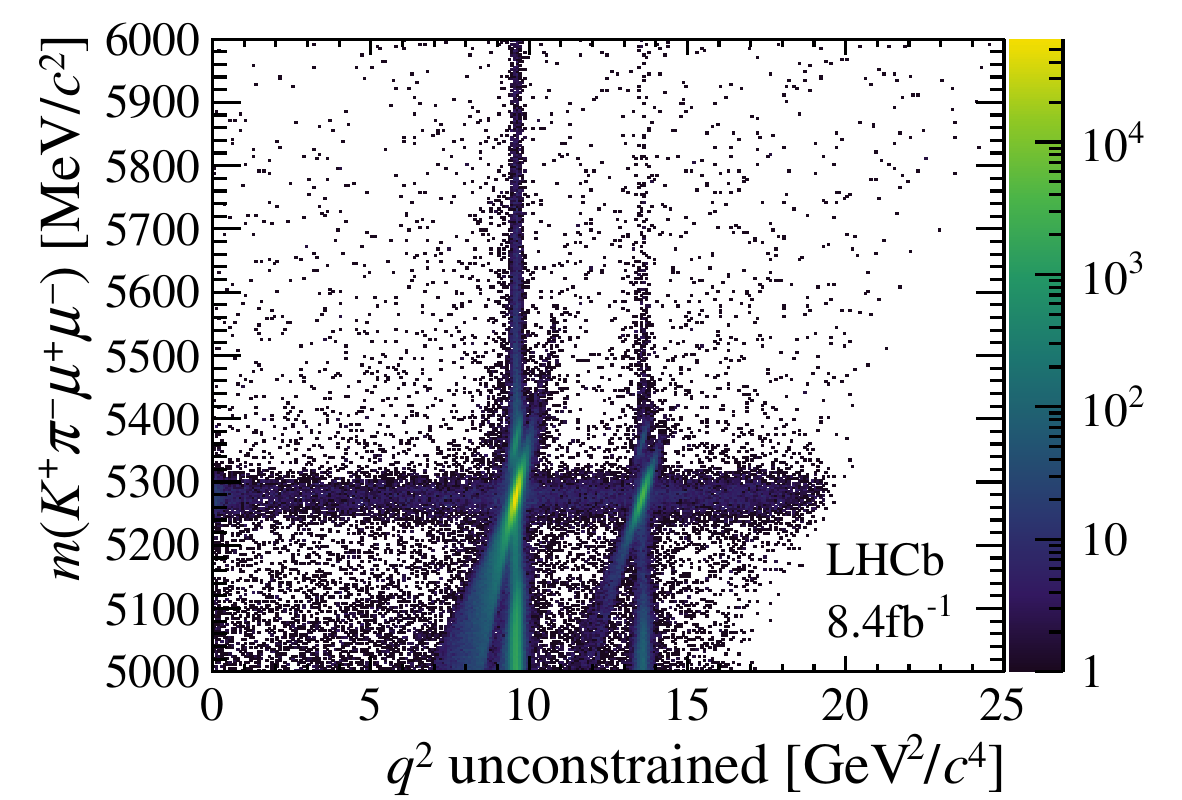}
    \caption{}
    \label{subfig:B0_mass_constraints_unconstrained}
    \end{subfigure}
    \begin{subfigure}{0.49\linewidth}
        \includegraphics[width=\linewidth]{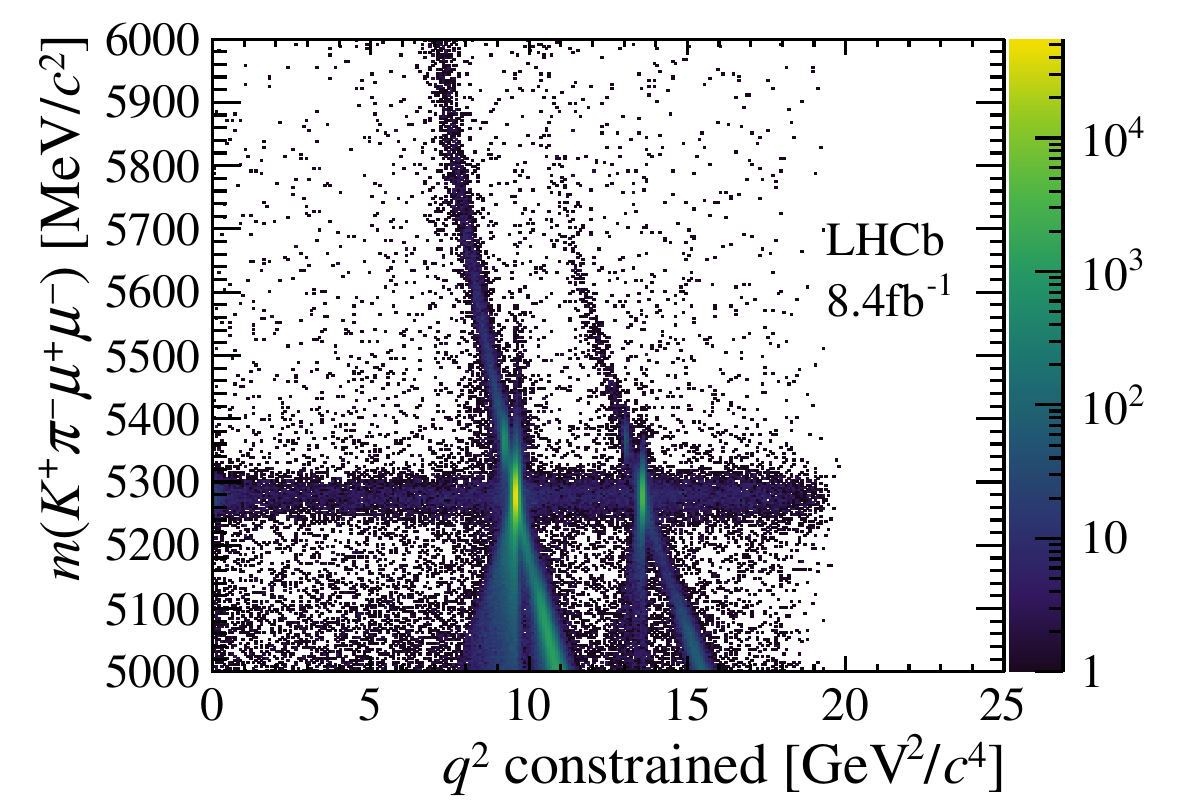}
    \caption{}
    \label{subfig:B0_mass_constraints_constrained}
  \end{subfigure}   
    \begin{subfigure}{0.49\linewidth}
        \includegraphics[width=\linewidth]{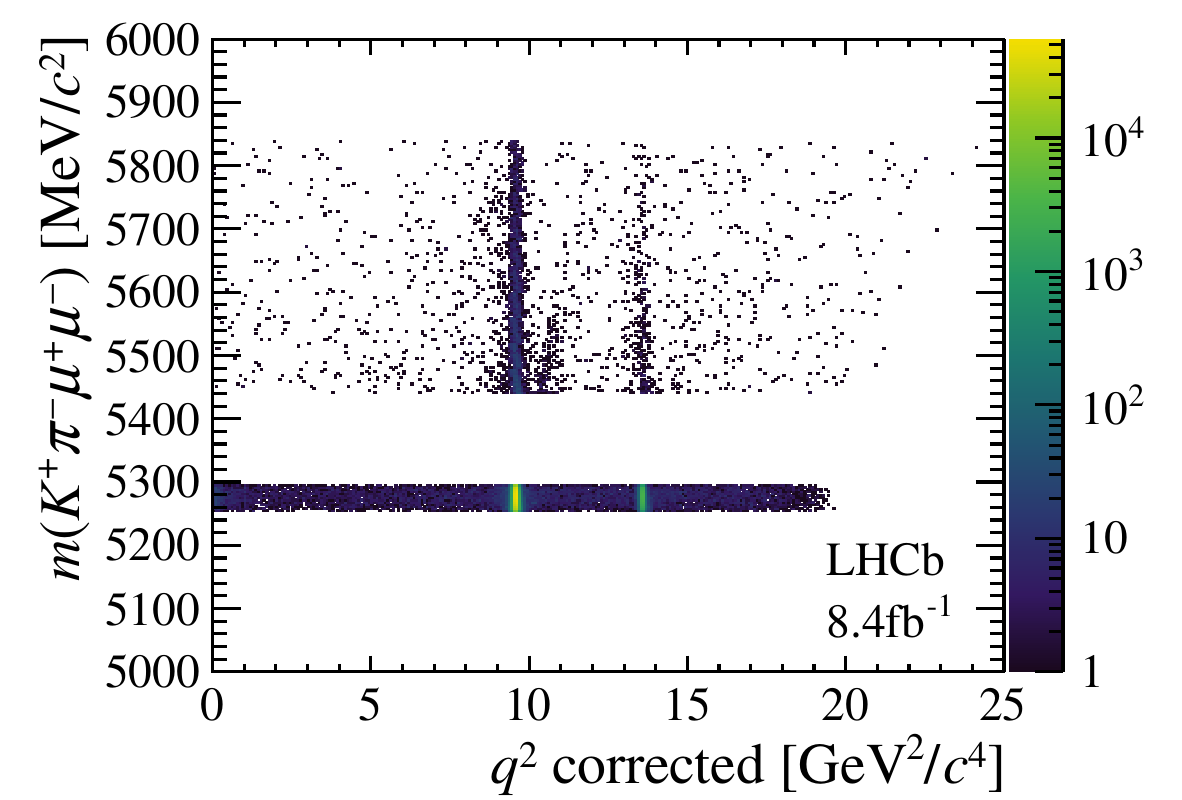}
    \caption{}
    \label{subfig:B0_mass_constraints_varying}
    \end{subfigure}
    \caption{Distributions of candidates in data with different treatments of the $B^0$ mass constraint when determining $q^2$. In (a) no constraint is applied, in (b) the final state is constrained to the $B^0$ mass, while in (c) the final-state mass is constrained to the centre of each of the $40\,\rm{MeV}/c^2$ wide signal and upper mass background windows. In all the plots, the signal regions correspond to the horizontal band. The diagonal lines in (a) that become vertical lines in (b) are the tails of poorly reconstructed $B^{0} \to J/\psi K^{*0}$ and $B^{0} \to \psi(2S) K^{*0}$ decays.}
    \label{fig:B0_mass_constraints}
\end{figure}

Following the determination of the signal fractions, a fit to the upper \Bd mass sideband ($5440 \leq m(\Kp\pim\mumu) \leq 5840 \mevcc$) is performed simultaneously in the three \qsq regions defined in Table~\ref{tab:WCFitCategories} in order to cleanly determine the combinatorial background angular distribution. The upper mass sideband is used since the combinatorial background is the only significant contribution in this region and the results can subsequently be extrapolated into the signal region as explained below. The lower mass sideband also contains combinatorial background events, but is not used for this purpose due to the presence of additional physical backgrounds from partially reconstructed decays and genuine low-mass signal events. The latter are particularly troublesome, as explained below. Special care is taken in the sideband fit to account for the use of a \Bd mass constraint in the fit to the signal region (see Sec.~\ref{sec:SignalFit} below). The combinatorial background events are not the decay products of a real \Bd meson; hence, the mass constraint causes a distortion of the background \qsq distribution which is correlated with the reconstructed \Bd mass. This effect can be observed by contrasting Fig.~\ref{subfig:B0_mass_constraints_unconstrained}, which shows the reconstructed \B mass as a function of \qsq without the mass constraint, and~\ref{subfig:B0_mass_constraints_constrained} showing the same with the mass constraint. The \qsq positions of the \jpsi- and \psitwos-dominated combinatorial peaks are observed to vary as a function of \mB in a way that is impractical to model. To remedy this, the upper \Bd mass sideband is divided into 10 windows, each of width 40\mevcc. In each window, the $K\pi\mu\mu$ mass is constrained to the centre of the region, so as to mimic the distortion of the background \qsq distribution that occurs in the signal region that also is 40\mevcc wide. As a result of this, the \jpsi and \psitwos peaks are aligned between the subregions and the signal region, as shown in Fig.~\ref{subfig:B0_mass_constraints_varying}. While this procedure mostly resolves issues arising from the \Bd mass constraint in the upper mass sideband, it further complicates matters in lower mass sideband. This is due to the fact that the lower mass sideband contains a significant portion of signal events which are wrongly mass constrained to the centre of the sideband region. The systematic uncertainty introduced by attempting to model these contributions outweighs any benefit of fitting the combinatorial background in the lower mass sideband.

In each of the ten \mB background windows, the background shape is modelled in the \ctl, \ctk, \phih, and \qsq dimensions, separately for the five \qsq regions defined in Table~\ref{tab:BMassFitCategories}. Those five \qsq regions are then reduced to the three regions of Table~\ref{tab:WCFitCategories} by adding the PDFs for the relevant contributions in each region. The PDF for the sideband fit is defined as
\begin{equation}
  \mathcal{P}(q^2, \vec{\Omega}) = \begin{cases} 
      \mathcal{P}_{\text{comb}}(q^2, \vec{\Omega}) & \text{low-} q^2 \\
      (1-f_\jpsi)\mathcal{P}_{\text{comb}}(q^2, \vec{\Omega}) + f_\jpsi\mathcal{P}_{\jpsi}(q^2, \vec{\Omega}) & \text{mid-} q^2 \\
      (1-f_\psitwos)\mathcal{P}_{\text{comb}}(q^2, \vec{\Omega}) + f_\psitwos\mathcal{P}_{\psitwos}(q^2, \vec{\Omega}) & \text{high-} q^2
   \end{cases}, 
   \label{eqn:SidebandFitPDF}
\end{equation}
where $f_\jpsi$ and $f_\psitwos$ represent the resonant background fractions relative to the fully combinatorial components in the mid- and high-\qsq regions, respectively.

Each of the PDFs in Eq.~\ref{eqn:SidebandFitPDF} is assumed to factorise completely such that each dimension can be modelled independently. For the \qsq dimension, the fully combinatorial contribution is modelled using the Weibull function~\cite{weibull1951statistical}; while the \jpsi- and \psitwos-dominated contributions are modelled using CB functions. The \ctl and \phih dimensions are modelled using second order Chebyshev polynomials; whilst the \ctk dimension is modelled with second order Bernstein polynomials. 

A complication arises due to the $\decay{\Bp}{\Kp\mumu}$ veto described in Sec.~\ref{sec:BkgComposition}. While the veto has no effect in the signal region, it causes a drop in the number of combinatorial background events in a $\cos{\theta_K}$ region of the phase space that depends on the reconstructed \Bd mass. If ignored, it leads to the wrong background shape extrapolated into the signal region. The solution employed is to exclude the affected $m(\Kp\pim\mumu)$ dependent $ \cos{\theta_K}$ interval in each of the ten sideband regions. This reduces the amount of events in the sideband fit by approximately 15\% but prevents any bias in the extrapolation.

The background fit in the upper mass sideband does not directly involve fitting the $m(\Kp\pim\mumu)$ dimension. Nevertheless, all parameters from the mass fit described in Sec.~\ref{sec:BMassFit} are allowed to vary again in the sideband fit. In this case, the parameters describing the $m(\Kp\pim\mumu)$ dimension enter the likelihood exclusively through a multivariate Gaussian constraint based on the full covariance matrix obtained from the direct mass fit. This allows the uncertainty in the signal fractions to be propagated through to the subsequent fitting stages.

To model the shape of the combinatorial background in the signal region, an extrapolation is made of the parameters describing the angular and \qsq distributions in each of the ten sideband bins. In this way, the angular distribution in the signal region is described by Eq.~\ref{eqn:SidebandFitPDF} as well. The parameters obtained from the extrapolation and their corresponding correlation matrix are used in the signal fit described below.

\subsection{Fit in the signal region}\label{sec:SignalFit}

The final step consists of an unbinned maximum likelihood fit to the four-dimensional (\ctl, \ctk, \phih, and \qsq) distribution in the signal region. 
The fit is performed simultaneously in the three \qsq regions shown in Table~\ref{tab:WCFitCategories}, with the total PDF in each region given by
\begin{align}
    \mathcal{P}_{\text{Total},i} (\vec{\Omega}, \qsq) &= f_{\text{Sig},i} \mathcal{P}_{\text{Sig},i} (\vec{\Omega}, \qsq) + \left( 1 - f_{\text{Sig},i} \right) \mathcal{P}_{\text{Bkg},i} (\vec{\Omega}, \qsq), \label{eqn:FinalPDF}
\end{align}
where $\mathcal{P}_{\text{Sig},i} (\vec{\Omega}, \qsq)$ is the full experimental signal PDF described in Eq.~\ref{eqn:fullSignalPDF}, and $\mathcal{P}_{\text{Bkg},i} (\vec{\Omega}, \qsq)$ is the corresponding background PDF from Eq.~\ref{eqn:SidebandFitPDF}. The  fractions $f_{\text{Sig},i}$ correspond to the fractions of signal events within each of the signal regions. They are not independent free parameters, rather, they are derived from the fitted signal fractions in Eq.~\ref{eqn:MassFitPDF}, which correspond to the full mass range and the five \qsq regions of Table~\ref{tab:BMassFitCategories}.
In the fit, the background shape is constrained using the results of the extrapolation from sideband fits and the background yield is constrained using signal fractions from the \mB mass fit as given in Table~\ref{tab:BMassFitCategories}. This is achieved through a single multivariate Gaussian constraint based on the full covariance matrix from the fit described in Sec.~\ref{sec:SidebandFit}. The signal angular distribution is modelled according to Sec.~\ref{sec:AcceptanceResolution}. The baseline fit configuration consists of 150 free parameters.  A summary of the parameters of the signal model is given in Appendix~\ref{app:fit_parameters}.

The complex amplitudes for each polarisation state of the nonlocal components $A_j^\lambda$ appearing in Eqs.~\ref{eqn:dy1p} and~\ref{eq:Y2P}  are determined from the fit to the data. For amplitudes that are expected to be significantly different from zero, the fit is performed in terms of the magnitude $|A_j^\lambda|$ and phase $\delta_j^\lambda$. In contrast, for components with a small expected amplitude the fit is performed in terms of the real $\Re{(A_j^\lambda)}$ and imaginary $\Im{(A_j^\lambda)}$ components. This ensures better stability of the fit.

The scale of both the $\BdToKstmm$ local and nonlocal amplitudes~are determined through the known value of the compound branching fraction \mbox{$\mathcal{B}(\BdToJpsiKst)\mathcal{B}(\jpsi\to\mumu)$} and by scaling the three polarisation amplitudes $A_{\jpsi}^{0,\parallel, \perp}$ such that
\begin{equation}
    \label{eq:normalisation}
    \mathcal{B}(B^0\to \jpsi \Kstarz)\mathcal{B}(\jpsi\to\mumu) = 
    | A_{\jpsi}^{0}|^{2} + | A_{\jpsi}^{\parallel} |^{2}+| A_{\jpsi}^{\perp} |^{2}.
\end{equation}
The branching fraction $\mathcal{B}(B^0\to \jpsi \Kstarz)$ is taken from Ref.~\cite{Belle:2014nuw} and the branching fraction $\mathcal{B}(\jpsi\to\mumu)$ is taken from Ref.~\cite{PDG2022}. In the fit this is implemented by calculating $|A_{\jpsi}^{0}|$ from Eq.~\ref{eq:normalisation} rather than having it as a free parameter. The uncertainty of the branching fraction $\mathcal{B}(B^0\to \jpsi \Kstarz)$ is a limiting source of uncertainty on numerous nonlocal parameters and is the largest systematic uncertainty on the Wilson Coefficients, as discussed in Sec.~\ref{sec:Systematics}.

The kinematically allowed $q^2$ region of $\decay{\Bd}{\Kstarz\mumu}$ decays ranges from $4m^{2}_{\mu}$ to $(m_B-m_{K\pi}^{\text{max}})^{2}$, where $m_{K\pi}^{\text{max}}$ denotes the maximum mass of the $\Kstarz\to K^+\pi^-$ system. In this analysis  $m_{K\pi}^{\text{max}}$ is set to $0.996\gevcc$ as discussed in Sec.~\ref{sec:DecayRate}. This results in a \qsq phase-space range of $0.044<\qsq<18.34~\gevgevcccc$. In order to reduce the model dependence of the \qsq resolution in regions where the decay rate varies rapidly with \qsq, the measurement is performed in the reconstructed \qsq range of $0.1<\qsq<18.0~\gevgevcccc$.

The analysis was performed in a blind fashion until finalised, by implementing an unknown offset to the values of the Wilson Coefficients $\mathcal{C}^{(')}_{9,10}$ and $\mathcal{C}_{9\tau}$.
Additionally, the signs of the differences between the Wilson Coefficients and their Standard Model values are switched randomly. Pseudoexperiments are performed following an identical procedure to that used for the data fit to validate the full analysis. The bias and error coverage obtained from the pull distributions for the Wilson Coefficients are listed in Table~\ref{tab:pull_vals}.
\begin{table}[!t]
    \centering
    \caption{The means and widths of the pull distributions in pseudoexperiments for the Wilson Coefficients. The bias is quoted as a fraction of the statistical uncertainty on the parameter.}
    \begin{tabular}{c c c }
\hline 
Variable & Mean (bias) & Width (coverage)\\
\hline 
\C9             & $-0.27 \pm 0.06$ & $ 1.00 \pm 0.04$  \\
\C{10}          & $-0.13 \pm 0.06$ & $ 0.94 \pm 0.04$  \\
\Cp9       & $-0.09 \pm 0.06$ & $ 1.05 \pm 0.04$  \\
\Cp{10}       & $-0.34 \pm 0.06$ & $0.99 \pm 0.04$  \\
$\mathcal{C}_{9\tau}$     & $-0.20 \pm 0.06$ & $ 1.03 \pm 0.04$  \\
\hline
    \end{tabular}
    \label{tab:pull_vals}
\end{table}
The observed biases are $\lesssim 30\%$ compared to the statistical uncertainty for all parameters, and are accounted for as corrections to the final results.

\section{Systematic uncertainties}
\label{sec:Systematics}
Several sources of systematic uncertainty are considered for this analysis, including those related to the modelling of the signal, backgrounds, detector effects, and the analysis method and implementation. The most significant effects are described in detail, followed by a brief overview of some additional effects which are considered but found to be subdominant or negligible. The final parameter uncertainties are obtained by combining the statistical covariance matrix from the likelihood fit with the total combined systematic covariance matrix accounting for all non-negligible effects. The systematic covariance matrices are obtained by performing fits to pseudoexperiments using alternative fit configurations and/or modifying the pseudoexperiments in a manner representative of the effect in question.

\subsection{Normalisation to the {\boldmath \BdToJpsiKst} branching fraction}
\label{sec:JpsiKstBFSyst}

The dominant source of systematic uncertainty on the parameters of interest
is found to arise from the normalisation to the \BdToJpsiKst branching fraction, which is only known with a relative uncertainty of 6.8\%~\cite{Belle:2014nuw}. Varying the known \BdToJpsiKst branching fraction within its uncertainties translates to an effect of the order 50\% of the statistical uncertainty for the \C9 , and 100\% for the \C{10} parameters. It should be noted that the reason this effect is so significant is due to the dramatically improved statistical precision in this analysis relative to the \belle measurement of \BF{(\BdToJpsiKst)}\cite{Belle:2014nuw}. This is currently an irreducible systematic uncertainty, but can be directly improved in the future with a new precise measurement of the \BdToJpsiKst branching fraction, \eg from the \belletwo experiment.

\subsection{Exotic charmonium-like states}
\label{sec:ExoticaSyst}

The presence of charmonium-like resonances in the $\jpsi\pi$ and $\psitwos\pi$ spectra, so-called exotic $T_{c\overline{c}1}$ states\footnote{Previously known as $Z_c$ states}, leads to the interference of the decay amplitudes $B\to T_{c\overline{c}1}(\to\psi\pi)K$ with both the rare decay and $B\to\psi K^{*0}$ final states. This analysis performs a fit across the full $\qsq$ spectrum, including the $\jpsi$ and $\psitwos$ regions, without accounting for these exotica contributions. The reason for not including these amplitude components in the fit is mainly due to computational efficiency; the $B\to T_{c\overline{c}1}(\to\psi\pi)K$ decays contribute with different functional angular dependencies, thus the decay rate no longer factorises into a simple sum of products of $J_i(\qsq)$ and $f(\Omega)$ terms as given in Eq.~\ref{eqn:diffdecayrate}.
To assess the impact of neglecting the exotica contributions, a correction is derived by generating pseudodata that contain all of the baseline local and nonlocal amplitudes added coherently to the exotica ones, following the procedure of Refs.~\cite{Gratrex:2015hna, Belle:2014nuw,Belle:2013shl}. The pseudoexperiments are fit back using the baseline model that neglects the exotic states. The exotic amplitudes are fixed in the generation of the pseudodata to the central values from measurements made by the Belle collaboration~\cite{Belle:2014nuw,Belle:2013shl}. The resulting average shift of the parameters from their generated values is taken as the correction. With the exception of the \jpsi and \psitwos magnitudes and phases, the exotica correction to the parameters of interest is small ($\lesssim 20\%$) relative to the statistical uncertainty.

A systematic uncertainty is derived for the correction by varying the exotic amplitudes within their measured 1$\sigma$ uncertainties and recalculating the correction. Again, with the exception of the \jpsi and \psitwos magnitudes and phases, the systematic uncertainty on the exotica correction is $\lesssim 20\%$ relative to the statistical uncertainty.

For the \jpsi and \psitwos parameters, the exotica correction and associated systematic uncertainty are large relative to the statistical uncertainty (from $100\%$ to $250\%$); however, the absolute effect remains small given the excellent statistical precision achieved on the resonance magnitudes and phases.

\subsection{Acceptance}
\label{sec:AcceptanceSyst}

The acceptance function described in Sec.~\ref{sec:AcceptanceResolution} is determined from simulated samples that are reweighted to agree with data. Corrections are applied specifically to ensure agreement in the hardware trigger and tracking efficiencies, the multiplicity of tracks in an event, and the distributions of PID and \Bd meson kinematic variables. 
The weights applied to the simulated samples have associated statistical and systematic uncertainties that propagate through to the eventual determination of the signal parameters. To assess the impact of these uncertainties, ensembles of pseudoexperiments are used in which alternative weights are derived and subsequently used to produce modified PDFs with alternative acceptance coefficients. Pseudoexperiments are generated from the alternative PDFs and then fitted back using both the baseline and alternative (true) PDFs.

The largest effect is found to come from the corrections to the \Bd meson kinematics. Relative to the respective statistical uncertainties, the effect is approximately 10\% for the Wilson Coefficients and 200\% for the magnitudes of the parallel and transverse \jpsi amplitudes.

The baseline weights for these kinematic corrections are derived by comparing the distributions of the \Bd meson transverse momentum, pseudorapidity, vertex quality, and impact parameter quality between \decay{\Bd}{\jpsi\Kstarz} decays in simulation and data. The alternative weights are derived by performing the same comparisons for \decay{\Bd}{\psitwos\Kstarz} decays in simulation and data.

\subsection{Two-particle open-charm constraint}
\label{sec:opencharm_constraint_systematic}

As described in Sec.~\ref{sec:nonlocalParam}, a Gaussian constraint is placed on the coefficients in Eq.~\ref{eq:Y2P} describing the size of the open-charm contributions to maintain the stability of the fit. The constraint restricts the real and imaginary parts, separately, of each two-particle open-charm state (\D\Dbar, \Dstar\Dbar and \Dstar\Dstarb) to be of a similar size to one another. A separate constraint is used for each polarisation amplitude. The constraint is given a conservative width of 1.0,\footnote{For context, the coherent sum of all the $D^{(*)}\overline{D}^{(*)}$ states would saturate the decay rate at around 0.22~\cite{Cornella:2020aoq}. } but could nonetheless cause biases in the open-charm contributions, as well as other parameters, if the components that are constrained have differences larger than this in data. To asses this bias, pseudoexperiments are generated with the difference between the open-charm components set to 1.5. These pseudoexperiments are then fitted twice, once with the baseline constraint, centered at zero, and once with an unbiased constraint centered at 1.5.
The difference in the fit results is assigned as a systematic uncertainty, and besides the open-charm parameters, the main affected parameters are $C_9$ and $C_{9\tau}$, with systematic uncertainties of 24\% and 29\% of the statistical uncertainty, respectively.

\subsection{Subdominant effects}

The experimental resolution in the angles \ctl, \ctk, and \phih is not explicitly accounted for in the signal model. Unlike the \qsq spectrum, however, the angular distributions contain no sharp peaks and are thus not greatly affected by the detector resolution. Ensembles of pseudoexperiments emulating the effects of the angular resolution are used to confirm that this has no significant effects on the signal parameters of interest. 

The \qsq resolution is accounted for in the baseline model as described in Sec.~\ref{sec:AcceptanceResolution}. As an approximation, the parameters of the resolution model are assumed to remain constant within each \qsq region. Pseudoexperiments investigating the effects of mismodelling the \qsq resolution are performed and no significant effects are observed to result from this assumption.

The mass of the $K^{*0}_{0}(700)$ scalar state has a large uncertainty. Varying the mass in the interval $0.680 < m_{K^{*0}_{0}(700)} < 0.900\gevgevcccc$ results in no significant change apart from the value of the effective form factor for the S-wave which is a nuisance parameter in the fit.

After the full selection has been applied, the fraction of events that contain more than one candidate is approximately 0.18\%. These events are unlikely to correspond to multiple true candidates and are not distributed evenly throughout the phase space. However, the distribution of events with multiple candidates is found to be well modelled in simulation, hence all candidates are retained in the subsequent analysis and a small systematic uncertainty related to their inclusion is determined from simulation.

\section{Results}
\label{sec:Results}
\begin{figure}[h]
    \centering
    \includegraphics[width=\linewidth]{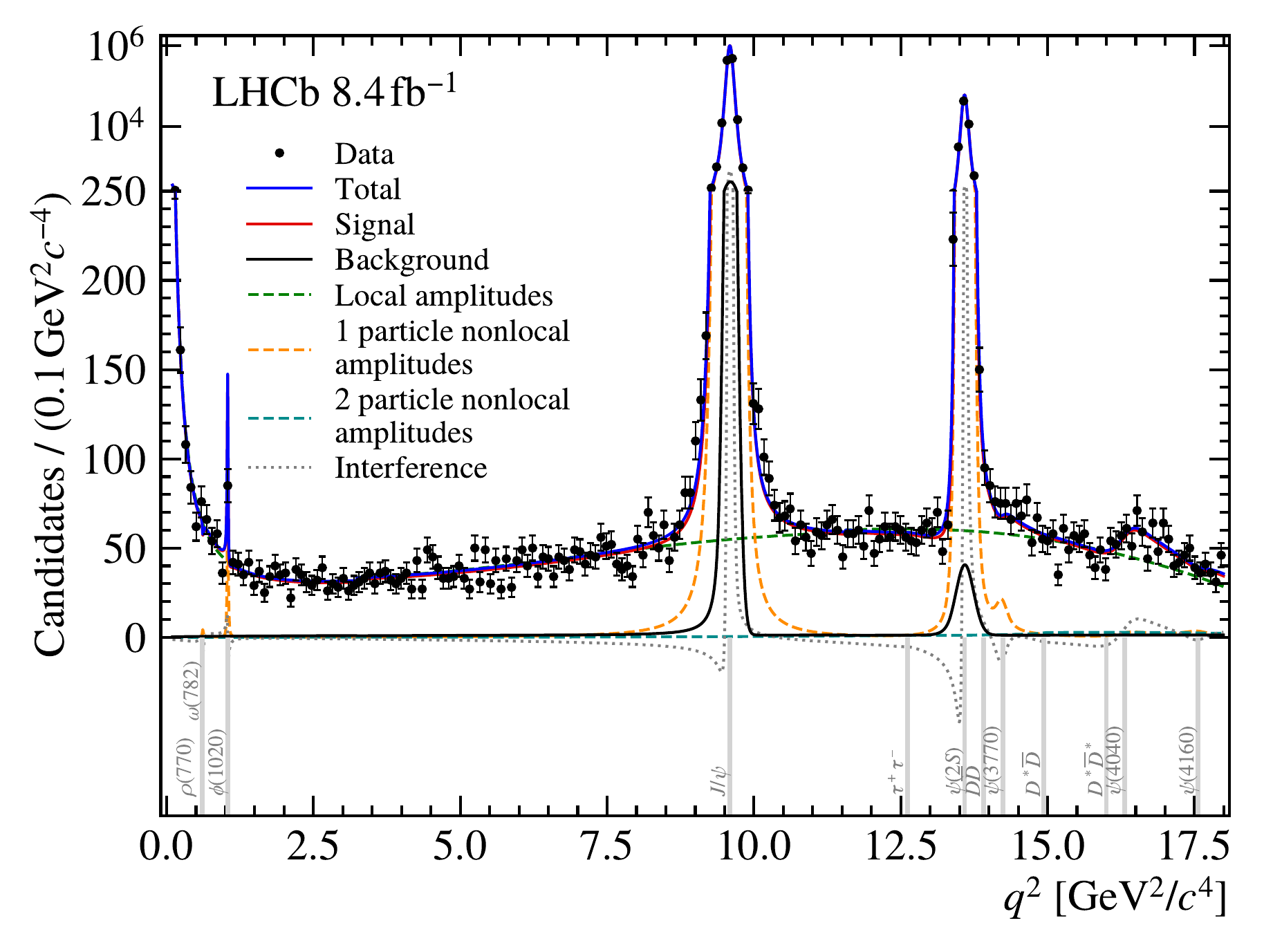}
    \caption{The $q^2$ distribution in the data, overlaid with the PDF projection from the baseline data fit. The total PDF is decomposed into signal and background components, with the signal contributions further decomposed into local and nonlocal contributions as described in Sec.~\ref{sec:nonlocalParam}. Note the hybrid linear/log scale to incorporate the very tall peaks from the charmonium states.}
    \label{fig:DataQ2ProjRes}
\end{figure}

The full \qsq spectrum resulting from the simultaneous fit is shown overlaid on the data in Fig.~\ref{fig:DataQ2ProjRes}. The total PDF is decomposed into signal and background components, and the signal component is further decomposed into the contributions from local amplitudes, one- and two-particle nonlocal amplitudes, and the interference between them. The same results are shown with alternative signal decompositions in Figs.~\ref{fig:DataQ2ProjHel} and~\ref{fig:DataQ2ProjLorentz} in Appendix~\ref{app:AltSigDecompQ2}. 
\input{tables/WCCorrectedFitResults}

\begin{figure}[h]
\centering
    \begin{subfigure}{0.44\linewidth}
        \centering
        \includegraphics[width=\linewidth]{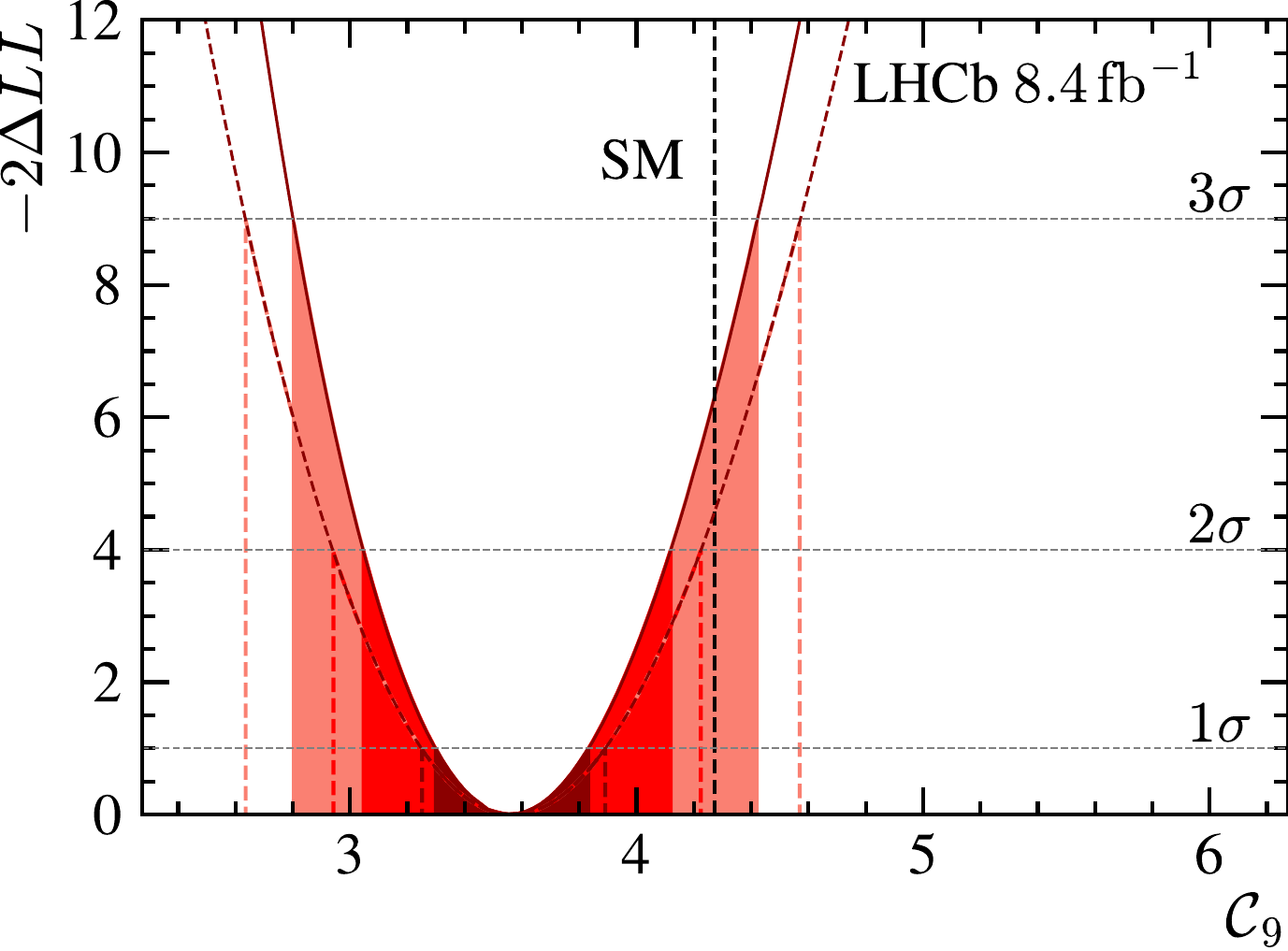}
        \label{fig:C9LH}
    \end{subfigure}
    \begin{subfigure}{0.44\linewidth}
        \centering
        \includegraphics[width=\linewidth]{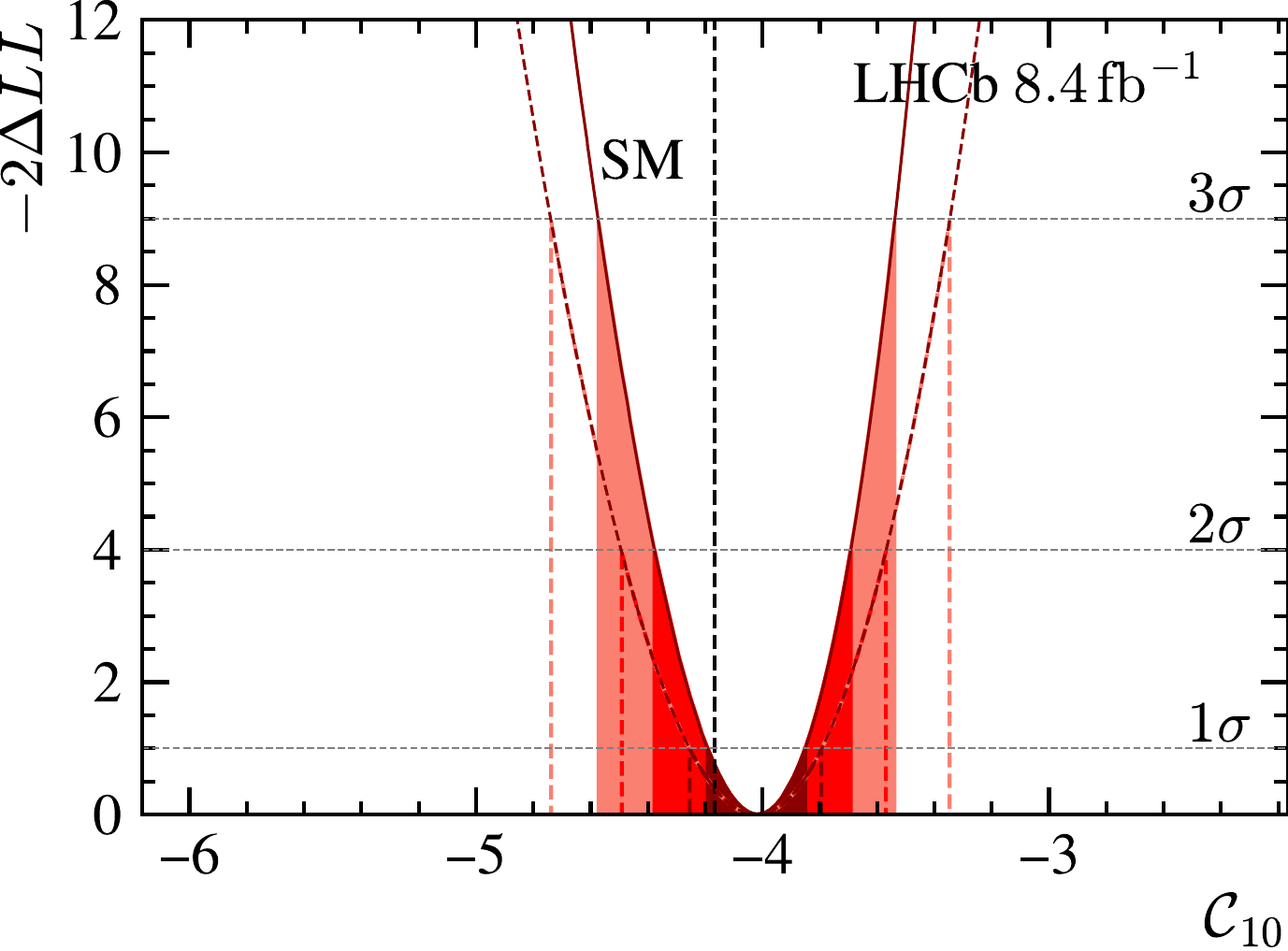}
        \label{fig:C10LH}
    \end{subfigure}
    \begin{subfigure}{0.44\linewidth}
        \centering
        \includegraphics[width=\linewidth]{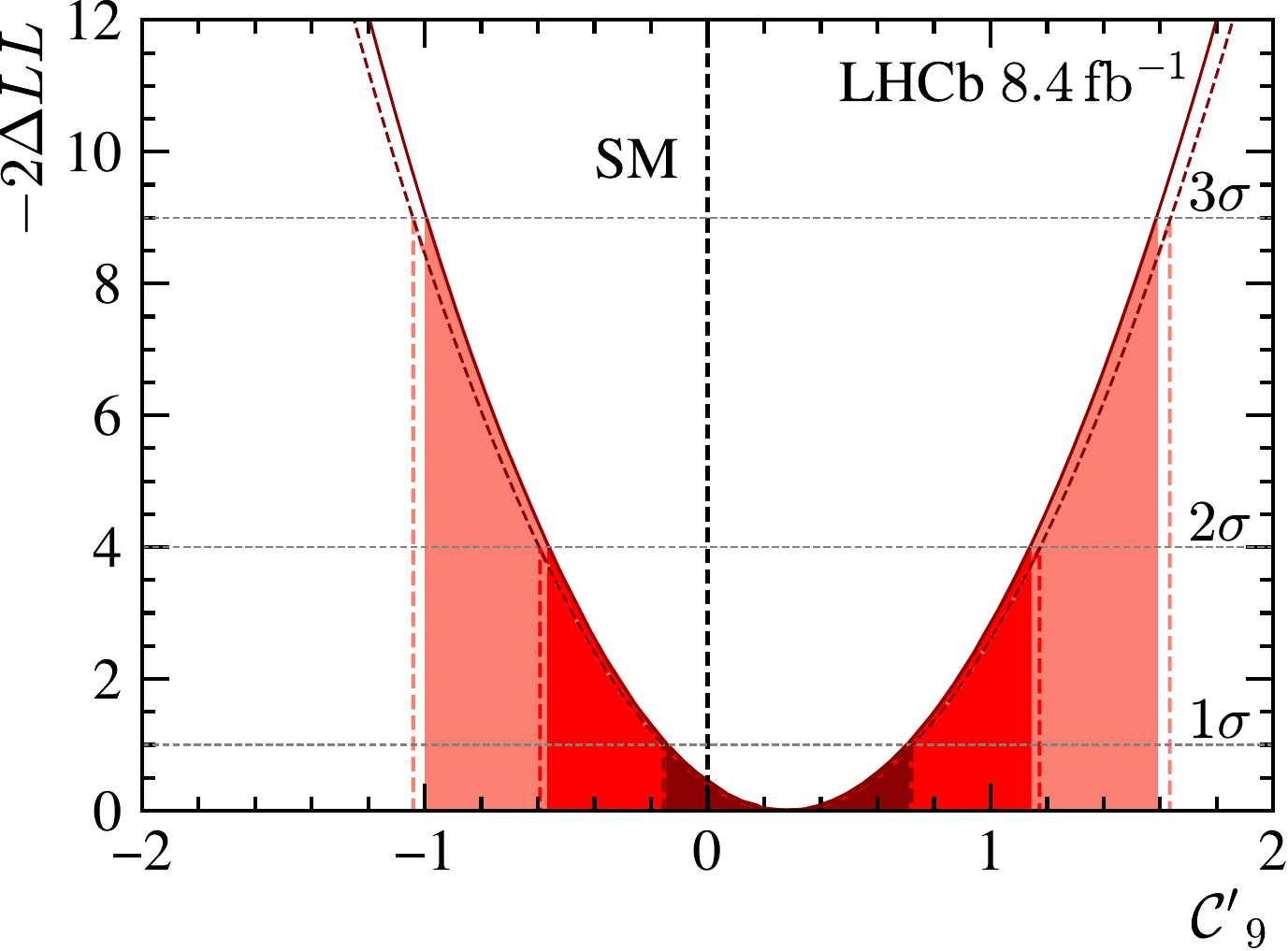}
        \label{fig:C9pLH}
    \end{subfigure}
    \begin{subfigure}{0.44\linewidth}
        \centering
        \includegraphics[width=\linewidth]{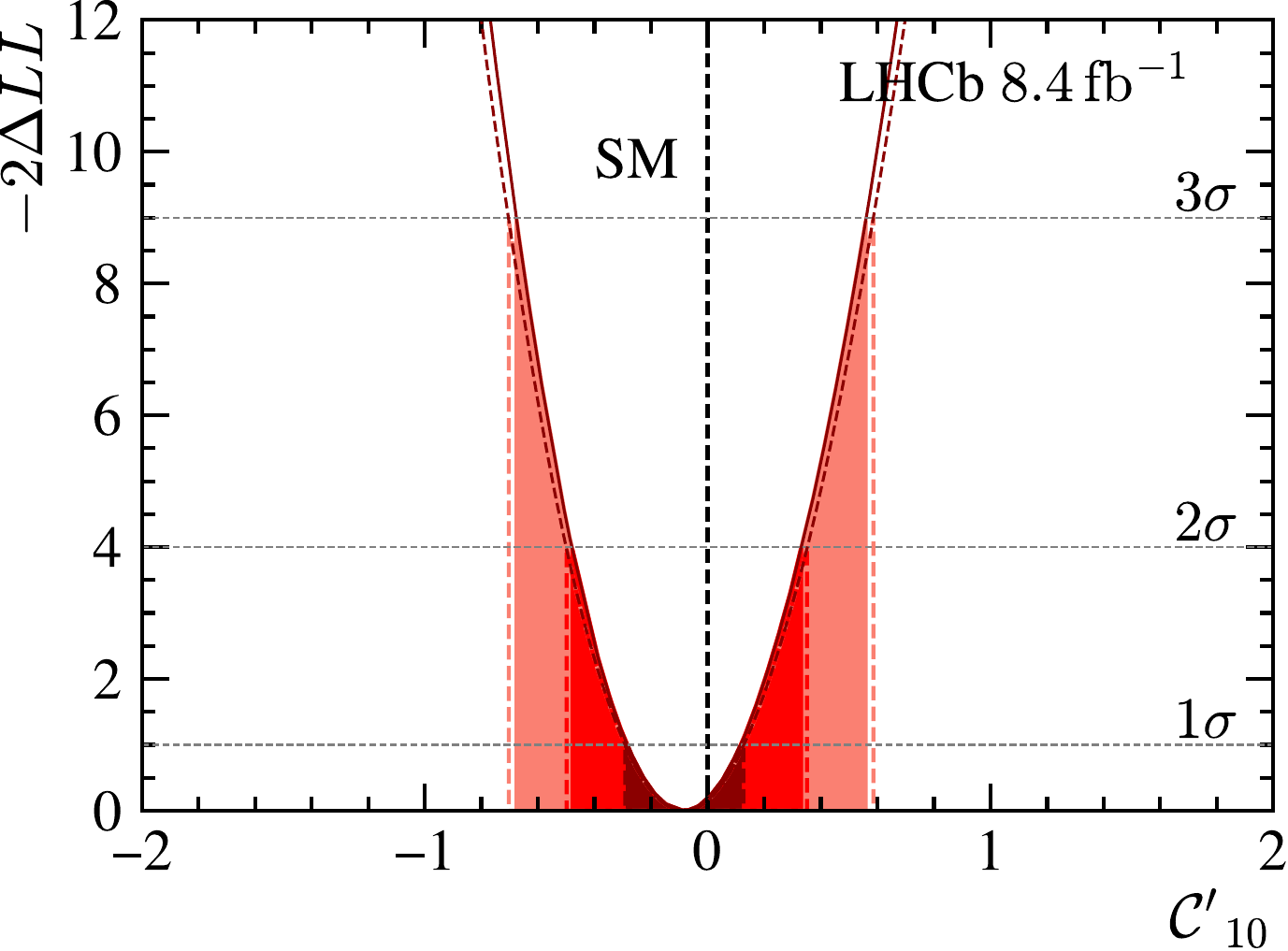}
        \label{fig:C10C10pLH}
    \end{subfigure}
    \begin{subfigure}{0.44\linewidth}
        \centering
        \includegraphics[width=\linewidth]{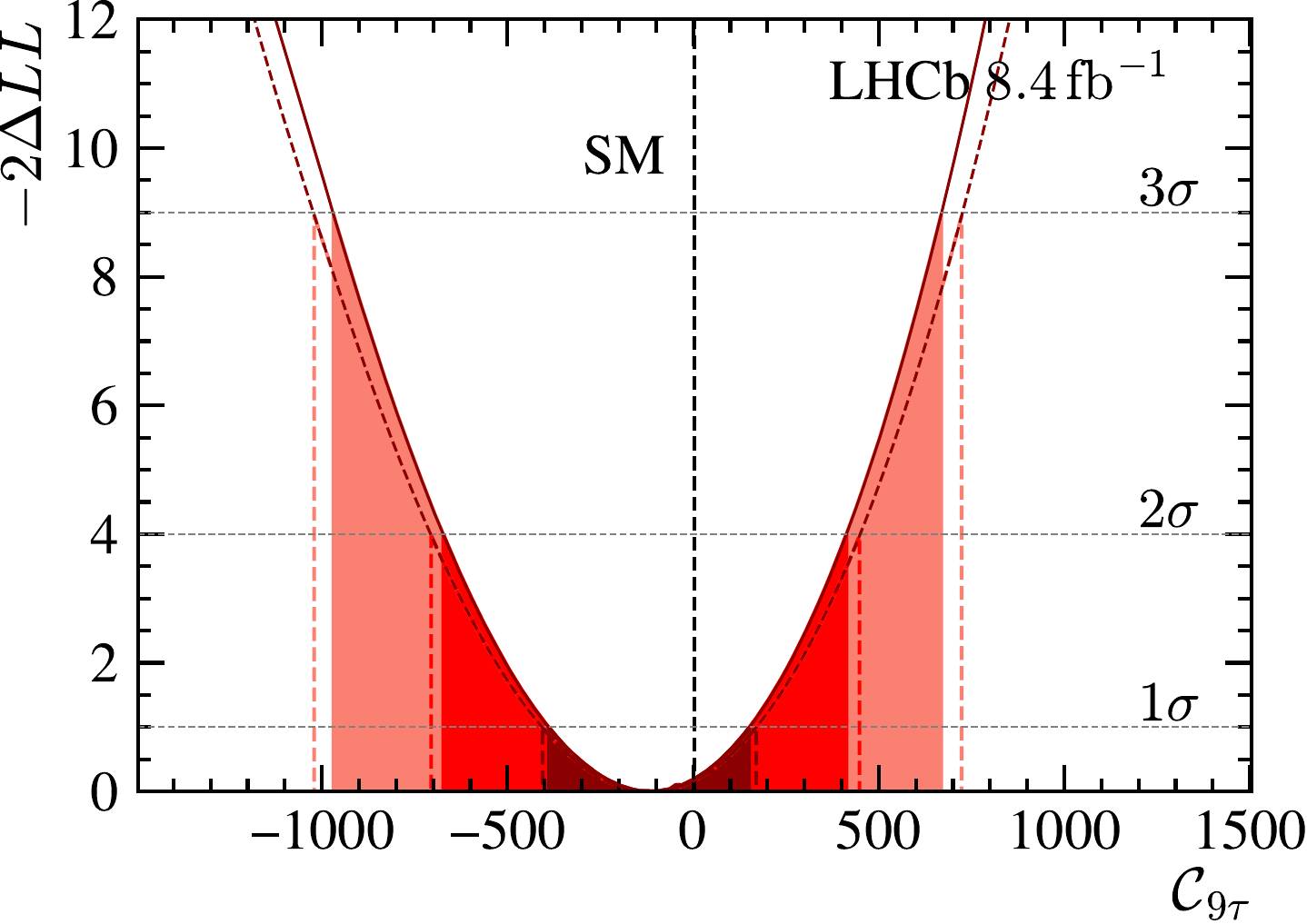}
        \label{fig:C9TAULH}
    \end{subfigure}
    \caption{One-dimensional confidence intervals for the Wilson Coefficients, obtained using a likelihood profile method. The shaded regions consider only statistical uncertainties, while the dashed vertical lines indicate the same regions with systematic uncertainties included. The vertical black dashed lines show the Standard Model values.}
    \label{fig:1DWCLHProfiles}
\end{figure}
\begin{figure}[h]
    \begin{subfigure}{0.5\linewidth}
        \centering
        \includegraphics[width=\linewidth]{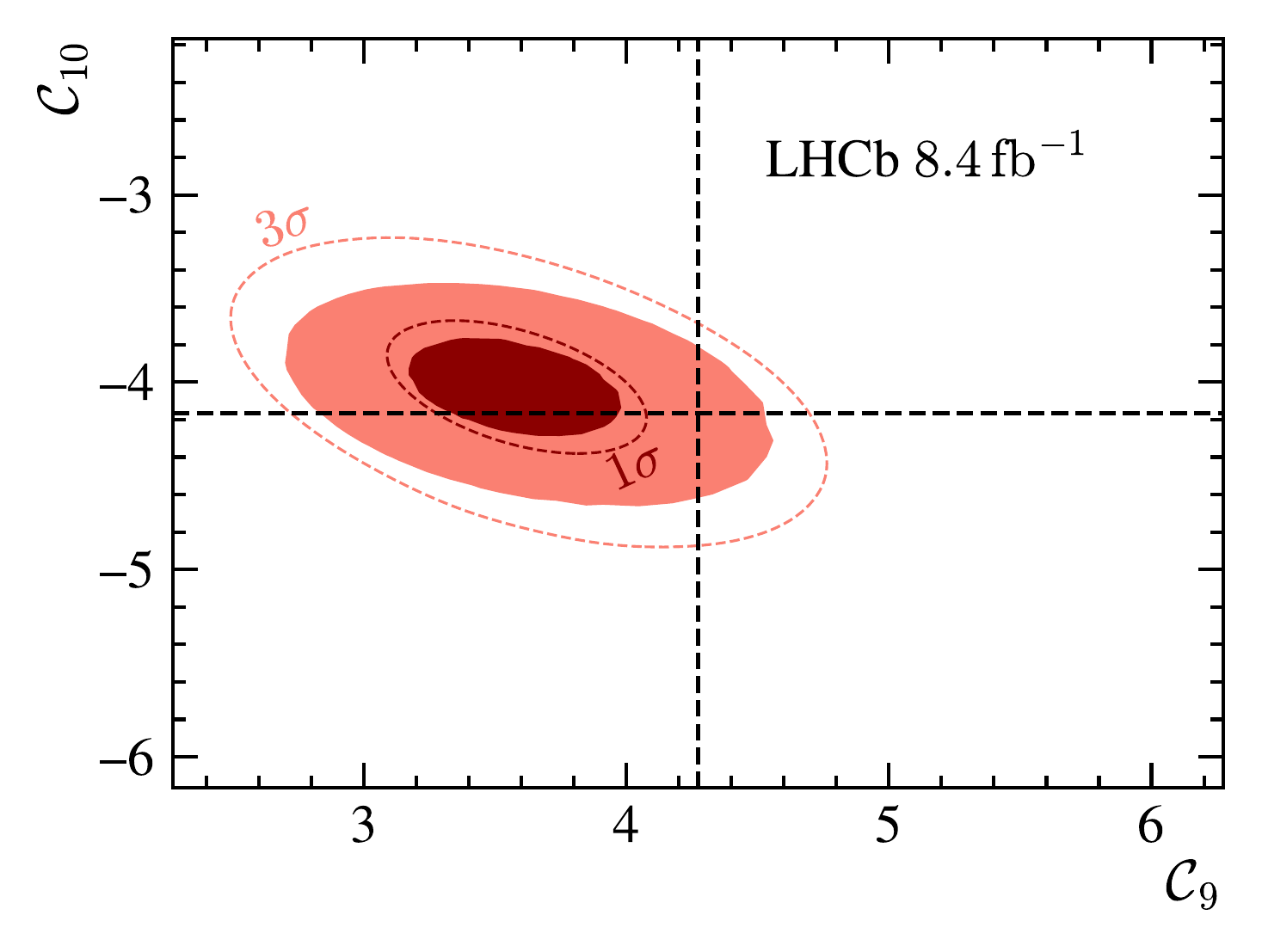}
    \end{subfigure}
    \begin{subfigure}{0.5\linewidth}
        \centering
        \includegraphics[width=\linewidth]{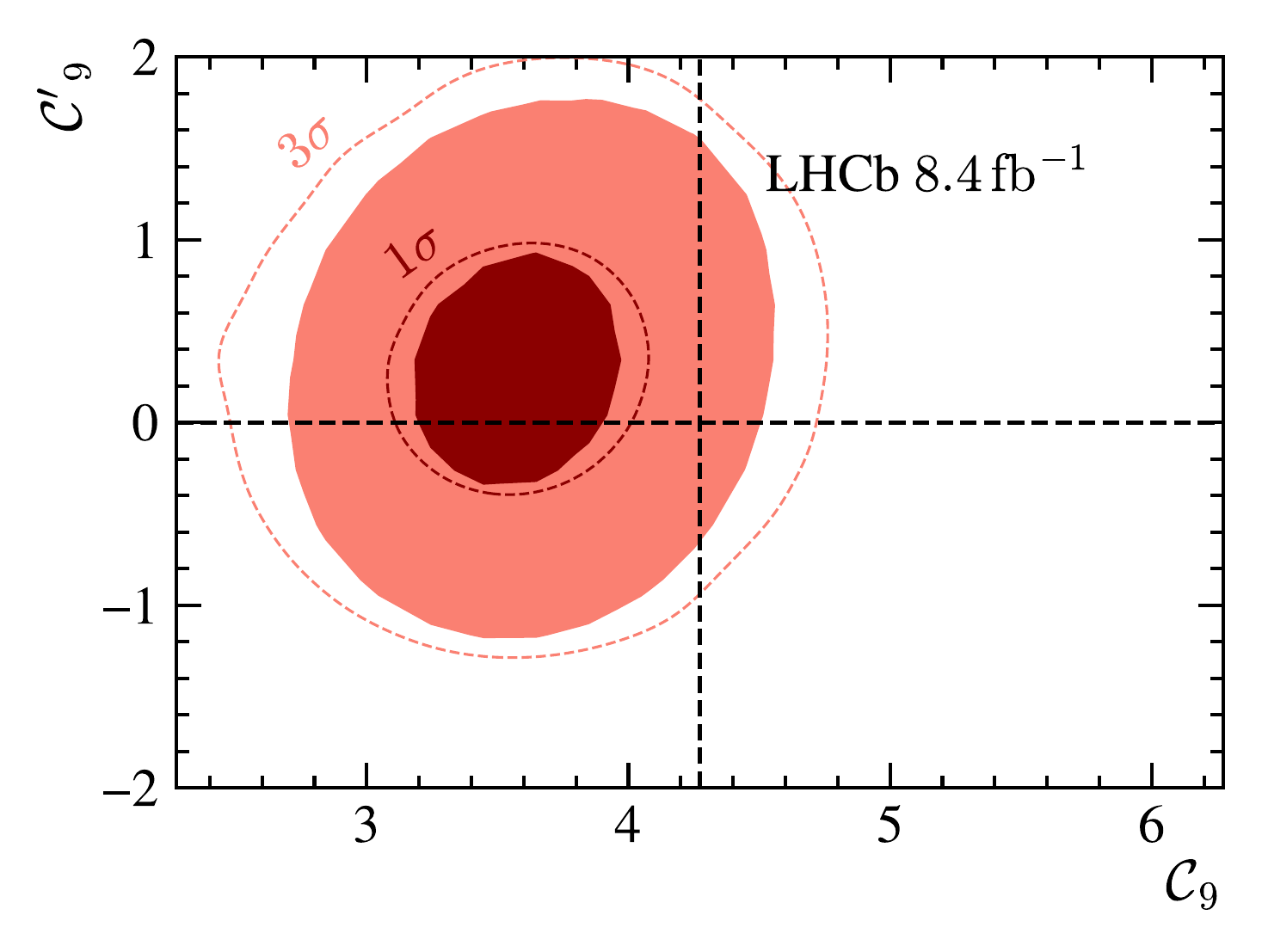}
    \end{subfigure}
    \begin{subfigure}{0.5\linewidth}
        \centering
        \includegraphics[width=\linewidth]{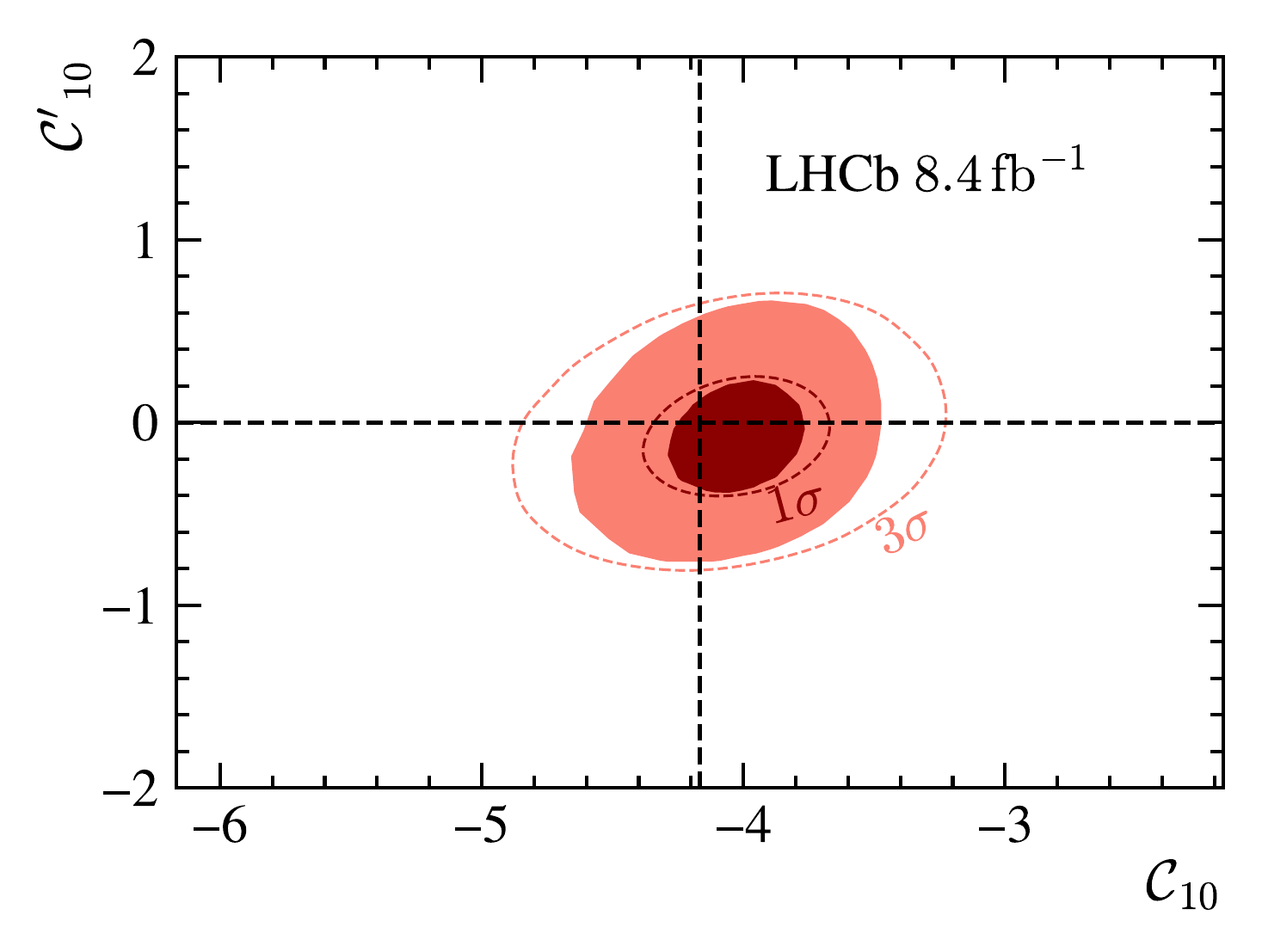}
    \end{subfigure}
    \begin{subfigure}{0.5\linewidth}
        \centering
        \includegraphics[width=\linewidth]{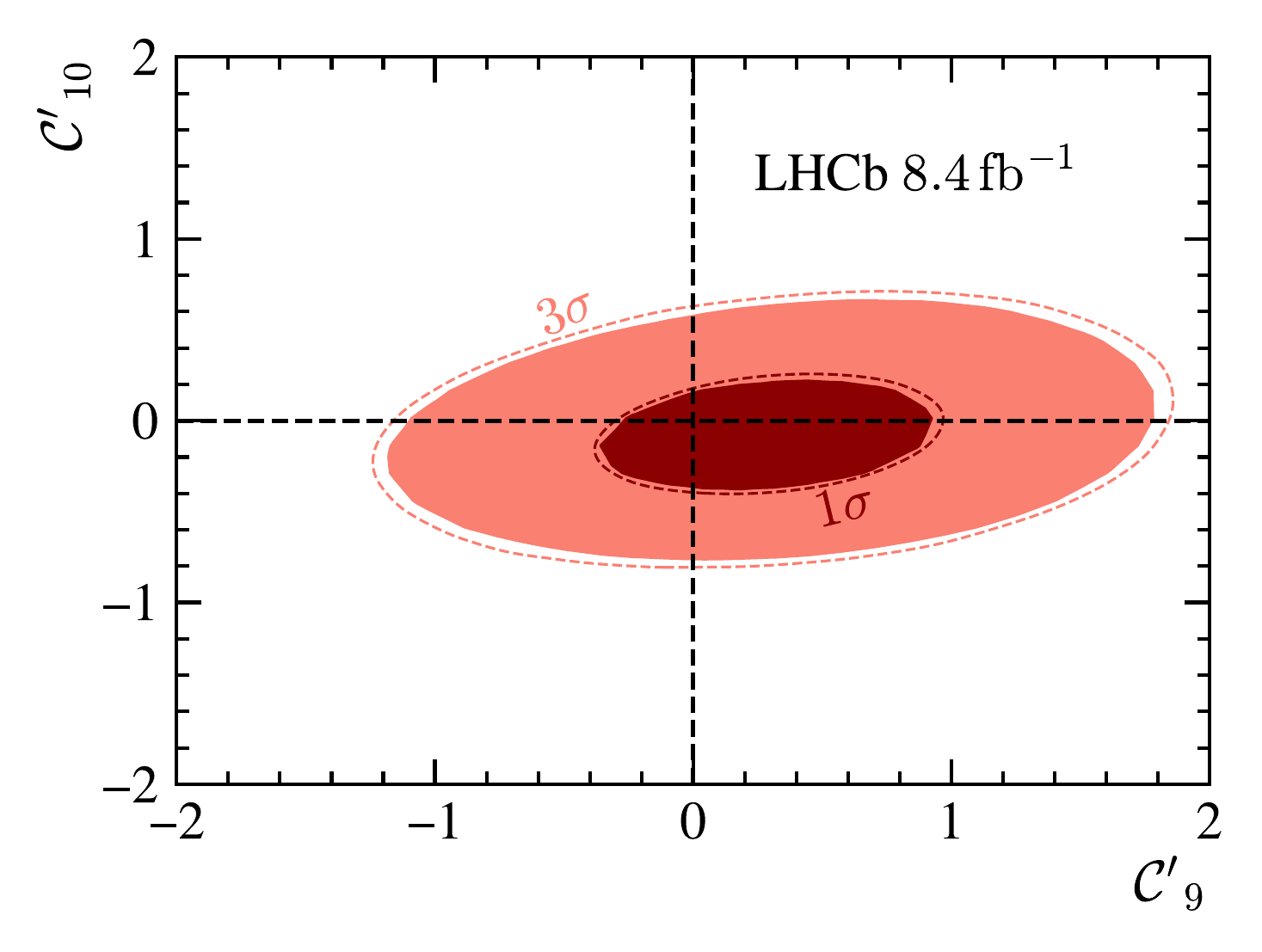}
    \end{subfigure}
    \caption{Two-dimensional confidence regions for selected combinations of the Wilson Coefficients, obtained using a likelihood profile method. The shaded regions indicate the $1\sigma$ and $3\sigma$ confidence regions considering only statistical uncertainties, while the dashed contours indicate the same regions with systematic uncertainties included. The horizontal and vertical dashed lines show the Standard Model values.}
    \label{fig:WCLHProfiles}
\end{figure}
The optimal values of the Wilson Coefficients $\mathcal{C}^{(')}_{9,10}$ and $\mathcal{C}_{9\tau}$ are listed in Table~\ref{tab:WCResults} along with $1\sigma$ statistical and systematic uncertainties. The quoted statistical uncertainties are obtained from the covariance matrix evaluated at the best fit point. The systematic uncertainties are evaluated as described in Sec.~\ref{sec:Systematics}. The corresponding one-dimensional likelihood profiles are shown in Fig.~\ref{fig:1DWCLHProfiles}, wherein the $1\sigma$, $2\sigma$, and $3\sigma$ confidence intervals are indicated considering both statistical and systematic uncertainties. The intervals obtained using the profile likelihood method are in excellent agreement with the parameter errors obtained from the covariance matrix. The SM values for the Wilson Coefficients obtained from Ref.~\cite{EOS, EOSAuthors:2021xpv} are also indicated in Fig.~\ref{fig:1DWCLHProfiles}, revealing a $2.1\sigma$ deviation in the \C9 fit result, and otherwise good agreement with SM. Two-dimensional likelihood profiles for $\mathcal{C}^{(')}_{9,10}$ are also obtained, as shown in Fig.~\ref{fig:WCLHProfiles}. The parameters of the dominant nonlocal contributions, \ie the one-particle resonance amplitudes, are listed in Tables~\ref{tab:NonLocalResults1PMagPhase} and~\ref{tab:NonLocalResults1PReIm}, and the two-particle and nonresonant contributions to the \C7 parameters are given in Table~\ref{tab:NonLocalResults2P}.
\input{tables/NonLocal1PCorrectedFitResults}
\input{tables/NonLocal2PAndDC7CorrectedFitResults}

The prior and posterior values for the local form factor parameters are given in Table~\ref{tab:LocalFFResults}. Projections of the fit on the angles as well as \qsq in the individual subregions can be found in Fig.~\ref{fig:signal_fit_result} in Appendix~\ref{app:FitProjInQ2Regions}. 

\input{tables/LocalFFCorrectedFitResults}

\section{Discussion}
\label{sec:Discussion}

The primary observation to be made based on the results of Sec.~\ref{sec:Results} is that while the nonlocal model used in this analysis shows that there is some contribution of nonlocal amplitudes in the \qsq regions used by previous binned analyses~\cite{LHCb-PAPER-2020-002}, it still prefers a value of \C9 that is shifted from the SM expectation. Based on a one-dimensional profile likelihood scan, shown in Fig.~\ref{fig:1DWCLHProfiles}, a shift of $\Delta \mathcal{C}^{\text{NP}}_{9} = -0.71 \pm 0.33$ is observed that corresponds to a $2.1\sigma$ deviation from the SM prediction of $\mathcal{C}^{\text{SM}}_{9} = 4.27$~\cite{EOS, EOSAuthors:2021xpv}, with both statistical and systematic uncertainties accounted for. The global significance of the deviation from the SM considering all of the Wilson Coefficients in Table~\ref{tab:WCResults} is reduced to $1.5\sigma$. This dilution of the statistical significance is due to the lack of a significant fit quality improvement when introducing the possibility of NP in Wilson Coefficients \C10 , \Cp10 , \Cp9 and $\C{9\tau} $, compared to only allowing for NP in the Wilson Coefficient \C9 .
No significant deviation in the Wilson Coefficient \C10 is observed, nor any evidence for the presence of right-handed currents.

This is the first direct measurement of the Wilson Coefficient $\mathcal{C}_{9\tau}$, and the value of \mbox{$\mathcal{C}_{9\tau} = (-1.2 \pm 2.6 \pm 1.0) \times 10^{2}$} is consistent with both zero and the SM expectation of lepton flavour universality $\mathcal{C}^\text{SM}_{9\tau} = 4.27$~\cite{EOS, EOSAuthors:2021xpv}. The uncertainty of the $\mathcal{C}_{9\tau}$ parameter is dominated by statistical effects. The largest systematic uncertainty, accounting for $\sim30\%$ of the total uncertainty, arises from the constraint on the relative size of the  $B^0 \to D^{(*)}\overline{D}^{(*)} K^{*0}$ contributions, as detailed in Sec.~\ref{sec:nonlocalParam}. The development of theory calculations that can be used to constrain the $B^0\to D^{(*)}\overline{D}^{(*)}(\to\mumu)\Kstarz$ amplitudes would help improve sensitivity to the Wilson Coefficient
$\mathcal{C}_{9\tau}$ in future measurements. 

The current best upper limit on the $\mathcal{B}(B^0 \to K^{*0}\tau^+\tau^-)$ branching fraction is $3.1\times 10^{-3}$ at 90\% Confidence Level~\cite{Belle:2021ecr} (CL), corresponding to an upper limit of $|\mathcal{C}_{9\tau}|<680$ at 90\% CL~(assuming no NP contribution in the $\mathcal{C}_{10\tau}$ coefficient) or $|\mathcal{C}_{9\tau}|<600$ (assuming the relation $\mathcal{C}_{10\tau} = - \mathcal{C}_{9\tau}$). The 90\% CL upper limit on the $|\mathcal{C}_{9\tau}|$ parameter from this work is $|\mathcal{C}_{9\tau}|<500$ ($|\mathcal{C}_{9\tau}|<600$ at 95\% CL). To convert the upper limits on the $B^0 \to K^{*0}\tau^+\tau^-$ branching fraction in Ref.~\cite{Belle:2021ecr} to upper limits on the parameter $|\mathcal{C}_{9\tau}|$, the  \texttt{flavio} package~\cite{Straub:2018kue} is used, with local \decay{B^{0}}{K^{*0}} form factors from Ref.~\cite{Gubernari:2022hxn} and subleading effects parameterised as in Ref.~\cite{Altmannshofer:2014rta}.  

A number of cross-checks are performed to validate the results of this analysis. The description of the dominant nonlocal amplitudes, \ie those of the \jpsi and \psitwos resonances, is validated by comparing the fitted amplitude parameters and resulting angular observables to those measured in previous analyses. To this end, the angular observables $F_L$, $S_3$, $S_4$, $S_8$, and $S_9$ are calculated at the \jpsi pole mass, and compared along with the magnitudes and phases $|A_{\parallel,\perp}^{\jpsi}|$ and $\delta_{\parallel, \perp} ^{\jpsi}$, to the results reported by \lhcb~\cite{LHCb-PAPER-2013-023}. Agreement within $1.5\sigma$ is observed between the two measurements for all observables, magnitudes, and phases. The measured magnitudes and phases of $\BdToJpsiKst$ and $\BdToPsitwosKst$ transitions are also in good agreement with previous amplitude analyses performed by \belle~\cite{Belle:2014nuw,Belle:2013shl}, once the systematic uncertainties due to the presence of $T_{c\overline{c}1}(4430)^+$ and $T_{c\overline{c}1}(4200)^+$ states are accounted for. 

In order to check that the model used in this analysis is complete regarding its description of the nonlocal contributions, an alternative fit is performed in which the values of coefficients \C9 and \C10 are allowed to carry a linear dependence on \qsq. Specifically, the following replacements are made,
\begin{align}
\mathcal{C}_{9}^{q^{2}} &= \mathcal{C}_{9}+\alpha(\qsq-q^2_{\text{mid}}), & & \mathcal{C}_{10}^{q^{2}} = \mathcal{C}_{10}+\beta(\qsq-q^2_{\text{mid}}),
\end{align}
where $q^2_{\text{mid}}=8.95~\gevgevcccc$ and denotes the middle of the fitted $q^2$ range. Statistically significant nonzero values of $\alpha$ and/or $\beta$ would imply an incorrect description of the nonlocal contributions since a \qsq dependent shift is not consistent with being of local origin. Allowing for this linear dependence in the fit does not significantly alter the values for \C9 and \C10 , and results in $\alpha=0.029\pm0.082$, $\beta=-0.058\pm 0.026$, where the uncertainties are statistical only. No evidence for an incorrect description of the nonlocal contributions to the \C9 parameter is observed, while for the parameter \C10, which receives only local contributions in the model, a $2.2\sigma$ deviation from zero is observed in the $\beta$ slope parameter. This could point to an inconsistency in form factors between the low and high \qsq regions but this is not explored and no systematic uncertainty is assigned due to this effect.

The results of the fit are also cross-checked for different choices of the dispersion relation subtraction point $q^2_0$ which serves as additional validation of the nonlocal model. The subtraction constant $Y_{c\bar{c}}(q^{2}_{0})$ enters Eq.~\ref{eqn:dispersionRelation} as a constant offset to \C9 and is degenerate with a NP contribution. In principle, the dispersion relation of Eq.~\ref{eqn:dispersionRelation} is exact and should be independent of the number and location of subtractions, provided the subtraction point is within the region in which $Y_{c\bar{c}}(q^{2}_{0})$ can be calculated reliably, \ie $q^2_0 < 0$. A deviation from this behaviour would reveal itself as a change in the \C9 fit results dependent upon the chosen subtraction point. This would indicate a problem in either the calculation of $Y_{c\bar{c}}(q^{2}_{0})$ or in the extrapolation to physical \qsq values via the dispersive integral --- that is, a problem with the parameterisation of the spectral densities used in this analysis. To check this, the fit is run twice with subtractions at $q^2_0 = -1 \gevgevcccc$ and $q^2_0 = -10 \gevgevcccc$ and the results are compared to the baseline fit with the subtraction at $q^2_0 = -4.6 \gevgevcccc$. The change in the \C9 parameter is found to be $\sim 0.1$ in both cases which is approximately $35\%$ of the statistical uncertainty. Therefore, within the precision of this measurement, the choice of subtraction point is found to have a negligible impact on the results.
 
\begin{figure}[t!]
    \centering
    \includegraphics[width=0.49\linewidth]{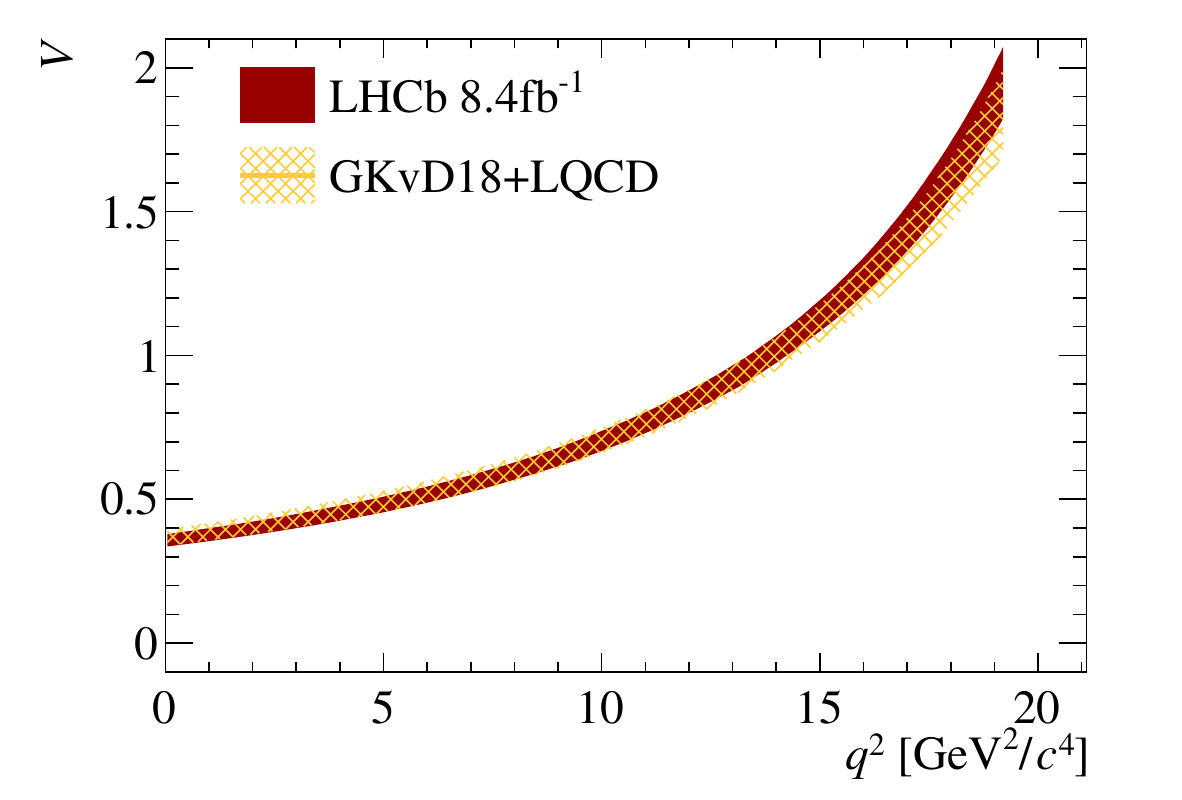}
    \includegraphics[width=0.49\linewidth]{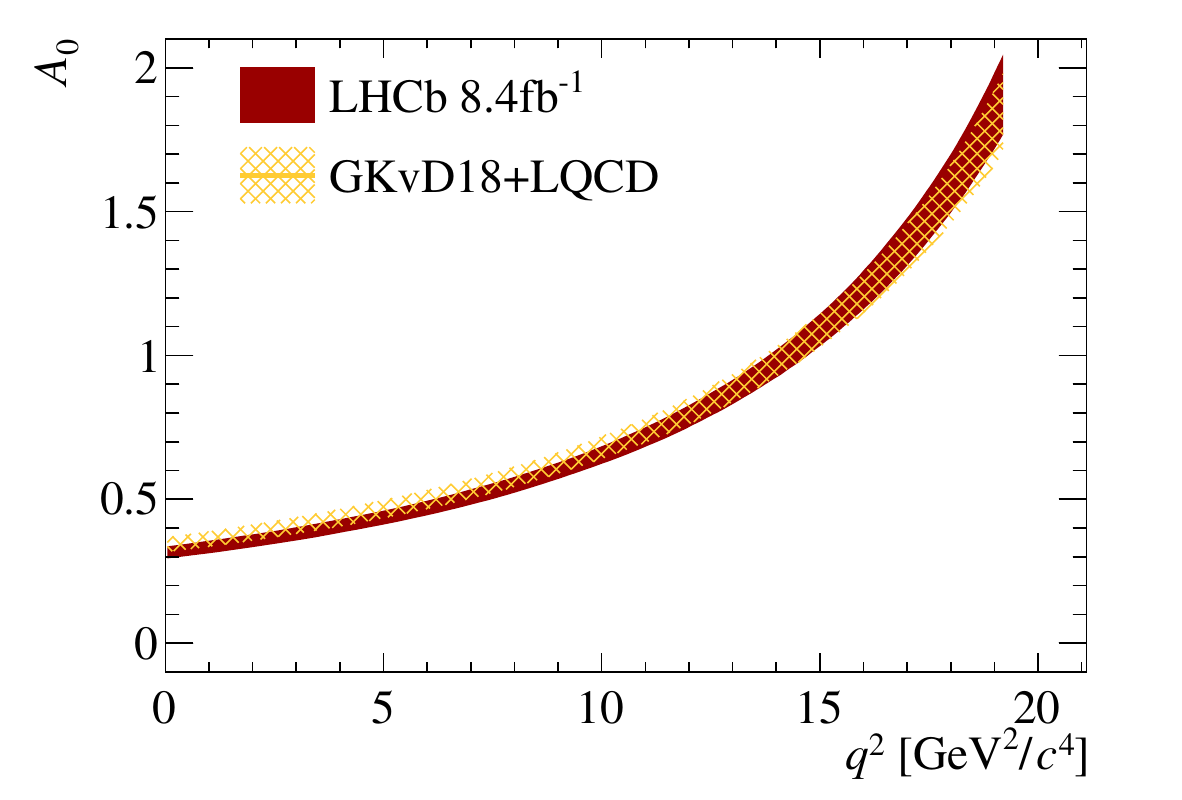}\\
    \includegraphics[width=0.49\linewidth]{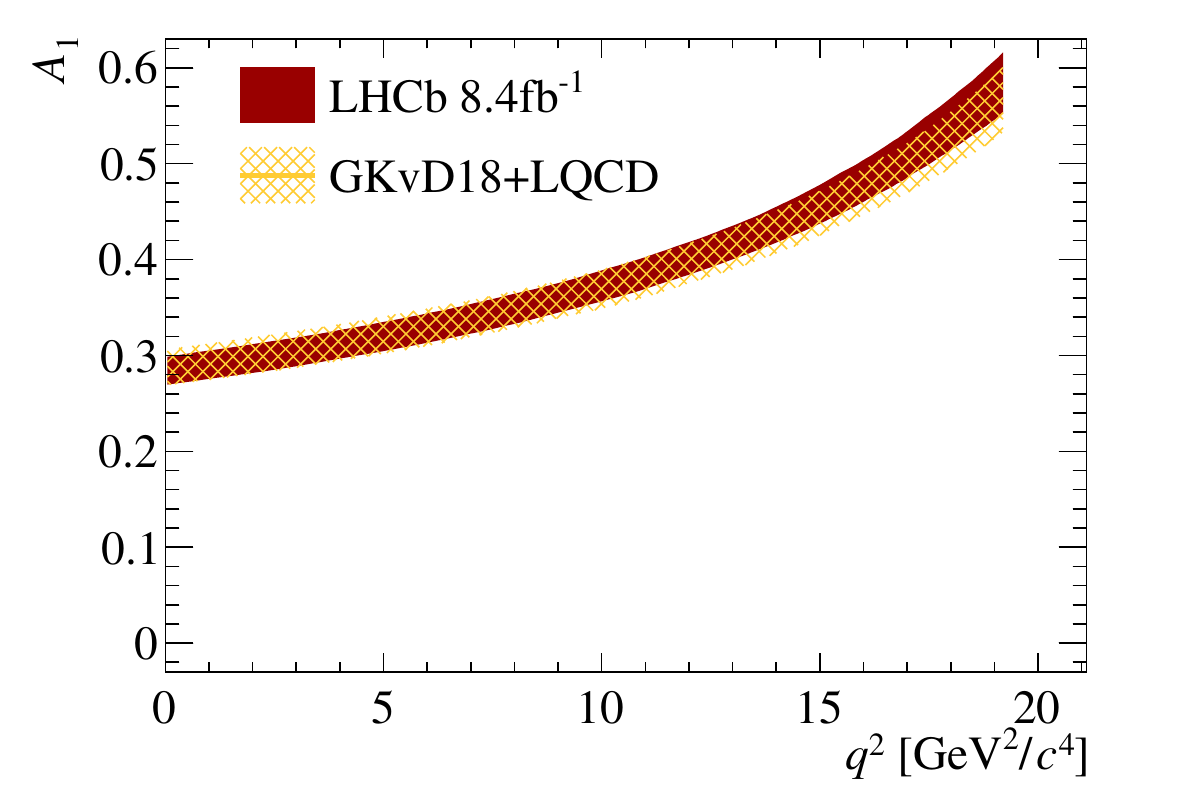}
    \includegraphics[width=0.49\linewidth]{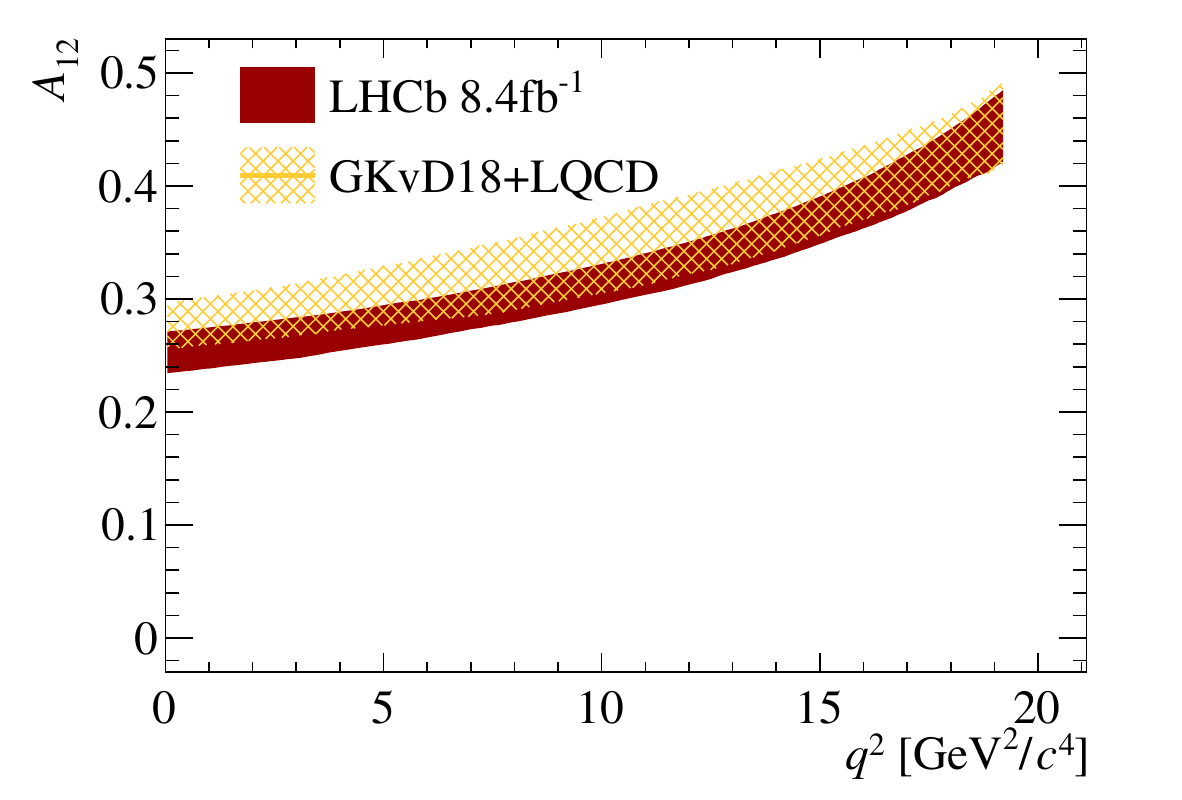}\\
    \includegraphics[width=0.49\linewidth]{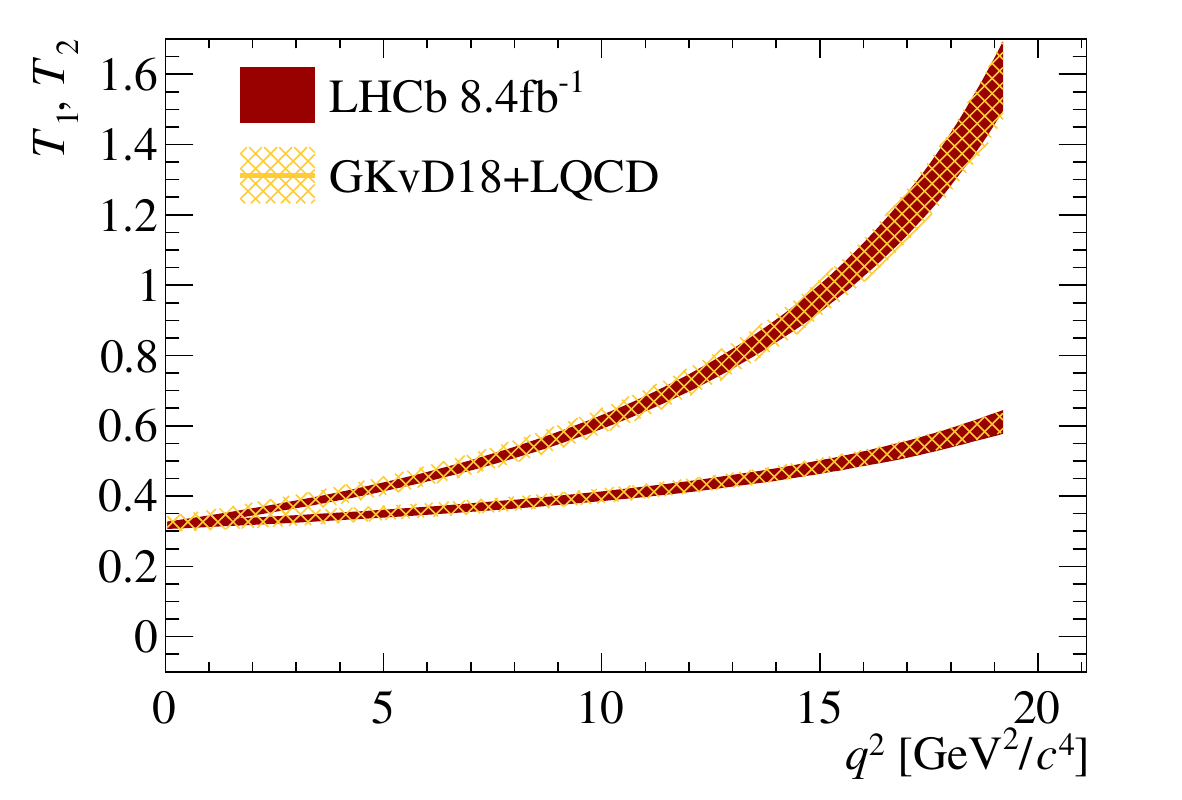}
    \includegraphics[width=0.49\linewidth]{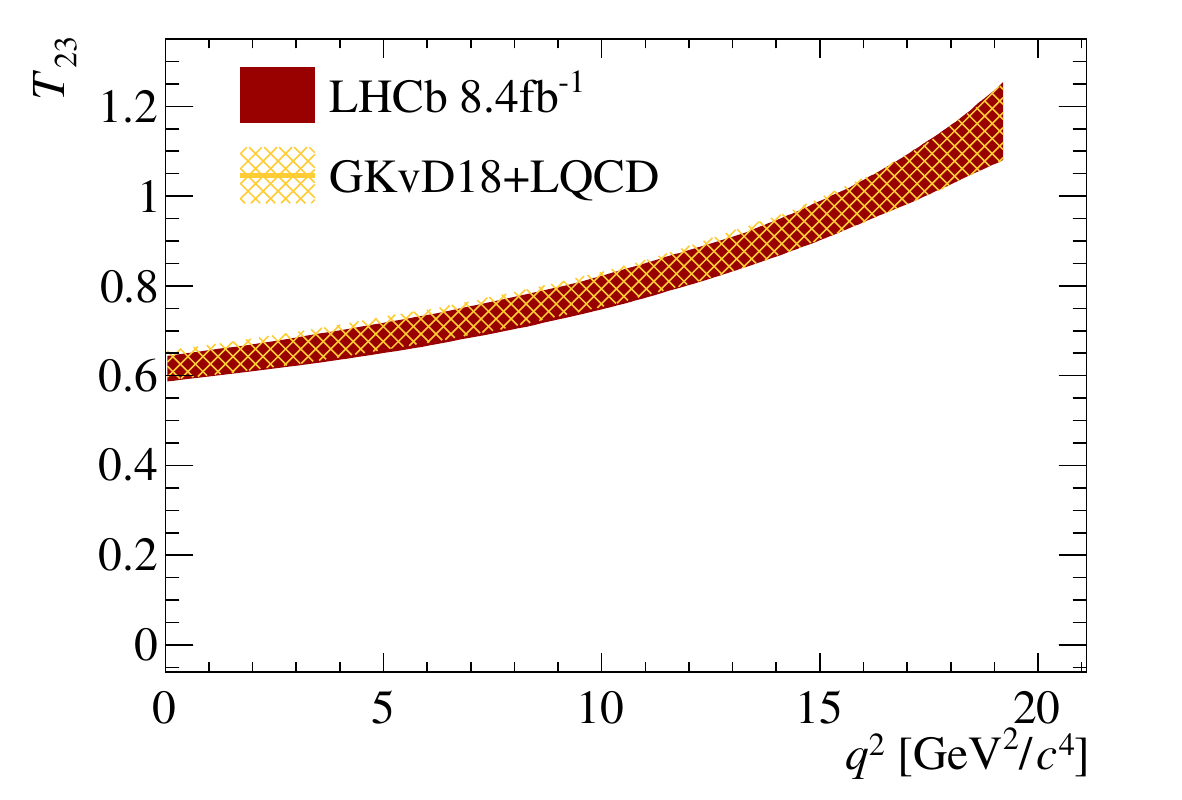}
    \caption{Comparison of form factors (orange) prefit and (maroon) postfit. The bands denote the 68\% intervals from varying the form factors according to the postfit and prefit covariance matrices, respectively. Only the statistical uncertainty is accounted for in the postfit intervals.}
    \label{fig:fits_ffs_post_fit}
\end{figure}

To investigate the sensitivity of the fit to the local form factor constraints, an alternative set of SM predictions from Ref.~\cite{Bharucha:2015bzk} is used to constrain the form factors. The main difference between the two sets of form-factor predictions are the LCSR inputs, leading to slight differences in the central values and widths of the constraints in this alternative fit. Modifying the constraint results in a non-negligible shift in the Wilson Coefficients. The effect is approximately 35\% of the statistical uncertainty for \C9 and 90\% for \C10. This difference is visible due to the improved precision of the measurement presented here. Further advances in the calculation of the local form factors are necessary to resolve these differences. Plots of the baseline local form factors as a function of \qsq are shown in Fig.~\ref{fig:fits_ffs_post_fit}, where only the statistical uncertainty is shown. The statistical precision of the data provides some mild overconstraining power and in some cases prefers slightly modified central values; however the global difference between the prefit and postfit form factors, evaluated using the change in $\chi^{2}$ of the Gaussian constraint, is negligible.

\begin{figure}[t!]
    \centering
    \includegraphics[width=0.49\linewidth]{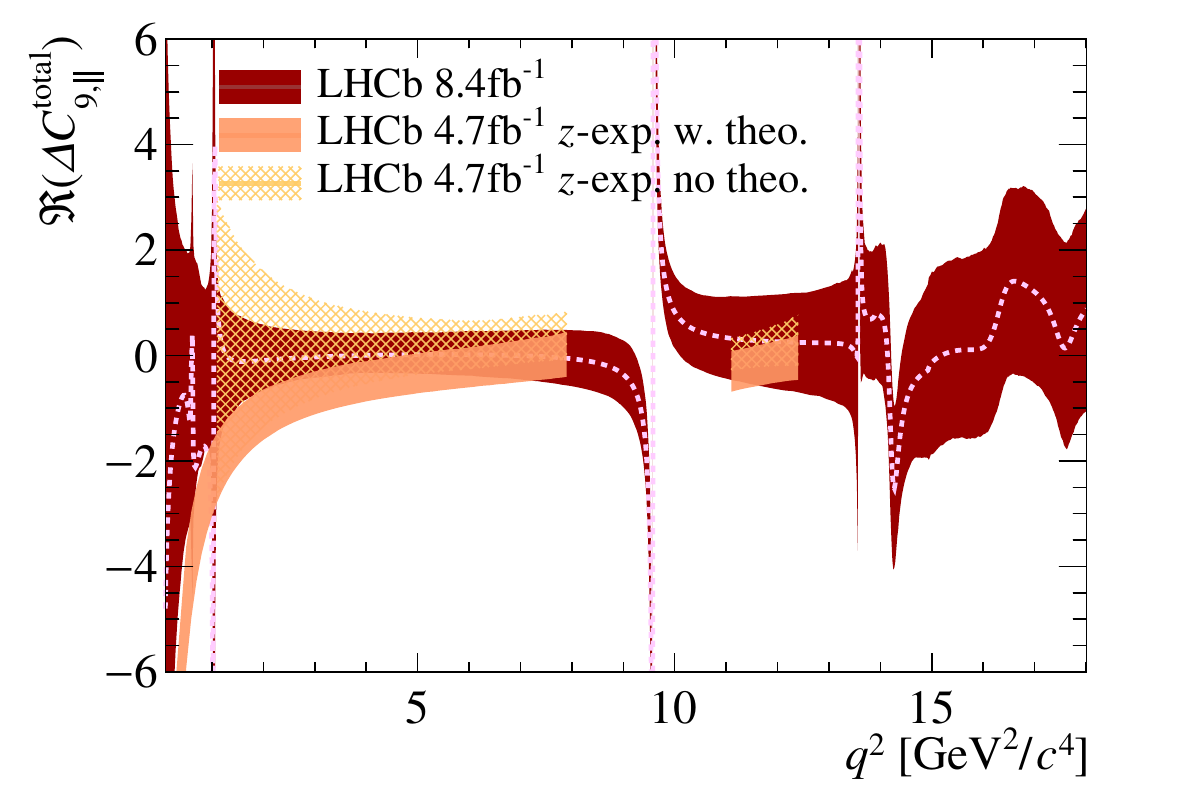}
    \includegraphics[width=0.49\linewidth]{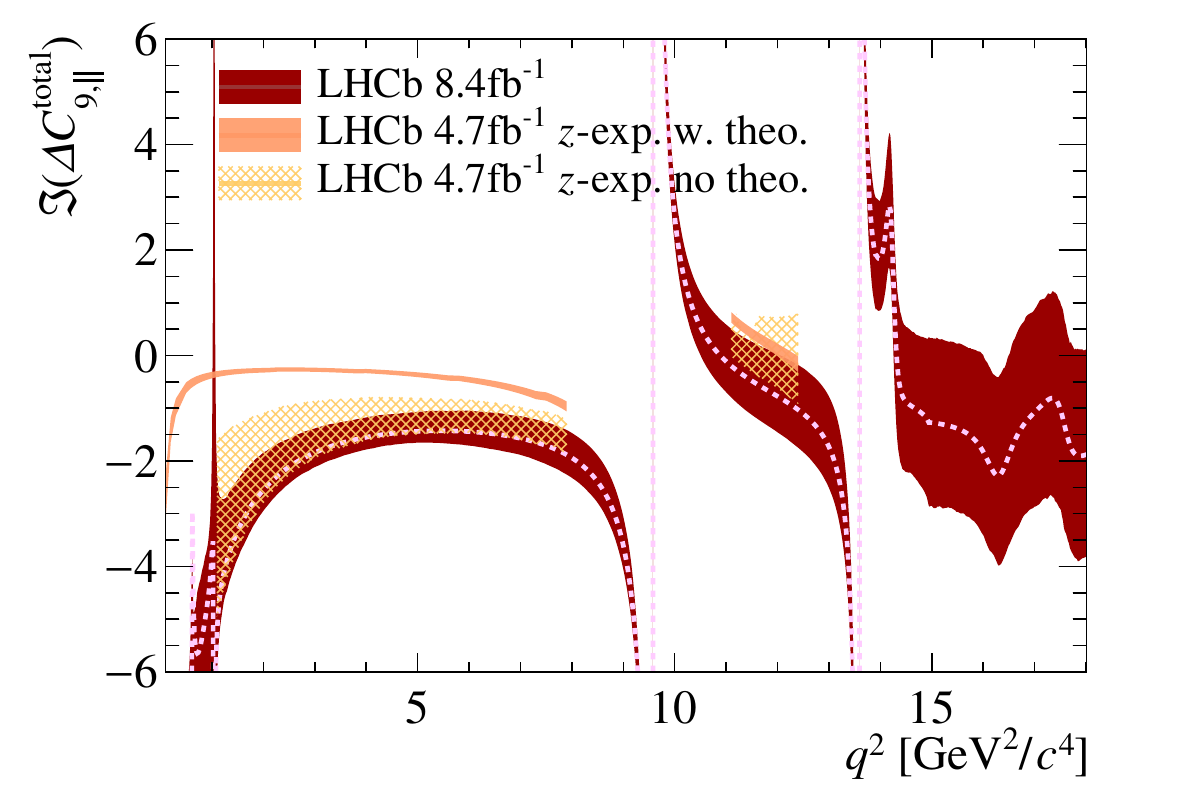}\\
    \includegraphics[width=0.49\linewidth]{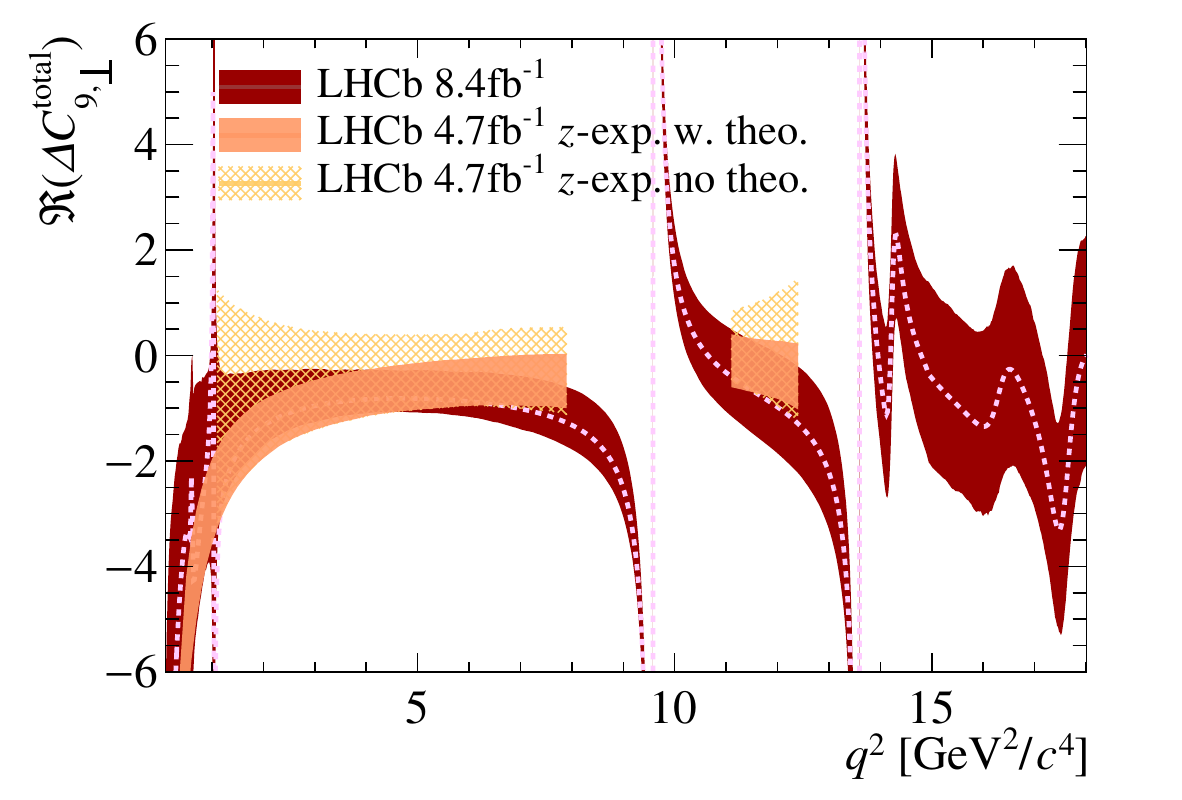}
    \includegraphics[width=0.49\linewidth]{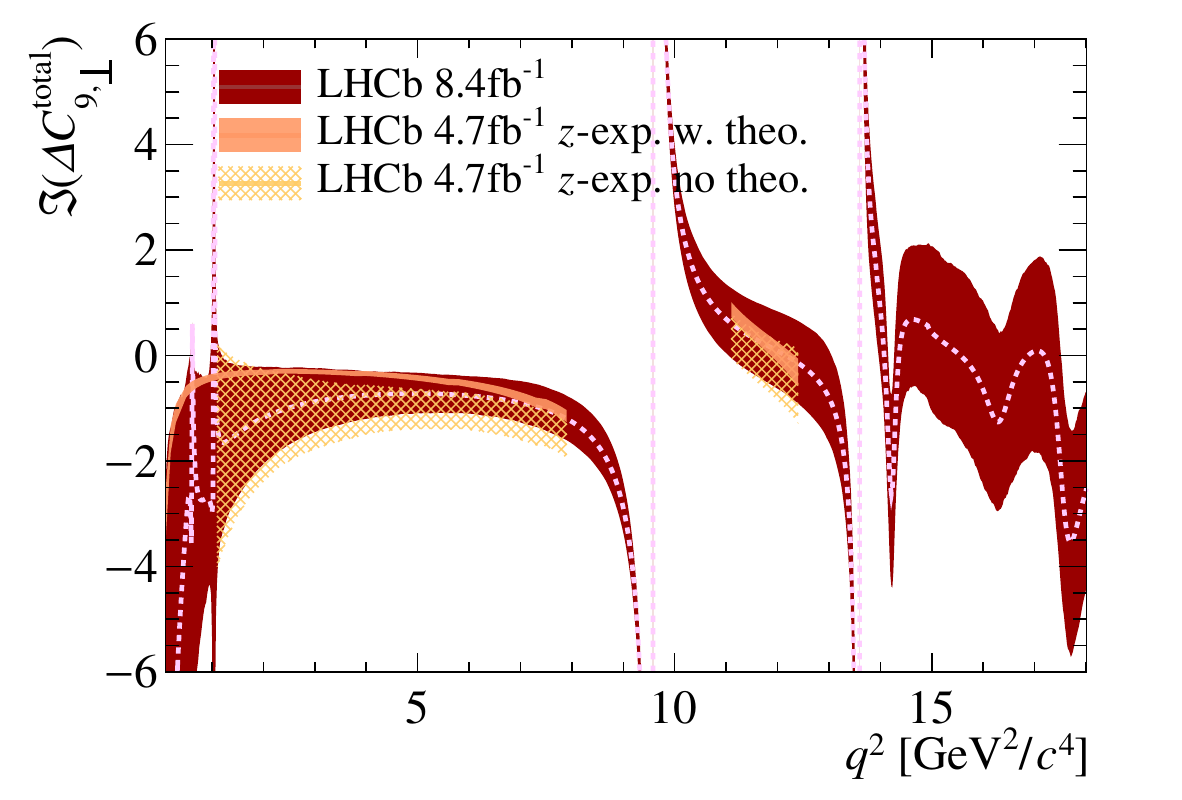}\\
    \includegraphics[width=0.49\linewidth]{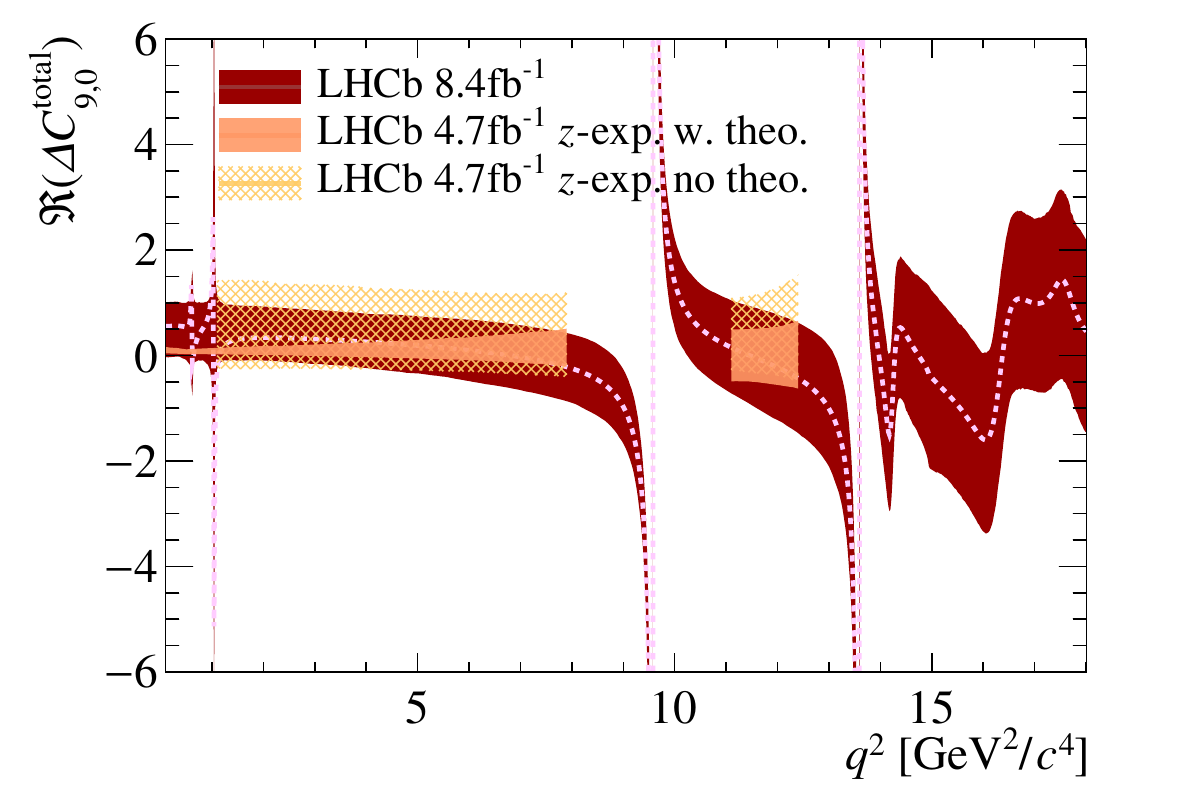}
    \includegraphics[width=0.49\linewidth]{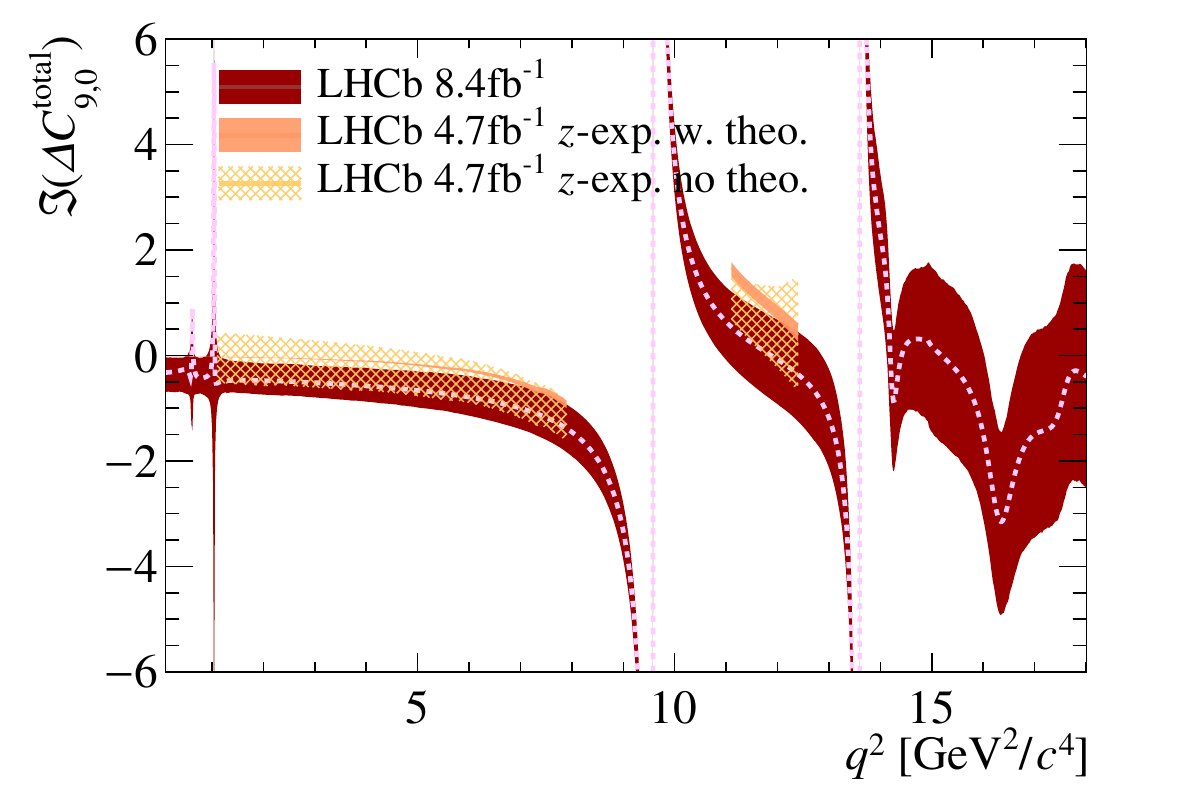}
    \caption{The nonlocal contributions from (maroon) this analysis that includes one- and two-particle hadronic amplitudes expressed as shifts to $\mathcal{C}_9$. The contributions from the $\Delta \mathcal{C}_7^{\lambda}$ terms are also included, but the tau-loop contribution is excluded. The shaded bands indicate 68\% confidence regions from varying the fit parameters according to the covariance matrix accounting for both statistical and systematic uncertainties. The results of $z$-expansion fits~\cite{Bobeth:2017vxj} from the $4.7\invfb$ LHCb analysis~\cite{LHCb-PAPER-2023-032} are also shown (pink) with and (yellow) without theory input from $\qsq<0$. See text for more detail.}
    \label{fig:fits_nonlocal_post_fit}
\end{figure}
The results of the nonlocal hadronic amplitudes, expressed as polarisation-dependent shifts to the \C9 parameter are shown in Fig.~\ref{fig:fits_nonlocal_post_fit}. A comparison is made to the nonlocal amplitudes measured using 4.7\invfb of \lhcb data~\cite{LHCb-PAPER-2023-032} that employed a polynomial expansion in the $z$ parameter, defined similarly to that shown in Eq.~\ref{eq:zexpansion} and relies on the analytical properties of these functions in the range $\qsq\in(1.1,8.0)\cup(11.0,12.5)\gevgevcccc$. In the measurement of Ref.~\cite{LHCb-PAPER-2023-032} two fits were considered. One fit that relied on a simultaneous fit to both \lhcb data in the region and theory calculations at $\qsq<0$ using an expansion up to fourth order in $z$, and another fit only to \lhcb data using an expansion up to second order in $z$. In contrast to the study of Ref.~\cite{LHCb-PAPER-2023-032}, the model used in this analysis gives access to the entire \qsq range of $\BdToKstmm$ decays. A good agreement is seen in the real part of nonlocal amplitudes between all three fit variations. However, it is clear that the data prefers large $\Im(\Delta \mathcal{C}_{9,\parallel}^{\rm total})$ contributions, that cannot be accommodated by the theory inputs at $\qsq<0$ for the $z$-expansion fit.

\begin{figure}[t!]
    \centering
    \begin{subfigure}{0.49\linewidth}
        \centering
        \includegraphics[width=\linewidth]{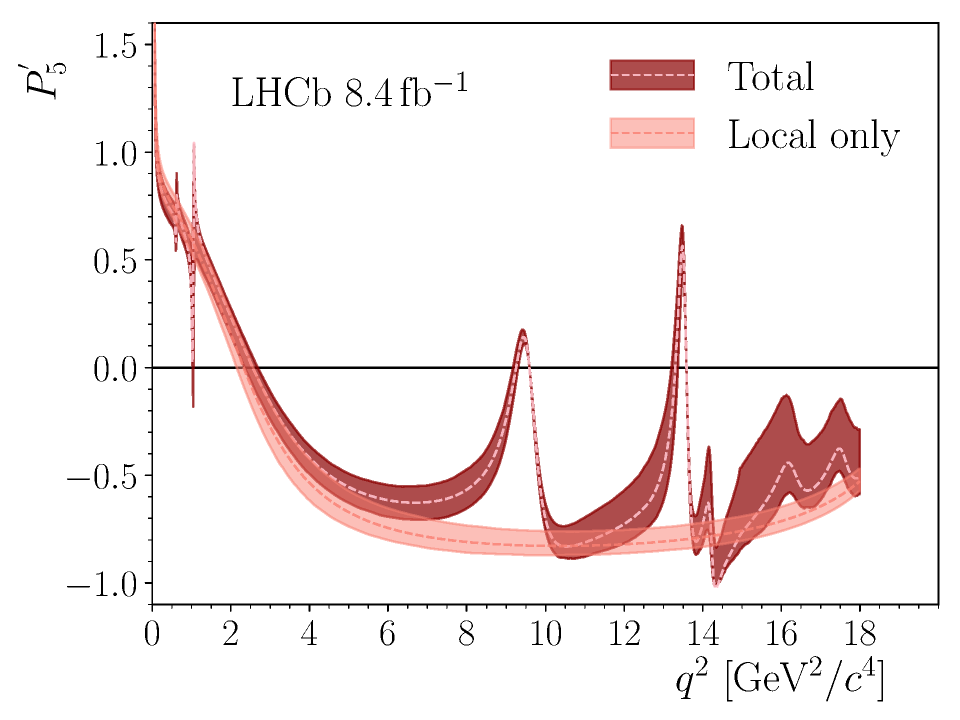}
        \caption{}
        \label{subfig:Run1Run2_Obs_unbinned}
    \end{subfigure}
    \begin{subfigure}{0.49\linewidth}
        \centering
        \includegraphics[width=\linewidth]{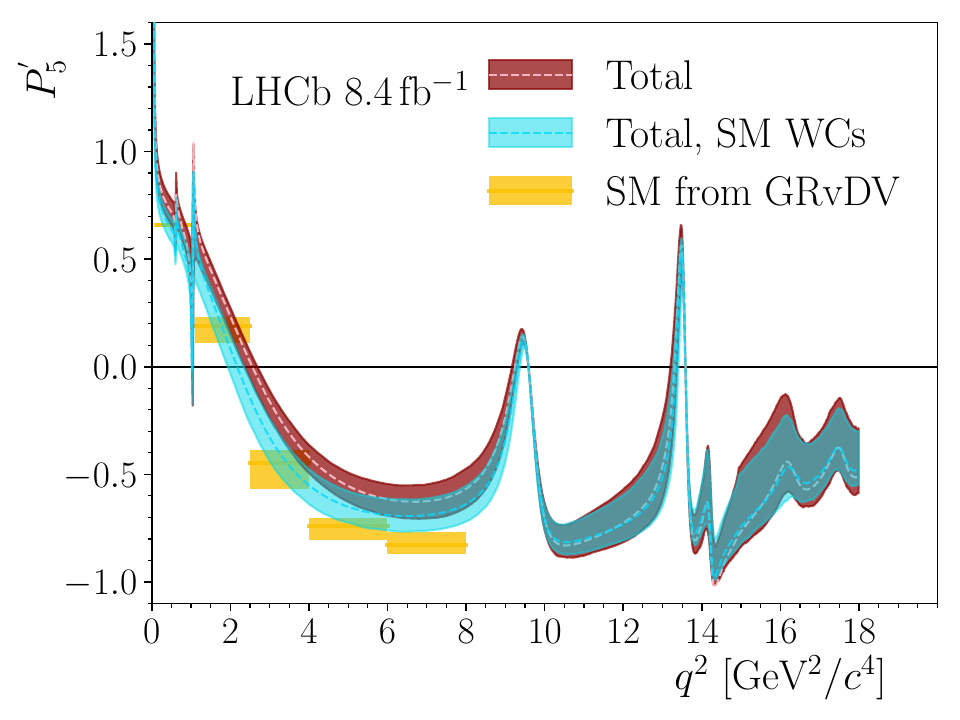}
        \caption{}
        \label{subfig:Run1Run2_Obs_binned}
    \end{subfigure}
    \caption{Distributions of the observable $P'_{5}$ constructed out of the signal parameters from the baseline fit to data. In~\subref{subfig:Run1Run2_Obs_unbinned} the distribution is shown both with and without the nonlocal contributions included in the amplitudes. In~\subref{subfig:Run1Run2_Obs_binned} the distribution is shown for the baseline fit to data, and with the Wilson Coefficients (WCs) set to their SM values. The shaded bands indicate 68\% confidence regions from varying the fit parameters according to the covariance matrix accounting for both statistical and systematic uncertainties. These are compared against SM predictions obtained from Ref.~\cite{Gubernari:2022hxn}.}
  \label{fig:Run1Run2_Obs}
\end{figure}

\begin{figure}[t!]
    \centering
    \begin{subfigure}{0.49\linewidth}
        \centering
        \includegraphics[width=\linewidth]{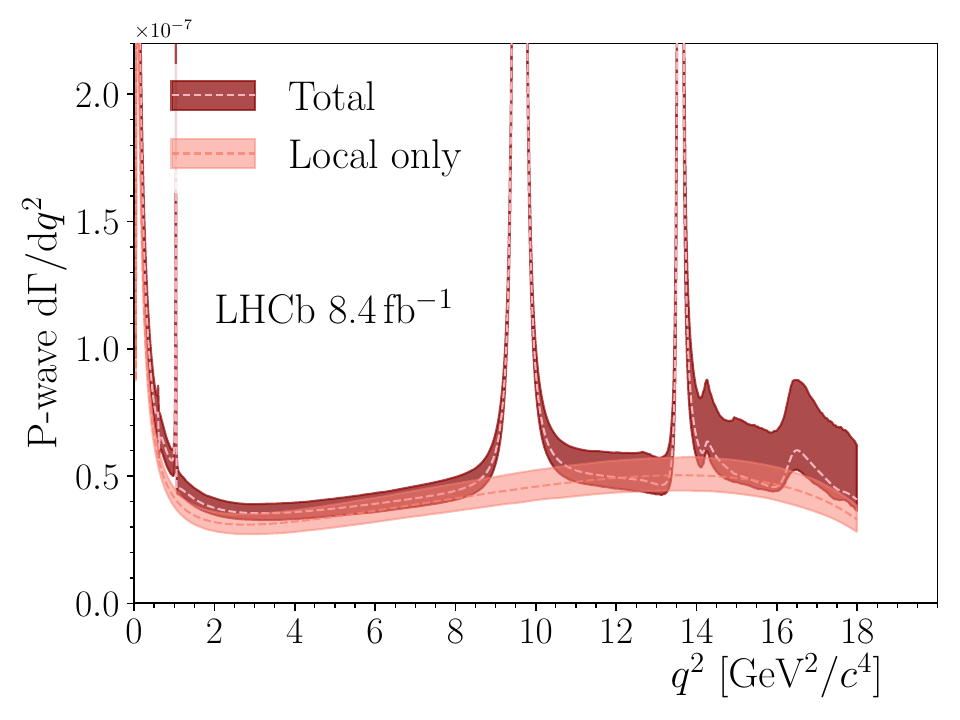}
        \caption{}
        \label{subfig:Run1Run2_dBdq2_unbinned}
    \end{subfigure}
    \begin{subfigure}{0.49\linewidth}
        \centering
        \includegraphics[width=\linewidth]{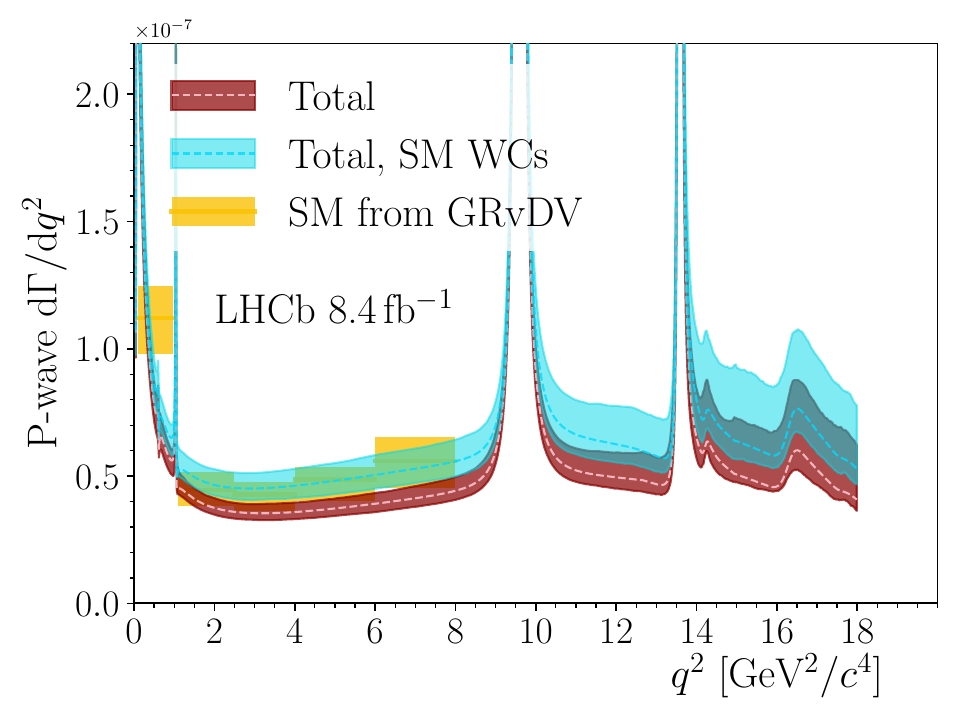}
        \caption{}
        \label{subfig:Run1Run2_dBdq2_binned}
    \end{subfigure}
    \caption{Distributions of the P-wave differential branching fraction $\deriv\Gamma/\deriv\qsq$ constructed out of the signal parameters from the baseline fit to data. In~\subref{subfig:Run1Run2_Obs_unbinned} the distribution is shown both with and without the nonlocal contributions included in the amplitudes. In~\subref{subfig:Run1Run2_Obs_binned} the distribution is shown for the baseline fit to data, and with the Wilson Coefficients~(WCs) set to their SM values. The shaded bands indicate 68\% confidence regions from varying the fit parameters according to the covariance matrix accounting for both statistical and systematic uncertainties. These are compared against SM predictions obtained from Ref.~\cite{Gubernari:2022hxn}. The shaded bands indicate $1\sigma$ confidence regions.}
  \label{fig:Run1Run2_dBdq2}
\end{figure}

Figures~\ref{fig:Run1Run2_Obs} and~\ref{fig:Run1Run2_dBdq2} show the role of the nonlocal contributions in the observable $P_{5}'$ and the differential branching fraction, respectively. The observables are plotted for only the signal component of the model with the effects of the detector resolution and acceptance removed. The nonlocal components are set to zero in the model when constructing the observables in order to plot only the local contributions, as shown in Fig.~\ref{subfig:Run1Run2_Obs_unbinned} for $P_{5}'$ and Fig.~\ref{subfig:Run1Run2_dBdq2_unbinned} for the differential branching fraction, $\deriv\Gamma/\deriv\qsq$. The local-only observables evidently differ from the total across much of the \qsq spectrum, including within the bins used in previous analyses~\cite{LHCb-PAPER-2020-002}. By setting the Wilson Coefficients to their SM values, SM ``postdictions'' of the angular observables can be computed from the signal parameters returned by the baseline fit to the data. The resulting observables are constructed using the nonlocal contributions derived from data in this analysis and can be compared to the formal SM predictions from Ref.~\cite{Gubernari:2022hxn}, as shown in Figs.~\ref{subfig:Run1Run2_Obs_binned} and~\ref{subfig:Run1Run2_dBdq2_binned}. The SM observable postdictions of this analysis have central values closer to those of the data fit results for the total observables, indicating that the data prefer larger nonlocal contributions than the formal SM computations. This is in agreement with the distributions of the nonlocal amplitudes shown in Fig.~\ref{fig:fits_nonlocal_post_fit}. Nevertheless, the SM postdictions also have different central values to the baseline fit that are closer to the SM predictions. The latter observation indicates that the nonlocal contributions, while important, are not sufficient to explain the deviation seen in the measured value of \C9.

Overall, this set of results is consistent with those reported in recent global analyses of \bsll decays~\cite{Alguero:2023jeh}, which favour lepton flavour universal NP contributions to the Wilson Coefficient \C9 . Moreover, they are consistent with the findings of other complementary analyses investigating the effect of the nonlocal contributions in \decay{\Bd}{K^{*}\ellell} decays~\cite{Bordone:2024hui,LHCb-PAPER-2023-033} which also found them to be of only minor importance.
\newpage

\section{Conclusion}
\label{sec:Conclusion}

An amplitude analysis of the decay \BdToKstmm in the reconstructed \qsq range of $0.1 \leq \qsq \leq 18.0 \gevgevcccc$ is performed for the first time using \lhcb data. The analysis employs a model of one- and two-particle nonlocal amplitudes to explicitly isolate the local and nonlocal contributions to the decay and capture the interference between them. In doing so, direct measurements of the $\bquark \to \squark \mumu$ Wilson Coefficients $\mathcal{C}^{(')}_{9,10}$ are obtained, as well as a first ever direct measurement of the Wilson Coefficient $\mathcal{C}_{9\tau}$. The values of \Cp9 , \C10 , \Cp10 , and $\mathcal{C}_{9\tau}$ are all found to be consistent with the SM, while a $2.1\sigma$ deviation is observed in the \C9 parameter. The observed shift in the value of \C9 is found to be independent of \qsq, but has a slight dependence on the local form factor constraints used. These results agree with the interpretations of previous binned angular analyses. Although the nonlocal contributions play a clear role in the angular distribution of \BdToKstmm decays, the tension in the measured value of \C9 persists. There is also agreement with the findings of the prior complementary analysis focusing on the effect of the nonlocal contributions in $\decay{\Bd}{K^{*0}\ellell}$ decays. The results of this analysis are obtained using all available information in the final state and cannot be combined with any other \lhcb measurement of the angular observables or the branching fraction of the same or partially the same dataset.

%% file: tables/WCCorrectedFitResults.tex
\begin{table}[t]
\centering
\caption{Results for the Wilson Coefficients. The first uncertainty is statistical, while the second is systematic.}
   \begin{tabular}{l  l }
       \hline
       \multicolumn{2}{c}{Wilson Coefficient results} \\
       \hline
\C9 & $\phantom{-}3.56 \pm 0.28 \pm 0.18$ \\
\C{10} & $-4.02 \pm 0.18 \pm 0.16$ \\
\Cp9 & $\phantom{-}0.28 \pm 0.41 \pm 0.12$ \\
\Cp{10} & $-0.09 \pm 0.21 \pm 0.06$ \\
\C{9\tau} & $(-1.2 \pm 2.6 \pm 1.0) \times 10^{2}$ \\
       \hline
   \end{tabular}
   \label{tab:WCResults}
\end{table}

%% file: tables/NonLocal1PCorrectedFitResults.tex
\begin{table}[h]
\centering
\caption{Results for the (left column) magnitudes and (right column) phases of the dominant one-particle nonlocal contributions. The first uncertainty is statistical, while the second is systematic. The magnitudes, $|A_{j}^{\lambda}|$, and phases, $\delta_{j}^{\lambda}$, are defined in Eq.~\ref{eqn:dy1p}. The values of amplitude parameters that are fixed in the fit to the data appear with a dash.}
   \label{tab:NonLocalResults1PMagPhase}
   \begin{tabular}[t]{l  r l r}
       \hline
       \multicolumn{4}{c}{Nonlocal parameter results} \\
       \hline
$| A_{J/\psi}^{\parallel} |$ & $(3.98 \pm 0.01 \pm 0.15) \times 10^{-3}$ & 
$\delta_{J/\psi}^{\parallel}$ & $0.23 \pm 0.01 \pm 0.01$ \\ 
$| A_{J/\psi}^{\perp} |$ & $(3.85 \pm 0.01 \pm 0.14) \times 10^{-3}$ & 
$\delta_{J/\psi}^{\perp}$ & $-0.21 \pm 0.00 \pm 0.01$ \\ 
$| A_{J/\psi}^{0} |$ & \multicolumn{1}{c}{--} & 
$\delta_{J/\psi}^{0}$ & $-1.92 \pm 0.05 \pm 0.02$ \\ 
$| A_{\psi(2S)}^{\parallel} |$ & $(9.59 \pm 0.28 \pm 0.82) \times 10^{-4}$ & 
$\delta_{\psi(2S)}^{\parallel}$ & $0.84 \pm 0.02 \pm 0.19$ \\ 
$| A_{\psi(2S)}^{\perp} |$ & $(8.38 \pm 0.27 \pm 0.62) \times 10^{-4}$ & 
$\delta_{\psi(2S)}^{\perp}$ & $-0.44 \pm 0.02 \pm 0.11$ \\ 
$| A_{\psi(2S)}^{0} |$ & $(13.4 \pm 0.4\phantom{0} \pm 1.1\phantom{0}) \times 10^{-4}$ & 
$\delta_{\psi(2S)}^{0}$ & $-2.54 \pm 0.13 \pm 0.12$ \\ 
$|A_{\rho(770)}^{0}|$ & \multicolumn{1}{c}{--} & 
$\delta_{\rho(770)}^{0}$ & $1.38 \pm 0.53 \pm 0.65$ \\ 
$|A_{\omega(782)}^{0}|$ & \multicolumn{1}{c}{--} & 
$\delta_{\omega(782)}^{0}$ & $-0.49 \pm 0.92 \pm 0.53$ \\ 
$|A_{\phi(1020)}^{0}|$ & \multicolumn{1}{c}{--} & 
$\delta_{\phi(1020)}^{0}$ & $0.10 \pm 0.82 \pm 0.78$ \\ 
       \hline
   \end{tabular}
\end{table}

\begin{table}[h]
\centering
\caption{Results for the (left column) real and (right column) imaginary parts of the higher charmonium resonance nonlocal amplitudes as defined in Eq.~\ref{eqn:dy1p}. The first uncertainty is statistical, while the second is systematic.}
   \begin{tabular}[t]{l  r l r}
       \hline
       \multicolumn{4}{c}{Nonlocal parameter results ($ \times 10^{-5}$)} \\
       \hline
$\Re(A_{\psi(3770)}^{\parallel})$ & $3.68 \pm 1.34 \pm 0.73$ & 
$\Im(A_{\psi(3770)}^{\parallel})$ & $2.87 \pm 1.88 \pm 0.49$ \\ 
$\Re(A_{\psi(3770)}^{\perp})$ & $-3.53 \pm 1.45 \pm 0.47$ & 
$\Im(A_{\psi(3770)}^{\perp})$ & $-0.86 \pm 1.56 \pm 0.53$ \\ 
$\Re(A_{\psi(3770)}^{0})$ & $-3.14 \pm 1.39 \pm 0.60$ & 
$\Im(A_{\psi(3770)}^{0})$ & $1.67 \pm 1.54 \pm 0.62$ \\ 
$\Re(A_{\psi(4040)}^{\parallel})$ & $-2.39 \pm 1.53 \pm 0.96$ & 
$\Im(A_{\psi(4040)}^{\parallel})$ & $-0.71 \pm 1.80 \pm 1.11$ \\ 
$\Re(A_{\psi(4040)}^{\perp})$ & $-2.01 \pm 1.47 \pm 0.59$ & 
$\Im(A_{\psi(4040)}^{\perp})$ & $0.35 \pm 1.49 \pm 0.82$ \\
$\Re(A_{\psi(4040)}^{0})$ & $-5.62  \pm 1.71  \pm 1.07$ & 
$\Im(A_{\psi(4040)}^{0})$ & $1.32 \pm 1.87 \pm 0.99$ \\ 
$\Re(A_{\psi(4160)}^{\parallel})$ & $0.04 \pm 1.72 \pm 0.56$ & 
$\Im(A_{\psi(4160)}^{\parallel})$ & $1.91  \pm 1.98  \pm 1.45$ \\ 
$\Re(A_{\psi(4160)}^{\perp})$ & $-2.81 \pm 1.75 \pm 0.61$ & 
$\Im(A_{\psi(4160)}^{\perp})$ & $0.32 \pm 0.15 \pm 0.09$ \\ 
$\Re(A_{\psi(4160)}^{0})$ & $1.03 \pm 1.77 \pm 0.39$ & 
$\Im(A_{\psi(4160)}^{0})$ & $-1.66  \pm 1.67  \pm 1.04 $ \\ 
       \hline
   \end{tabular}
   \label{tab:NonLocalResults1PReIm}
\end{table}

%% file: tables/NonLocal2PAndDC7CorrectedFitResults.tex
\begin{table}[h]
\centering
\caption{Results for the parameters of the two-particle and nonresonant nonlocal contributions for the (left) real and (right) imaginary components as defined in Eqs.~\ref{eq:Y2P} and Sec.~\ref{sec:delta_c7}. The first uncertainty is statistical, while the second is systematic.}
   \begin{tabular}[t]{l  r l r}
       \hline
       \multicolumn{4}{c}{Nonlocal parameter results} \\
       \hline
$\Re(A_{D^{0}\bar{D}^{0}}^{\parallel})$ & $-0.07 \pm 0.93 \pm 0.69$ & 
$\Im(A_{D^{0}\bar{D}^{0}}^{\parallel})$ & $-0.44 \pm 0.71 \pm 0.73$ \\ 
$\Re(A_{D^{0}\bar{D}^{0}}^{\perp})$ & $-0.12 \pm 0.83 \pm 0.71$ & 
$\Im(A_{D^{0}\bar{D}^{0}}^{\perp})$ & $0.02 \pm 0.80 \pm 0.74$ \\ 
$\Re(A_{D^{0}\bar{D}^{0}}^{0})$ & $-0.33 \pm 0.91 \pm 0.70$ & 
$\Im(A_{D^{0}\bar{D}^{0}}^{0})$ & $-0.27 \pm 0.77 \pm 0.81$ \\ 
$\Re(A_{D^{*0}\bar{D}^{*0}}^{\parallel})$ & $-0.06 \pm 0.96 \pm 0.63$ & 
$\Im(A_{D^{*0}\bar{D}^{*0}}^{\parallel})$ & $-0.25 \pm 0.79 \pm 0.67$ \\ 
$\Re(A_{D^{*0}\bar{D}^{*0}}^{\perp})$ & $-0.16 \pm 0.91 \pm 0.66$ & 
$\Im(A_{D^{*0}\bar{D}^{*0}}^{\perp})$ & $-0.03 \pm 0.85 \pm 0.70$ \\ 
$\Re(A_{D^{*0}\bar{D}^{*0}}^{0})$ & $-0.17 \pm 0.95 \pm 0.66$ & 
$\Im(A_{D^{*0}\bar{D}^{*0}}^{0})$ & $-0.28 \pm 0.85 \pm 0.78$ \\ 
$\Re(A_{D^{*0}\bar{D}^{0}}^{\parallel})$ & $0.02 \pm 0.42 \pm 0.66$ & 
$\Im(A_{D^{*0}\bar{D}^{0}}^{\parallel})$ & $-0.46 \pm 0.32 \pm 0.58$ \\ 
$\Re(A_{D^{*0}\bar{D}^{0}}^{\perp})$ & $-0.24 \pm 0.42 \pm 0.70$ & 
$\Im(A_{D^{*0}\bar{D}^{0}}^{\perp})$ & $-0.11 \pm 0.39 \pm 0.61$ \\ 
$\Re(A_{D^{*0}\bar{D}^{0}}^{0})$ & $-0.51 \pm 0.41 \pm 0.68$ & 
$\Im(A_{D^{*0}\bar{D}^{0}}^{0})$ & $0.12 \pm 0.35 \pm 0.58$ \\ 
$\Re(\Delta \mathcal{C}^\parallel_7)$ & $0.00 \pm 0.03 \pm 0.02$ & 
$\Im(\Delta \mathcal{C}^\parallel_7)$ & $-0.10 \pm 0.03 \pm 0.01$ \\ 
$\Re(\Delta \mathcal{C}^\perp_7)$ & $-0.05 \pm 0.03 \pm 0.02$ & 
$\Im(\Delta \mathcal{C}^\perp_7)$ & $-0.04 \pm 0.04 \pm 0.01$ \\ 
$\Re(\Delta \mathcal{C}^0_7)$ & $0.33 \pm 0.33 \pm 0.09$ & 
$\Im(\Delta \mathcal{C}^0_7)$ & $-0.19 \pm 0.20 \pm 0.09$ \\ 
       \hline
   \end{tabular}
   \label{tab:NonLocalResults2P}
\end{table}

%% file: tables/LocalFFCorrectedFitResults.tex
    \begin{table}[h]
\centering
   \caption{Results for the local form factors. The first uncertainty is statistical, while the second is systematic. The dashed entries represent the parameters being fixed in the fit due to their degeneracy with the nonlocal $\Delta \mathcal{C}_{7}^{\perp,0}$ parameters.}
   \begin{tabular}{c  r  r}
       \hline
       \multicolumn{3}{c}{Local form-factor results} \\
       \hline
       Parameter & \multicolumn{1}{c}{Prior~\cite{Gubernari:2022hxn}} & \multicolumn{1}{c}{Posterior} \\
       \hline
$\alpha_{1}^{A_0}$ & $-1.12 \pm 0.20$ & $-1.21 \pm 0.19 \pm 0.02$ \\
$\alpha_{2}^{A_0}$ & $2.18 \pm 1.76$ & $3.23 \pm 1.69 \pm 0.18$ \\
$\alpha_{0}^{A_1}$ & $0.29 \pm 0.02$ & $0.29 \pm 0.01 \pm 0.00$ \\
$\alpha_{1}^{A_1}$ & $0.46 \pm 0.13$ & $0.40 \pm 0.10 \pm 0.01$ \\
$\alpha_{2}^{A_1}$ & $1.22 \pm 0.73$ & $1.21 \pm 0.69 \pm 0.10$ \\
$\alpha_{0}^{A_{12}}$ & $0.28 \pm 0.02$ & $0.26 \pm 0.02 \pm 0.00$ \\
$\alpha_{1}^{A_{12}}$ & $0.55 \pm 0.34$ & $0.47 \pm 0.22 \pm 0.04$ \\
$\alpha_{2}^{A_{12}}$ & $0.58 \pm 2.08$ & $0.53 \pm 1.26 \pm 0.17$ \\
$\alpha_{0}^{V}$ & $0.36 \pm 0.03$ & $0.36 \pm 0.02 \pm 0.00$ \\
$\alpha_{1}^{V}$ & $-1.09 \pm 0.17$ & $-1.09 \pm 0.17 \pm 0.01$ \\
$\alpha_{2}^{V}$ & $2.73 \pm 1.99$ & $3.93 \pm 1.74 \pm 0.25$ \\
$\alpha_{1}^{T_1}$ & $-0.95 \pm 0.14$ & $-0.94 \pm 0.14 \pm 0.01$ \\
$\alpha_{2}^{T_1}$ & $2.11 \pm 1.28$ & $2.07 \pm 1.16 \pm 0.05$ \\
$\alpha_{0}^{T_2}$ & $0.32 \pm 0.02$ & \multicolumn{1}{c}{--} \\
$\alpha_{1}^{T_2}$ & $0.60 \pm 0.18$ & $0.61 \pm 0.16 \pm 0.01$ \\
$\alpha_{2}^{T_2}$ & $1.70 \pm 0.99$ & $1.78 \pm 0.98 \pm 0.03$ \\
$\alpha_{0}^{T_{23}}$ & $0.62 \pm 0.03$ & \multicolumn{1}{c}{--} \\
$\alpha_{1}^{T_{23}}$ & $0.97 \pm 0.32$ & $0.95 \pm 0.30 \pm 0.01$ \\
$\alpha_{2}^{T_{23}}$ & $1.81 \pm 2.45$ & $1.68 \pm 2.15 \pm 0.04$ \\
       \hline
   \end{tabular}
   \label{tab:LocalFFResults}
\end{table}

%% file: acknowledgements.tex
\section*{Acknowledgements}
%
%
\noindent We are immensely grateful to Danny van Dyk and Javier Virto for the useful discussions.
We express our gratitude to our colleagues in the CERN
accelerator departments for the excellent performance of the LHC. We
thank the technical and administrative staff at the LHCb
institutes.
We acknowledge support from CERN and from the national agencies:
CAPES, CNPq, FAPERJ and FINEP (Brazil); 
MOST and NSFC (China); 
CNRS/IN2P3 (France); 
BMBF, DFG and MPG (Germany); 
INFN (Italy); 
NWO (Netherlands); 
MNiSW and NCN (Poland); 
MCID/IFA (Romania); 
MICIU and AEI (Spain);
SNSF and SER (Switzerland); 
NASU (Ukraine); 
STFC (United Kingdom); 
DOE NP and NSF (USA).
We acknowledge the computing resources that are provided by CERN, IN2P3
(France), KIT and DESY (Germany), INFN (Italy), SURF (Netherlands),
PIC (Spain), GridPP (United Kingdom), 
CSCS (Switzerland), IFIN-HH (Romania), CBPF (Brazil),
and Polish WLCG (Poland).
We are indebted to the communities behind the multiple open-source
software packages on which we depend.
Individual groups or members have received support from
ARC and ARDC (Australia);
Key Research Program of Frontier Sciences of CAS, CAS PIFI, CAS CCEPP, 
Fundamental Research Funds for the Central Universities, 
and Sci. \& Tech. Program of Guangzhou (China);
Minciencias (Colombia);
EPLANET, Marie Sk\l{}odowska-Curie Actions, ERC and NextGenerationEU (European Union);
A*MIDEX, ANR, IPhU and Labex P2IO, and R\'{e}gion Auvergne-Rh\^{o}ne-Alpes (France);
AvH Foundation (Germany);
ICSC (Italy); 
Severo Ochoa and Mar\'ia de Maeztu Units of Excellence, GVA, XuntaGal, GENCAT, InTalent-Inditex and Prog. ~Atracci\'on Talento CM (Spain);
SRC (Sweden);
the Leverhulme Trust, the Royal Society
 and UKRI (United Kingdom).

%% file: appendix.tex

\newpage
\section*{Appendices}

\appendix

\section{Angular observables and spherical harmonics}
\label{app:DecayRateFunctions}

For a P-wave $\Kp\pim$ system, the explicit forms of the angular terms in Eq.\eqref{eqn:diffdecayrate} are given by
\begin{equation}
\begin{split}
f_{1s}(\cos{\theta_\ell},\cos{\theta_K},\phi) &= \sin^2\theta_{K,}\\
f_{1c}(\cos{\theta_\ell},\cos{\theta_K},\phi) &= \cos^2\theta_{K},\\
f_{2s}(\cos{\theta_\ell},\cos{\theta_K},\phi) &= \sin^2\theta_{K}\cos2\theta_{\ell},\\
f_{2c}(\cos{\theta_\ell},\cos{\theta_K},\phi) &= \cos^2\theta_{K}\cos2\theta_{\ell},\\
f_3   (\cos{\theta_\ell},\cos{\theta_K},\phi) &= \sin^2\theta_{K}\sin^2\theta_{\ell}\cos2\phi,\\
f_4   (\cos{\theta_\ell},\cos{\theta_K},\phi) &= \sin 2\theta_{K}\sin 2\theta_{\ell}\cos\phi,\\
f_5   (\cos{\theta_\ell},\cos{\theta_K},\phi) &= \sin 2\theta_{K}\sin\theta_{\ell}\cos\phi,\\
f_{6s}(\cos{\theta_\ell},\cos{\theta_K},\phi) &= \sin^2\theta_{K}\cos\theta_{\ell},\\
f_7   (\cos{\theta_\ell},\cos{\theta_K},\phi) &= \sin 2\theta_{K}\sin \theta_{\ell}\sin\phi,\\
f_8   (\cos{\theta_\ell},\cos{\theta_K},\phi) &= \sin 2\theta_{K}\sin 2\theta_{\ell}\sin\phi,\\
f_9   (\cos{\theta_\ell},\cos{\theta_K},\phi) &= \sin^2\theta_{K}\sin^2 \theta_{\ell}\sin2\phi.
\end{split}
\end{equation}
and for the S-wave, they are given by
\begin{equation}
\begin{split}
f^{S'}_{1c}(\cos{\theta_\ell},\cos{\theta_K},\phi) &= 1,\\
f^{S'}_{2c}(\cos{\theta_\ell},\cos{\theta_K},\phi) &= \cos2\theta_{\ell},\\
f_{1c}'(\cos{\theta_\ell},\cos{\theta_K},\phi) &= \cos\theta_{K},\\
f_{2c}'(\cos{\theta_\ell},\cos{\theta_K},\phi) &= \cos\theta_{K}\cos2\theta_{\ell},\\
f_4'   (\cos{\theta_\ell},\cos{\theta_K},\phi) &= \sin \theta_{K}\sin 2\theta_{\ell}\cos\phi,\\
f_5'   (\cos{\theta_\ell},\cos{\theta_K},\phi) &= \sin \theta_{K}\sin\theta_{\ell}\cos\phi,\\
f_7'   (\cos{\theta_\ell},\cos{\theta_K},\phi) &= \sin \theta_{K}\sin \theta_{\ell}\sin\phi,\\
f_8'   (\cos{\theta_\ell},\cos{\theta_K},\phi) &= \sin \theta_{K}\sin 2\theta_{\ell}\sin\phi.
\end{split}
\end{equation}

\newcommand{\Am}{\mathcal{A}}

The corresponding angular coefficients $J_i(q^2)$ can be constructed using the transversity amplitudes $\Am_0^{L,R}$,  $\Am_\parallel^{L,R}$, $\Am_\perp^{L,R}$ and $\Am_t$ described in Sec.~\ref{sec:TransversityAmplitudes}, where contributions from scalar amplitudes are assumed to be zero. The explicit forms of the P-wave angular coefficients are given by
\begin{equation}
    \begin{split}
    J_{1s}(q^2) &=  \frac{2+ \beta_\ell^2}{4}\left( |\Am_\perp^L|^2+|\Am_\parallel^L|^2+|\Am_\perp^R|^2+|\Am_\parallel^R|^2\right) + \frac{4m_\ell^2}{q^2} \operatorname{Re}\left(\Am_\perp^L \Am_\perp^{R*}+\Am_\parallel^L \Am_\parallel^{R*} \right), \\
    J_{1c}(q^2) &= |\Am_0^L|^2 + |\Am_0^R|^2 + \frac{4m_\ell^2}{q^2}\left( |\Am_t|^2 + 2 \operatorname{Re}\left(\Am_0^L \Am_0^{R*}\right) \right),\\
    J_{2s}(q^2) &= \frac{\beta_\ell^2}{4}\left( |\Am_\perp^L|^2 +|\Am_\parallel^L|^2+|\Am_\perp^R|^2 +|\Am_\parallel^R|^2 \right),\\
    J_{2c}(q^2) &= -\beta_\ell^2 \left( |\Am_0^L|^2 +|\Am_0^R|^2 \right),\\
    J_{3}(q^2)&=\frac{\beta_\ell^2}{2} \left( |\Am_\perp^L|^2 -|\Am_\parallel^L|^2 + |\Am_\perp^R|^2 -|\Am_\parallel^R|^2 \right),\\
    J_{4}(q^2) &= -\frac{\beta_\ell^2}{\sqrt{2}} \operatorname{Re} \left( \Am_0^L\Am_\parallel^{L*} +\Am_0^R\Am_\parallel^{R*} \right),\\
    J_{5}(q^2) &= \sqrt{2}\beta_\ell \operatorname{Re} \left(\Am_0^L\Am_\perp^{L*} -\Am_0^R\Am_\perp^{R*} \right),\\
    J_{6s}(q^2) &= -2\beta_\ell \operatorname{Re} \left( \Am_\parallel^{L} \Am_\perp^{L*} - \Am_\parallel^{R} \Am_\perp^{R*} \right),\\
    J_{7}(q^2) & = -\sqrt{2}\beta_\ell\operatorname{Im}\left(\Am_0^{L}\Am_\parallel^{L*} - \Am_0^{R}\Am_\parallel^{R*} \right),\\
    J_{8}(q^2) & = \frac{\beta_\ell^2}{\sqrt{2}}\operatorname{Im}\left( \Am_0^L \Am_\perp^{L*} +\Am_0^R \Am_\perp^{R*} \right) ,\\
    J_{9}(q^2) & = -\beta_\ell^2 \operatorname{Im} \left( \Am_\parallel^{L*}\Am_\perp^L + \Am_\parallel^{R*}\Am_\perp^{R}\right), 
    \end{split}
\end{equation}
where the parameter $\beta_\ell$ is given by $\beta_\ell = \sqrt{1 - \frac{4m_\ell^2}{q^2}}$. The S-wave angular coefficients also involve the S-wave transversity amplitude $\Am_{00}^{L,R}$ in addition to the P-wave amplitudes, and are given by
\begin{equation} 
\begin{split}
    J^{S'}_{1c}(q^2) &= \frac{1}{3}\left( \left(|\Am_{00}^{L}|^2+|\Am_{00}^{R}|^2\right) + \frac{4m_\ell^2}{q^2}2 \operatorname{Re}(\Am^L_{00} \Am^{R*}_{00})\right),\\
    J^{S'}_{2c}(q^2) &= -\frac{1}{3}\beta_\ell^2\left(|\Am_{00}^{L}|^2+|\Am_{00}^{R}|^2\right),\\
    J_{1c}'(q^2) &= \frac{2}{\sqrt{3}}\operatorname{Re}\left( \Am_{00}^{L} \Am_{0}^{L*} +\Am_{00}^{R}\Am_{0}^{R*} + \frac{4m_\ell^2}{q^2}\left( \Am_{00}^{L} \Am_{0}^{R*} + \Am_{0}^{L} \Am_{00}^{R*} \right)   \right),\\
    J_{2c}'(q^2) &= -\frac{2}{\sqrt{3}} \beta_\ell^2 \operatorname{Re}\left( \Am_{00}^{L}A_{0}^{L*} + \Am_{00}^{R} \Am_{0}^{R*}\right),\\
    J_{4}'(q^2)  &= -\sqrt{\frac{2}{3}}\beta_\ell^2 \left( \operatorname{Re}\left( \Am_{00}^{L}\Am_{\parallel}^{L*}\right)+\operatorname{Re}\left( \Am_{00}^{R}\Am_{\parallel}^{R*}\right) \right),\\
    J_{5}'(q^2)  &= 2\sqrt{\frac{2}{3}}\beta_\ell^2 \left( \operatorname{Re}\left( \Am_{00}^{L}\Am_{\perp}^{L*}\right)+\operatorname{Re}\left( \Am_{00}^{R}\Am_{\perp}^{R*}\right) \right),\\
    J_{7}'(q^2)  &= -2\sqrt{\frac{2}{3}}\beta_\ell^2\left( \operatorname{Re}\left( \Am_{00}^{L}\Am_{\parallel}^{L*}\right)-\operatorname{Re}\left( \Am_{00}^{R}\Am_{\parallel}^{R*}\right) \right),\\
    J_{8}'(q^2)  &= \sqrt{\frac{2}{3}}\beta_\ell^2 \left( \operatorname{Re}\left( \Am_{00}^{L}\Am_{\perp}^{L*}\right)-\operatorname{Re}\left( \Am_{00}^{R}\Am_{\perp}^{R*}\right) \right).
\end{split}
\end{equation}

%% file: appendix-fit-parameters.tex
\newpage
\section{Fit parameters}
\label{app:fit_parameters}
The parameters of the signal model and their behaviour in the fit are shown in Table~\ref{tab:SigPars}.
\input{tables/SignalParameters.tex}

%% file: tables/SignalParameters.tex
\begin{longtable}[c]{c c c c c c}
\caption{A summary of the parameters of the signal model, including whether the parameters are freely varied, constrained, or fixed in the maximum likelihood fit. For the fixed and constrained parameters, more details can be found in the text.}
\label{tab:SigPars} \\

\endfirsthead

\endhead

\multicolumn{6}{r}{ $\hookrightarrow$ Continued on next page} \\
\endfoot

\endlastfoot

\toprule
\multicolumn{6}{ c }{Wilson Coefficients} \\
\midrule
\C9  &  \C{10}  & $\C{9\tau} $ & \Cp9  & \Cp{10}  &  \\
\midrule
\multicolumn{6}{ c }{Local form factors --- constrained (by Ref.~\cite{Gubernari:2022hxn})} \\
\midrule
${\alpha_{0}^{A_0}}$ & ${\alpha_{1}^{A_0}}$ & ${\alpha_{2}^{A_0}}$ &
${\alpha_{0}^{A_1}}$    & ${\alpha_{1}^{A_1}}$    & ${\alpha_{2}^{A_1}}$ \\ 
${\alpha_{0}^{A_{12}}}$ & ${\alpha_{1}^{A_{12}}}$ & ${\alpha_{2}^{A_{12}}}$ &
${\alpha_{0}^{V}}$      & ${\alpha_{1}^{V}}$      & ${\alpha_{2}^{V}}$  \\
${\alpha_{0}^{T_1}}$    & ${\alpha_{1}^{T_1}}$    & ${\alpha_{2}^{T_1}}$ &
${\alpha_{0}^{T_2}}$    & ${\alpha_{1}^{T_2}}$    & ${\alpha_{2}^{T_2}}$ \\
${\alpha_{0}^{T_23}}$    & ${\alpha_{1}^{T_{23}}}$ & ${\alpha_{2}^{T_{23}}}$ & & \\
\midrule
\multicolumn{6}{ c }{Nonlocal (one-particle)} \\
\midrule
$| A^{\parallel}_{J/\psi} |$    &  $| A^{\perp}_{J/\psi} |$      & $| A^{0}_{J/\psi} |$                     &
$\delta^{\parallel} _{J/\psi}$  &  $\delta^{\perp} _{J/\psi}$    & $\delta^{0} _{J/\psi}$\\

$| A^{\parallel}_{\psi(2S)} |$  &   $| A^{\perp}_{\psi(2S)} |$   & $| A^{0}_{\psi(2S)} |$ &
$\delta^{\parallel}_{\psi(2S)}$ &  $\delta^{\perp} _{\psi(2S)}$  & $\delta^{0} _{\psi(2S)}$\\
$| A^{\parallel}_{\rho(770)} |$ & $| A^{\perp}_{\rho(770)} |$ & $| A^{0}_{\rho(770)} |$ &
$\delta^{\parallel} _{\rho(770)}$  &  $\delta^{\perp} _{\rho(770)}$    &
$\delta^{0}  _{\rho(770)}$  \\
$| A^{\parallel}_{\omega(782)} |$ & $| A^{\perp}_{\omega(782)} |$ & $| A^{0}_{\omega(782)} |$ &
$\delta^{\parallel} _{\omega(782)}$  &  $\delta^{\perp} _{\omega(782)}$    &  $\delta^{0}  _{\omega(782)}$  \\ 
$| A^{\parallel}_{\phi(1020)} |$ & $| A^{\perp}_{\phi(1020)} |$ & $| A^{0}_{\phi(1020)} |$ &
$\delta^{\parallel} _{\phi(1020)}$  &  $\delta^{\perp} _{\phi(1020)}$    & $\delta^{0}  _{\phi(1020)}$  \\

$\Re(A^{\parallel} _{\psi(3770)})$ & $\Re(A^{\perp} _{\psi(3770)})$  & $\Re(A^{0} _{\psi(3770)})$ & 
$\Im(A^{\parallel}_{\psi(3770)})$  & $\Im(A^{\perp} _{\psi(3770)})$  & $\Im(A^{0} _{\psi(3770)})$  \\

$\Re(A^{\parallel} _{\psi(4040)})$ & $\Re(A^{\perp} _{\psi(4040)})$  & $\Re(A^{0} _{\psi(4040)})$ &
$\Im(A^{\parallel} _{\psi(4040)})$ & $\Im(A^{\perp} _{\psi(4040)})$  & $\Im(A^{0} _{\psi(4040)})$\\  

$\Re(A^{\parallel}_{\psi(4160)})$  &$\Re(A^{\perp}_{\psi(4160)})$ & $\Re(A^{0}_{\psi(4160)})$ &
$\Im(A^{\parallel}_{\psi(4160)})$  &$\Im(A^{\perp}_{\psi(4160)})$ & $\Im(A^{0}_{\psi(4160)})$\\
$m_\jpsi$ & $m_\psitwos$ & & & & \\
\midrule
\multicolumn{6}{ c }{Nonlocal (two-particle) --- constrained (see Sec.~\ref{sec:opencharm_constraint_systematic})} \\
\midrule
$\Re(A^{\parallel}_{D^{*0}\overline{D}^{0}})$ & $\Re(A^{\perp}_{D^{*0}\overline{D}^{0}})$ &  ${\Re(A^{0}_{D^{*0}\overline{D}^{0}})}$ &
$\Im(A^{\parallel}_{D^{*0}\overline{D}^{0}})$ & $\Im(A^{\perp}_{D^{*0}\overline{D}^{0}})$   & ${\Im(A^{0}_{D^{*0}\overline{D}^{0}})}$\\

$\Re(A^{\parallel}_{D^{0}\overline{D}^{0}})$ & $\Re(A^{\perp}_{D^{0}\overline{D}^{0}})$ &  ${\Re(A^{0}_{D^{0}\overline{D}^{0}})}$ &
$\Im(A^{\parallel}_{D^{0}\overline{D}^{0}})$ & $\Im(A^{\perp}_{D^{0}\overline{D}^{0}})$   & ${\Im(A^{0}_{D^{0}\overline{D}^{0}})}$\\

$\Re(A^{\parallel}_{D^{*0}\overline{D}^{*0}})$ & $\Re(A^{\perp}_{D^{*0}\overline{D}^{*0}})$ &  ${\Re(A^{0}_{D^{*0}\overline{D}^{*0}})}$ &
$\Im(A^{\parallel}_{D^{*0}\overline{D}^{*0}})$ & $\Im(A^{\perp}_{D^{*0}\overline{D}^{*0}})$   & ${\Im(A^{0}_{D^{*0}\overline{D}^{*0}})}$\\
\midrule
\multicolumn{6}{ c }{Nonlocal (nonresonant)} \\
\midrule
$\Re(\Delta C^\parallel_7)$      & $\Re(\Delta C^\perp_7)$         & $\Re(\Delta C^0_7)$ &
$\Im(\Delta C^\parallel_7)$      & $\Im(\Delta C^\perp_7)$         & $\Im(\Delta C^0_7)$ \\
\midrule
\multicolumn{6}{ c }{S-wave} \\
\midrule
$\mathcal{C}_7^S$ & $\mathcal{C}_9^S$ & $\mathcal{C}_{10}^S$ & 
$F(0)$ & $\alpha_F$ & $\beta_F$ \\
$| A^{00}_{J/\psi} |$ & $\delta^{00} _{J/\psi}$ & $| A^{00}_{\psitwos} |$ &
$\delta^{00} _{\psitwos}$ & & \\
\bottomrule
\pagebreak

\toprule
\multicolumn{6}{ c }{Resolution} \\
\midrule
$\alpha_{l,1}$ & $n_{l,1}$ & $\sigma_{C,1}$ &
$\alpha_{u,1}$ & $n_{u,1}$ & $\sigma_{G,1}$ \\
$\alpha_{l,2}$ & $n_{l,2}$ & $\sigma_{C,2}$ &
$\alpha_{u,2}$ & $n_{u,2}$ & $\sigma_{G,2}$ \\
$\alpha_{l,3}$ & $n_{l,3}$ & $\sigma_{C,3}$ &
$\alpha_{u,3}$ & $n_{u,3}$ & $\sigma_{G,3}$ \\
$f_{G,1}$ & $f_{G,2}$ & $f_{G,3}$ & & \\
\midrule
\multicolumn{6}{ c }{Fixed parameters} \\
\midrule
\C7      & \Cp7         &  ${\alpha_{0}^{A_0}}$  & ${\alpha_{0}^{T_1}}$  & ${\alpha_{0}^{T_2}}$  & ${\alpha_{0}^{T_{23}}}$ \\ 
$\mathcal{C}_7^S$ & $\mathcal{C}_9^S$ & $\alpha_F$ & $\alpha_{l,1}$ & $n_{l,1}$ & $\sigma_{C,1}$ \\
$\alpha_{u,1}$ & $n_{u,1}$ & $\sigma_{G,1}$ & $f_{G,1}$ & $| A^{0}_{J/\psi} |$ & $| A^{\parallel}_{\rho(770)} |$ \\
$| A^{\perp}_{\rho(770)} |$ & $| A^{0}_{\rho(770)} |$ & $\delta^{\parallel} _{\rho(770)}$  & $\delta^{\perp} _{\rho(770)}$    & $| A^{\parallel}_{\omega(782)} |$ & $| A^{\perp}_{\omega(782)} |$ \\
$| A^{0}_{\omega(782)} |$ & $\delta^{\parallel} _{\omega(782)}$  &  $\delta^{\perp} _{\omega(782)}$    & $| A^{\parallel}_{\phi(1020)} |$ &  $| A^{\perp}_{\phi(1020)} |$ & $| A^{0}_{\phi(1020)} |$ \\ 
$\delta^{\parallel} _{\phi(1020)}$  &  $\delta^{\perp} _{\phi(1020)}$ & & & \\
\bottomrule
\end{longtable}

%% file: appendix-plots.tex
\newpage
\section{Plots of the full set of angular observables}
\begin{figure}
    \centering
    \includegraphics[width=0.47\linewidth]{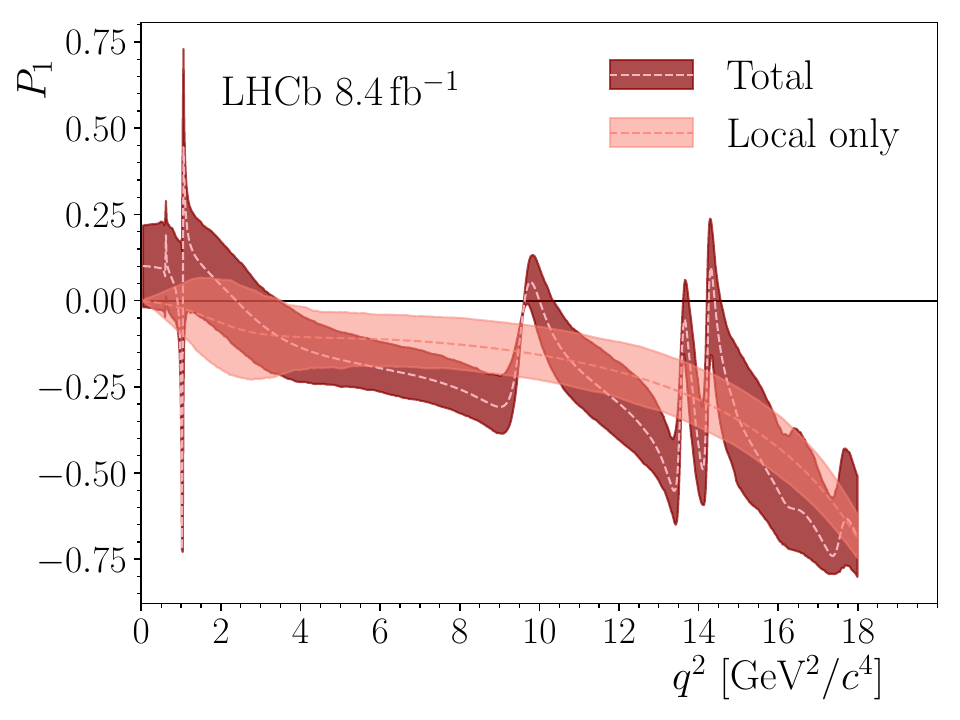}
    \includegraphics[width=0.47\linewidth]{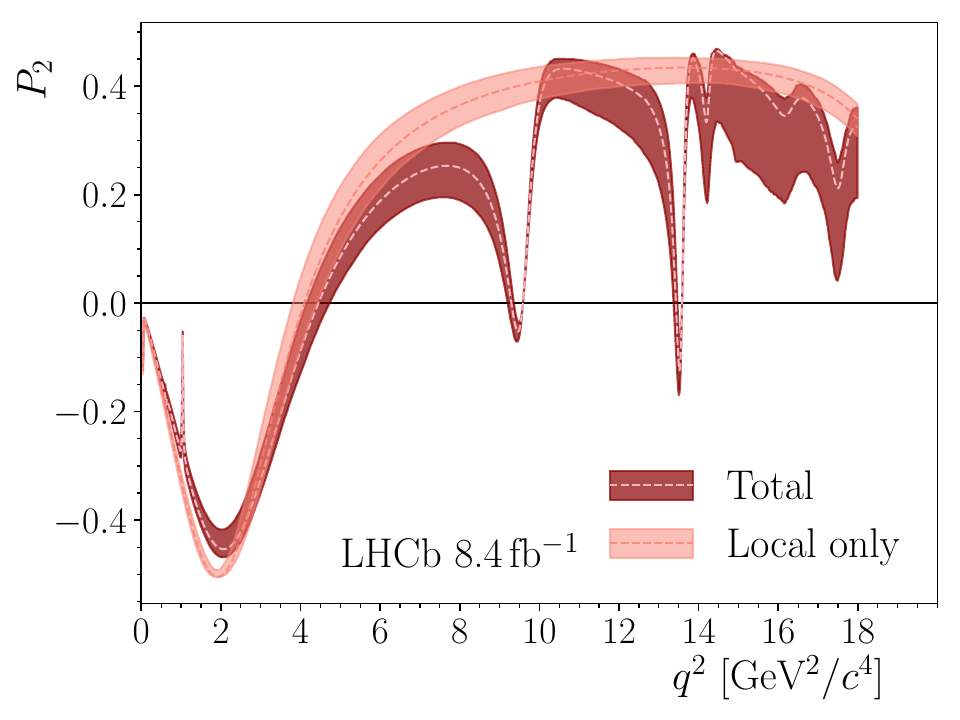}\\
    \includegraphics[width=0.47\linewidth]{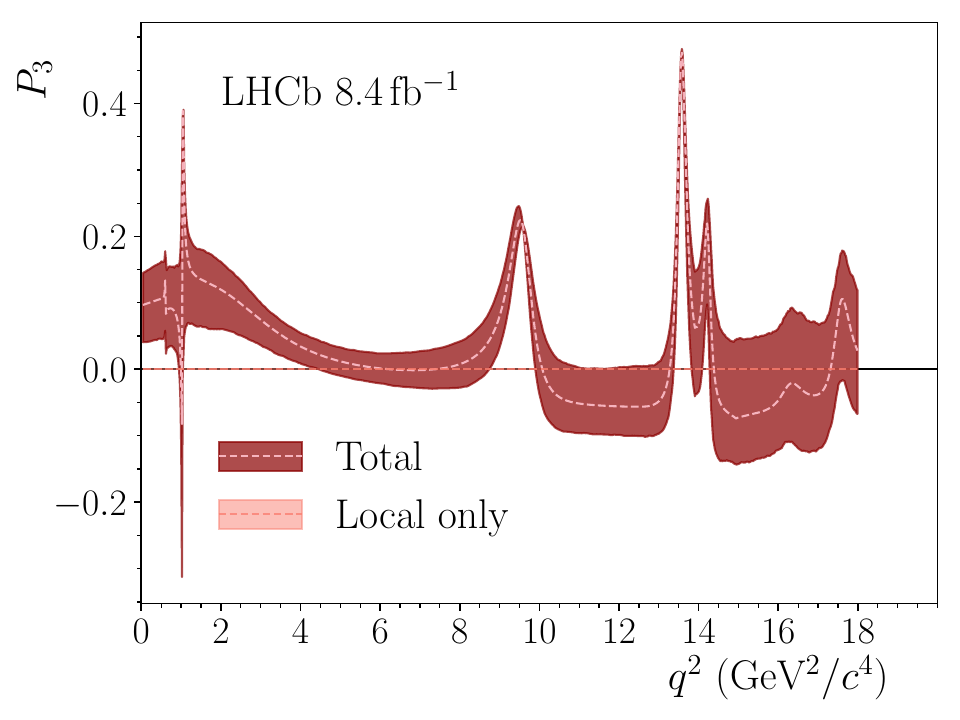}
    \includegraphics[width=0.47\linewidth]{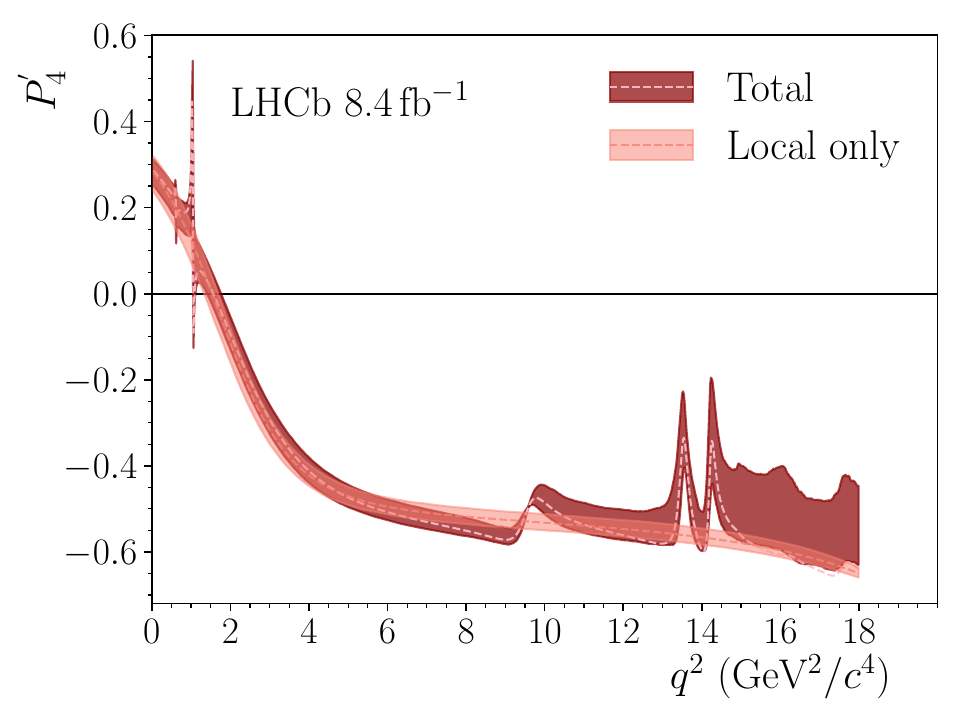}\\
    \includegraphics[width=0.47\linewidth]{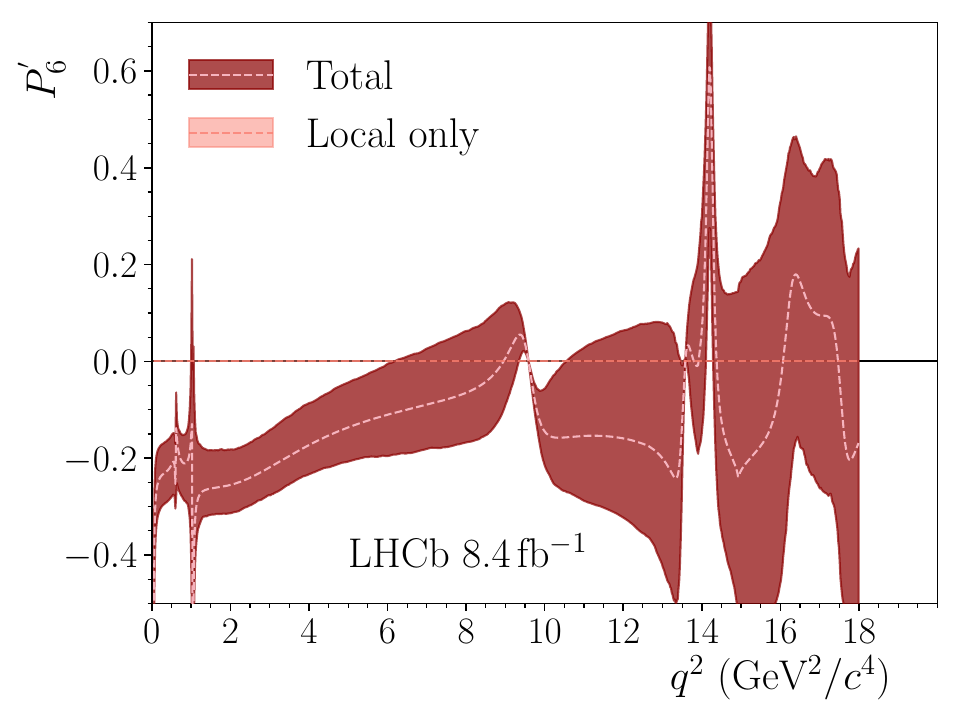}
    \includegraphics[width=0.47\linewidth]{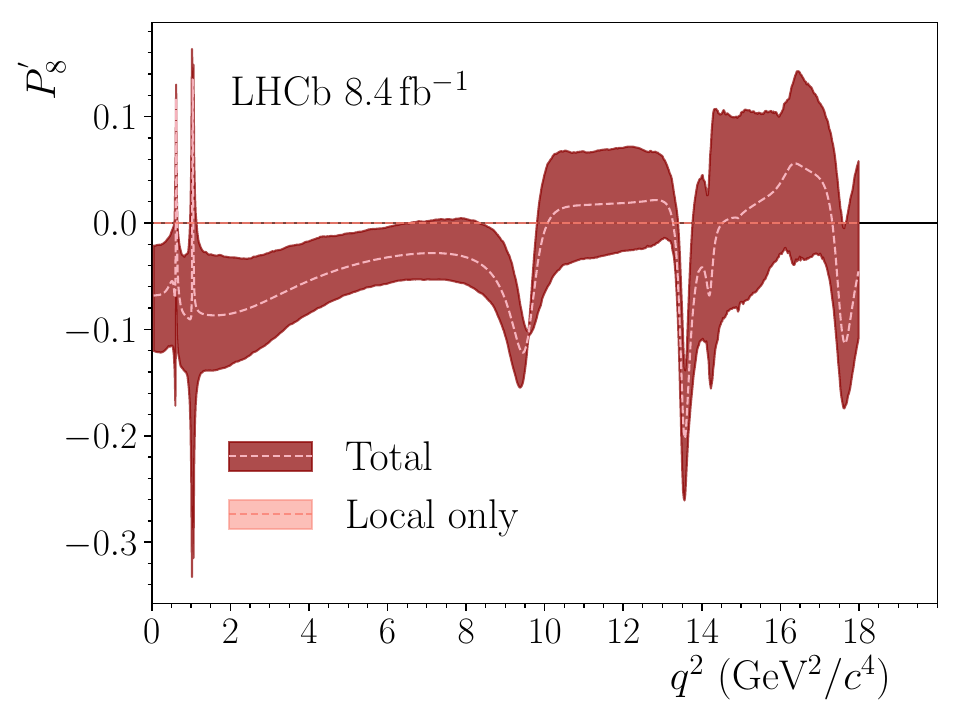}
    \caption{Plots of the angular observables in the optimised basis showing both the total and the contributions from local amplitudes only.}
    \label{fig:UnbinnedPObs}
\end{figure}

\begin{figure}
    \centering
    \includegraphics[width=0.47\linewidth]{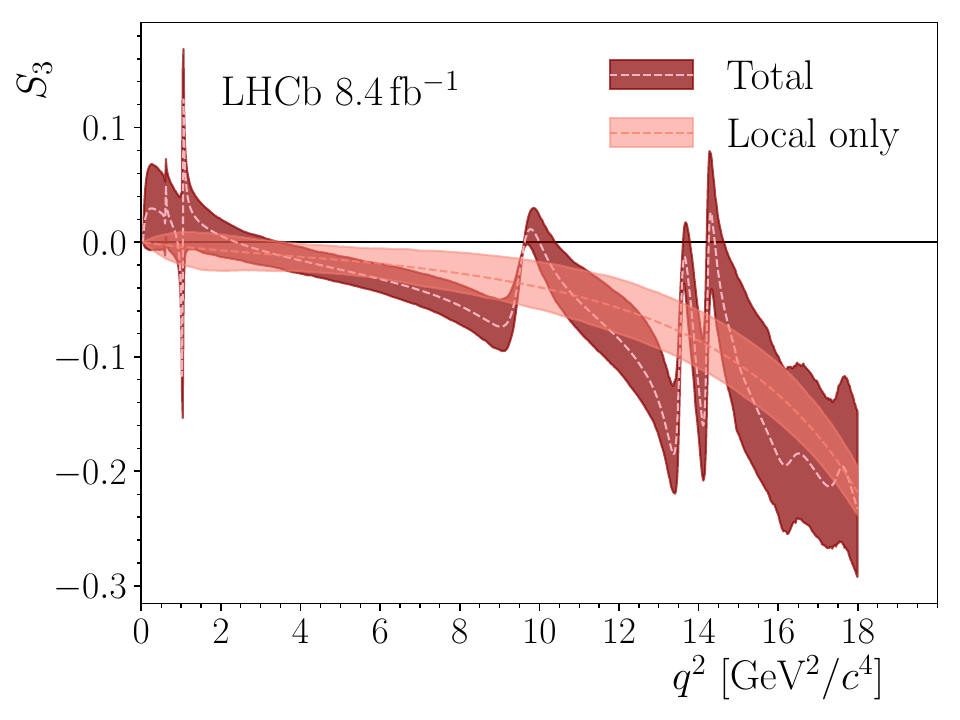}
    \includegraphics[width=0.47\linewidth]{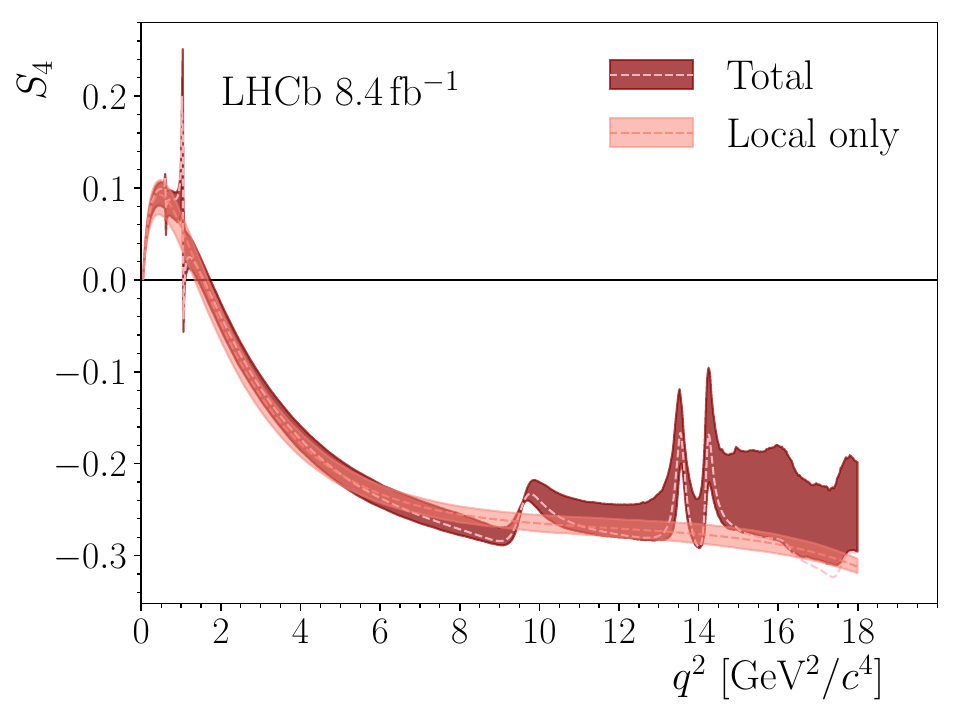}\\
    \includegraphics[width=0.47\linewidth]{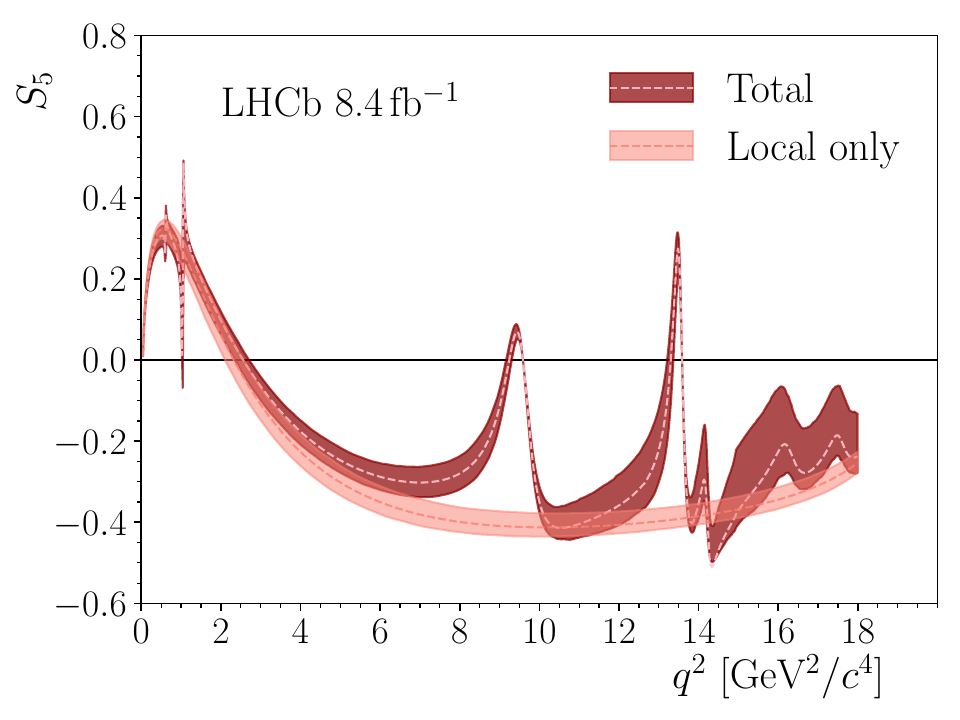}
    \includegraphics[width=0.47\linewidth]{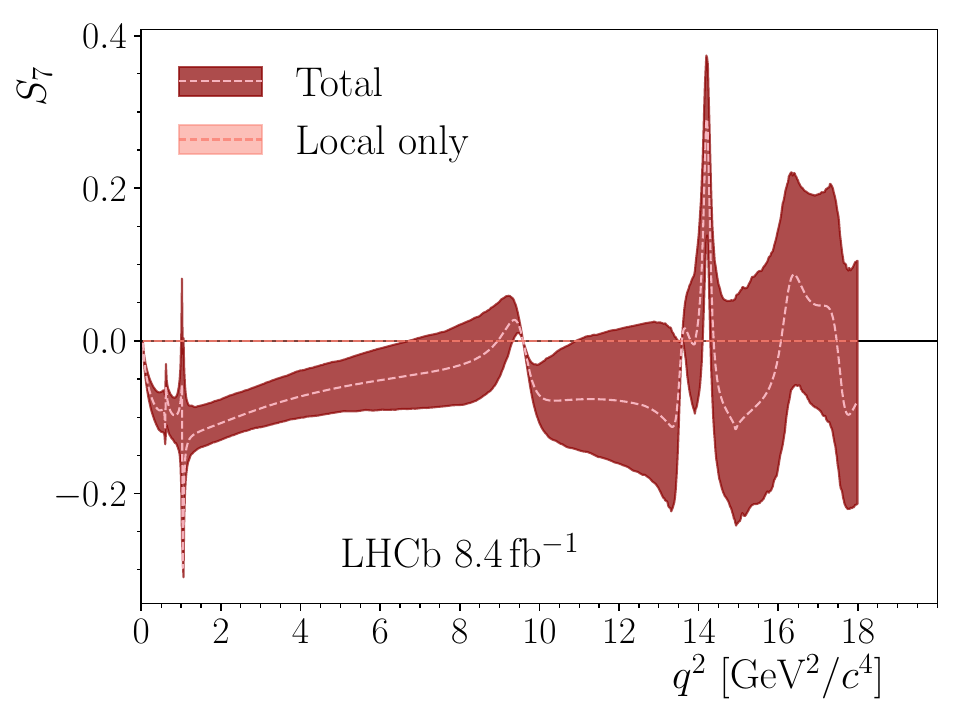}\\
    \includegraphics[width=0.47\linewidth]{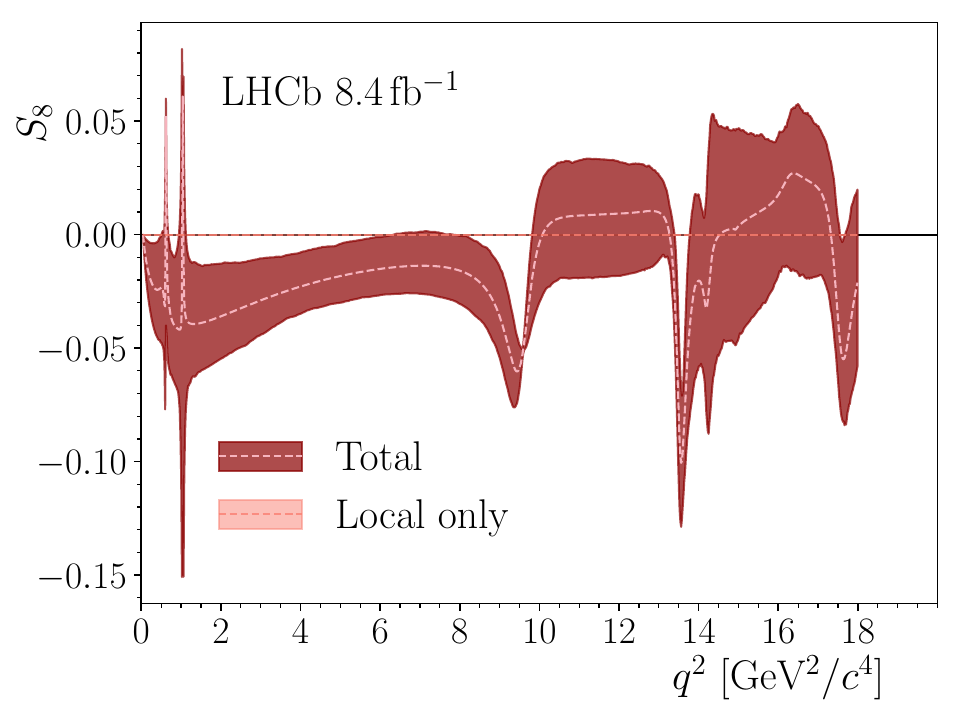}
    \includegraphics[width=0.47\linewidth]{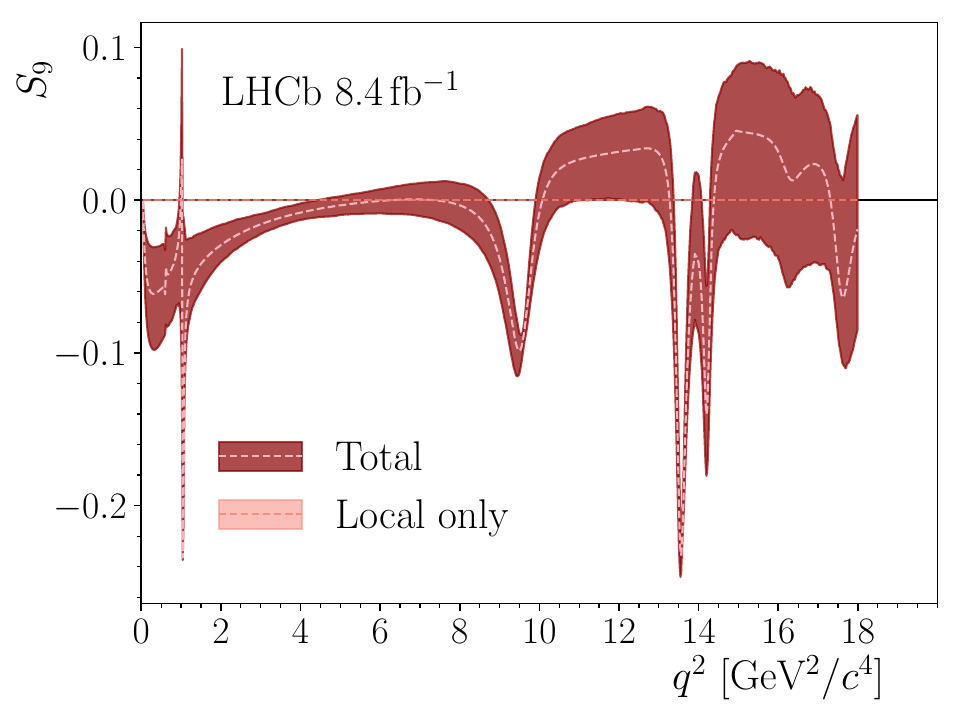}\\
    \includegraphics[width=0.47\linewidth]{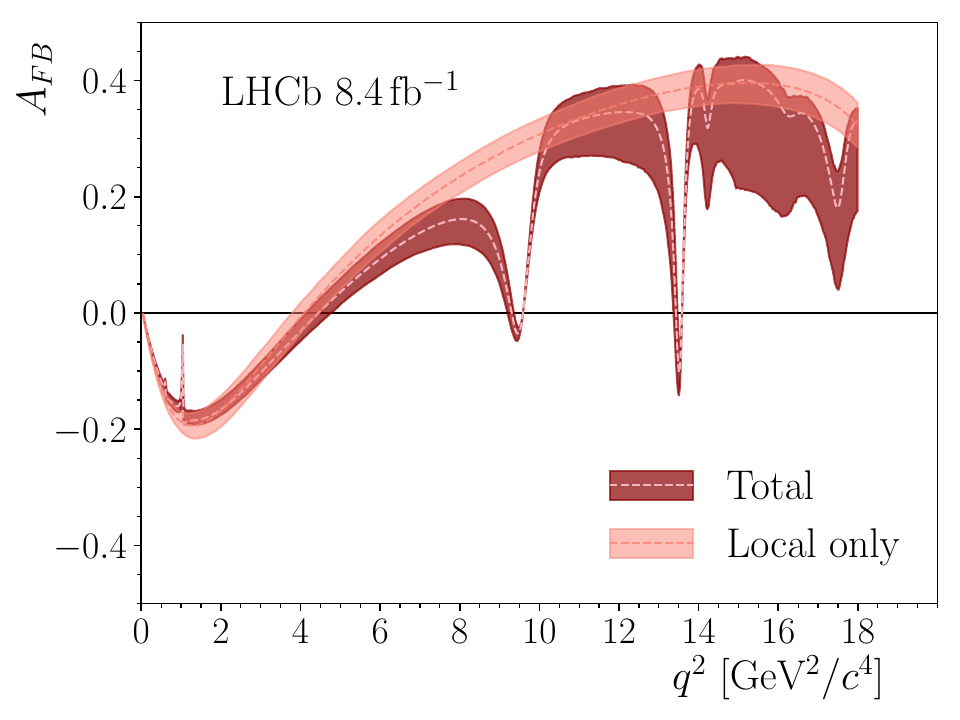}
    \includegraphics[width=0.47\linewidth]{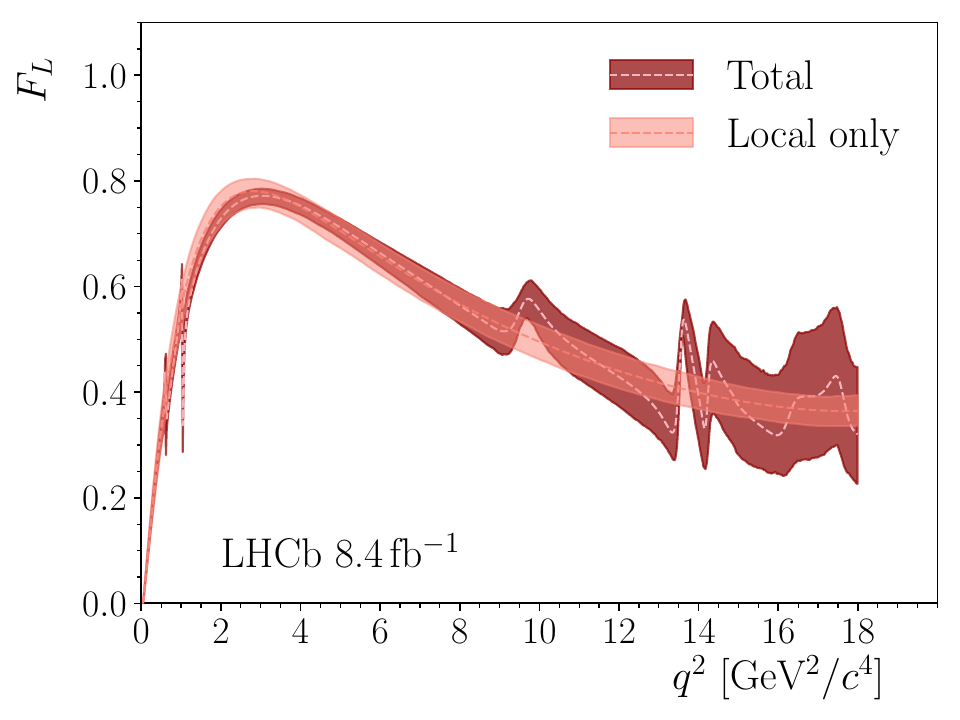}
    \caption{Plots of the angular observables in the standard basis showing both the total and the contributions from local amplitudes only.}
    \label{fig:UnbinnedSObs}
\end{figure}

\begin{figure}
    \centering
    \includegraphics[width=0.47\linewidth]{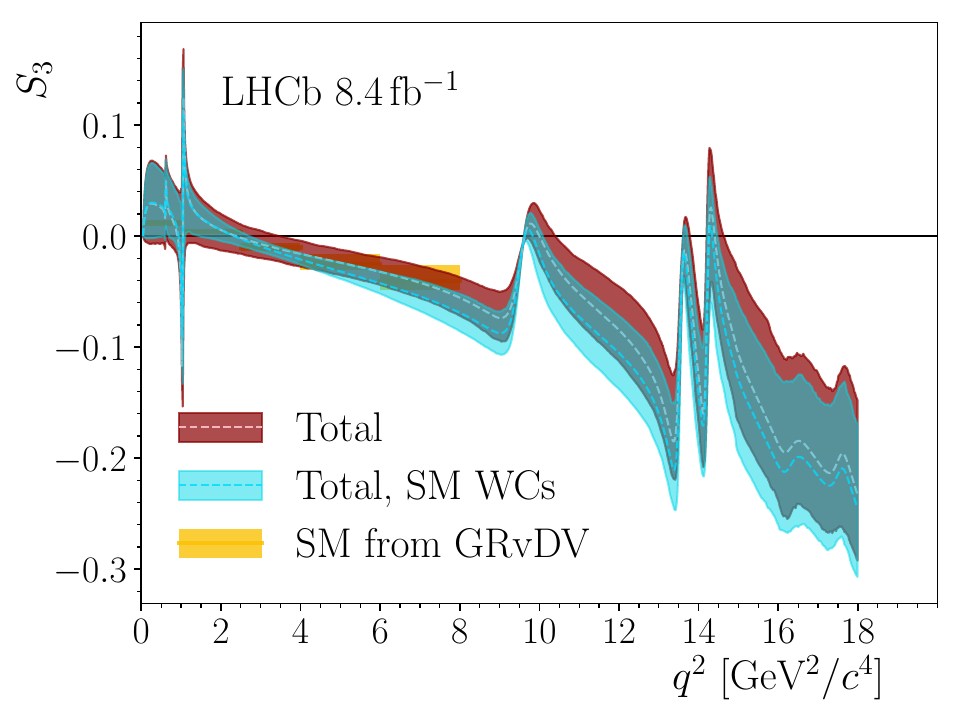}
    \includegraphics[width=0.47\linewidth]{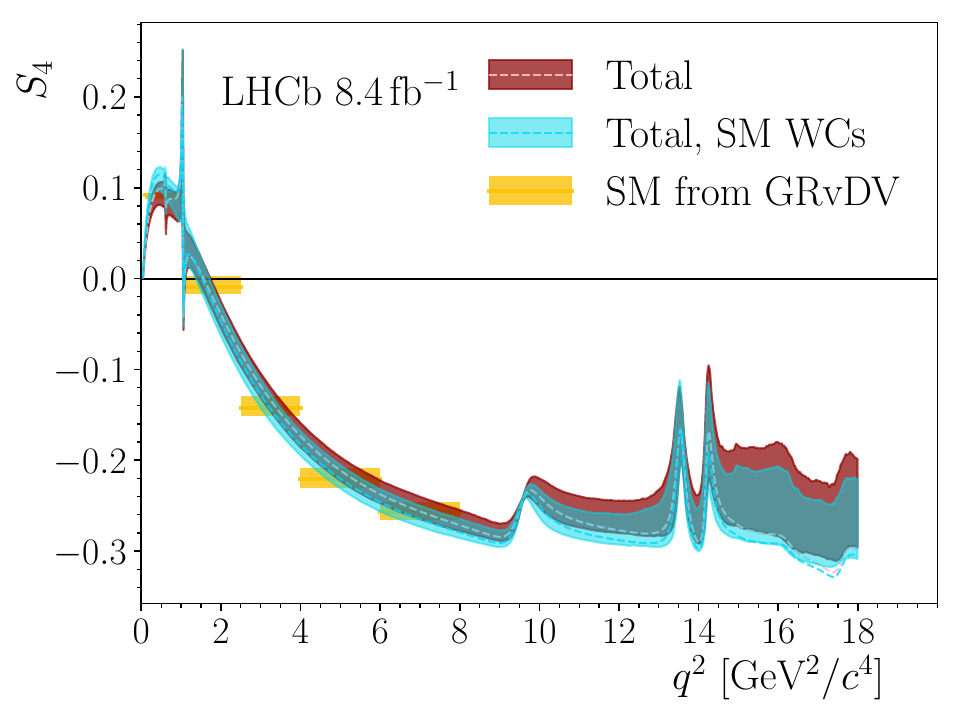}\\
    \includegraphics[width=0.47\linewidth]{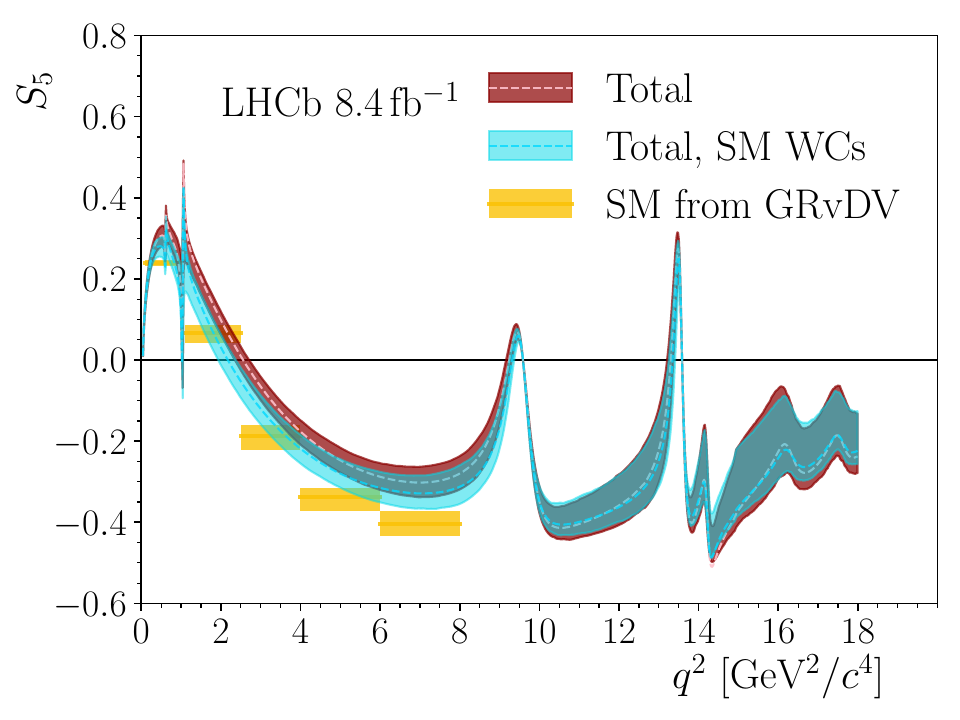}
    \includegraphics[width=0.47\linewidth]{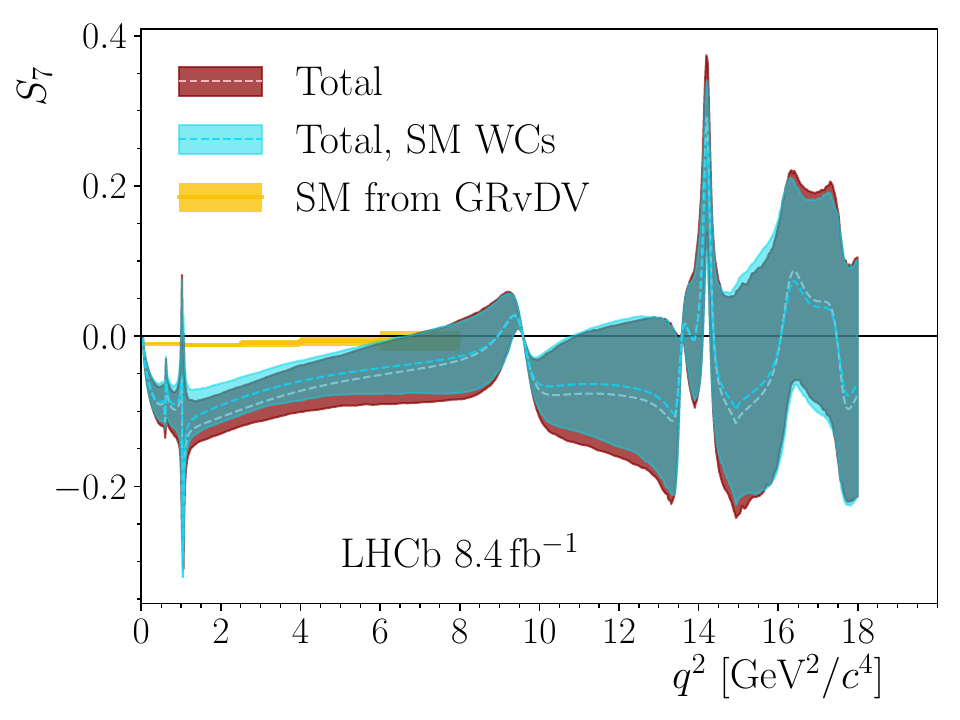}\\
    \includegraphics[width=0.47\linewidth]{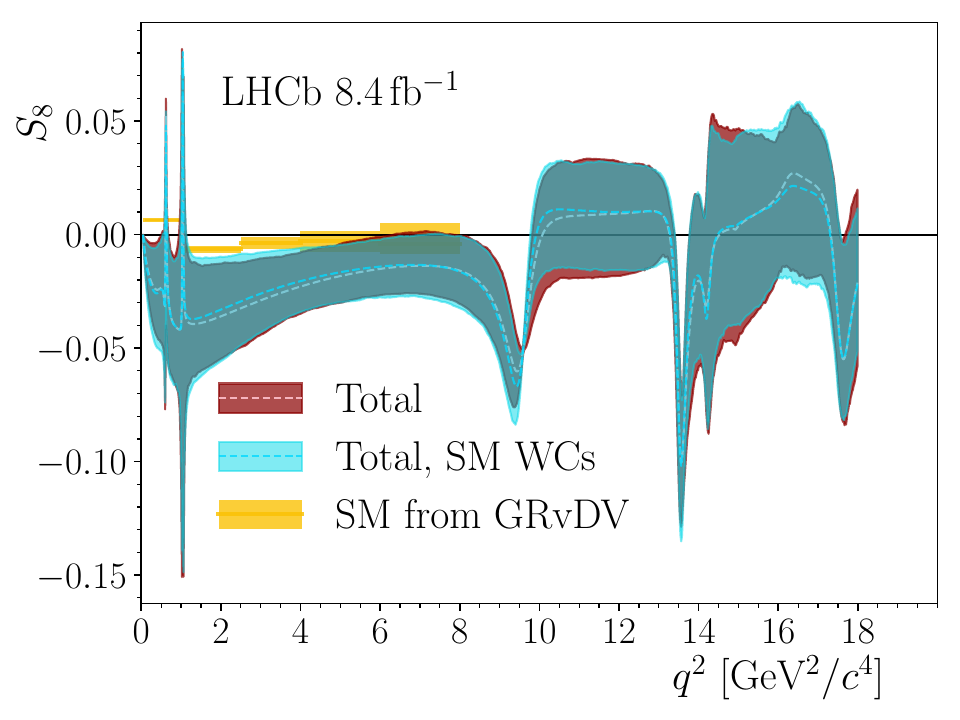}
    \includegraphics[width=0.47\linewidth]{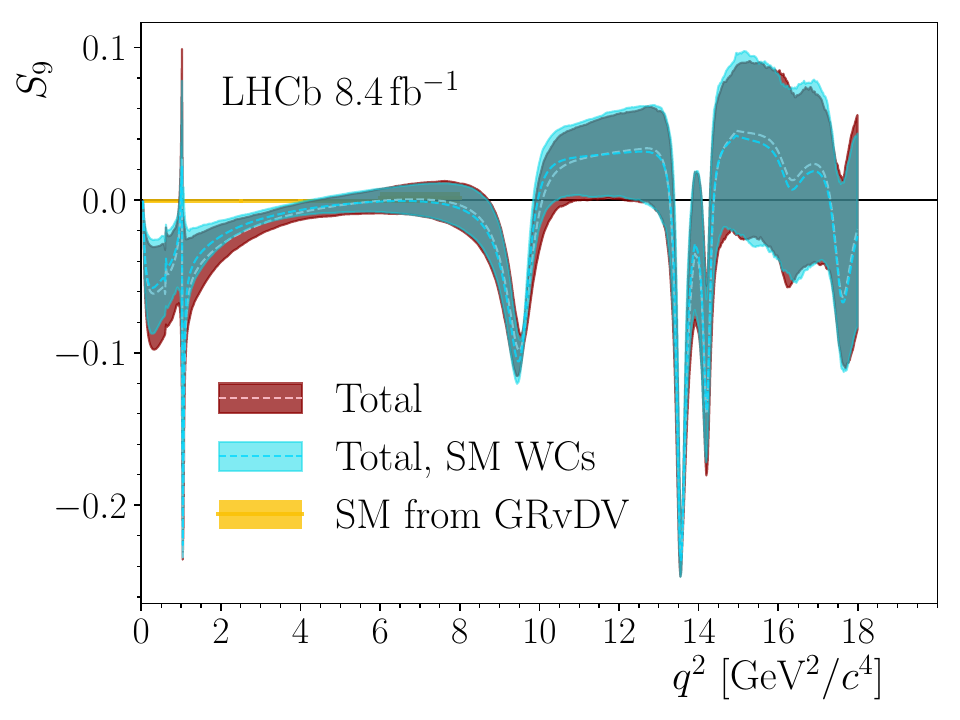}\\
    \includegraphics[width=0.47\linewidth]{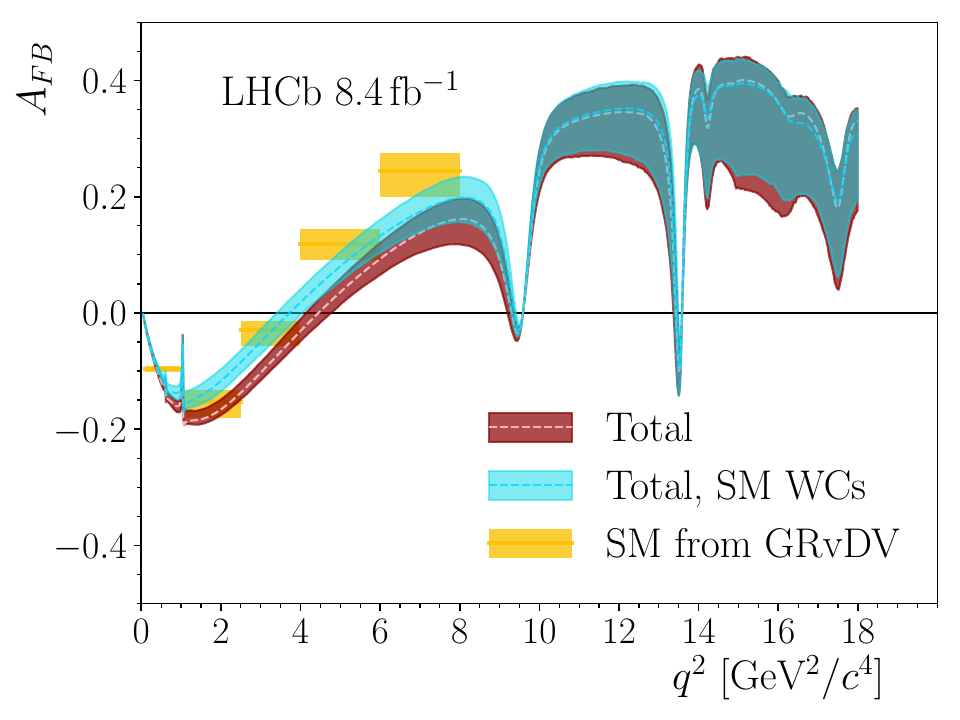}
    \includegraphics[width=0.47\linewidth]{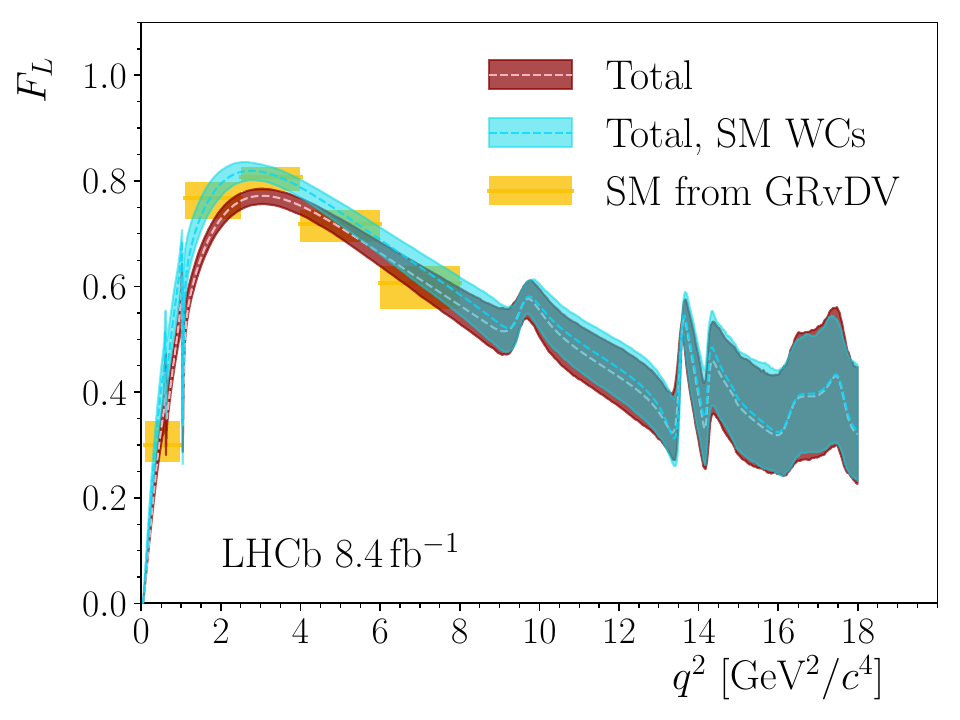}
    \caption{Plots of the unbinned angular observables in the standard basis shown for the both the baseline fit to data, and with the Wilson Coefficients~(WCs) set to their Standard Model (SM) values. These are compared against SM predictions from Ref.~\cite{Gubernari:2022hxn}.}
    \label{fig:BinnedSObsSMcomp}
\end{figure}

As the overall fit calculates all amplitudes, the observables of previous measurements can be plotted for both the optimised (Fig.~\ref{fig:UnbinnedPObs}) and the standard (Fig.~\ref{fig:UnbinnedSObs}) basis. In all of the plots, the observables are shown when calculated from both the full amplitudes as well as from the local amplitudes only. In this way the contribution of the nonlocal amplitudes is made clear. Finally, in Fig.~\ref{fig:BinnedSObsSMcomp} there are plots in the standard basis compared against Standard Model predictions from Ref.~\cite{Gubernari:2022hxn}.

\clearpage

%% file: appendix-projections.tex
\section{Fit projections in \boldmath{\qsq} sub-regions}
\label{app:FitProjInQ2Regions}

\begin{figure}[h]
    \centering
    \includegraphics[width=0.32\linewidth]{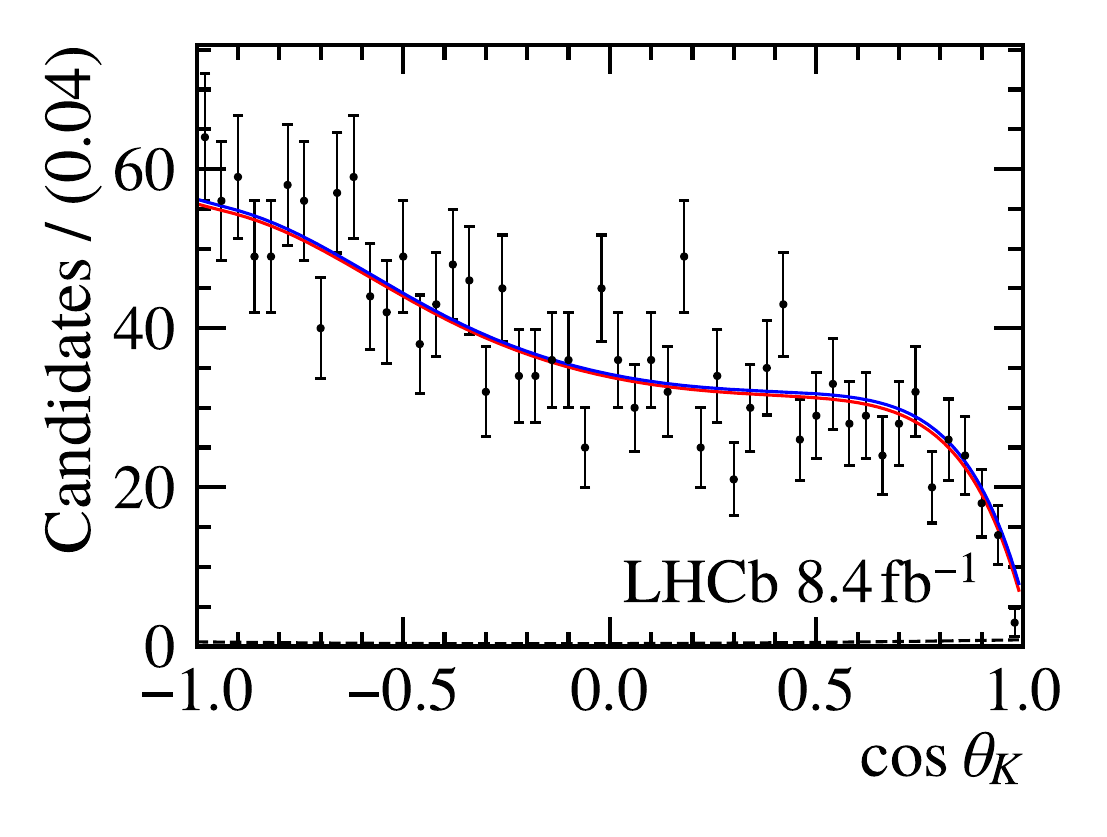}
    \includegraphics[width=0.32\linewidth]{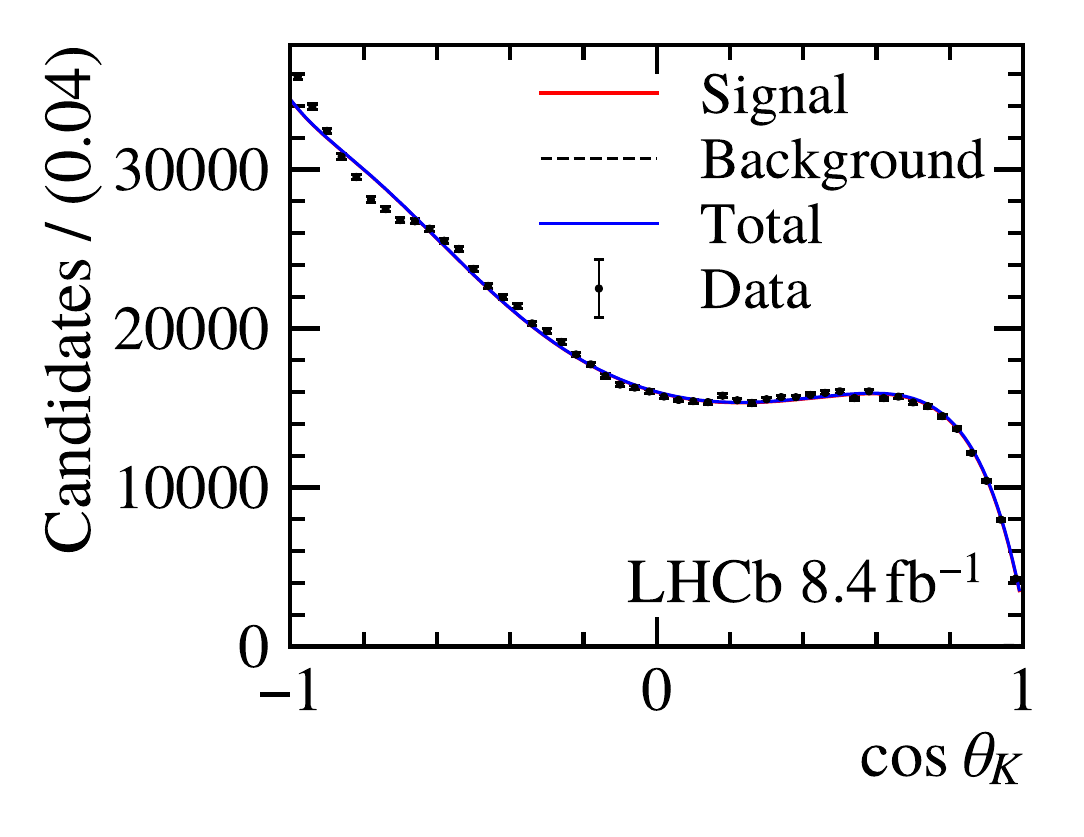}
    \includegraphics[width=0.32\linewidth]{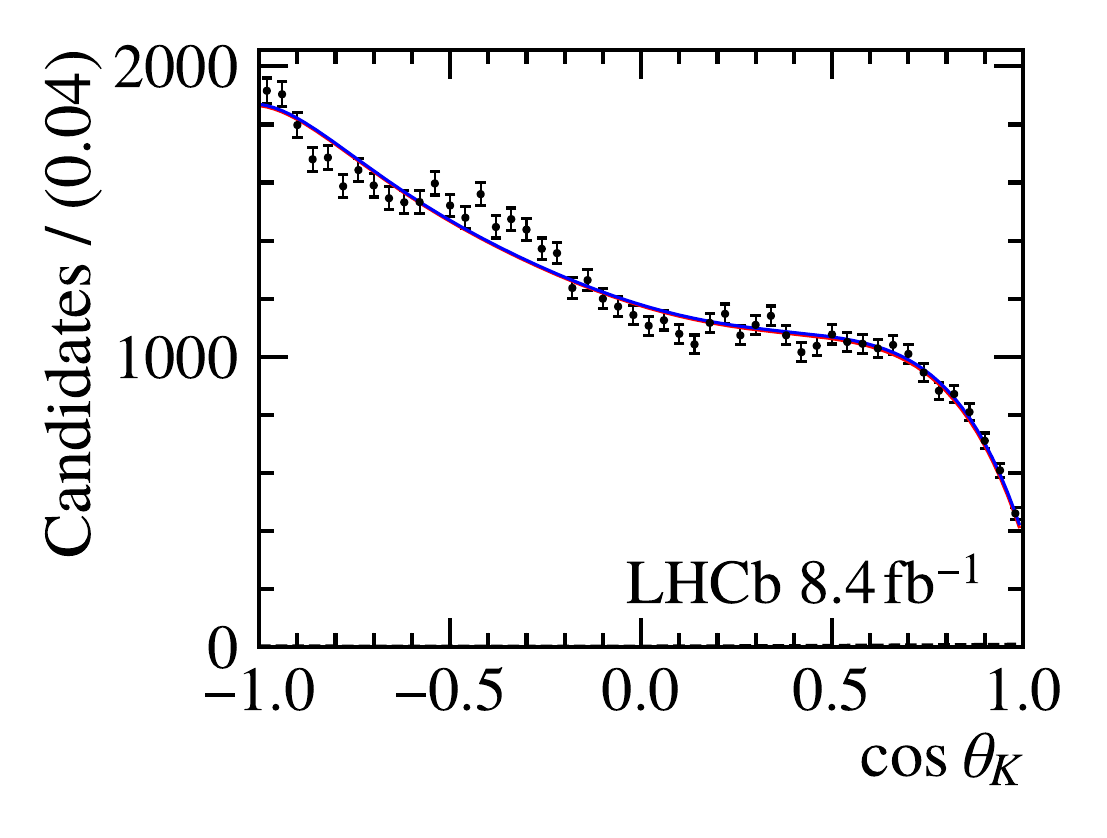}
    
    \includegraphics[width=0.32\linewidth]{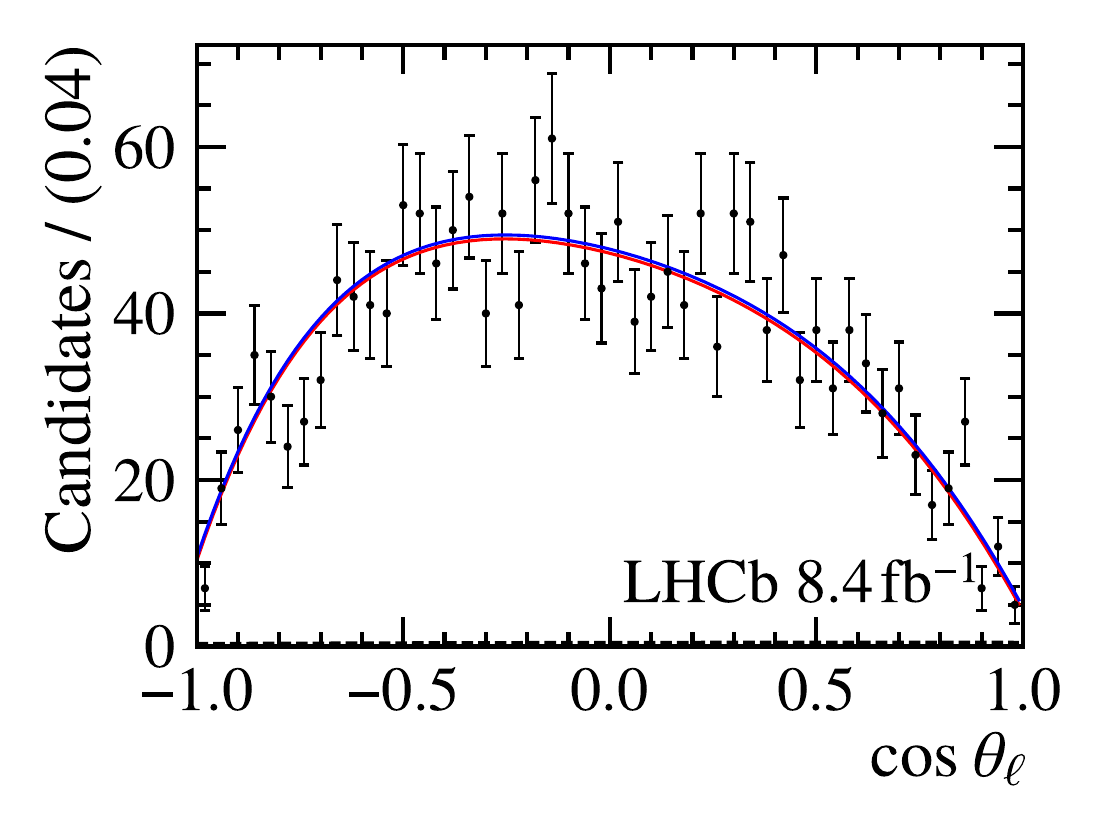}
    \includegraphics[width=0.32\linewidth]{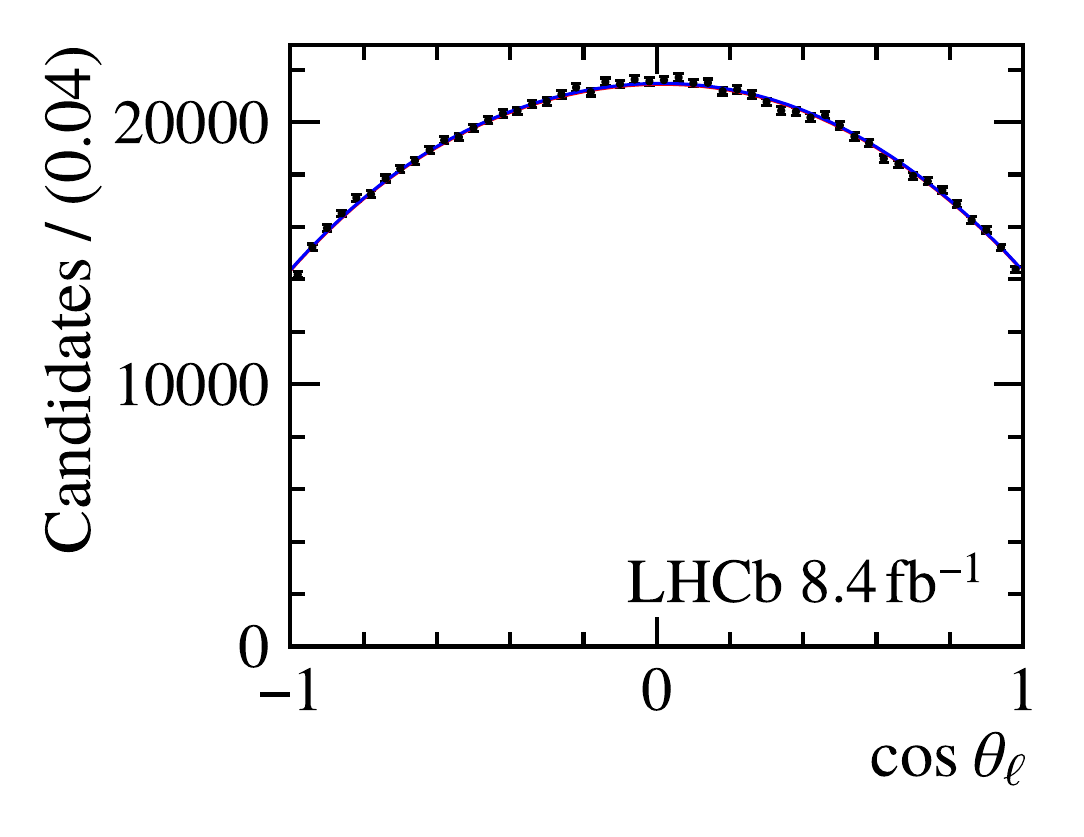}
    \includegraphics[width=0.32\linewidth]{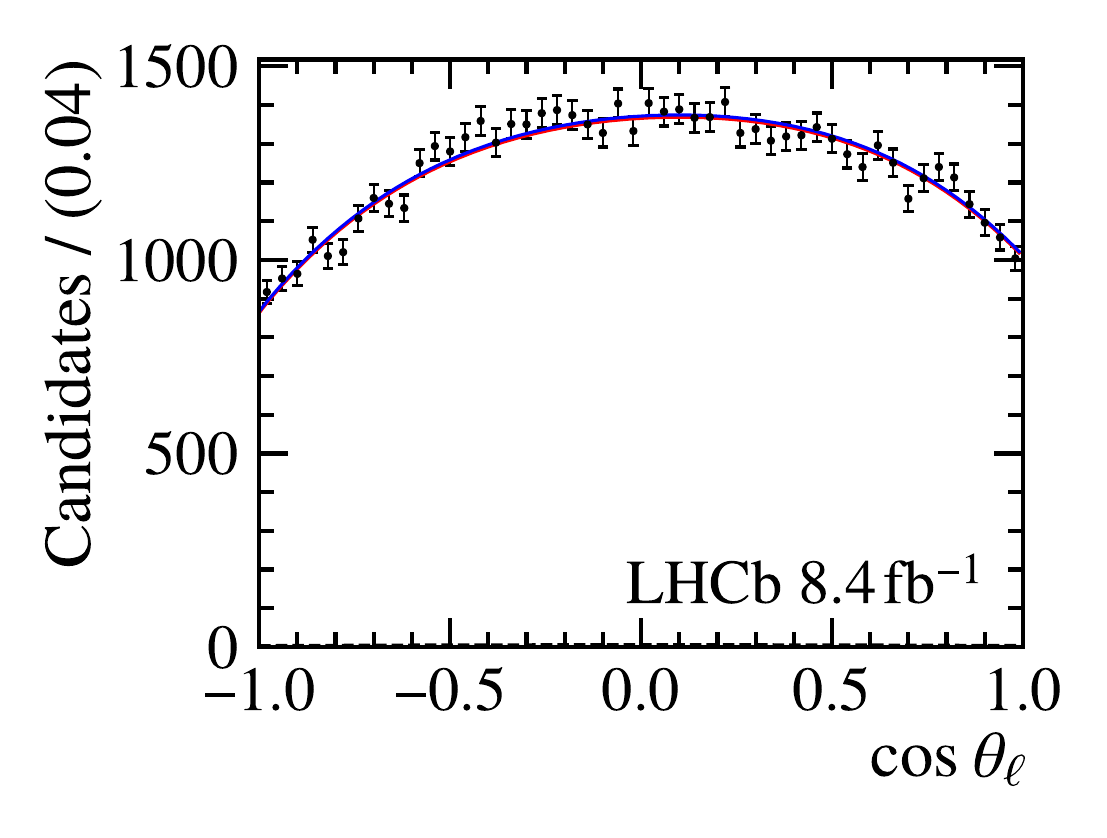}
    
    \includegraphics[width=0.32\linewidth]{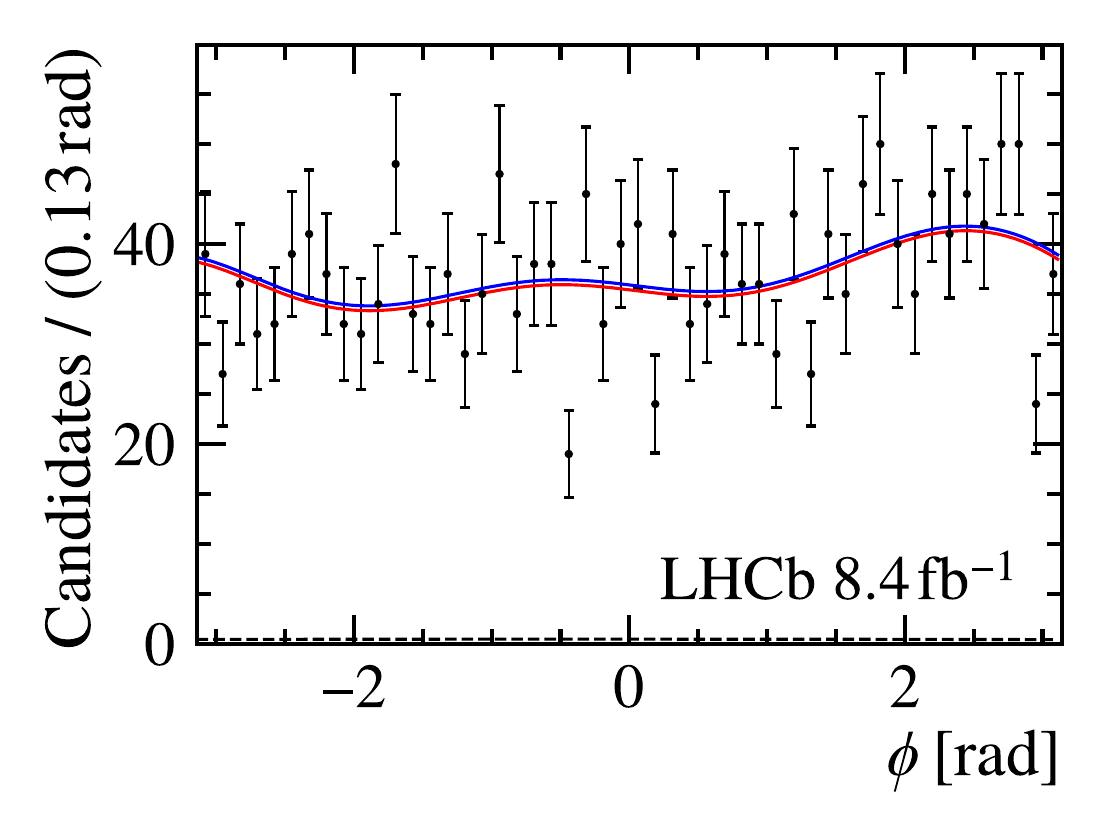}
    \includegraphics[width=0.32\linewidth]{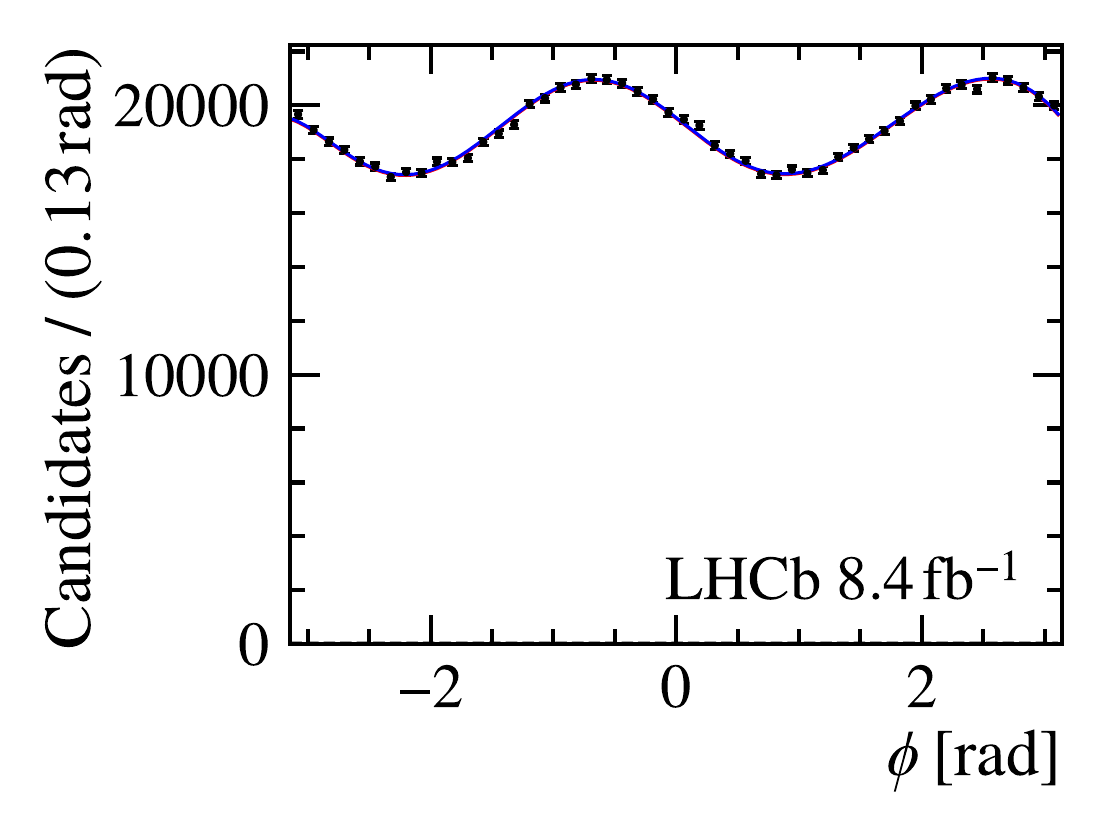}
    \includegraphics[width=0.32\linewidth]{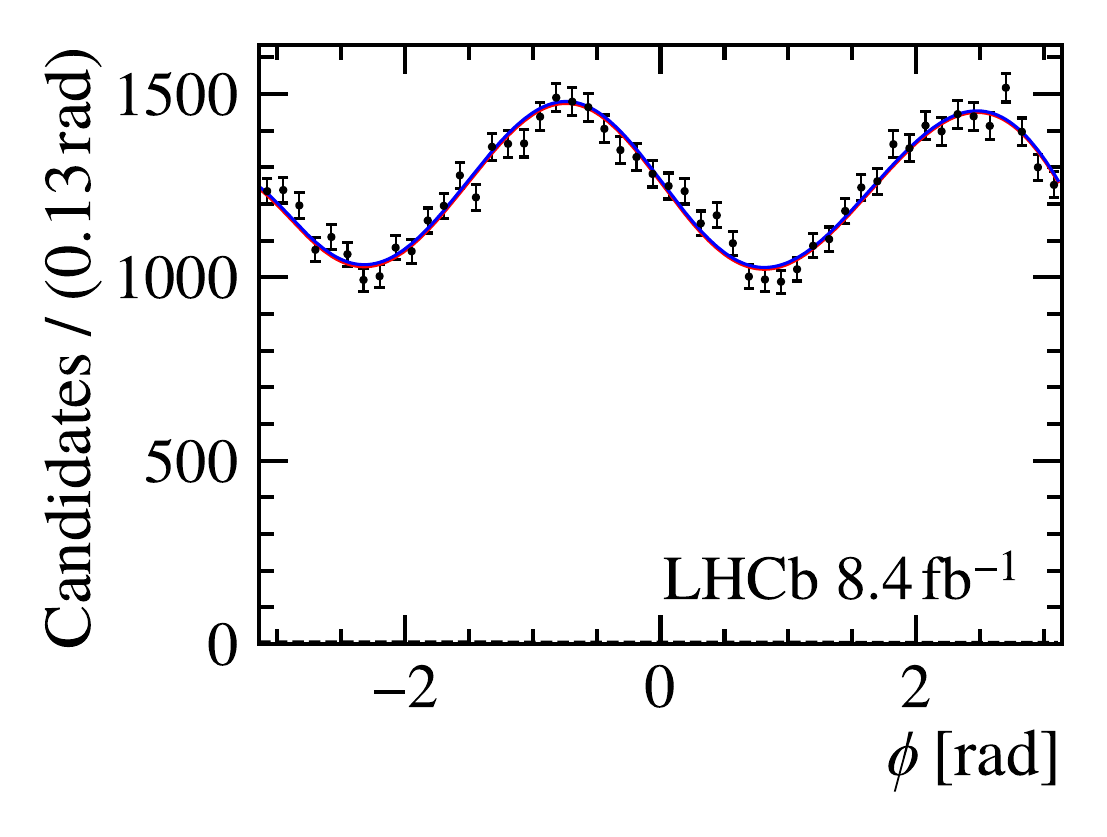}
    
    \includegraphics[width=0.32\linewidth]{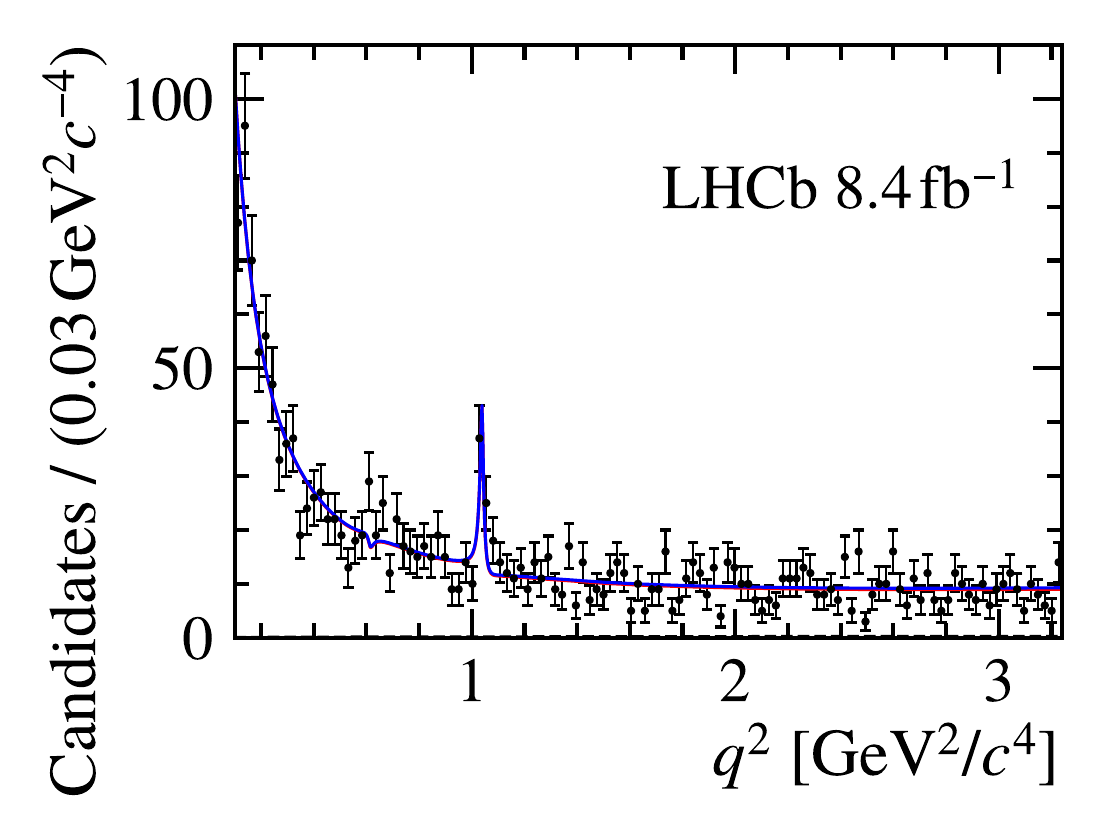}
    \includegraphics[width=0.32\linewidth]{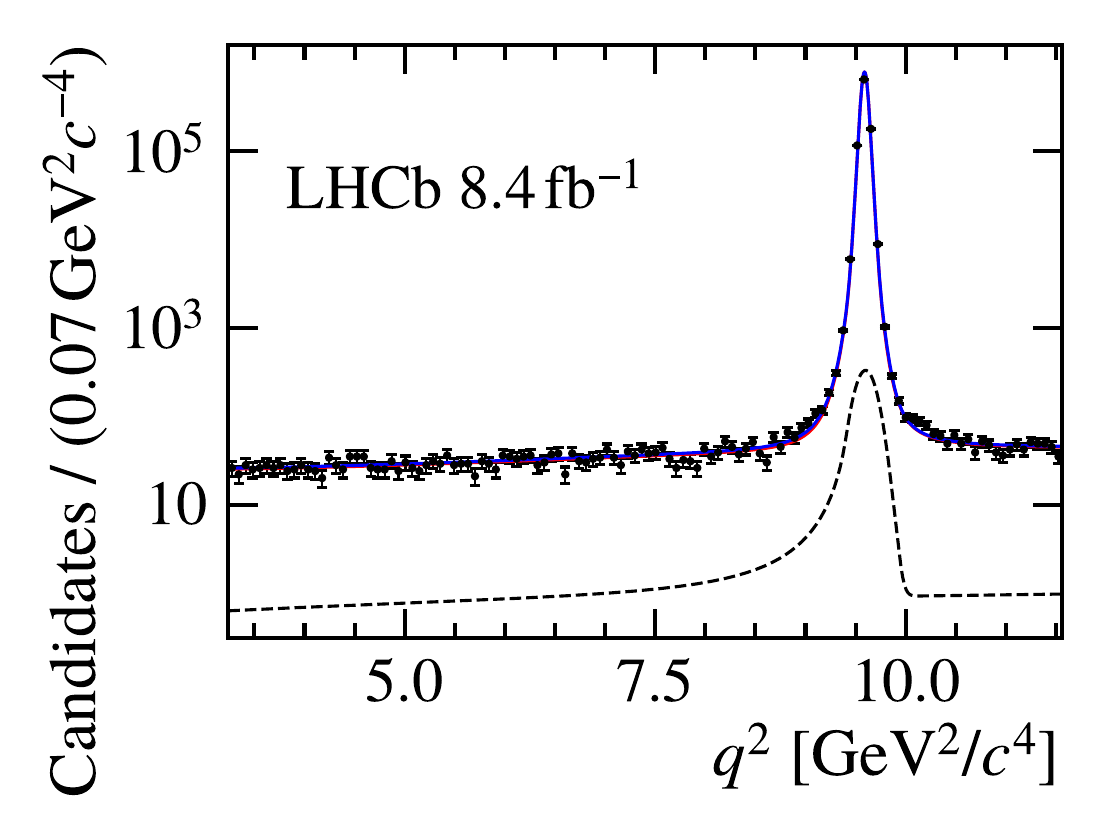}
    \includegraphics[width=0.32\linewidth]{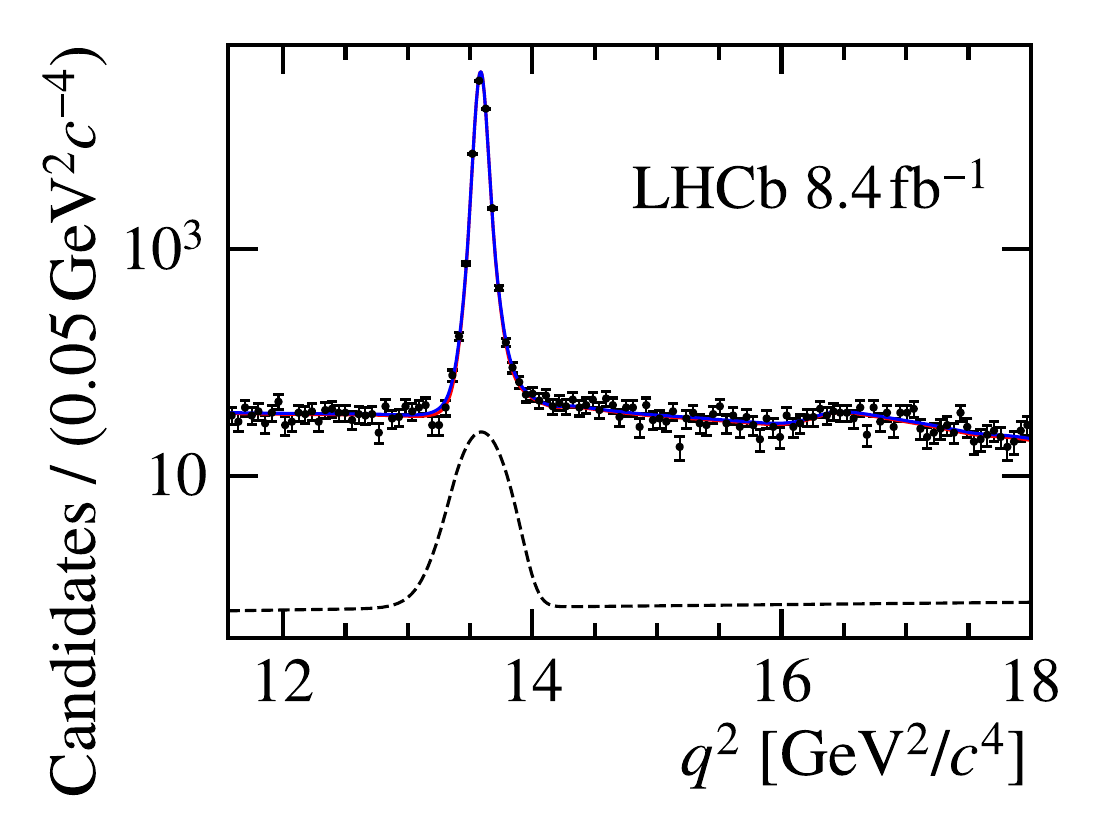}
    
    \caption{Result of the fit to candidates in the signal mass region. The four rows correspond to the distributions of $\cos{\theta_{K}}$, $\cos{\theta_{\ell}}$, $\phi$ and $q^2$. The three columns correspond to the low-, mid- and high-$q^2$ regions.   The total PDF is shown in blue, the signal PDF in red and the background PDF in dotted black. The impact of the neglected exotic states is visible in the $\cos{\theta_{K}}$ distributions. }
    \label{fig:signal_fit_result}
\end{figure}

The four-dimensional maximum-likelihood fit to the signal region is performed simultaneously in three \qsq regions, as described in Sec.~\ref{sec:AcceptanceResolution}. The results of the fits to the \ctk, \ctl, \phih, and \qsq distributions within each of the three regions are shown in Fig.~\ref{fig:signal_fit_result}.

\subsection{Projections of the \qsq spectrum with alternative signal decompositions}
\label{app:AltSigDecompQ2}

\begin{figure}[h]
    \centering
    \includegraphics[width=\linewidth]{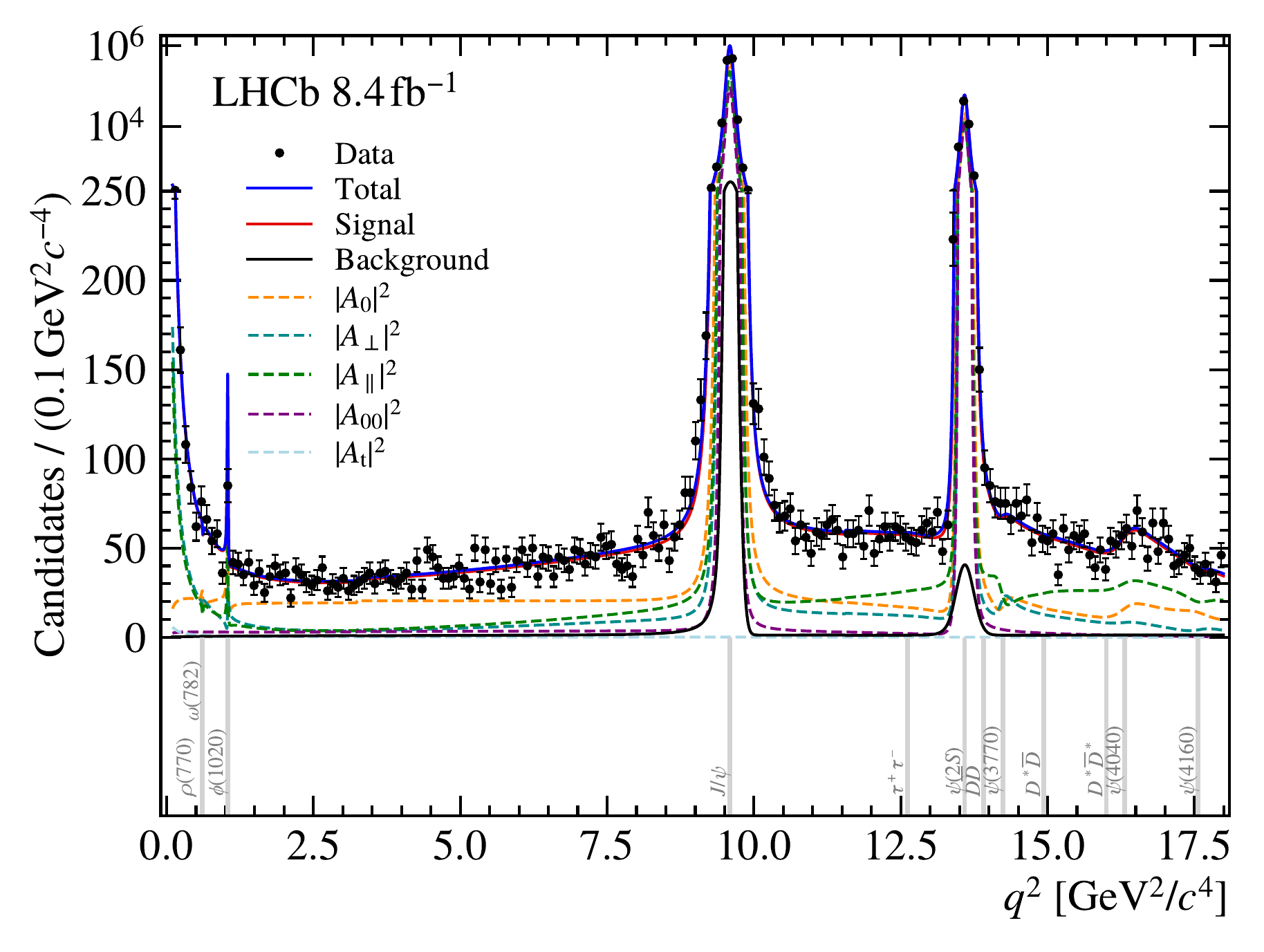}
    \caption{The $q^2$ distribution in the data, overlaid with the PDF projection from the baseline data fit. The total PDF is decomposed into signal and background components, with the signal contributions further decomposed into contributions from the different transversity amplitudes.}
    \label{fig:DataQ2ProjHel}
\end{figure}

\begin{figure}[h]
    \centering
    \includegraphics[width=\linewidth]{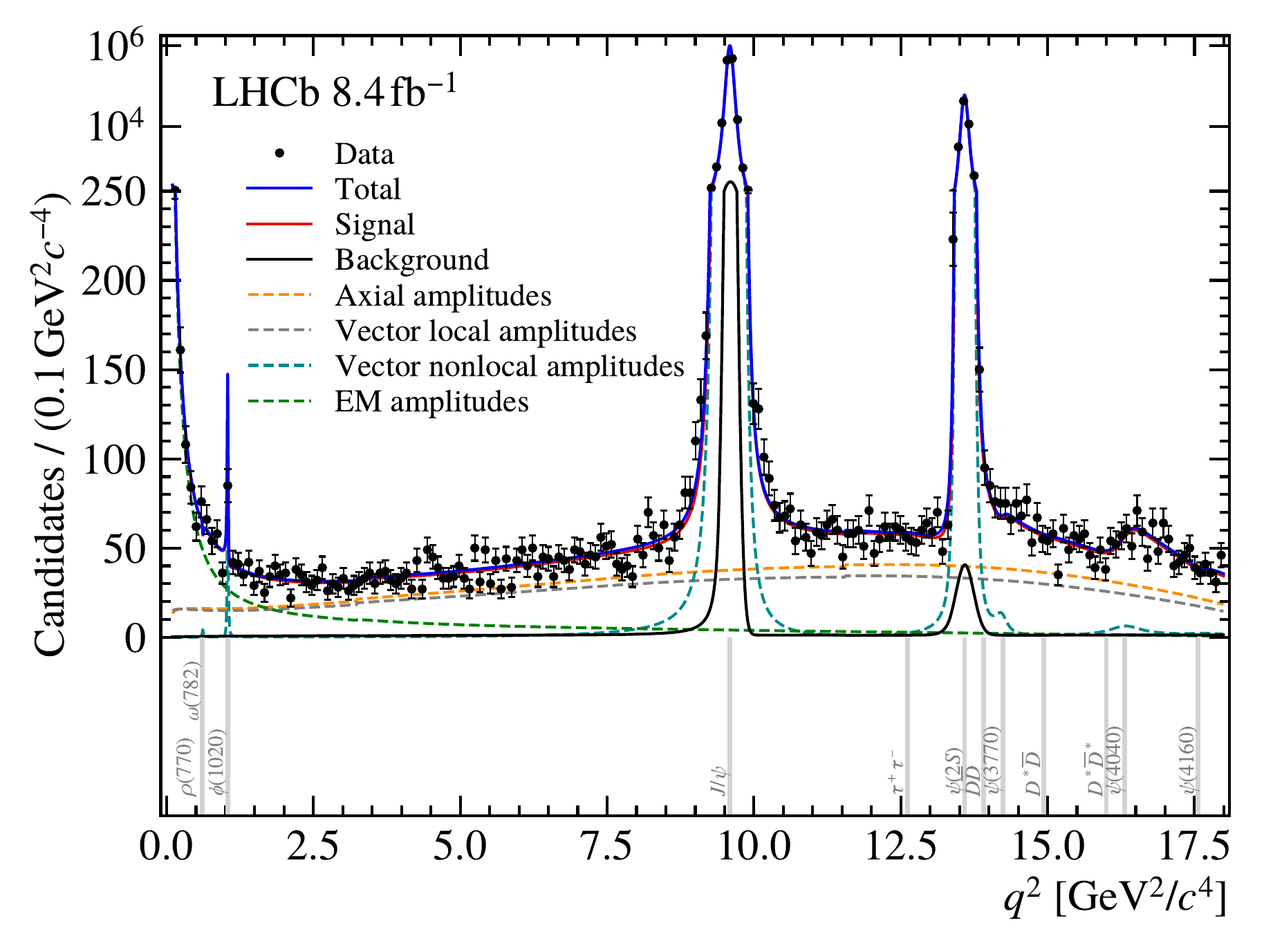}
    \caption{The $q^2$ distribution in the data, overlaid with the PDF projection from the baseline data fit. The total PDF is decomposed into signal and background components, with the signal contributions further decomposed into contributions from different Lorentz structures.}
    \label{fig:DataQ2ProjLorentz}
\end{figure}

The signal contributions can be decomposed in various ways --- for example, in terms of the local and nonlocal contributions, as done in Fig.~\ref{fig:DataQ2ProjRes} in the main text. Alternatively, they can be decomposed into contributions from different transversity amplitudes (see Sec.~\ref{sec:TransversityAmplitudes}) as shown in Fig.~\ref{fig:DataQ2ProjHel}, or into contributions from different Lorentz structures as shown in Fig.~\ref{fig:DataQ2ProjLorentz}.

\section{Comparison of observables to previous analyses}
\label{app:ObsComp}

\begin{figure}[h]
 \centering
  \includegraphics[width=0.40\linewidth]{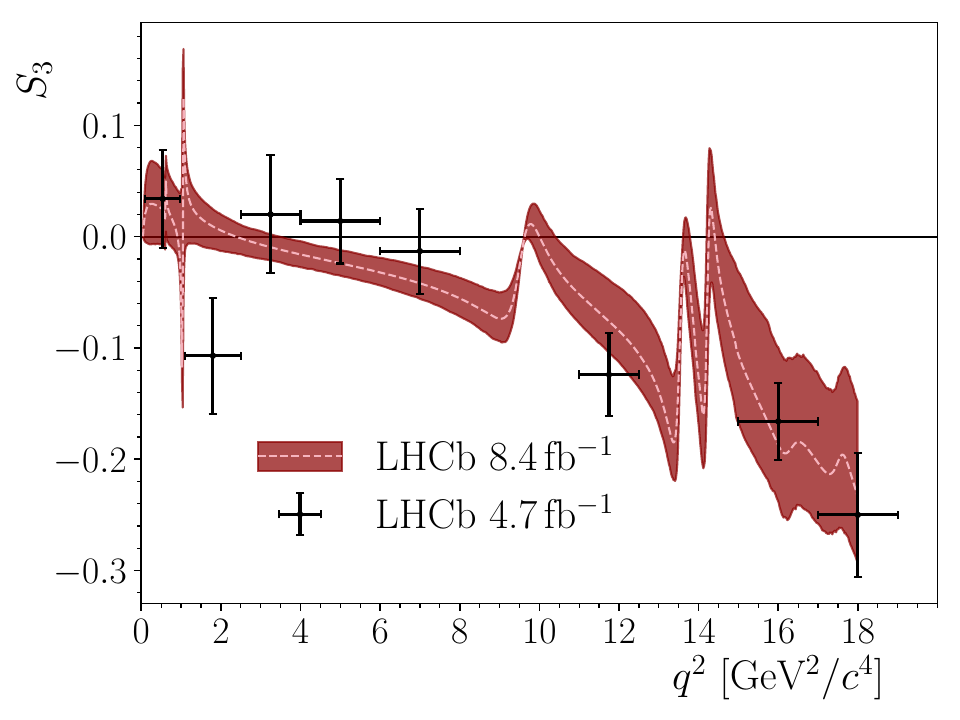}
  \includegraphics[width=0.40\linewidth]{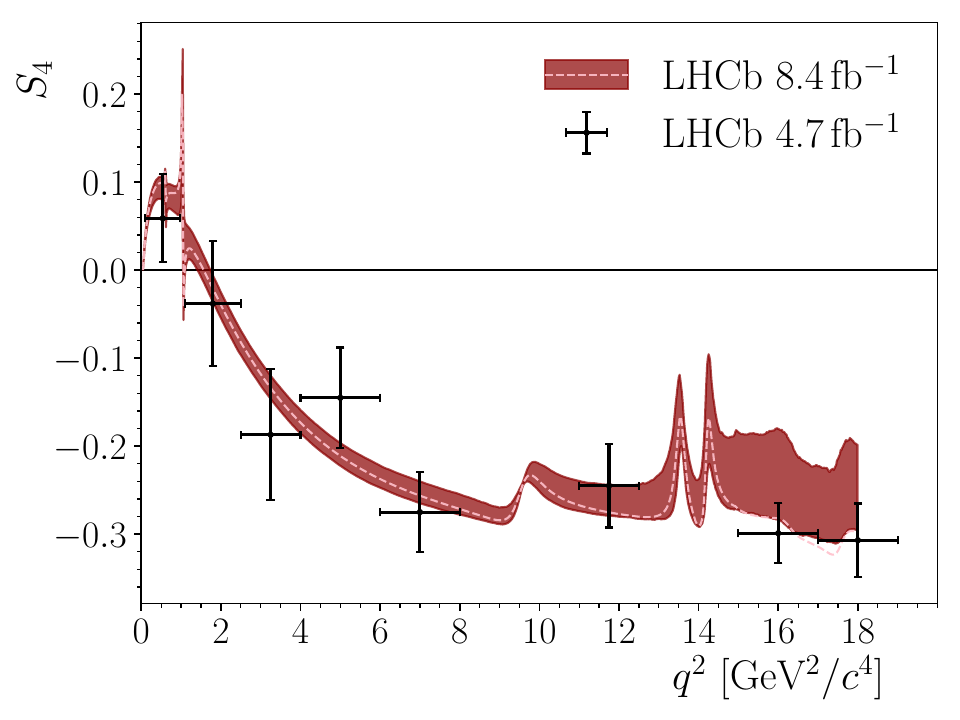}\\
  \includegraphics[width=0.40\linewidth]{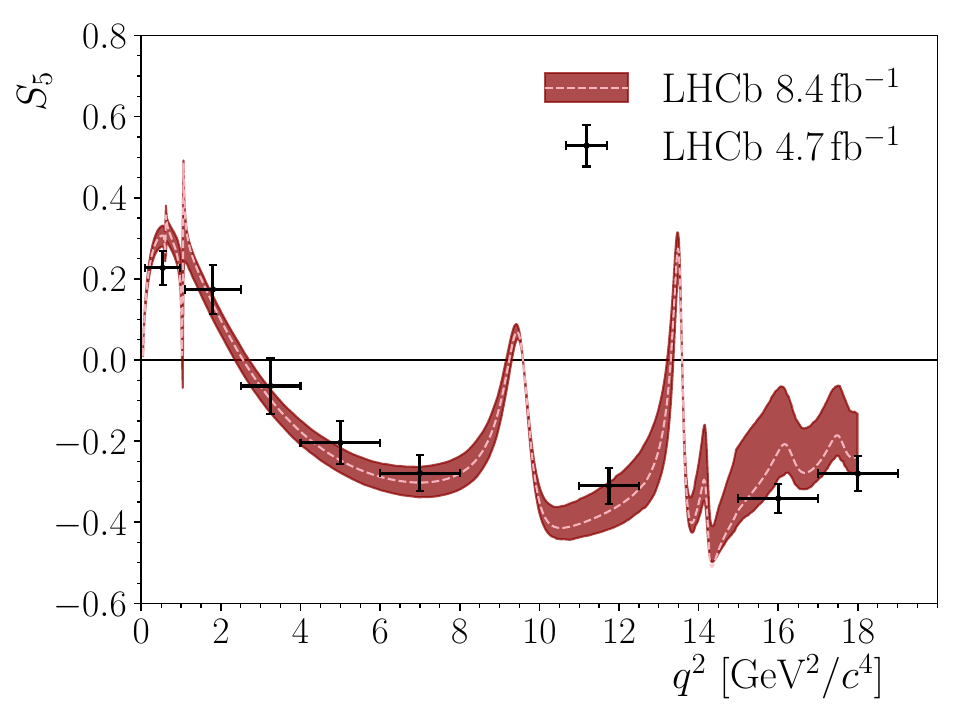}
  \includegraphics[width=0.40\linewidth]{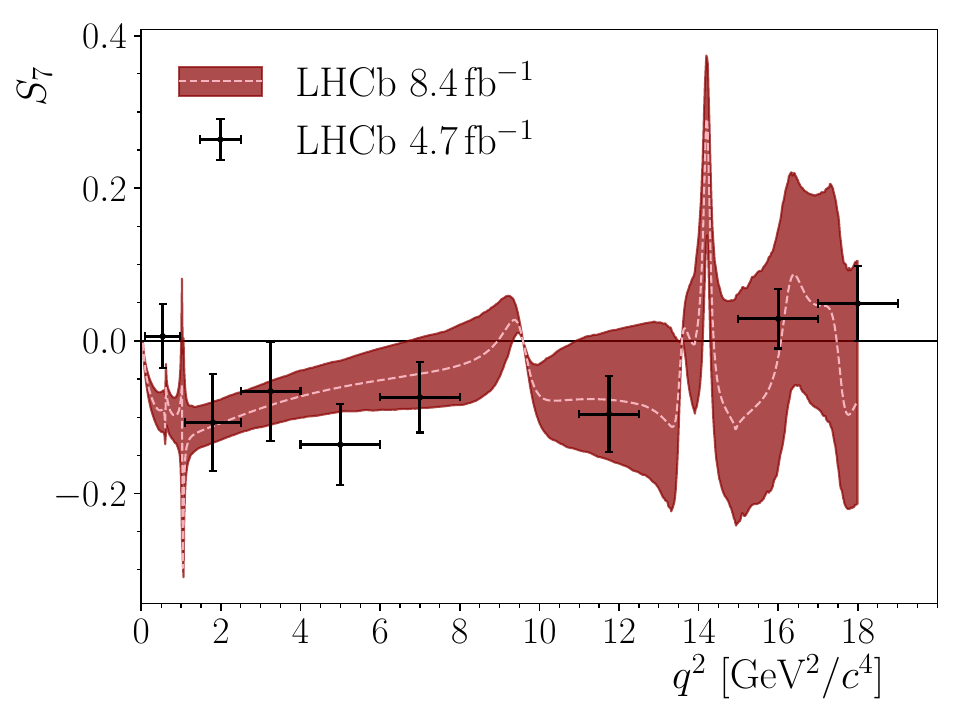}\\
  \includegraphics[width=0.40\linewidth]{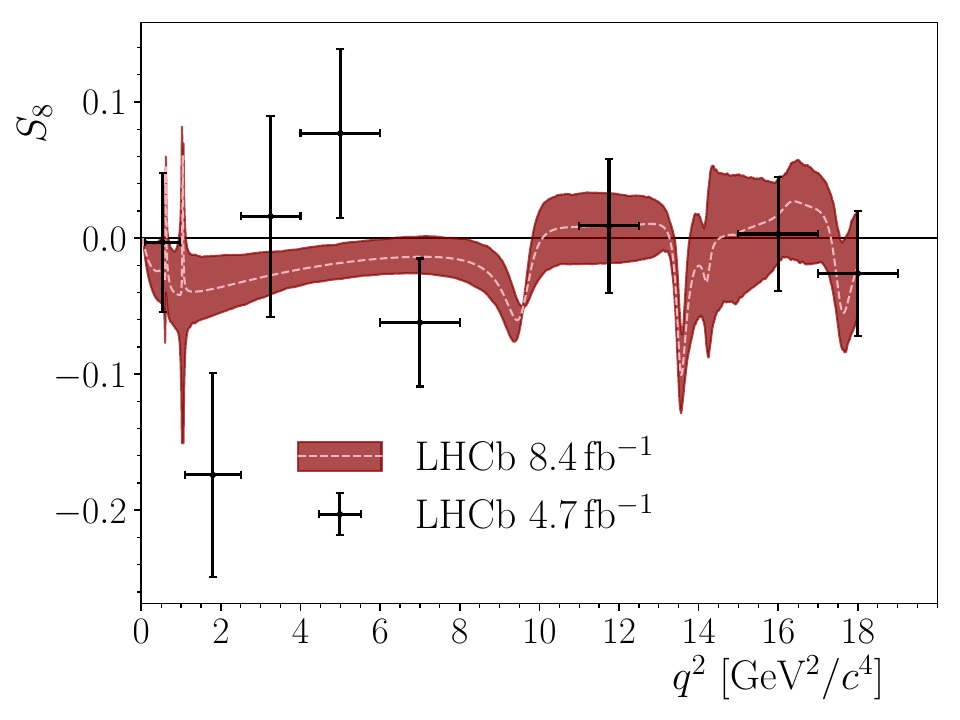}
  \includegraphics[width=0.40\linewidth]{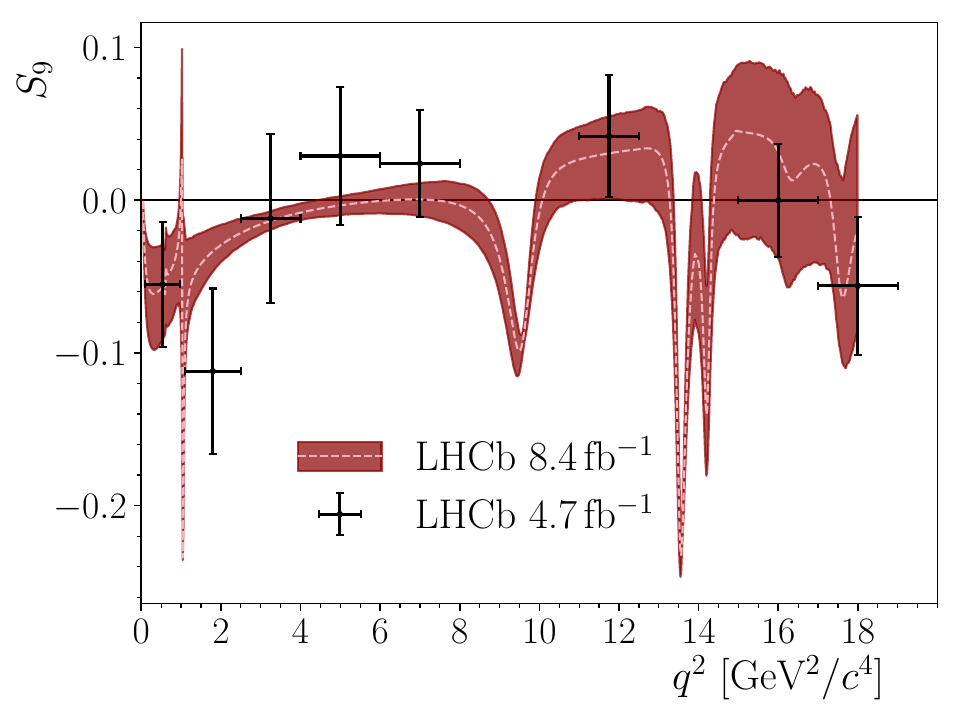}\\
  \includegraphics[width=0.40\linewidth]{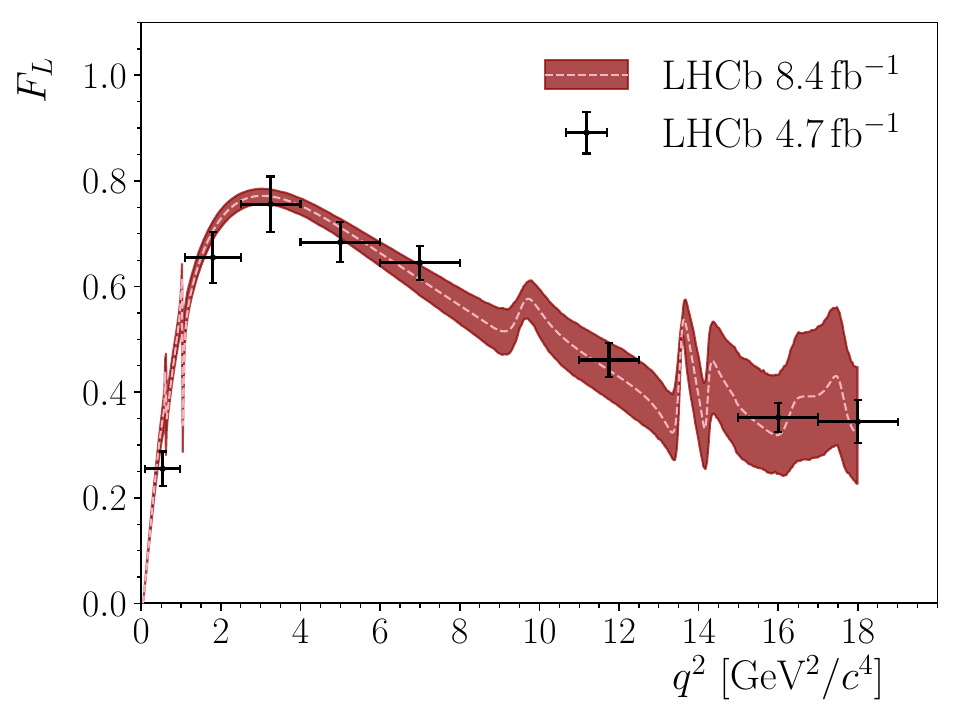}
  \includegraphics[width=0.40\linewidth]{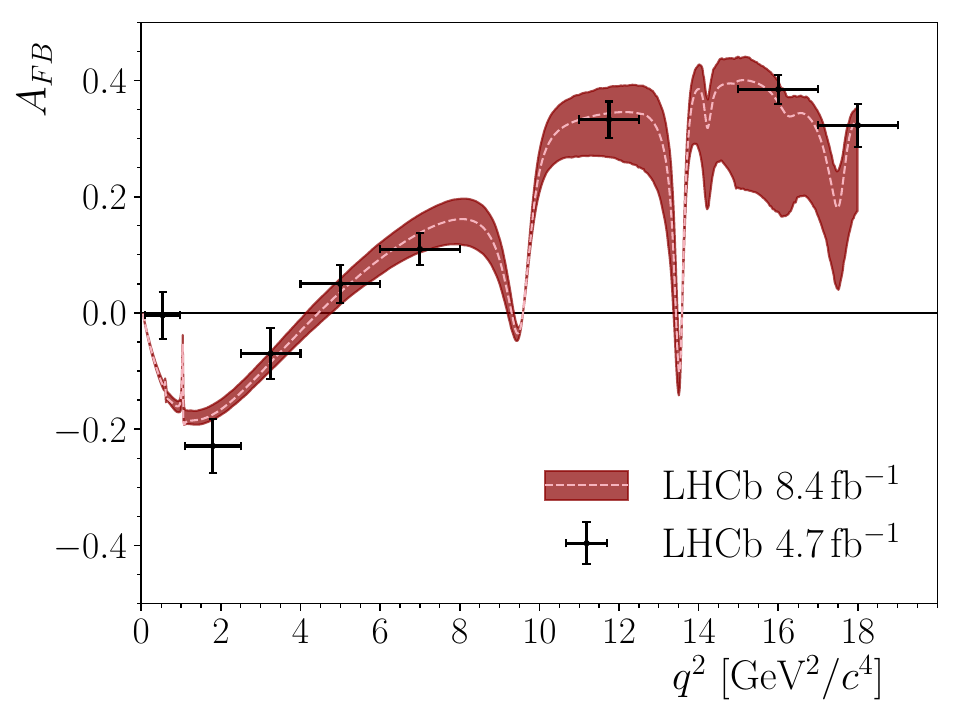}\\
  \caption{Comparison of unbinned observables constructed out of the signal parameters with the measurements from the dedicated LHCb binned analyses~\cite{LHCb-PAPER-2020-002,LHCb-PAPER-2016-012} that used $4.7\invfb$ of data for the angular analysis (black). The shaded bands indicate 68\% confidence regions from varying the fit parameters according to the covariance matrix accounting for both statistical and systematic uncertainties.
  \label{fig:preunblind_run1run2016_obs}}
\end{figure}

\begin{figure}[h]
 \centering
  \includegraphics[width=0.40\linewidth]{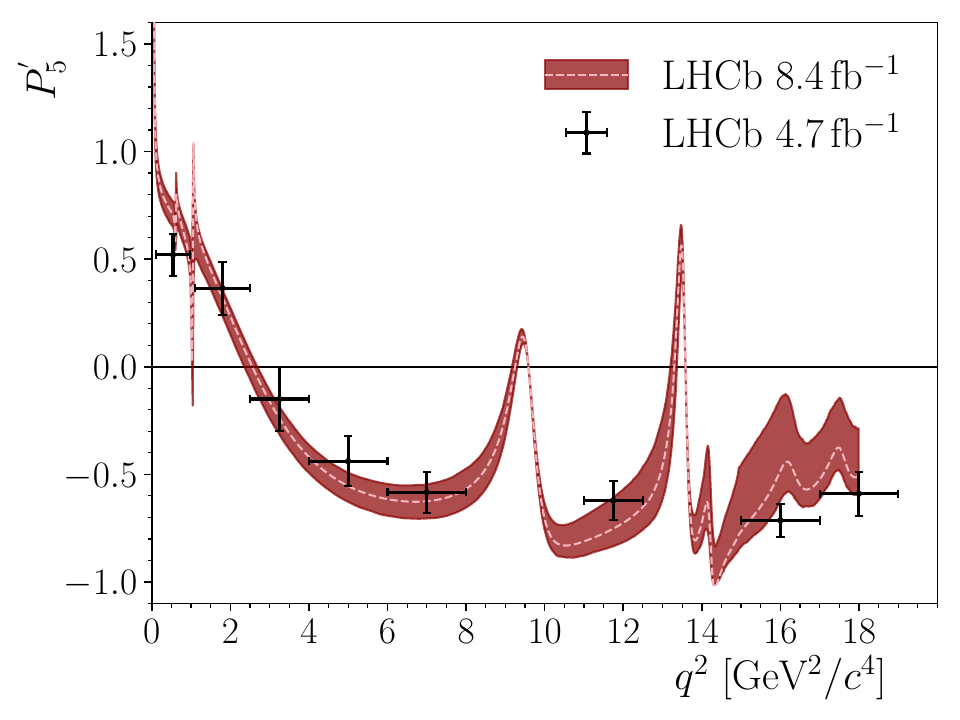} 
  \includegraphics[width=0.40\linewidth]{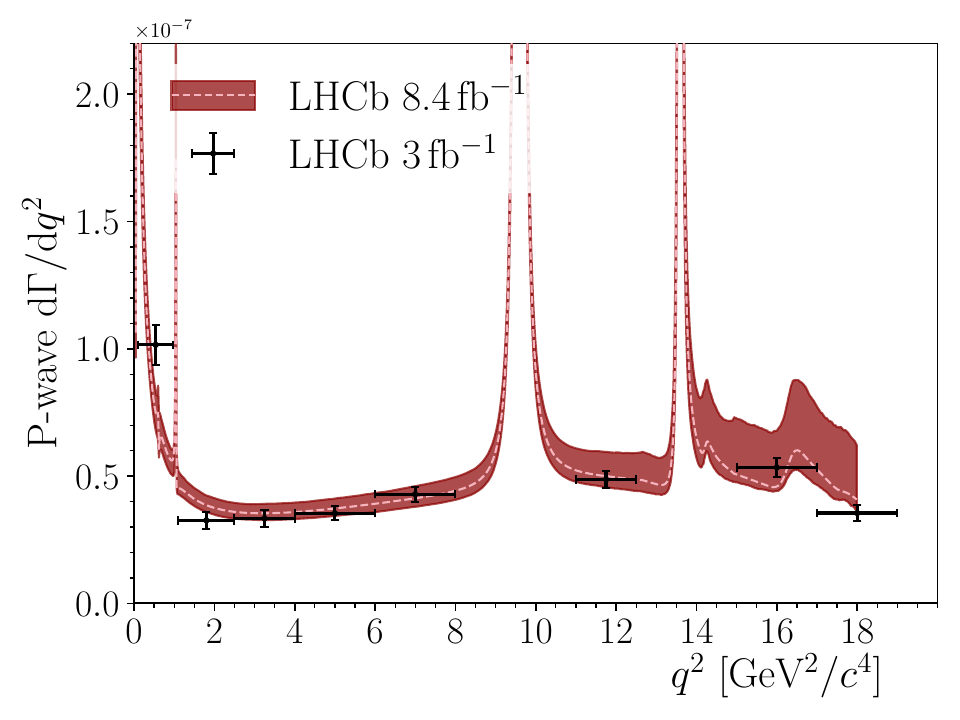}
  \caption{Comparison of unbinned observables constructed out of the signal parameters with the measurements from the dedicated LHCb binned analyses~\cite{LHCb-PAPER-2020-002,LHCb-PAPER-2016-012} that used $4.7\invfb$ of data for the angular analysis and $3\invfb$ for the branching fraction (black). The shaded bands indicate 68\% confidence regions from varying the fit parameters according to the covariance matrix accounting for both statistical and systematic uncertainties.
  \label{fig:preunblind_run1run2016_obs_again}}
\end{figure}

The total angular observables obtained from the signal parameters can be compared to previous \lhcb measurements. The binned angular observables were measured in Ref.~\cite{LHCb-PAPER-2020-002} using the Run 1 and 2016 data samples, corresponding to 4.7\invfb. The decay rate was measured in Ref.~\cite{LHCb-PAPER-2016-012} using Run 1 data, corresponding to 3\invfb. The comparison is shown in Figs.~\ref{fig:preunblind_run1run2016_obs} and~\ref{fig:preunblind_run1run2016_obs_again}. The shaded bands indicate 68\% confidence regions from varying the fit parameters according to the covariance matrix accounting for both statistical and systematic uncertainties. It is worth remarking that, in this analysis, the uncertainty on the observables is heavily correlated across the \qsq spectrum, given the unbinned nature of the measurement. In contrast, the measurements in each \qsq bin in Refs.~\cite{LHCb-PAPER-2020-002,LHCb-PAPER-2016-012} are statistically independent from other bins. The comparatively large uncertainty at high-\qsq in this analysis is a result of the large number of nonlocal contributions present in this region which are not easily separable at the current experimental sensitivity.

\clearpage

\clearpage

%% file: Authorship_LHCb-PAPER-2024-011.tex
\centerline
{\large\bf LHCb collaboration}
\begin
{flushleft}
\small
R.~Aaij$^{36}$\lhcborcid{0000-0003-0533-1952},
A.S.W.~Abdelmotteleb$^{55}$\lhcborcid{0000-0001-7905-0542},
C.~Abellan~Beteta$^{49}$,
F.~Abudin{\'e}n$^{55}$\lhcborcid{0000-0002-6737-3528},
T.~Ackernley$^{59}$\lhcborcid{0000-0002-5951-3498},
A. A. ~Adefisoye$^{67}$\lhcborcid{0000-0003-2448-1550},
B.~Adeva$^{45}$\lhcborcid{0000-0001-9756-3712},
M.~Adinolfi$^{53}$\lhcborcid{0000-0002-1326-1264},
P.~Adlarson$^{79}$\lhcborcid{0000-0001-6280-3851},
C.~Agapopoulou$^{13}$\lhcborcid{0000-0002-2368-0147}
% ...

{\footnotesize \it

% ...
$^{13}$Universit{\'e} Paris-Saclay, CNRS/IN2P3, IJCLab, Orsay, France\\
% ...
$^{36}$Nikhef National Institute for Subatomic Physics, Amsterdam, Netherlands\\
% ...
$^{45}$Instituto Galego de F{\'\i}sica de Altas Enerx{\'\i}as (IGFAE), Universidade de Santiago de Compostela, Santiago de Compostela, Spain\\
% ...
$^{49}$Physik-Institut, Universit{\"a}t Z{\"u}rich, Z{\"u}rich, Switzerland\\
% ...
$^{53}$H.H. Wills Physics Laboratory, University of Bristol, Bristol, United Kingdom\\
% ...
$^{55}$Department of Physics, University of Warwick, Coventry, United Kingdom\\
% ...
$^{58}$School of Physics and Astronomy, University of Glasgow, Glasgow, United Kingdom\\
$^{59}$Oliver Lodge Laboratory, University of Liverpool, Liverpool, United Kingdom\\
% ...
$^{67}$Syracuse University, Syracuse, NY, United States\\
% ...
$^{79}$Department of Physics and Astronomy, Uppsala University, Uppsala, Sweden, associated to $^{58}$\\
% ...
}
\end{flushleft}

%% file: main.bbl
\ifx\mcitethebibliography\mciteundefinedmacro
\PackageError{LHCb.bst}{mciteplus.sty has not been loaded}
{This bibstyle requires the use of the mciteplus package.}\fi
\providecommand{\href}[2]{#2}
\begin{mcitethebibliography}{10}
\mciteSetBstSublistMode{n}
\mciteSetBstMaxWidthForm{subitem}{\alph{mcitesubitemcount})}
\mciteSetBstSublistLabelBeginEnd{\mcitemaxwidthsubitemform\space}
{\relax}{\relax}

\bibitem{Kruger:1999xa}
F.~Kruger, L.~M. Sehgal, N.~Sinha, and R.~Sinha,
  \ifthenelse{\boolean{articletitles}}{\emph{{Angular distribution and CP
  asymmetries in the decays $\bar B \to K^- \pi^+ e^- e^+$ and $\bar B \to
  \pi^- \pi^+ e^- e^+$}},
  }{}\href{https://doi.org/10.1103/PhysRevD.61.114028}{Phys.\ Rev.\ D
  \textbf{61} (2000) 114028},
  \href{http://arxiv.org/abs/hep-ph/9907386}{{\normalfont\ttfamily
  arXiv:hep-ph/9907386}}, [Erratum: Phys. Rev. D 63, 019901 (2001)]\relax
\mciteBstWouldAddEndPuncttrue
\mciteSetBstMidEndSepPunct{\mcitedefaultmidpunct}
{\mcitedefaultendpunct}{\mcitedefaultseppunct}\relax
\EndOfBibitem
\bibitem{LHCb-PAPER-2013-019}
LHCb collaboration, R.~Aaij {\em et~al.},
  \ifthenelse{\boolean{articletitles}}{\emph{{Differential branching fraction
  and angular analysis of the decay \mbox{\decay{\Bz}{\Kstarz\mumu}}}},
  }{}\href{https://doi.org/10.1007/JHEP08(2013)131}{JHEP \textbf{08} (2013)
  131}, \href{http://arxiv.org/abs/1304.6325}{{\normalfont\ttfamily
  arXiv:1304.6325}}\relax
\mciteBstWouldAddEndPuncttrue
\mciteSetBstMidEndSepPunct{\mcitedefaultmidpunct}
{\mcitedefaultendpunct}{\mcitedefaultseppunct}\relax
\EndOfBibitem
\bibitem{LHCb-PAPER-2013-037}
LHCb collaboration, R.~Aaij {\em et~al.},
  \ifthenelse{\boolean{articletitles}}{\emph{{Measurement of
  form-factor-independent observables in the decay
  \mbox{\decay{\Bz}{\Kstarz\mumu}}}},
  }{}\href{https://doi.org/10.1103/PhysRevLett.111.191801}{Phys.\ Rev.\ Lett.\
  \textbf{111} (2013) 191801},
  \href{http://arxiv.org/abs/1308.1707}{{\normalfont\ttfamily
  arXiv:1308.1707}}\relax
\mciteBstWouldAddEndPuncttrue
\mciteSetBstMidEndSepPunct{\mcitedefaultmidpunct}
{\mcitedefaultendpunct}{\mcitedefaultseppunct}\relax
\EndOfBibitem
\bibitem{LHCb-PAPER-2015-051}
LHCb collaboration, R.~Aaij {\em et~al.},
  \ifthenelse{\boolean{articletitles}}{\emph{{Angular analysis of the
  \mbox{\decay{\Bz}{\Kstarz\mumu}} decay using $3\invfb$ of integrated
  luminosity}}, }{}\href{https://doi.org/10.1007/JHEP02(2016)104}{JHEP
  \textbf{02} (2016) 104},
  \href{http://arxiv.org/abs/1512.04442}{{\normalfont\ttfamily
  arXiv:1512.04442}}\relax
\mciteBstWouldAddEndPuncttrue
\mciteSetBstMidEndSepPunct{\mcitedefaultmidpunct}
{\mcitedefaultendpunct}{\mcitedefaultseppunct}\relax
\EndOfBibitem
\bibitem{LHCb-PAPER-2020-002}
LHCb collaboration, R.~Aaij {\em et~al.},
  \ifthenelse{\boolean{articletitles}}{\emph{{Measurement of \CP-averaged
  observables in the \mbox{\decay{\Bz}{\Kstarz\mumu}} decay}},
  }{}\href{https://doi.org/10.1103/PhysRevLett.125.011802}{Phys.\ Rev.\ Lett.\
  \textbf{125} (2020) 011802},
  \href{http://arxiv.org/abs/2003.04831}{{\normalfont\ttfamily
  arXiv:2003.04831}}\relax
\mciteBstWouldAddEndPuncttrue
\mciteSetBstMidEndSepPunct{\mcitedefaultmidpunct}
{\mcitedefaultendpunct}{\mcitedefaultseppunct}\relax
\EndOfBibitem
\bibitem{LHCb-PAPER-2023-033}
LHCb collaboration, R.~Aaij {\em et~al.},
  \ifthenelse{\boolean{articletitles}}{\emph{{Amplitude analysis of the $\Bz
  \to \Kstarz\mup\mun$ decay}},
  }{}\href{https://doi.org/10.1103/PhysRevLett.132.131801}{Phys.\ Rev.\ Lett.\
  \textbf{132} (2024) 131801},
  \href{http://arxiv.org/abs/2312.09115}{{\normalfont\ttfamily
  arXiv:2312.09115}}\relax
\mciteBstWouldAddEndPuncttrue
\mciteSetBstMidEndSepPunct{\mcitedefaultmidpunct}
{\mcitedefaultendpunct}{\mcitedefaultseppunct}\relax
\EndOfBibitem
\bibitem{ATLAS:2018gqc}
ATLAS collaboration, M.~Aaboud {\em et~al.},
  \ifthenelse{\boolean{articletitles}}{\emph{{Angular analysis of $B^0_d
  \rightarrow K^{*}\mu^+\mu^-$ decays in $pp$ collisions at $\sqrt{s}= 8$ TeV
  with the ATLAS detector}},
  }{}\href{https://doi.org/10.1007/JHEP10(2018)047}{JHEP \textbf{10} (2018)
  047}, \href{http://arxiv.org/abs/1805.04000}{{\normalfont\ttfamily
  arXiv:1805.04000}}\relax
\mciteBstWouldAddEndPuncttrue
\mciteSetBstMidEndSepPunct{\mcitedefaultmidpunct}
{\mcitedefaultendpunct}{\mcitedefaultseppunct}\relax
\EndOfBibitem
\bibitem{BaBar:2006tnv}
BaBar collaboration, B.~Aubert {\em et~al.},
  \ifthenelse{\boolean{articletitles}}{\emph{{Measurements of branching
  fractions, rate asymmetries, and angular distributions in the rare decays $B
  \to K \ell^{+} \ell^{-}$ and \mbox{$B \to K^{*} \ell^{+} \ell^{-}$}}},
  }{}\href{https://doi.org/10.1103/PhysRevD.73.092001}{Phys.\ Rev.\
  \textbf{D73} (2006) 092001},
  \href{http://arxiv.org/abs/hep-ex/0604007}{{\normalfont\ttfamily
  arXiv:hep-ex/0604007}}\relax
\mciteBstWouldAddEndPuncttrue
\mciteSetBstMidEndSepPunct{\mcitedefaultmidpunct}
{\mcitedefaultendpunct}{\mcitedefaultseppunct}\relax
\EndOfBibitem
\bibitem{BaBar:2015wkg}
BaBar collaboration, J.~P. Lees {\em et~al.},
  \ifthenelse{\boolean{articletitles}}{\emph{{Measurement of angular
  asymmetries in the decays $B \to K^*\ell^+\ell^-$}},
  }{}\href{https://doi.org/10.1103/PhysRevD.93.052015}{Phys.\ Rev.\
  \textbf{D93} (2016) 052015},
  \href{http://arxiv.org/abs/1508.07960}{{\normalfont\ttfamily
  arXiv:1508.07960}}\relax
\mciteBstWouldAddEndPuncttrue
\mciteSetBstMidEndSepPunct{\mcitedefaultmidpunct}
{\mcitedefaultendpunct}{\mcitedefaultseppunct}\relax
\EndOfBibitem
\bibitem{Belle:2009zue}
Belle collaboration, J.-T. Wei {\em et~al.},
  \ifthenelse{\boolean{articletitles}}{\emph{{Measurement of the differential
  branching fraction and forward-backward asymmetry for $B \to
  K^{(*)}\ell^+\ell^-$}},
  }{}\href{https://doi.org/10.1103/PhysRevLett.103.171801}{Phys.\ Rev.\ Lett.\
  \textbf{103} (2009) 171801},
  \href{http://arxiv.org/abs/0904.0770}{{\normalfont\ttfamily
  arXiv:0904.0770}}\relax
\mciteBstWouldAddEndPuncttrue
\mciteSetBstMidEndSepPunct{\mcitedefaultmidpunct}
{\mcitedefaultendpunct}{\mcitedefaultseppunct}\relax
\EndOfBibitem
\bibitem{Belle:2016fev}
Belle collaboration, S.~Wehle {\em et~al.},
  \ifthenelse{\boolean{articletitles}}{\emph{{Lepton-flavor-dependent angular
  analysis of $B\to K^\ast \ell^+\ell^-$}},
  }{}\href{https://doi.org/10.1103/PhysRevLett.118.111801}{Phys.\ Rev.\ Lett.\
  \textbf{118} (2017) 111801},
  \href{http://arxiv.org/abs/1612.05014}{{\normalfont\ttfamily
  arXiv:1612.05014}}\relax
\mciteBstWouldAddEndPuncttrue
\mciteSetBstMidEndSepPunct{\mcitedefaultmidpunct}
{\mcitedefaultendpunct}{\mcitedefaultseppunct}\relax
\EndOfBibitem
\bibitem{CDF:2011tds}
CDF collaboration, T.~Aaltonen {\em et~al.},
  \ifthenelse{\boolean{articletitles}}{\emph{{Measurements of the angular
  distributions in the decays $B \to K^{(*)} \mu^+ \mu^-$ at CDF}},
  }{}\href{https://doi.org/10.1103/PhysRevLett.108.081807}{Phys.\ Rev.\ Lett.\
  \textbf{108} (2012) 081807},
  \href{http://arxiv.org/abs/1108.0695}{{\normalfont\ttfamily
  arXiv:1108.0695}}\relax
\mciteBstWouldAddEndPuncttrue
\mciteSetBstMidEndSepPunct{\mcitedefaultmidpunct}
{\mcitedefaultendpunct}{\mcitedefaultseppunct}\relax
\EndOfBibitem
\bibitem{CMS:2015bcy}
CMS collaboration, V.~Khachatryan {\em et~al.},
  \ifthenelse{\boolean{articletitles}}{\emph{{Angular analysis of the decay
  \mbox{$B^0 \to K^{*0} \mu^+ \mu^-$} from pp collisions at $\sqrt s = 8$
  TeV}}, }{}\href{https://doi.org/10.1016/j.physletb.2015.12.020}{Phys.\ Lett.\
   \textbf{B753} (2016) 424},
  \href{http://arxiv.org/abs/1507.08126}{{\normalfont\ttfamily
  arXiv:1507.08126}}\relax
\mciteBstWouldAddEndPuncttrue
\mciteSetBstMidEndSepPunct{\mcitedefaultmidpunct}
{\mcitedefaultendpunct}{\mcitedefaultseppunct}\relax
\EndOfBibitem
\bibitem{CMS:2017rzx}
CMS collaboration, A.~M. Sirunyan {\em et~al.},
  \ifthenelse{\boolean{articletitles}}{\emph{{Measurement of angular parameters
  from the decay $\mathrm{B}^0 \to \mathrm{K}^{*0} \mu^+ \mu^-$ in
  proton-proton collisions at $\sqrt{s} = $ 8 TeV}},
  }{}\href{https://doi.org/10.1016/j.physletb.2018.04.030}{Phys.\ Lett.\
  \textbf{B781} (2018) 517},
  \href{http://arxiv.org/abs/1710.02846}{{\normalfont\ttfamily
  arXiv:1710.02846}}\relax
\mciteBstWouldAddEndPuncttrue
\mciteSetBstMidEndSepPunct{\mcitedefaultmidpunct}
{\mcitedefaultendpunct}{\mcitedefaultseppunct}\relax
\EndOfBibitem
\bibitem{Altmannshofer:2008dz}
W.~Altmannshofer {\em et~al.},
  \ifthenelse{\boolean{articletitles}}{\emph{{Symmetries and asymmetries of $B
  \to K^{*} \mu^{+} \mu^{-}$ decays in the Standard Model and beyond}},
  }{}\href{https://doi.org/10.1088/1126-6708/2009/01/019}{JHEP \textbf{01}
  (2009) 019}, \href{http://arxiv.org/abs/0811.1214}{{\normalfont\ttfamily
  arXiv:0811.1214}}\relax
\mciteBstWouldAddEndPuncttrue
\mciteSetBstMidEndSepPunct{\mcitedefaultmidpunct}
{\mcitedefaultendpunct}{\mcitedefaultseppunct}\relax
\EndOfBibitem
\bibitem{EOSAuthors:2021xpv}
EOS collaboration, D.~van Dyk {\em et~al.},
  \ifthenelse{\boolean{articletitles}}{\emph{{EOS: a software for flavor
  physics phenomenology}},
  }{}\href{https://doi.org/10.1140/epjc/s10052-022-10177-4}{Eur.\ Phys.\ J.\
  \textbf{C82} (2022) 569},
  \href{http://arxiv.org/abs/2111.15428}{{\normalfont\ttfamily
  arXiv:2111.15428}}\relax
\mciteBstWouldAddEndPuncttrue
\mciteSetBstMidEndSepPunct{\mcitedefaultmidpunct}
{\mcitedefaultendpunct}{\mcitedefaultseppunct}\relax
\EndOfBibitem
\bibitem{EOS}
D.~van Dyk {\em et~al.}, \ifthenelse{\boolean{articletitles}}{\emph{{eos/eos:
  EOS Version 1.0.11}}, }{} 2024.
\newblock
  doi:~\href{https://doi.org/10.5281/zenodo.10600399}{10.5281/zenodo.10600399}\relax
\mciteBstWouldAddEndPuncttrue
\mciteSetBstMidEndSepPunct{\mcitedefaultmidpunct}
{\mcitedefaultendpunct}{\mcitedefaultseppunct}\relax
\EndOfBibitem
\bibitem{Descotes-Genon:2012isb}
S.~Descotes-Genon, J.~Matias, M.~Ramon, and J.~Virto,
  \ifthenelse{\boolean{articletitles}}{\emph{{Implications from clean
  observables for the binned analysis of $B \to K^*\mu^+\mu^-$ at large
  recoil}}, }{}\href{https://doi.org/10.1007/JHEP01(2013)048}{JHEP \textbf{01}
  (2013) 048}, \href{http://arxiv.org/abs/1207.2753}{{\normalfont\ttfamily
  arXiv:1207.2753}}\relax
\mciteBstWouldAddEndPuncttrue
\mciteSetBstMidEndSepPunct{\mcitedefaultmidpunct}
{\mcitedefaultendpunct}{\mcitedefaultseppunct}\relax
\EndOfBibitem
\bibitem{Altmannshofer:2014rta}
W.~Altmannshofer and D.~M. Straub,
  \ifthenelse{\boolean{articletitles}}{\emph{{New physics in $b\rightarrow s$
  transitions after LHC run 1}},
  }{}\href{https://doi.org/10.1140/epjc/s10052-015-3602-7}{Eur.\ Phys.\ J.\
  \textbf{C75} (2015) 382},
  \href{http://arxiv.org/abs/1411.3161}{{\normalfont\ttfamily
  arXiv:1411.3161}}\relax
\mciteBstWouldAddEndPuncttrue
\mciteSetBstMidEndSepPunct{\mcitedefaultmidpunct}
{\mcitedefaultendpunct}{\mcitedefaultseppunct}\relax
\EndOfBibitem
\bibitem{Capdevila:2017bsm}
B.~Capdevila {\em et~al.}, \ifthenelse{\boolean{articletitles}}{\emph{{Patterns
  of New Physics in $b\to s\ell^+\ell^-$ transitions in the light of recent
  data}}, }{}\href{https://doi.org/10.1007/JHEP01(2018)093}{JHEP \textbf{01}
  (2018) 093}, \href{http://arxiv.org/abs/1704.05340}{{\normalfont\ttfamily
  arXiv:1704.05340}}\relax
\mciteBstWouldAddEndPuncttrue
\mciteSetBstMidEndSepPunct{\mcitedefaultmidpunct}
{\mcitedefaultendpunct}{\mcitedefaultseppunct}\relax
\EndOfBibitem
\bibitem{Beaujean:2013soa}
F.~Beaujean, C.~Bobeth, and D.~van Dyk,
  \ifthenelse{\boolean{articletitles}}{\emph{{Comprehensive Bayesian analysis
  of rare (semi)leptonic and radiative $B$ decays}},
  }{}\href{https://doi.org/10.1140/epjc/s10052-014-2897-0}{Eur.\ Phys.\ J.\
  \textbf{C74} (2014) 2897},
  \href{http://arxiv.org/abs/1310.2478}{{\normalfont\ttfamily
  arXiv:1310.2478}}, [Erratum: Eur. Phys. J. C74, 3179 (2014)]\relax
\mciteBstWouldAddEndPuncttrue
\mciteSetBstMidEndSepPunct{\mcitedefaultmidpunct}
{\mcitedefaultendpunct}{\mcitedefaultseppunct}\relax
\EndOfBibitem
\bibitem{Descotes-Genon:2013wba}
S.~Descotes-Genon, J.~Matias, and J.~Virto,
  \ifthenelse{\boolean{articletitles}}{\emph{{Understanding the \mbox{$B\to
  K^*\mu^+\mu^-$ anomaly}}},
  }{}\href{https://doi.org/10.1103/PhysRevD.88.074002}{Phys.\ Rev.\
  \textbf{D88} (2013) 074002},
  \href{http://arxiv.org/abs/1307.5683}{{\normalfont\ttfamily
  arXiv:1307.5683}}\relax
\mciteBstWouldAddEndPuncttrue
\mciteSetBstMidEndSepPunct{\mcitedefaultmidpunct}
{\mcitedefaultendpunct}{\mcitedefaultseppunct}\relax
\EndOfBibitem
\bibitem{Alguero:2021anc}
M.~Alguer\'o {\em et~al.},
  \ifthenelse{\boolean{articletitles}}{\emph{{$b\rightarrow s\ell ^+\ell ^-$
  global fits after $R_{K_S}$ and $R_{K^{*+}}$}},
  }{}\href{https://doi.org/10.1140/epjc/s10052-022-10231-1}{Eur.\ Phys.\ J.\
  \textbf{C82} (2022) 326},
  \href{http://arxiv.org/abs/2104.08921}{{\normalfont\ttfamily
  arXiv:2104.08921}}\relax
\mciteBstWouldAddEndPuncttrue
\mciteSetBstMidEndSepPunct{\mcitedefaultmidpunct}
{\mcitedefaultendpunct}{\mcitedefaultseppunct}\relax
\EndOfBibitem
\bibitem{Alguero:2023jeh}
M.~Alguer\'o {\em et~al.}, \ifthenelse{\boolean{articletitles}}{\emph{{To (b)e
  or not to (b)e: no electrons at LHCb}},
  }{}\href{https://doi.org/10.1140/epjc/s10052-023-11824-0}{Eur.\ Phys.\ J.\
  \textbf{C83} (2023) 648},
  \href{http://arxiv.org/abs/2304.07330}{{\normalfont\ttfamily
  arXiv:2304.07330}}\relax
\mciteBstWouldAddEndPuncttrue
\mciteSetBstMidEndSepPunct{\mcitedefaultmidpunct}
{\mcitedefaultendpunct}{\mcitedefaultseppunct}\relax
\EndOfBibitem
\bibitem{London:2021lfn}
D.~London and J.~Matias, \ifthenelse{\boolean{articletitles}}{\emph{{$B$
  flavour anomalies: 2021 theoretical status report}},
  }{}\href{https://doi.org/10.1146/annurev-nucl-102020-090209}{Ann.\ Rev.\
  Nucl.\ Part.\ Sci.\  \textbf{72} (2022) 37},
  \href{http://arxiv.org/abs/2110.13270}{{\normalfont\ttfamily
  arXiv:2110.13270}}\relax
\mciteBstWouldAddEndPuncttrue
\mciteSetBstMidEndSepPunct{\mcitedefaultmidpunct}
{\mcitedefaultendpunct}{\mcitedefaultseppunct}\relax
\EndOfBibitem
\bibitem{Horgan:2013hoa}
R.~R. Horgan, Z.~Liu, S.~Meinel, and M.~Wingate,
  \ifthenelse{\boolean{articletitles}}{\emph{{Lattice QCD calculation of form
  factors describing the rare decays $B \to K^* \ell^+ \ell^-$ and $B_s \to
  \phi \ell^+ \ell^-$}},
  }{}\href{https://doi.org/10.1103/PhysRevD.89.094501}{Phys.\ Rev.\
  \textbf{D89} (2014) 094501},
  \href{http://arxiv.org/abs/1310.3722}{{\normalfont\ttfamily
  arXiv:1310.3722}}\relax
\mciteBstWouldAddEndPuncttrue
\mciteSetBstMidEndSepPunct{\mcitedefaultmidpunct}
{\mcitedefaultendpunct}{\mcitedefaultseppunct}\relax
\EndOfBibitem
\bibitem{Horgan:2015vla}
R.~R. Horgan, Z.~Liu, S.~Meinel, and M.~Wingate,
  \ifthenelse{\boolean{articletitles}}{\emph{{Rare $B$ decays using lattice QCD
  form factors}}, }{}\href{https://doi.org/10.22323/1.214.0372}{PoS
  \textbf{LATTICE2014} (2015) 372},
  \href{http://arxiv.org/abs/1501.00367}{{\normalfont\ttfamily
  arXiv:1501.00367}}\relax
\mciteBstWouldAddEndPuncttrue
\mciteSetBstMidEndSepPunct{\mcitedefaultmidpunct}
{\mcitedefaultendpunct}{\mcitedefaultseppunct}\relax
\EndOfBibitem
\bibitem{Bharucha:2015bzk}
A.~Bharucha, D.~M. Straub, and R.~Zwicky,
  \ifthenelse{\boolean{articletitles}}{\emph{{$B\to V\ell^+\ell^-$ in the
  Standard Model from light-cone sum rules}},
  }{}\href{https://doi.org/10.1007/JHEP08(2016)098}{JHEP \textbf{08} (2016)
  098}, \href{http://arxiv.org/abs/1503.05534}{{\normalfont\ttfamily
  arXiv:1503.05534}}\relax
\mciteBstWouldAddEndPuncttrue
\mciteSetBstMidEndSepPunct{\mcitedefaultmidpunct}
{\mcitedefaultendpunct}{\mcitedefaultseppunct}\relax
\EndOfBibitem
\bibitem{Gubernari:2018wyi}
N.~Gubernari, A.~Kokulu, and D.~van Dyk,
  \ifthenelse{\boolean{articletitles}}{\emph{{$B\to P$ and $B\to V$ form
  factors from $B$-Meson Light-Cone Sum Rules beyond leading twist}},
  }{}\href{https://doi.org/10.1007/JHEP01(2019)150}{JHEP \textbf{01} (2019)
  150}, \href{http://arxiv.org/abs/1811.00983}{{\normalfont\ttfamily
  arXiv:1811.00983}}\relax
\mciteBstWouldAddEndPuncttrue
\mciteSetBstMidEndSepPunct{\mcitedefaultmidpunct}
{\mcitedefaultendpunct}{\mcitedefaultseppunct}\relax
\EndOfBibitem
\bibitem{Khodjamirian:2010vf}
A.~Khodjamirian, T.~Mannel, A.~A. Pivovarov, and Y.-M. Wang,
  \ifthenelse{\boolean{articletitles}}{\emph{{Charm-loop effect in $B \to
  K^{(*)} \ell^{+} \ell^{-}$ and $B\to K^*\gamma$}},
  }{}\href{https://doi.org/10.1007/JHEP09(2010)089}{JHEP \textbf{09} (2010)
  089}, \href{http://arxiv.org/abs/1006.4945}{{\normalfont\ttfamily
  arXiv:1006.4945}}\relax
\mciteBstWouldAddEndPuncttrue
\mciteSetBstMidEndSepPunct{\mcitedefaultmidpunct}
{\mcitedefaultendpunct}{\mcitedefaultseppunct}\relax
\EndOfBibitem
\bibitem{Korchin:2012kz}
A.~Y. Korchin and V.~A. Kovalchuk,
  \ifthenelse{\boolean{articletitles}}{\emph{{Contribution of vector resonances
  to the ${\bar B}_d^0 \to {\bar K}^{*0} \mu^+ \mu^-$ decay}},
  }{}\href{https://doi.org/10.1140/epjc/s10052-012-2155-2}{Eur.\ Phys.\ J.\ C
  \textbf{72} (2012) 2155},
  \href{http://arxiv.org/abs/1205.3683}{{\normalfont\ttfamily
  arXiv:1205.3683}}\relax
\mciteBstWouldAddEndPuncttrue
\mciteSetBstMidEndSepPunct{\mcitedefaultmidpunct}
{\mcitedefaultendpunct}{\mcitedefaultseppunct}\relax
\EndOfBibitem
\bibitem{Lyon:2014hpa}
J.~Lyon and R.~Zwicky, \ifthenelse{\boolean{articletitles}}{\emph{{Resonances
  gone topsy turvy - the charm of QCD or new physics in $b \to s \ell^+
  \ell^-$?}}, }{}\href{http://arxiv.org/abs/1406.0566}{{\normalfont\ttfamily
  arXiv:1406.0566}}\relax
\mciteBstWouldAddEndPuncttrue
\mciteSetBstMidEndSepPunct{\mcitedefaultmidpunct}
{\mcitedefaultendpunct}{\mcitedefaultseppunct}\relax
\EndOfBibitem
\bibitem{Ciuchini:2015qxb}
M.~Ciuchini {\em et~al.}, \ifthenelse{\boolean{articletitles}}{\emph{{$B\to K^*
  \ell^+ \ell^-$ decays at large recoil in the Standard Model: a theoretical
  reappraisal}}, }{}\href{https://doi.org/10.1007/JHEP06(2016)116}{JHEP
  \textbf{06} (2016) 116},
  \href{http://arxiv.org/abs/1512.07157}{{\normalfont\ttfamily
  arXiv:1512.07157}}\relax
\mciteBstWouldAddEndPuncttrue
\mciteSetBstMidEndSepPunct{\mcitedefaultmidpunct}
{\mcitedefaultendpunct}{\mcitedefaultseppunct}\relax
\EndOfBibitem
\bibitem{Ciuchini:2022wbq}
M.~Ciuchini {\em et~al.},
  \ifthenelse{\boolean{articletitles}}{\emph{{Constraints on lepton
  universality violation from rare B decays}},
  }{}\href{https://doi.org/10.1103/PhysRevD.107.055036}{Phys.\ Rev.\ D
  \textbf{107} (2023) 055036},
  \href{http://arxiv.org/abs/2212.10516}{{\normalfont\ttfamily
  arXiv:2212.10516}}\relax
\mciteBstWouldAddEndPuncttrue
\mciteSetBstMidEndSepPunct{\mcitedefaultmidpunct}
{\mcitedefaultendpunct}{\mcitedefaultseppunct}\relax
\EndOfBibitem
\bibitem{Gubernari:2020eft}
N.~Gubernari, D.~van Dyk, and J.~Virto,
  \ifthenelse{\boolean{articletitles}}{\emph{{Non-local matrix elements in
  \mbox{$B_{(s)}\to \{K^{(*)},\phi\}\ell^+\ell^-$}}},
  }{}\href{https://doi.org/10.1007/JHEP02(2021)088}{JHEP \textbf{02} (2021)
  088}, \href{http://arxiv.org/abs/2011.09813}{{\normalfont\ttfamily
  arXiv:2011.09813}}\relax
\mciteBstWouldAddEndPuncttrue
\mciteSetBstMidEndSepPunct{\mcitedefaultmidpunct}
{\mcitedefaultendpunct}{\mcitedefaultseppunct}\relax
\EndOfBibitem
\bibitem{Gubernari:2022hxn}
N.~Gubernari, M.~Reboud, D.~van Dyk, and J.~Virto,
  \ifthenelse{\boolean{articletitles}}{\emph{{Improved theory predictions and
  global analysis of exclusive $b \to s\mu^+\mu^-$ processes}},
  }{}\href{https://doi.org/10.1007/JHEP09(2022)133}{JHEP \textbf{09} (2022)
  133}, \href{http://arxiv.org/abs/2206.03797}{{\normalfont\ttfamily
  arXiv:2206.03797}}\relax
\mciteBstWouldAddEndPuncttrue
\mciteSetBstMidEndSepPunct{\mcitedefaultmidpunct}
{\mcitedefaultendpunct}{\mcitedefaultseppunct}\relax
\EndOfBibitem
\bibitem{Ciuchini:2020gvn}
M.~Ciuchini {\em et~al.}, \ifthenelse{\boolean{articletitles}}{\emph{{Lessons
  from the $B^{0,+}\to K^{*0,+}\mu^+\mu^-$ angular analyses}},
  }{}\href{https://doi.org/10.1103/PhysRevD.103.015030}{Phys.\ Rev.\ D
  \textbf{103} (2021) 015030},
  \href{http://arxiv.org/abs/2011.01212}{{\normalfont\ttfamily
  arXiv:2011.01212}}\relax
\mciteBstWouldAddEndPuncttrue
\mciteSetBstMidEndSepPunct{\mcitedefaultmidpunct}
{\mcitedefaultendpunct}{\mcitedefaultseppunct}\relax
\EndOfBibitem
\bibitem{Blake:2017fyh}
T.~Blake {\em et~al.}, \ifthenelse{\boolean{articletitles}}{\emph{{An empirical
  model to determine the hadronic resonance contributions to $\overline{B}{} ^0
  \!\rightarrow \overline{K}{} ^{*0} \mu ^+ \mu ^- $ transitions}},
  }{}\href{https://doi.org/10.1140/epjc/s10052-018-5937-3}{Eur.\ Phys.\ J.\
  \textbf{C78} (2018) 453},
  \href{http://arxiv.org/abs/1709.03921}{{\normalfont\ttfamily
  arXiv:1709.03921}}\relax
\mciteBstWouldAddEndPuncttrue
\mciteSetBstMidEndSepPunct{\mcitedefaultmidpunct}
{\mcitedefaultendpunct}{\mcitedefaultseppunct}\relax
\EndOfBibitem
\bibitem{Bobeth:2017vxj}
C.~Bobeth, M.~Chrzaszcz, D.~van Dyk, and J.~Virto,
  \ifthenelse{\boolean{articletitles}}{\emph{{Long-distance effects in
  $B\rightarrow K^*\ell \ell $ from analyticity}},
  }{}\href{https://doi.org/10.1140/epjc/s10052-018-5918-6}{Eur.\ Phys.\ J.\
  \textbf{C78} (2018) 451},
  \href{http://arxiv.org/abs/1707.07305}{{\normalfont\ttfamily
  arXiv:1707.07305}}\relax
\mciteBstWouldAddEndPuncttrue
\mciteSetBstMidEndSepPunct{\mcitedefaultmidpunct}
{\mcitedefaultendpunct}{\mcitedefaultseppunct}\relax
\EndOfBibitem
\bibitem{LHCb-PAPER-2016-025}
LHCb collaboration, R.~Aaij {\em et~al.},
  \ifthenelse{\boolean{articletitles}}{\emph{{Differential branching fraction
  and angular analysis of the decay \mbox{\decay{\Bz}{\Kp\pim\mumu}} in the
  $K^{*}_{0,2}(1430)^0$ region}},
  }{}\href{https://doi.org/10.1007/JHEP12(2016)065}{JHEP \textbf{12} (2016)
  065}, \href{http://arxiv.org/abs/1609.04736}{{\normalfont\ttfamily
  arXiv:1609.04736}}\relax
\mciteBstWouldAddEndPuncttrue
\mciteSetBstMidEndSepPunct{\mcitedefaultmidpunct}
{\mcitedefaultendpunct}{\mcitedefaultseppunct}\relax
\EndOfBibitem
\bibitem{Belle:2014nuw}
Belle collaboration, K.~Chilikin {\em et~al.},
  \ifthenelse{\boolean{articletitles}}{\emph{{Observation of a new charged
  charmoniumlike state in $\bar{B}^0 \to J/\psi K^-\pi^+$ decays}},
  }{}\href{https://doi.org/10.1103/PhysRevD.90.112009}{Phys.\ Rev.\
  \textbf{D90} (2014) 112009},
  \href{http://arxiv.org/abs/1408.6457}{{\normalfont\ttfamily
  arXiv:1408.6457}}\relax
\mciteBstWouldAddEndPuncttrue
\mciteSetBstMidEndSepPunct{\mcitedefaultmidpunct}
{\mcitedefaultendpunct}{\mcitedefaultseppunct}\relax
\EndOfBibitem
\bibitem{Belle:2013shl}
Belle collaboration, K.~Chilikin {\em et~al.},
  \ifthenelse{\boolean{articletitles}}{\emph{{Experimental constraints on the
  spin and parity of the $Z$(4430)$^+$}},
  }{}\href{https://doi.org/10.1103/PhysRevD.88.074026}{Phys.\ Rev.\
  \textbf{D88} (2013) 074026},
  \href{http://arxiv.org/abs/1306.4894}{{\normalfont\ttfamily
  arXiv:1306.4894}}\relax
\mciteBstWouldAddEndPuncttrue
\mciteSetBstMidEndSepPunct{\mcitedefaultmidpunct}
{\mcitedefaultendpunct}{\mcitedefaultseppunct}\relax
\EndOfBibitem
\bibitem{LHCb-PAPER-2023-032}
LHCb collaboration, R.~Aaij {\em et~al.},
  \ifthenelse{\boolean{articletitles}}{\emph{{Determination of short- and
  long-distance contributions in $\Bz \to \Kstarz\mup\mun$ decays}},
  }{}\href{https://doi.org/10.1103/PhysRevD.109.052009}{Phys.\ Rev.\
  \textbf{D109} (2024) 052009},
  \href{http://arxiv.org/abs/2312.09102}{{\normalfont\ttfamily
  arXiv:2312.09102}}\relax
\mciteBstWouldAddEndPuncttrue
\mciteSetBstMidEndSepPunct{\mcitedefaultmidpunct}
{\mcitedefaultendpunct}{\mcitedefaultseppunct}\relax
\EndOfBibitem
\bibitem{PDG2022}
Particle Data Group, R.~L. Workman {\em et~al.},
  \ifthenelse{\boolean{articletitles}}{\emph{{\href{http://pdg.lbl.gov/}{Review
  of particle physics}}}, }{}\href{https://doi.org/10.1093/ptep/ptac097}{Prog.\
  Theor.\ Exp.\ Phys.\  \textbf{2022} (2022) 083C01}\relax
\mciteBstWouldAddEndPuncttrue
\mciteSetBstMidEndSepPunct{\mcitedefaultmidpunct}
{\mcitedefaultendpunct}{\mcitedefaultseppunct}\relax
\EndOfBibitem
\bibitem{Becirevic:2012dp}
D.~Becirevic and A.~Tayduganov,
  \ifthenelse{\boolean{articletitles}}{\emph{{Impact of $B\to K^\ast_0
  \ell^+\ell^-$ on the New Physics search in $B\to K^\ast \ell^+\ell^-$
  decay}}, }{}\href{https://doi.org/10.1016/j.nuclphysb.2012.11.016}{Nucl.\
  Phys.\  \textbf{B868} (2013) 368},
  \href{http://arxiv.org/abs/1207.4004}{{\normalfont\ttfamily
  arXiv:1207.4004}}\relax
\mciteBstWouldAddEndPuncttrue
\mciteSetBstMidEndSepPunct{\mcitedefaultmidpunct}
{\mcitedefaultendpunct}{\mcitedefaultseppunct}\relax
\EndOfBibitem
\bibitem{Doring:2013wka}
M.~D\"oring, U.-G. Mei\ss{}ner, and W.~Wang,
  \ifthenelse{\boolean{articletitles}}{\emph{{Chiral dynamics and S-wave
  contributions in semileptonic B decays}},
  }{}\href{https://doi.org/10.1007/JHEP10(2013)011}{JHEP \textbf{10} (2013)
  011}, \href{http://arxiv.org/abs/1307.0947}{{\normalfont\ttfamily
  arXiv:1307.0947}}\relax
\mciteBstWouldAddEndPuncttrue
\mciteSetBstMidEndSepPunct{\mcitedefaultmidpunct}
{\mcitedefaultendpunct}{\mcitedefaultseppunct}\relax
\EndOfBibitem
\bibitem{Descotes-Genon:2019bud}
S.~Descotes-Genon, A.~Khodjamirian, and J.~Virto,
  \ifthenelse{\boolean{articletitles}}{\emph{{Light-cone sum rules for $B\to
  K\pi$ form factors and applications to rare decays}},
  }{}\href{https://doi.org/10.1007/JHEP12(2019)083}{JHEP \textbf{12} (2019)
  083}, \href{http://arxiv.org/abs/1908.02267}{{\normalfont\ttfamily
  arXiv:1908.02267}}\relax
\mciteBstWouldAddEndPuncttrue
\mciteSetBstMidEndSepPunct{\mcitedefaultmidpunct}
{\mcitedefaultendpunct}{\mcitedefaultseppunct}\relax
\EndOfBibitem
\bibitem{Khodjamirian:2012rm}
A.~Khodjamirian, T.~Mannel, and Y.~M. Wang,
  \ifthenelse{\boolean{articletitles}}{\emph{{$B \to K \ell^{+}\ell^{-}$ decay
  at large hadronic recoil}},
  }{}\href{https://doi.org/10.1007/JHEP02(2013)010}{JHEP \textbf{02} (2013)
  010}, \href{http://arxiv.org/abs/1211.0234}{{\normalfont\ttfamily
  arXiv:1211.0234}}\relax
\mciteBstWouldAddEndPuncttrue
\mciteSetBstMidEndSepPunct{\mcitedefaultmidpunct}
{\mcitedefaultendpunct}{\mcitedefaultseppunct}\relax
\EndOfBibitem
\bibitem{Bordone:2024hui}
M.~Bordone, G.~Isidori, S.~M\"achler, and A.~Tinari,
  \ifthenelse{\boolean{articletitles}}{\emph{{Short- vs. long-distance physics
  in $B\rightarrow K^{(*)} \ell ^+\ell ^-$: a data-driven analysis}},
  }{}\href{https://doi.org/10.1140/epjc/s10052-024-12869-5}{Eur.\ Phys.\ J.\ C
  \textbf{84} (2024) 547},
  \href{http://arxiv.org/abs/2401.18007}{{\normalfont\ttfamily
  arXiv:2401.18007}}\relax
\mciteBstWouldAddEndPuncttrue
\mciteSetBstMidEndSepPunct{\mcitedefaultmidpunct}
{\mcitedefaultendpunct}{\mcitedefaultseppunct}\relax
\EndOfBibitem
\bibitem{Isidori:2024lng}
G.~Isidori, Z.~Polonsky, and A.~Tinari,
  \ifthenelse{\boolean{articletitles}}{\emph{{An explicit estimate of charm
  rescattering in $B^0 \to K^0 \bar{\ell} \ell$}},
  }{}\href{http://arxiv.org/abs/2405.17551}{{\normalfont\ttfamily
  arXiv:2405.17551}}\relax
\mciteBstWouldAddEndPuncttrue
\mciteSetBstMidEndSepPunct{\mcitedefaultmidpunct}
{\mcitedefaultendpunct}{\mcitedefaultseppunct}\relax
\EndOfBibitem
\bibitem{Cornella:2020aoq}
C.~Cornella {\em et~al.}, \ifthenelse{\boolean{articletitles}}{\emph{{Hunting
  for $B^+\rightarrow K^+ \tau ^+\tau ^-$ imprints on the $B^+ \rightarrow K^+
  \mu ^+\mu ^-$ dimuon spectrum}},
  }{}\href{https://doi.org/10.1140/epjc/s10052-020-08674-5}{Eur.\ Phys.\ J.\
  \textbf{C80} (2020) 1095},
  \href{http://arxiv.org/abs/2001.04470}{{\normalfont\ttfamily
  arXiv:2001.04470}}\relax
\mciteBstWouldAddEndPuncttrue
\mciteSetBstMidEndSepPunct{\mcitedefaultmidpunct}
{\mcitedefaultendpunct}{\mcitedefaultseppunct}\relax
\EndOfBibitem
\bibitem{LHCb-PAPER-2013-023}
LHCb collaboration, R.~Aaij {\em et~al.},
  \ifthenelse{\boolean{articletitles}}{\emph{{Measurement of the polarization
  amplitudes in \mbox{\decay{\Bz}{\jpsi\Kstar(892)^0}} decays}},
  }{}\href{https://doi.org/10.1103/PhysRevD.88.052002}{Phys.\ Rev.\
  \textbf{D88} (2013) 052002},
  \href{http://arxiv.org/abs/1307.2782}{{\normalfont\ttfamily
  arXiv:1307.2782}}\relax
\mciteBstWouldAddEndPuncttrue
\mciteSetBstMidEndSepPunct{\mcitedefaultmidpunct}
{\mcitedefaultendpunct}{\mcitedefaultseppunct}\relax
\EndOfBibitem
\bibitem{BaBar:2007rbr}
BaBar collaboration, B.~Aubert {\em et~al.},
  \ifthenelse{\boolean{articletitles}}{\emph{{Measurement of decay amplitudes
  of \mbox{$B \to J/\psi K^{*}, \psi(2S) K^{*}$}, and $\chi_{c1} K^{*}$ with an
  angular analysis}},
  }{}\href{https://doi.org/10.1103/PhysRevD.76.031102}{Phys.\ Rev.\
  \textbf{D76} (2007) 031102},
  \href{http://arxiv.org/abs/0704.0522}{{\normalfont\ttfamily
  arXiv:0704.0522}}\relax
\mciteBstWouldAddEndPuncttrue
\mciteSetBstMidEndSepPunct{\mcitedefaultmidpunct}
{\mcitedefaultendpunct}{\mcitedefaultseppunct}\relax
\EndOfBibitem
\bibitem{LHCb-PAPER-2018-042}
LHCb collaboration, R.~Aaij {\em et~al.},
  \ifthenelse{\boolean{articletitles}}{\emph{{Study of the
  \mbox{\decay{\Bz}{\rho(770)^0 K^*(892)^0}} decay with an amplitude analysis
  of \mbox{\decay{\Bz}{(\pip\pim) (\Kp\pim)}} decays}},
  }{}\href{https://doi.org/10.1007/JHEP05(2019)026}{JHEP \textbf{05} (2019)
  026}, \href{http://arxiv.org/abs/1812.07008}{{\normalfont\ttfamily
  arXiv:1812.07008}}\relax
\mciteBstWouldAddEndPuncttrue
\mciteSetBstMidEndSepPunct{\mcitedefaultmidpunct}
{\mcitedefaultendpunct}{\mcitedefaultseppunct}\relax
\EndOfBibitem
\bibitem{LHCb-PAPER-2014-005}
LHCb collaboration, R.~Aaij {\em et~al.},
  \ifthenelse{\boolean{articletitles}}{\emph{{Measurement of polarization
  amplitudes and \CP asymmetries in \mbox{\decay{\Bz}{\phiz\Kstar(892)^0}}}},
  }{}\href{https://doi.org/10.1007/JHEP05(2014)069}{JHEP \textbf{05} (2014)
  069}, \href{http://arxiv.org/abs/1403.2888}{{\normalfont\ttfamily
  arXiv:1403.2888}}\relax
\mciteBstWouldAddEndPuncttrue
\mciteSetBstMidEndSepPunct{\mcitedefaultmidpunct}
{\mcitedefaultendpunct}{\mcitedefaultseppunct}\relax
\EndOfBibitem
\bibitem{LHCb-PAPER-2016-045}
LHCb collaboration, R.~Aaij {\em et~al.},
  \ifthenelse{\boolean{articletitles}}{\emph{{Measurement of the phase
  difference between the short- and long-distance amplitudes in the
  \mbox{\decay{\Bp}{\Kp\mumu}} decay}},
  }{}\href{https://doi.org/10.1140/epjc/s10052-017-4703-2}{Eur.\ Phys.\ J.\
  \textbf{C77} (2017) 161},
  \href{http://arxiv.org/abs/1612.06764}{{\normalfont\ttfamily
  arXiv:1612.06764}}\relax
\mciteBstWouldAddEndPuncttrue
\mciteSetBstMidEndSepPunct{\mcitedefaultmidpunct}
{\mcitedefaultendpunct}{\mcitedefaultseppunct}\relax
\EndOfBibitem
\bibitem{Asatrian:2019kbk}
H.~M. Asatrian, C.~Greub, and J.~Virto,
  \ifthenelse{\boolean{articletitles}}{\emph{{Exact NLO matching and
  analyticity in $b\to s\ell\ell$}},
  }{}\href{https://doi.org/10.1007/JHEP04(2020)012}{JHEP \textbf{04} (2020)
  012}, \href{http://arxiv.org/abs/1912.09099}{{\normalfont\ttfamily
  arXiv:1912.09099}}\relax
\mciteBstWouldAddEndPuncttrue
\mciteSetBstMidEndSepPunct{\mcitedefaultmidpunct}
{\mcitedefaultendpunct}{\mcitedefaultseppunct}\relax
\EndOfBibitem
\bibitem{Paul:2016urs}
A.~Paul and D.~M. Straub,
  \ifthenelse{\boolean{articletitles}}{\emph{{Constraints on new physics from
  radiative $B$ decays}},
  }{}\href{https://doi.org/10.1007/JHEP04(2017)027}{JHEP \textbf{04} (2017)
  027}, \href{http://arxiv.org/abs/1608.02556}{{\normalfont\ttfamily
  arXiv:1608.02556}}\relax
\mciteBstWouldAddEndPuncttrue
\mciteSetBstMidEndSepPunct{\mcitedefaultmidpunct}
{\mcitedefaultendpunct}{\mcitedefaultseppunct}\relax
\EndOfBibitem
\bibitem{LHCb-PAPER-2020-020}
LHCb collaboration, R.~Aaij {\em et~al.},
  \ifthenelse{\boolean{articletitles}}{\emph{{Strong constraints on the $b \to
  s\gamma$ photon polarisation from $B^0 \to K^{\ast 0} e^+ e^-$ decays}},
  }{}\href{https://doi.org/10.1007/JHEP12(2020)081}{JHEP \textbf{12} (2020)
  081}, \href{http://arxiv.org/abs/2010.06011}{{\normalfont\ttfamily
  arXiv:2010.06011}}\relax
\mciteBstWouldAddEndPuncttrue
\mciteSetBstMidEndSepPunct{\mcitedefaultmidpunct}
{\mcitedefaultendpunct}{\mcitedefaultseppunct}\relax
\EndOfBibitem
\bibitem{LHCb-DP-2008-001}
LHCb collaboration, A.~A. Alves~Jr.\ {\em et~al.},
  \ifthenelse{\boolean{articletitles}}{\emph{{The \lhcb detector at the LHC}},
  }{}\href{https://doi.org/10.1088/1748-0221/3/08/S08005}{JINST \textbf{3}
  (2008) S08005}\relax
\mciteBstWouldAddEndPuncttrue
\mciteSetBstMidEndSepPunct{\mcitedefaultmidpunct}
{\mcitedefaultendpunct}{\mcitedefaultseppunct}\relax
\EndOfBibitem
\bibitem{LHCb-DP-2014-002}
LHCb collaboration, R.~Aaij {\em et~al.},
  \ifthenelse{\boolean{articletitles}}{\emph{{LHCb detector performance}},
  }{}\href{https://doi.org/10.1142/S0217751X15300227}{Int.\ J.\ Mod.\ Phys.\
  \textbf{A30} (2015) 1530022},
  \href{http://arxiv.org/abs/1412.6352}{{\normalfont\ttfamily
  arXiv:1412.6352}}\relax
\mciteBstWouldAddEndPuncttrue
\mciteSetBstMidEndSepPunct{\mcitedefaultmidpunct}
{\mcitedefaultendpunct}{\mcitedefaultseppunct}\relax
\EndOfBibitem
\bibitem{LHCb-DP-2012-004}
R.~Aaij {\em et~al.}, \ifthenelse{\boolean{articletitles}}{\emph{{The \lhcb
  trigger and its performance in 2011}},
  }{}\href{https://doi.org/10.1088/1748-0221/8/04/P04022}{JINST \textbf{8}
  (2013) P04022}, \href{http://arxiv.org/abs/1211.3055}{{\normalfont\ttfamily
  arXiv:1211.3055}}\relax
\mciteBstWouldAddEndPuncttrue
\mciteSetBstMidEndSepPunct{\mcitedefaultmidpunct}
{\mcitedefaultendpunct}{\mcitedefaultseppunct}\relax
\EndOfBibitem
\bibitem{LHCb-DP-2019-001}
R.~Aaij {\em et~al.}, \ifthenelse{\boolean{articletitles}}{\emph{{Design and
  performance of the LHCb trigger and full real-time reconstruction in Run 2 of
  the LHC}}, }{}\href{https://doi.org/10.1088/1748-0221/14/04/P04013}{JINST
  \textbf{14} (2019) P04013},
  \href{http://arxiv.org/abs/1812.10790}{{\normalfont\ttfamily
  arXiv:1812.10790}}\relax
\mciteBstWouldAddEndPuncttrue
\mciteSetBstMidEndSepPunct{\mcitedefaultmidpunct}
{\mcitedefaultendpunct}{\mcitedefaultseppunct}\relax
\EndOfBibitem
\bibitem{Sjostrand:2007gs}
T.~Sj\"{o}strand, S.~Mrenna, and P.~Skands,
  \ifthenelse{\boolean{articletitles}}{\emph{{A brief introduction to PYTHIA
  8.1}}, }{}\href{https://doi.org/10.1016/j.cpc.2008.01.036}{Comput.\ Phys.\
  Commun.\  \textbf{178} (2008) 852},
  \href{http://arxiv.org/abs/0710.3820}{{\normalfont\ttfamily
  arXiv:0710.3820}}\relax
\mciteBstWouldAddEndPuncttrue
\mciteSetBstMidEndSepPunct{\mcitedefaultmidpunct}
{\mcitedefaultendpunct}{\mcitedefaultseppunct}\relax
\EndOfBibitem
\bibitem{LHCb-PROC-2010-056}
I.~Belyaev {\em et~al.}, \ifthenelse{\boolean{articletitles}}{\emph{{Handling
  of the generation of primary events in Gauss, the LHCb simulation
  framework}}, }{}\href{https://doi.org/10.1088/1742-6596/331/3/032047}{J.\
  Phys.\ Conf.\ Ser.\  \textbf{331} (2011) 032047}\relax
\mciteBstWouldAddEndPuncttrue
\mciteSetBstMidEndSepPunct{\mcitedefaultmidpunct}
{\mcitedefaultendpunct}{\mcitedefaultseppunct}\relax
\EndOfBibitem
\bibitem{Lange:2001uf}
D.~J. Lange, \ifthenelse{\boolean{articletitles}}{\emph{{The EvtGen particle
  decay simulation package}},
  }{}\href{https://doi.org/10.1016/S0168-9002(01)00089-4}{Nucl.\ Instrum.\
  Meth.\  \textbf{A462} (2001) 152}\relax
\mciteBstWouldAddEndPuncttrue
\mciteSetBstMidEndSepPunct{\mcitedefaultmidpunct}
{\mcitedefaultendpunct}{\mcitedefaultseppunct}\relax
\EndOfBibitem
\bibitem{davidson2015photos}
N.~Davidson, T.~Przedzinski, and Z.~Was,
  \ifthenelse{\boolean{articletitles}}{\emph{{PHOTOS interface in C++:
  Technical and physics documentation}},
  }{}\href{https://doi.org/https://doi.org/10.1016/j.cpc.2015.09.013}{Comp.\
  Phys.\ Comm.\  \textbf{199} (2016) 86},
  \href{http://arxiv.org/abs/1011.0937}{{\normalfont\ttfamily
  arXiv:1011.0937}}\relax
\mciteBstWouldAddEndPuncttrue
\mciteSetBstMidEndSepPunct{\mcitedefaultmidpunct}
{\mcitedefaultendpunct}{\mcitedefaultseppunct}\relax
\EndOfBibitem
\bibitem{Allison:2006ve}
Geant4 collaboration, J.~Allison {\em et~al.},
  \ifthenelse{\boolean{articletitles}}{\emph{{Geant4 developments and
  applications}}, }{}\href{https://doi.org/10.1109/TNS.2006.869826}{IEEE
  Trans.\ Nucl.\ Sci.\  \textbf{53} (2006) 270}\relax
\mciteBstWouldAddEndPuncttrue
\mciteSetBstMidEndSepPunct{\mcitedefaultmidpunct}
{\mcitedefaultendpunct}{\mcitedefaultseppunct}\relax
\EndOfBibitem
\bibitem{Agostinelli:2002hh}
Geant4 collaboration, S.~Agostinelli {\em et~al.},
  \ifthenelse{\boolean{articletitles}}{\emph{{Geant4: A simulation toolkit}},
  }{}\href{https://doi.org/10.1016/S0168-9002(03)01368-8}{Nucl.\ Instrum.\
  Meth.\  \textbf{A506} (2003) 250}\relax
\mciteBstWouldAddEndPuncttrue
\mciteSetBstMidEndSepPunct{\mcitedefaultmidpunct}
{\mcitedefaultendpunct}{\mcitedefaultseppunct}\relax
\EndOfBibitem
\bibitem{LHCb-PROC-2011-006}
M.~Clemencic {\em et~al.}, \ifthenelse{\boolean{articletitles}}{\emph{{The
  \lhcb simulation application, Gauss: Design, evolution and experience}},
  }{}\href{https://doi.org/10.1088/1742-6596/331/3/032023}{J.\ Phys.\ Conf.\
  Ser.\  \textbf{331} (2011) 032023}\relax
\mciteBstWouldAddEndPuncttrue
\mciteSetBstMidEndSepPunct{\mcitedefaultmidpunct}
{\mcitedefaultendpunct}{\mcitedefaultseppunct}\relax
\EndOfBibitem
\bibitem{LHCb-DP-2018-004}
D.~M{\"u}ller, M.~Clemencic, G.~Corti, and M.~Gersabeck,
  \ifthenelse{\boolean{articletitles}}{\emph{{ReDecay: A novel approach to
  speed up the simulation at LHCb}},
  }{}\href{https://doi.org/10.1140/epjc/s10052-018-6469-6}{Eur.\ Phys.\ J.\
  \textbf{C78} (2018) 1009},
  \href{http://arxiv.org/abs/1810.10362}{{\normalfont\ttfamily
  arXiv:1810.10362}}\relax
\mciteBstWouldAddEndPuncttrue
\mciteSetBstMidEndSepPunct{\mcitedefaultmidpunct}
{\mcitedefaultendpunct}{\mcitedefaultseppunct}\relax
\EndOfBibitem
\bibitem{Breiman}
L.~Breiman, J.~H. Friedman, R.~A. Olshen, and C.~J. Stone, {\em Classification
  and regression trees}, Wadsworth international group, Belmont, California,
  USA, 1984\relax
\mciteBstWouldAddEndPuncttrue
\mciteSetBstMidEndSepPunct{\mcitedefaultmidpunct}
{\mcitedefaultendpunct}{\mcitedefaultseppunct}\relax
\EndOfBibitem
\bibitem{AdaBoost}
Y.~Freund and R.~E. Schapire, \ifthenelse{\boolean{articletitles}}{\emph{A
  decision-theoretic generalization of on-line learning and an application to
  boosting}, }{}\href{https://doi.org/10.1006/jcss.1997.1504}{J.\ Comput.\
  Syst.\ Sci.\  \textbf{55} (1997) 119}\relax
\mciteBstWouldAddEndPuncttrue
\mciteSetBstMidEndSepPunct{\mcitedefaultmidpunct}
{\mcitedefaultendpunct}{\mcitedefaultseppunct}\relax
\EndOfBibitem
\bibitem{Pivk:2004ty}
M.~Pivk and F.~R. Le~Diberder,
  \ifthenelse{\boolean{articletitles}}{\emph{{sPlot: A statistical tool to
  unfold data distributions}},
  }{}\href{https://doi.org/10.1016/j.nima.2005.08.106}{Nucl.\ Instrum.\ Meth.\
  \textbf{A555} (2005) 356},
  \href{http://arxiv.org/abs/physics/0402083}{{\normalfont\ttfamily
  arXiv:physics/0402083}}\relax
\mciteBstWouldAddEndPuncttrue
\mciteSetBstMidEndSepPunct{\mcitedefaultmidpunct}
{\mcitedefaultendpunct}{\mcitedefaultseppunct}\relax
\EndOfBibitem
\bibitem{Skwarnicki:1986xj}
T.~Skwarnicki, {\em {A study of the radiative cascade transitions between the
  Upsilon-prime and Upsilon resonances}}, PhD thesis, Institute of Nuclear
  Physics, Krakow, 1986,
  {\href{http://inspirehep.net/record/230779/}{DESY-F31-86-02}}\relax
\mciteBstWouldAddEndPuncttrue
\mciteSetBstMidEndSepPunct{\mcitedefaultmidpunct}
{\mcitedefaultendpunct}{\mcitedefaultseppunct}\relax
\EndOfBibitem
\bibitem{Hulsbergen:2005pu}
W.~D. Hulsbergen, \ifthenelse{\boolean{articletitles}}{\emph{{Decay chain
  fitting with a Kalman filter}},
  }{}\href{https://doi.org/10.1016/j.nima.2005.06.078}{Nucl.\ Instrum.\ Meth.\
  \textbf{A552} (2005) 566},
  \href{http://arxiv.org/abs/physics/0503191}{{\normalfont\ttfamily
  arXiv:physics/0503191}}\relax
\mciteBstWouldAddEndPuncttrue
\mciteSetBstMidEndSepPunct{\mcitedefaultmidpunct}
{\mcitedefaultendpunct}{\mcitedefaultseppunct}\relax
\EndOfBibitem
\bibitem{weibull1951statistical}
W.~Weibull, \ifthenelse{\boolean{articletitles}}{\emph{A statistical
  distribution function of wide applicability}, }{}Journal of applied mechanics
  (1951)\relax
\mciteBstWouldAddEndPuncttrue
\mciteSetBstMidEndSepPunct{\mcitedefaultmidpunct}
{\mcitedefaultendpunct}{\mcitedefaultseppunct}\relax
\EndOfBibitem
\bibitem{Gratrex:2015hna}
J.~Gratrex, M.~Hopfer, and R.~Zwicky,
  \ifthenelse{\boolean{articletitles}}{\emph{{Generalised helicity formalism,
  higher moments and the $B \to K_{J_K}(\to K \pi) \bar{\ell}_1 \ell_2$ angular
  distributions}}, }{}\href{https://doi.org/10.1103/PhysRevD.93.054008}{Phys.\
  Rev.\  \textbf{D93} (2016) 054008},
  \href{http://arxiv.org/abs/1506.03970}{{\normalfont\ttfamily
  arXiv:1506.03970}}\relax
\mciteBstWouldAddEndPuncttrue
\mciteSetBstMidEndSepPunct{\mcitedefaultmidpunct}
{\mcitedefaultendpunct}{\mcitedefaultseppunct}\relax
\EndOfBibitem
\bibitem{Belle:2021ecr}
Belle collaboration, T.~V. Dong {\em et~al.},
  \ifthenelse{\boolean{articletitles}}{\emph{{Search for the decay $B^0 \to
  K^{*0}\tau^+\tau^-$ at the Belle experiment}},
  }{}\href{https://doi.org/10.1103/PhysRevD.108.L011102}{Phys.\ Rev.\
  \textbf{D108} (2023) L011102},
  \href{http://arxiv.org/abs/2110.03871}{{\normalfont\ttfamily
  arXiv:2110.03871}}\relax
\mciteBstWouldAddEndPuncttrue
\mciteSetBstMidEndSepPunct{\mcitedefaultmidpunct}
{\mcitedefaultendpunct}{\mcitedefaultseppunct}\relax
\EndOfBibitem
\bibitem{Straub:2018kue}
D.~M. Straub, \ifthenelse{\boolean{articletitles}}{\emph{{flavio: a Python
  package for flavour and precision phenomenology in the Standard Model and
  beyond}}, }{}\href{http://arxiv.org/abs/1810.08132}{{\normalfont\ttfamily
  arXiv:1810.08132}}\relax
\mciteBstWouldAddEndPuncttrue
\mciteSetBstMidEndSepPunct{\mcitedefaultmidpunct}
{\mcitedefaultendpunct}{\mcitedefaultseppunct}\relax
\EndOfBibitem
\bibitem{LHCb-PAPER-2016-012}
LHCb collaboration, R.~Aaij {\em et~al.},
  \ifthenelse{\boolean{articletitles}}{\emph{{Measurements of the S-wave
  fraction in \mbox{\decay{\Bz}{\Kp\pim\mumu}} decays and the
  \mbox{\decay{\Bz}{\Kstar(892)^0\mumu}} differential branching fraction}},
  }{}\href{https://doi.org/10.1007/JHEP11(2016)047}{JHEP \textbf{11} (2016)
  047}, Erratum \href{https://doi.org/10.1007/JHEP04(2017)142}{ibid.\
  \textbf{04} (2017) 142},
  \href{http://arxiv.org/abs/1606.04731}{{\normalfont\ttfamily
  arXiv:1606.04731}}\relax
\mciteBstWouldAddEndPuncttrue
\mciteSetBstMidEndSepPunct{\mcitedefaultmidpunct}
{\mcitedefaultendpunct}{\mcitedefaultseppunct}\relax
\EndOfBibitem
\end{mcitethebibliography}
